\documentclass[School=Harvard]{Dissertate}

\graphicspath{ {figures/} }

\def\deg{$^{\circ}$N}
\def\U{\mathsf{U}}
\def\V{\mathsf{V}}
\def\Z{\mathsf{Z}}
\def\Uv{\boldsymbol{U}}
\def\Rv{\boldsymbol{R}}
\def\Vv{\boldsymbol{V}}

\def\C{\bm{\mathsf{C}}}
\def\I{\bm{\mathsf{I}}}

\def\L{\bm{\mathsf{L}}}

\def\R{{\bf R}}

\def\D{{\bf D}}

\def\D{\bm{\mathsf{D}}}
\def\A{\bm{\mathsf{A}}}
\def\F{\bm{\mathsf{F}}}

\def\Q{\bm{\mathsf{Q}}}
\def\M{\bm{\mathsf{M}}}

\newcounter{saveeqn}%

\newcommand{\be}{\begin{equation}}
\newcommand{\ee}{\end{equation}}
\newcommand{\bdm}{\begin{equation*}}
\newcommand{\edm}{\end{equation*}}
\newcommand{\bea}{\begin{eqnarray}}
\newcommand{\eea}{\end{eqnarray}}

\newcommand{\partialf}[2]
{
 \ifthenelse{\equal{#1}{}}{\frac{\partial}{\partial #2}}{\frac{\partial #1}{\partial #2}}
}

\newcommand{\sgn}{\mathop{\mathrm{sgn}}}
\newcommand{\real}{\mathop{\mathrm{Re}}}
\newcommand{\imag}{\mathop{\mathrm{Im}}}
\newcommand{\vecd}{\mathop{\mathrm{vecd}}}

\newcommand{\diag}{\mathop{\mathrm{diag}}}
\newcommand{\Tr}{\mathop{\mathrm{Tr}}}
\renewcommand{\(}{\left(}
\renewcommand{\)}{\right)}
\renewcommand{\[}{\left[}
\renewcommand{\]}{\right]}
\newcommand{\<}{\left\langle}
\renewcommand{\>}{\right\rangle}

\newcommand{\Del}{\Delta}
\newcommand{\DDel}{\bm{\mathsf{\Del}}}
\renewcommand{\d}{\delta}
\newcommand{\Df}{\text{D}}
\newcommand{\df}{\mathrm{d}}
\renewcommand{\l}{\ell}
\newcommand{\la}{\lambda}
\newcommand{\s}{\sigma}
\renewcommand{\b}{\beta}
\renewcommand{\a}{\alpha}
\newcommand{\rU}{r_{\textrm{m}}}

\newcommand{\z}{\zeta}

\newcommand{\zv}{\mbox{\boldmath$\z$}}

\newcommand{\Psiv}{\mbox{\boldmath$\Psi$}}

\renewcommand{\i}{\mathrm{i}}

\newsavebox{\astrutbox}
\sbox{\astrutbox}{\rule[-5pt]{0pt}{20pt}}

\newcommand{\transp}{\textrm{T}}

\def\bit{\vphantom{\dot{W}}}

\def\x{\chi}
\def\om{\omega}
\def\Om{\Omega}

\def\thet{\theta}

\def\Gcal{\mathcal{G}}
\def\Dcal{\mathcal{D}}
\def\Db{\Dcal_\b}
\def\Db{\Dcal_2}
\def\Dbj{\Dcal_{2,j}}
\def\Ocal{\mathcal{O}}
\def\Fcal{\mathcal{F}}
\def\Acal{\mathcal{A}}
\def\Tcal{\mathcal{T}}
\def\Lcal{\mathcal{L}}
\def\Rcal{\mathcal{R}}

\def\FcalR{\mathcal{F}^{(R)}}
\def\FcalR{\mathcal{F}}
\def\Ncal{\mathcal{N}}
\def\deg{^\circ}

\def\xv{\mathbf{x}}
\def\cv{\mathbf{c}}
\def\av{\mathbf{a}}
\def\rv{\mathbf{r}}
\def\Uv{\mathbf{U}}
\def\Zv{\mathbf{Z}}
\def\uv{\mathbf{u}}

\def\nv{\mathbf{n}}
\def\kv{\mathbf{k}}
\def\av{\mathbf{a}}
\def\e{\varepsilon}

\def\qv{\mathbf{q}}
\def\pv{\mathbf{p}}
\def\bv{\bm\beta}
\def\alphav{\bm\alpha}

\def\Omv{\bm\Omega}

\def\ecz{\varepsilon_{c,\textrm{z}}}
\def\ecnz{\varepsilon_{c,\textrm{nz}}}

\def\spa{\vphantom{\frac1{(2\pi)^2}\frac{\Ncal\,\Dcal_0}{ \Dcal_0^2 + \b^2\,\Db^2 }}}
\def\spb{\vphantom{\int\limits_{\thet_{j-1}+\d}^{\thet_j-\d\thet} }}

\def\thet{\theta}

\def\betav{\bm{\b}}
\def\nablav{\bm{\nabla}}

\def\tol{\textrm{tol}}

\newcommand{\tav}[1]{\Tcal\[\,#1\,\]}

\def\dU{\delta\tilde{U}}

\def\dZ{\delta\tilde{Z}}
\def\dC{\delta\tilde{C}}

\def\pv{\mathbf{p}}

\def\ecz{\varepsilon_{c,\textrm{z}}}
\def\ecnz{\varepsilon_{c,\textrm{nz}}}

\def\spa{\vphantom{\frac1{(2\pi)^2}\frac{\Ncal\,\Dcal_0}{ \Dcal_0^2 + \b^2\,\Db^2 }}}
\def\spb{\vphantom{\int\limits_{\thet_{j-1}+\d}^{\thet_j-\d\thet} }}

\def\f{f}
\def\fr{f_r}

\def\zhat{\hat{\mathbf{z}}}

\makeatletter
\newcommand{\upmax}{\def\blx@maxcitenames{99}}
\newcommand{\dnmax}{\def\blx@maxcitenames{2}}
\makeatother

\def\topline{\hline\hline\vrule height 10pt depth4pt width0pt\relax}
\def\midline{\hline\vrule height 10pt width0pt\relax}
\def\botline{\hline}

\def\NLinv{NL$_\text{inv}$}

\newcommand{\textcite}{\citet}
\newcommand{\parencite}{\citep}

\begin{document}

% the front matter

%% Some details about the dissertation.
%\title{Title of the dissertation}
%\author{Firstname M. Lastname}
%\advisor{Bigname Scientist}
%
%% ... about the degree.
%\degree{Doctor of Philosophy}
%\field{Psychology}
%\degreeyear{2024}
%\degreemonth{May}
%\department{Psychology}
%
%% ... about the candidate's previous degrees.
%\pdOneName{B.S.}
%\pdOneSchool{Boston University}
%\pdOneYear{2018}
%
%\pdTwoName{M.A.}
%\pdTwoSchool{Monster's Univeristy}
%\pdTwoYear{2021}

% some details about the thesis
\title{Formation of large-scale structures by turbulence in rotating planets}
\author{Navid C. Constantinou}
\advisor{Petros J. Ioannou}

%\titlegreek{Σχηματισμός δομών μεγάλης κλίμακας από την τύρβη\\ σε περιστρεφόμενους πλανήτες}
%\authorgreek{Ναβίτ Κ. Κωνσταντίνου}
%\advisorgreek{Πέτρος Ι. Ιωάννου}

% about the degree
\degree{Doctor of Philosophy}
\field{Physics}
\degreeyear{2015}
\degreemonth{February}

% about the university
\department{Department of Physics}
\university{National and Kapodistrian University of Athens}
\universitycity{Athens}
\universitystate{Greece}

\maketitle
  
%

%\coloronline
\copyrightpage
\signaturespage
\abstractpage
\tableofcontents	
%%\authorlist
%%\listoffigures

	\chapter*{Acknowledgments}
	\noindent
	% !TEX root = ../thesis.tex

% the acknowledgments section

%\newthought{I would like to thank} everyone. Also the Pope of Rome and his fabulous cat.
In 2004, when I was in my second year of undergraduate studies, I took the Newtonian Mechanics course, jointly taught by Profs.~Petros Ioannou and Theocharis Apostolatos. Their inspiring, lively and often entertaining lectures changed the way I viewed physics. During the last 10 years both of them have influenced my scientific way of thinking immensely.

Now that my doctoral studies are complete I would like to thank first and foremost my thesis advisor Prof.~Petros Ioannou. He has been a tutor and a mentor for me during these years. Besides teaching me everything I know regarding atmospheric dynamics, he has taught me how to really understand something and how to approach a new problem by decomposing it to simpler pieces. His clear physical and mathematical intuition along with his immense passion and enthusiasm for attacking physical problems have both been sources of inspiration for me. I hope that some day I will be able to convey but a fraction of his passion and clarity to any future student that I might have.

Special thanks should go to Prof.~Brian Farrell and to Prof.~Theocharis Apostolatos.
Their attitude towards science has often helped me remember that science is fun, even when I found myself stuck in a particular problem. I thank both of them for being readily available to discuss anything regarding my research. 

Next I want to thank Nikos Bakas. He has been my academic ``older brother'' and, for most of these years, my officemate. I thank him for the multitude of discussions we had regarding the scientific problems and also for the many times he has helped me out when I was in trouble with some specific calculation. 

I want to thank Prof.~Fokion Hatziioannou. The afternoon discussion sessions that were organized by him when I was an undergraduate and in the early years of my graduate studies helped me become acquainted with subjects that were not approached in the lectures and also revisit subjects that I had been taught in the lectures but from a completely different perspective.  Also, for the organization of the famous ``Physics Department excursions'' and the long hikes in mountains all over Greece. The majority of friends I made from the Physics Department I met them in one of these field trips, or, if I had already met them before, it was in these excursions that we really came close to each other for the first time.

I would like to thank Prof.~Javier Jim\'enez for inviting me to the First Multiflow Summer Workshop during June 2013 in Madrid.  Also I would like to thank the IPAM (Institute for Pure and Applied Mathematics), UCLA for the invitation to the Mathematics of Turbulence Long program (September to December 2014) as a core participant and their hospitality. A large part of this manuscript was written during my time at IPAM.

It is my pleasure to acknowledge the Alexander S. Onassis Public Benefit Foundation for the financial support they have provided me during the period 2009-2014. Also, my friend A. Koudounas for partially supporting me during the last six months of my doctoral studies.

Special thanks goes to my parents Ntinos \& Ntina and the rest of my family for their continuous support, love and understanding the past eleven years of both my undergraduate and post-graduate studies.

Last but not least I would like to take this opportunity to collaboratively thank all of my friends who have been close supporters of me throughout these years. Indicatively I should mention: 
Giannis M. for helping me pack and move all my stuff away from my apartment before departing for the IPAM workshop in L.A. and also for his hospitality in Lefkosia in January 2015; 
Savvas~K. for his hospitality in Archanes during COMECAP 2014 conference;
Haris \& Kyriaki for their hospitality in Lefkada during both summers of 2011 and 2014 where we enjoyed Lefkada's crystal clear waters and also parts of this manuscript were written;
Petros for joining me to innumerable bicycle rides up the mountains around Athens;
Theodore~D. for his company during the IPAM workshop and for reading and commenting on parts of this manuscript;
Savvas~S. for his hospitality in Lefkosia both in September 2011 and December 2012;
Andreas~K. and Mairi for their hospitality in Lefkosia in January 2013;
Andreas~A. who except from being a friend also was an extremely helpful neighbor in Exarchia;
Andreas~S. for the great moments we had in the university in Athens, for introducing me to Pandelis gastronomical experience, and for his support even after he left Athens towards the colder Scandinavian latitudes. Finally, and perhaps most importantly, my Antigoni for caring.

%
%\vspace{2em}
%
%\begin{flushright}
%\emph{Navid\\
%Athens, Feb. 2015}
%\end{flushright}
%

	\vspace*{\fill} \newpage
	\setcounter{page}{1}
	\pagenumbering{arabic}

	\newpage \thispagestyle{fancy} \vspace*{\fill}
	\scshape \noindent % dedication text

To my friends
	\vspace*{\fill} \newpage \rm

\setcounter{page}{1}	\pagenumbering{arabic}

\doublespacing
\onehalfspacing

% include each chapter...
%\setcounter{chapter}{-1}  % start chapter numbering at 0

% !TEX root = ../thesis.tex

%\begin{savequote}[110mm] 
%When you're thinking about something that you don't understand you get this terrible unhappy feeling called `confusion', it's a very difficult and unhappy business, and so most of the time you're rather unhappy -- actually -- with this confusion. You can't penetrate this thing. And the reason for this confusion is that we're all some kind of apes -- and a kind of stupid -- trying to figure out how to put the two sticks together to reach the banana; and we can't quite make it. And I feel like this all the time. I'm some kind of ape trying to put two sticks together; so I always feel stupid. Once in a while though the sticks go together and I reach the banana.
%\qauthor{R. P. Feynman} 
%\end{savequote}

\chapter{Introduction}
\label{ch:intro}

\begin{figure}[h!]
\centering
\includegraphics[height=2.0in]{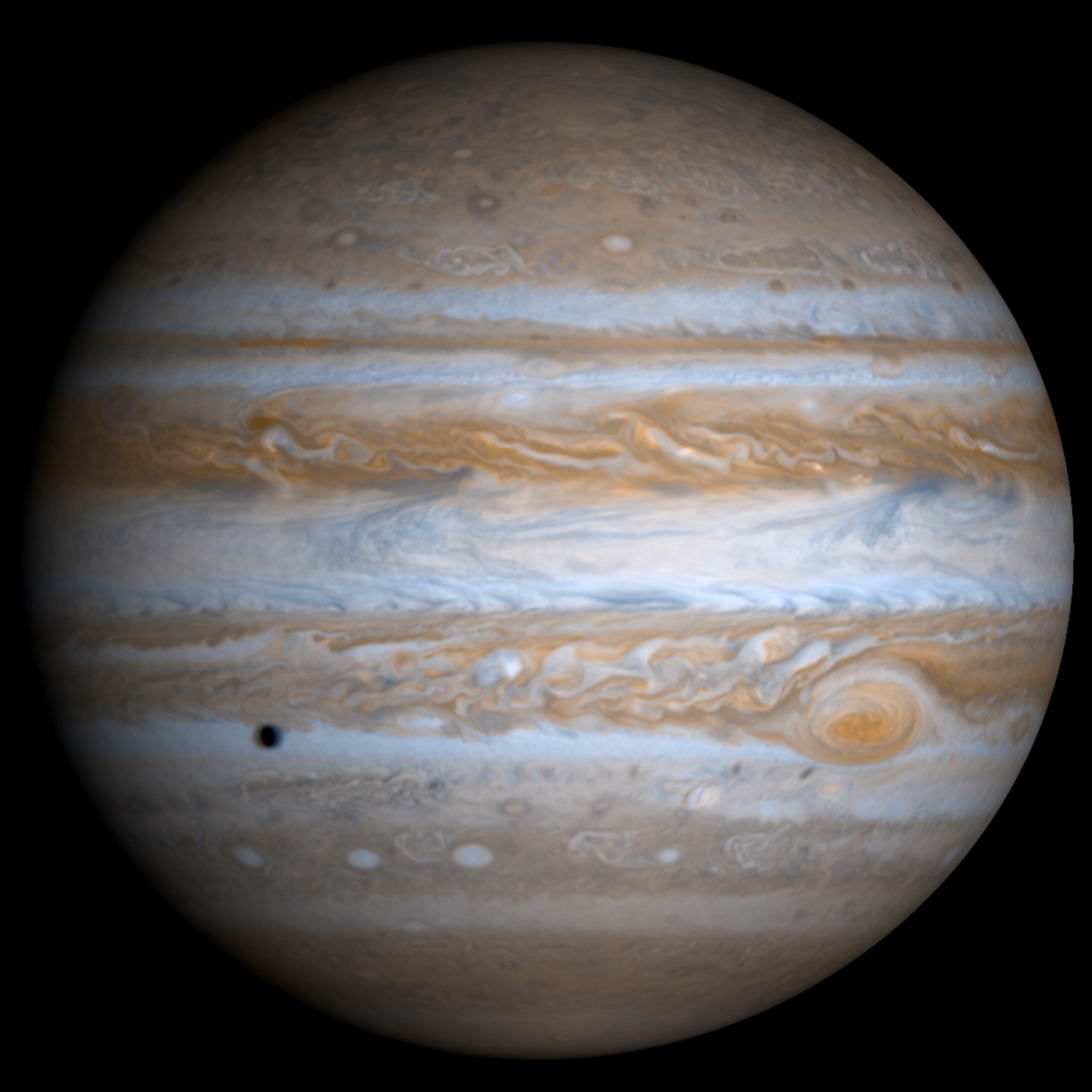}\hspace{-.5mm}\includegraphics[height=2.0in]{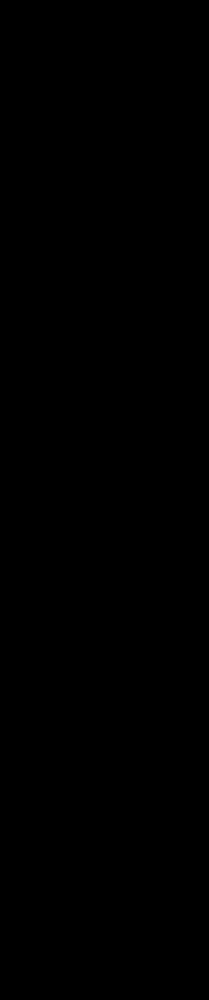}\hspace{-.5mm}\includegraphics[height=2.0in]{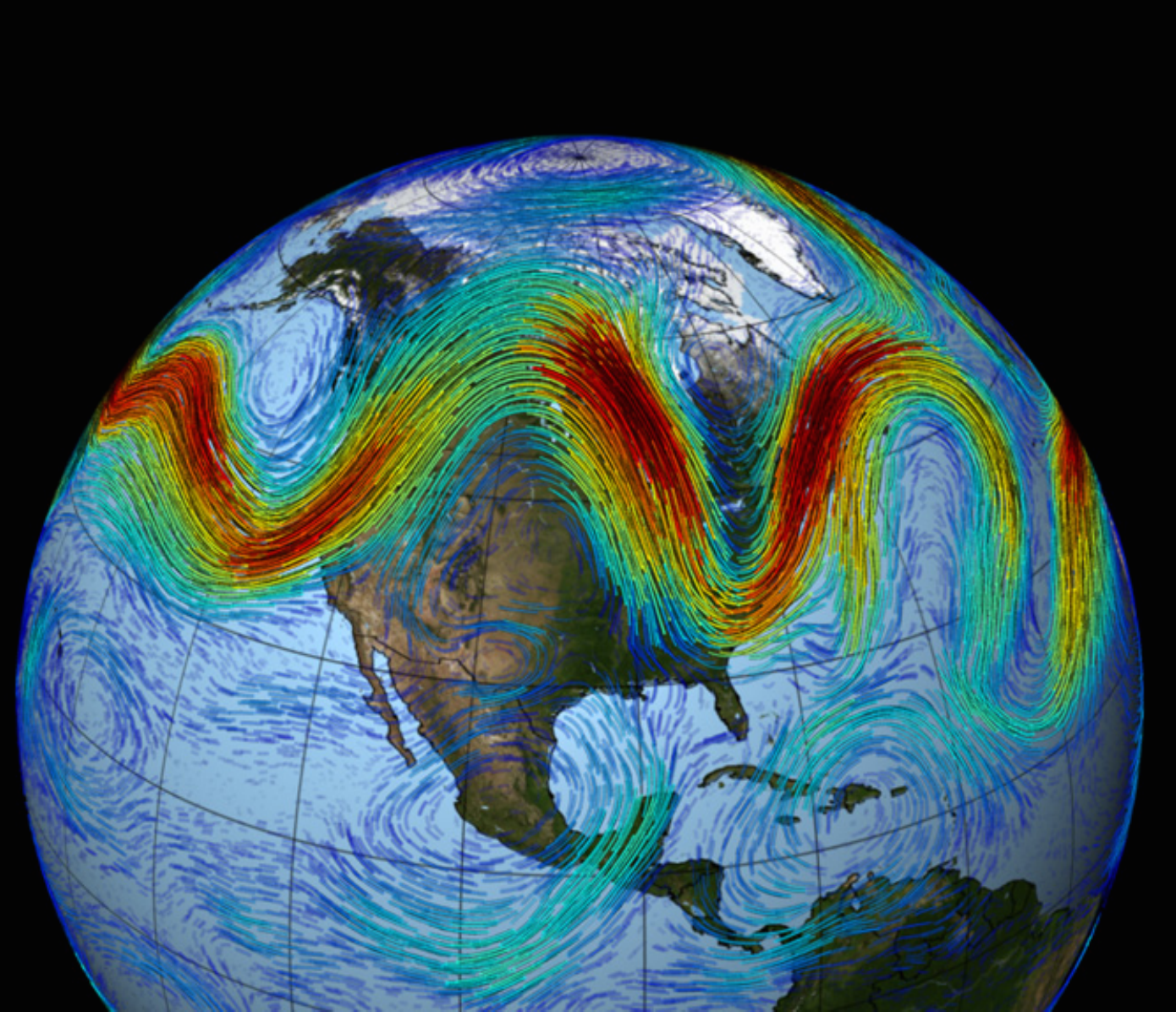}
\caption{\label{fig:Jupiter_Earth} The atmospheres of planets Jupiter and Earth are turbulent. The turbulence is anisotropic and inhomogeneous and the kinetic energy of the flow is concentrated in large-scale zonal or wavy jets and large-scale vortices which persist in the flow enhancing the large-scale long time range predictability of the flow. Credits: NASA/JPL and NASA/GSFC.}
\end{figure}

\section{Jets on Earth and Jupiter}
\label{sec:1.1}

Turbulent atmospheric flows in rotating planets are observed to self-organize into large-scale structures. These structures vary at a time scale much larger compared to the turbulent eddy motions with which they coexist. Prominent characteristic examples are the Earth's subtropical and polar jet streams or the zonal winds in Jupiter and its Great Red Spot (see~Fig.~\ref{fig:Jupiter_Earth}). Changes in the structure or the position of the Earth's jet streams may induce dramatic changes in regional weather patterns. Recent such examples are the 2003 and 2010 European heat waves and the 2013-14 North American cold wave that were caused by shifts in the position of the jets. 

%An insightful understanding on how this large-scale structures are organized and also maintained by turbulence is therefore necessary for a better understanding of climate dynamics and climate variability. However, the way that these structures are formed and maintain is in many cases still poorly understood. 
\begin{figure}
\centering
\begin{minipage}[c]{.48\textwidth}
\includegraphics[width=.95\linewidth,trim = 0mm 0mm 1mm 0mm, clip]{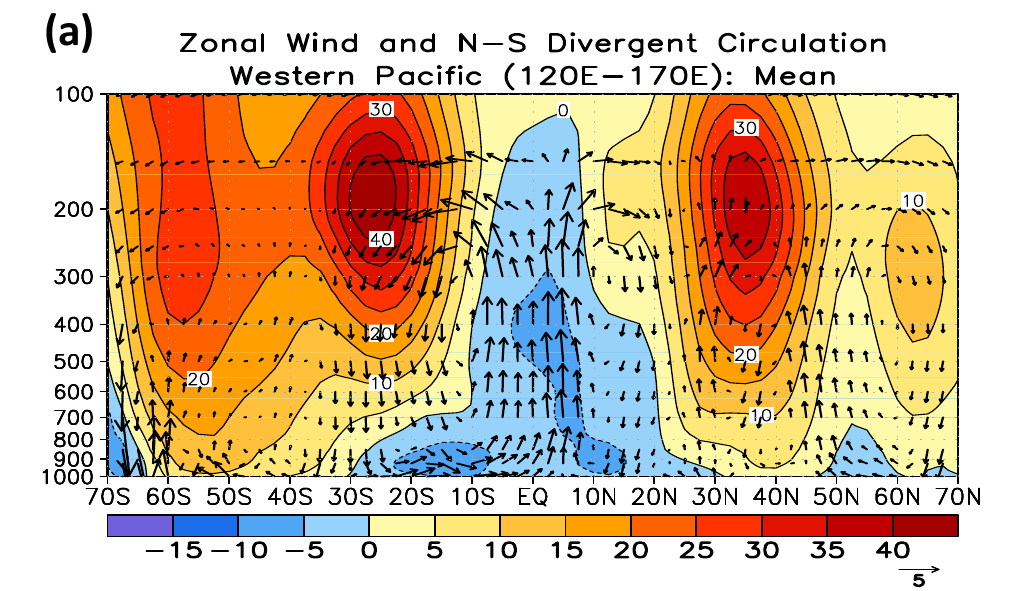}\\
\includegraphics[width=.95\linewidth,trim = 0mm 0mm 1mm 0mm, clip]{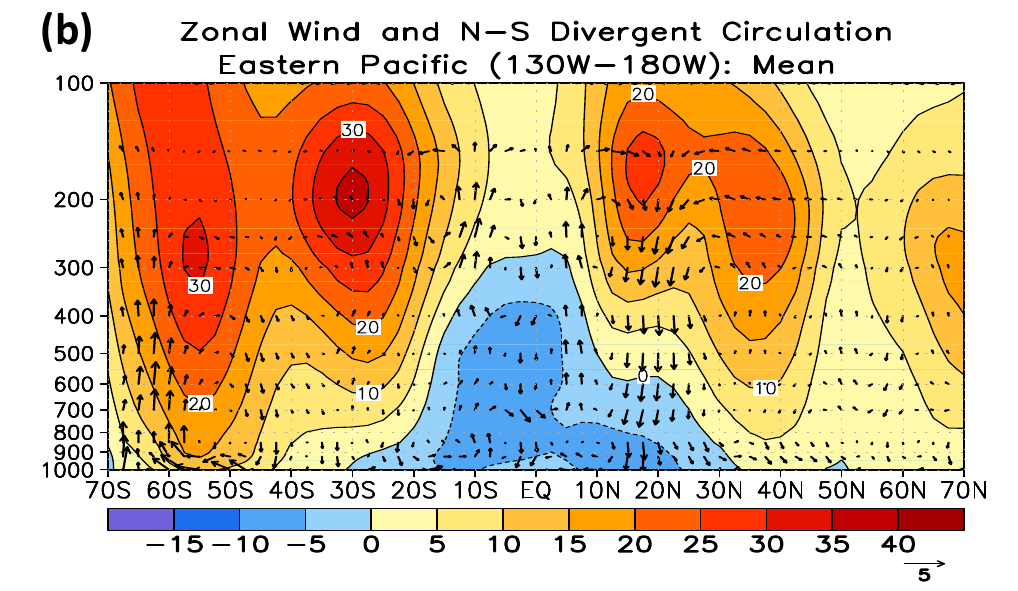}
\end{minipage}%
\noindent\begin{minipage}[c]{.48\textwidth}
\includegraphics[width=.95\linewidth]{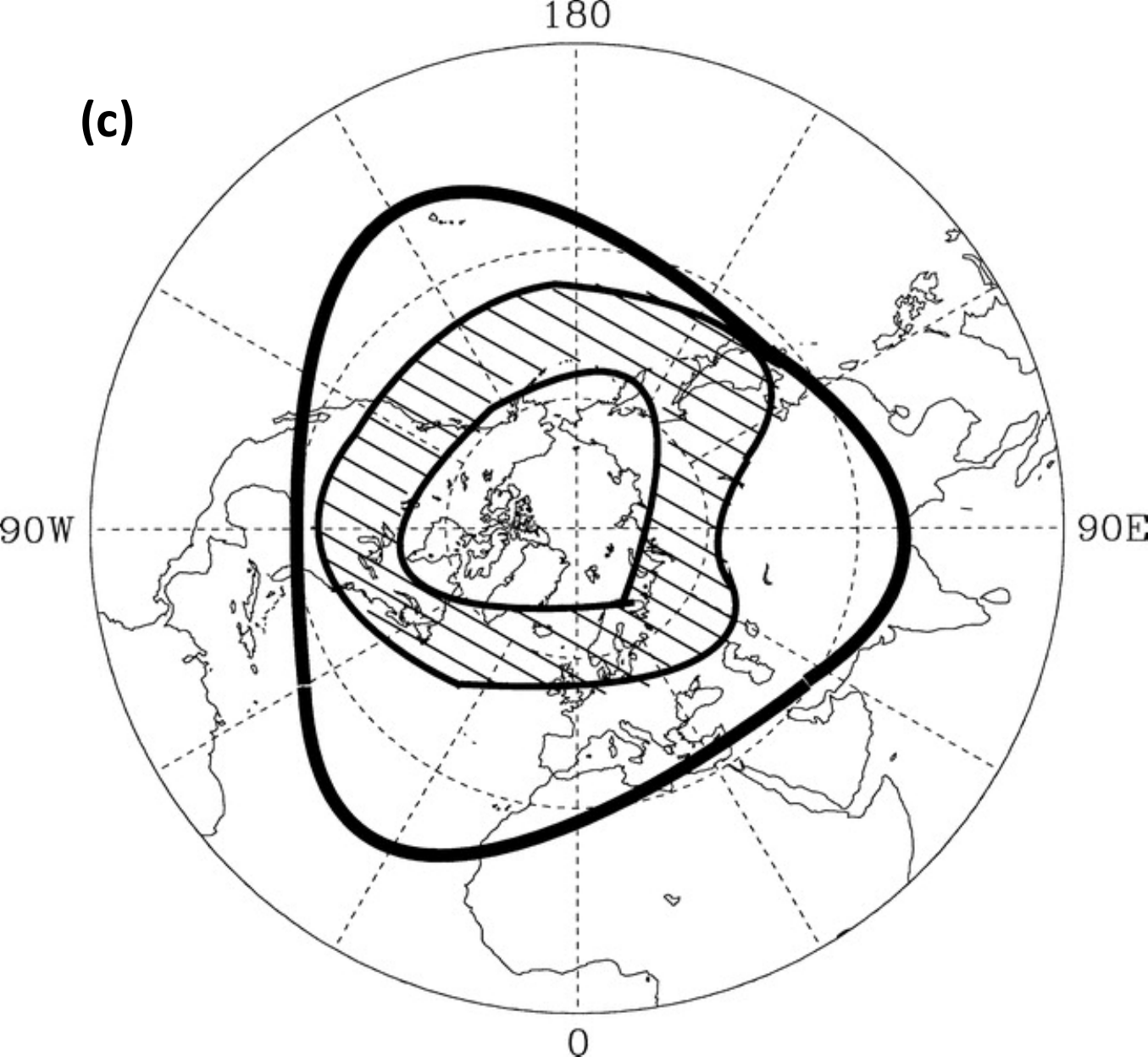}
\end{minipage}
\caption{\label{fig:EarthJets} Earth's jet streams: Shown are pressure (mb)-latitude sections of the mean zonal wind ($\textrm{m}\,\textrm{s}^{-1}$) and divergent circulation averaged over (a) the west Pacific (120$\deg$E-170$\deg$E) and (b) the east Pacific (130$\deg$W-180$\deg$W). Shown are time averages for May 2014. Divergent circulation is represented by vectors of combined pressure vertical velocity and the divergent component of the meridional wind.) Red (blue) shading and solid (dashed) contours denote eastward (westward) zonal wind. The eastward wind maxima at around $30\deg$N/S correspond to the subtropical jet streams while the maxima at higher latitudes to the eddy-driven polar jet streams. (Credit: NWW, Climate Prediction Center.) (c) The mean position of the subtropical jet (thick solid line) and the region (shaded) of principal polar jet stream activity for the northern hemisphere (Taken from \textcite{Riehl-1962}).}
\vspace{1em}
%\end{figure}
%
%\begin{figure}
%\centering
\includegraphics[width=.65\textwidth]{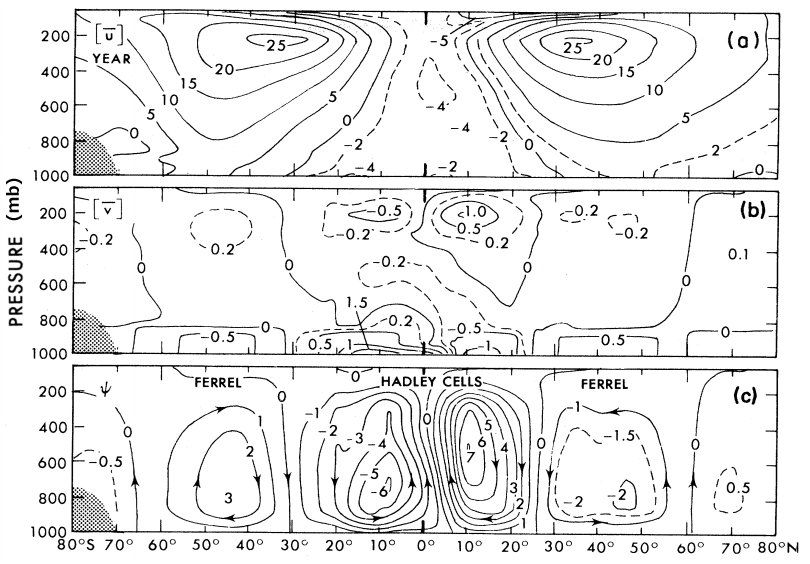}
\caption{\label{fig:annual_winds} Zonal mean cross sections of (a) the zonal wind component, $\overline{u}$, ($\textrm{m}\,\textrm{s}^{-1}$), (b) the meridional wind component, $\overline{v}$, ($\textrm{m}\,\textrm{s}^{-1}$) and the inferred flow of atmospheric mass ($10^{10}\ \textrm{kg}\,\textrm{s}^{-1}$) for annual-mean and zonally averaged conditions. (Brackets denote time average.) (Taken from \textcite{Peixoto-Oort-1984}.)}
\end{figure}

Jet streams (or jets) are strong and narrow quasi-rectilinear air currents found in the atmospheres of some planets. In the Earth there are two jets in each hemisphere that flow eastwards. The typical structure of the mean zonal winds over the Pacific ocean reveals the double jet structure in each hemisphere (cf.~Figs.~\ref{fig:EarthJets}\hyperref[fig:EarthJets]{a,b}). In a frame rotating with the Earth the jets have typical speeds of $40\ \textrm{m}\,\textrm{s}^{-1}$ ($145\ \textrm{km}\,\textrm{h}^{-1}$) and may reach $70\ \textrm{m}\,\textrm{s}^{-1}$ ($250\ \textrm{km}\,\textrm{h}^{-1}$) in the winter of each hemisphere. The wind maximum of the subtropical jet is located at around 30$\deg$N/S and at a height of $10\textrm{-}16\ \textrm{km}$ (or at a pressure of $200\ \textrm{mb}$). The polar jet is located at 40$\deg$-60$\deg$N/S and at a height of about $10\ \textrm{km}$ ($300\ \textrm{mb}$). The weaker subtropical jet is much more axisymmetric while the stronger polar jet has a pronounced slowly translating non-zonal wave component, especially in the Northern Hemisphere, as shown in Fig.~\ref{fig:Jupiter_Earth}, with the jet maxima distributed over an annular region as depicted in the schematic of~\textcite{Riehl-1962} in Fig.~\ref{fig:EarthJets}\hyperref[fig:EarthJets]{c}. Due to its spatial and temporal variation the polar jet stream does not appear as a prominent feature in plots of the annual mean zonal velocity. For example, in Fig.~\ref{fig:annual_winds}\hyperref[fig:annual_winds]{a} only the subtropical jets appear.

\begin{figure}
\centering
\includegraphics[width=\textwidth]{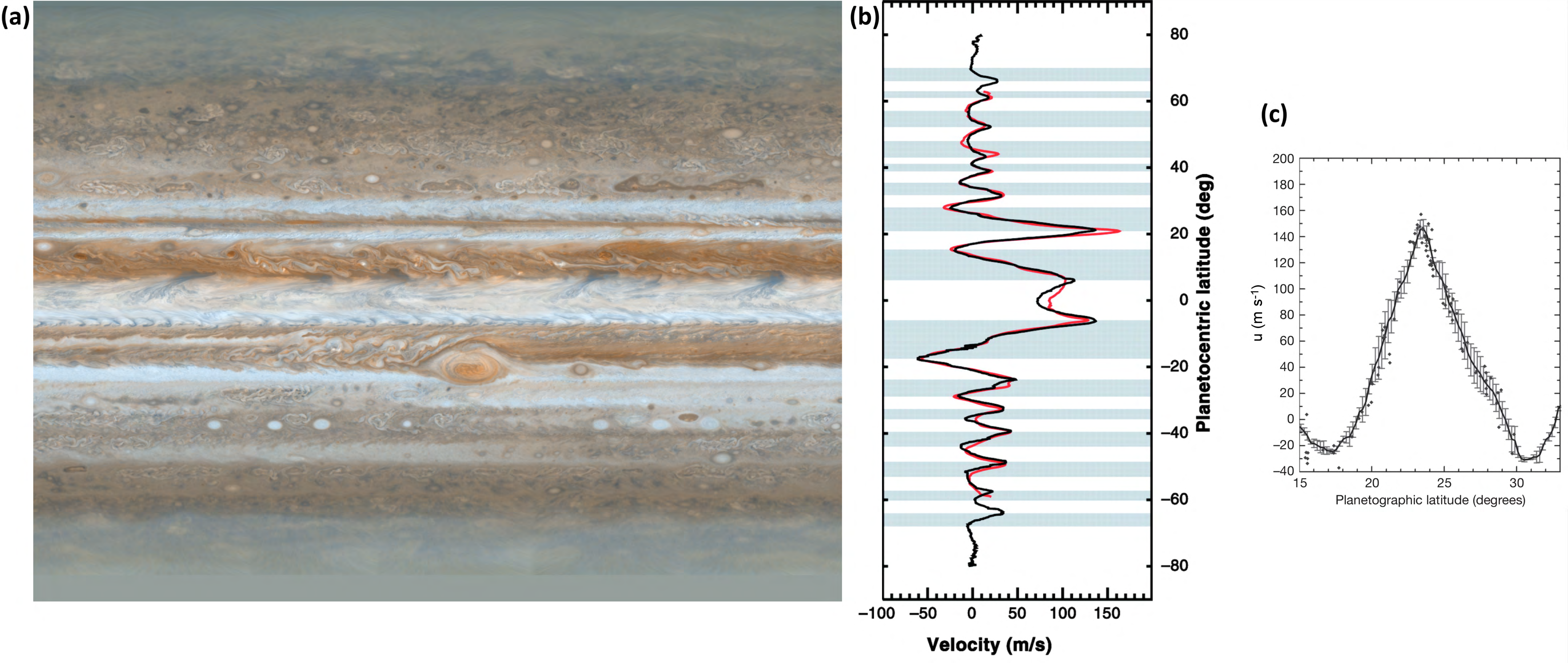}
%\begin{minipage}[c]{.72\textwidth}
%\includegraphics[height=2.2in]{jovian_structure_jets.pdf}
%\end{minipage}%
%\noindent\begin{minipage}[c]{.26\textwidth}
%\includegraphics[height=1.5in]{23Njet.pdf}
%\end{minipage}
%\includegraphics[height=2.2in]{jovian_structure_jets.pdf}\includegraphics[height=1.6in]{23Njet.pdf}
\caption{\label{fig:JupiterJets} Jupiter's jets: (a) A cylindrical projections of Jupiter from \emph{Cassini} images (Credits: NASA/JPL/Space Science Institute). (b) The zonal wind speeds measured by~\textcite{Limaye-1986} on \emph{Voyager 2} images taken on 1979 (red curve) and by~\textcite{Porco-etal-2003} on \emph{Cassini} images taken on 2000 (black curve). (Taken from \textcite{Porco-etal-2003}.) (c) A zoom at the eastward jet found at $24\deg$N. (Taken from \textcite{Sanchez-etal-2008}.)}
\end{figure}

While little information is known about the vertical structure of the winds on Jupiter\footnote{Preliminary evidence suggests that the jets increase below the clouds \parencite{Atkinson-etal-1997}. Definitive answers are expected from analysis of the gravitometric measurements that will be collected by space probe \emph{Juno} \parencite{Kaspi-2013,Read-2013}.} there is a lot of information about the latitudinal structure of the winds at cloud level (about $700\ \textrm{mb}$) obtained from involved cloud tracking techniques. These measurements revealed the existence of an alternating jet structure consisting of 15 eastward and 15 westward jets located at the latitudes that separate the colored belts of the planet (cf. Fig.~\ref{fig:JupiterJets}\hyperref[fig:JupiterJets]{a,b}). Near the equator the speed of the eastward zonal jet exceeds by $120\ \textrm{m}\,\textrm{s}^{-1}$ ($430\ \textrm{km}\,\textrm{h}^{-1}$) the rotational speed of the planet, indicating that the Jovian atmosphere has an appreciable superrotation at the equator. Moreover, the jets of Jupiter seem to vary very slowly despite being embedded in strong turbulent flow. This remarkable fact was discovered when the wind measurements made by \emph{Voyager 2} and the \emph{Cassini} space probes were analyzed and were found to produce nearly identical winds, although 20 years intervened between the measurements. The respective wind measurements made by the two probes are shown in Fig.~\ref{fig:JupiterJets}\hyperref[fig:JupiterJets]{b}. Also, the jets have a very special shape: the jet maxima (the superrotating flow) are very pointed while the westward jets (the subrotating flows) are weaker and blunted (cf.~Fig.~\ref{fig:JupiterJets}\hyperref[fig:JupiterJets]{c}). This thesis will provide an explanation for both the constancy of the zonal winds and their shape.
% and the Great Red Spot persists, in more or less the same form, for more than 300 years now, despite living in a flow in which the radiative time scale is of the order of $10$~years~\parencite{Marcus-1993}.

\begin{figure}
\centering
\includegraphics[height=1.1in]{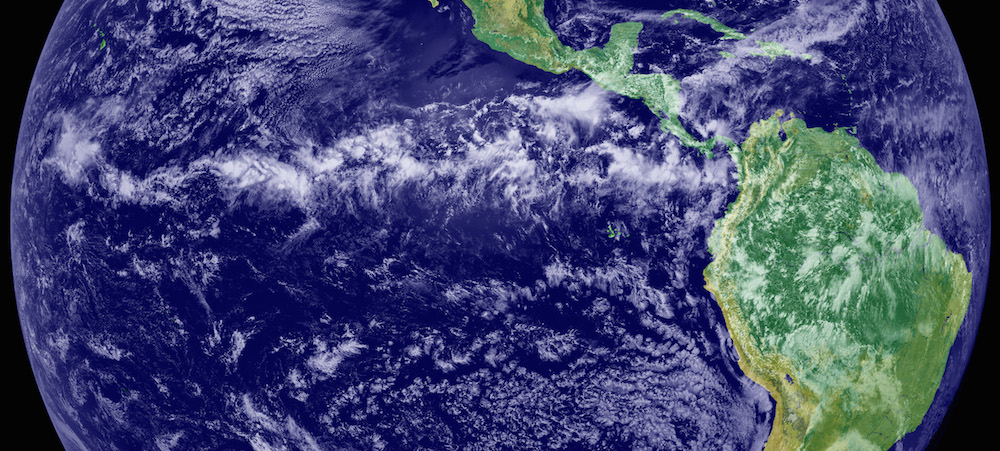}
\caption{\label{fig:ITCZ} The Intertropical Convergence Zone (ITCZ) over the eastern Pacific on 12 July 2000 becomes visible 
by the nearly zonal cloud band. The major convective activity that drives the upwelling of the Hadley cell has been organized to occur in this narrow zone. %Surface trade winds SW (NW) orientation north (south) of the equator converge and give rise to hot humid air from the tropics forming clouds.
Credit: NASA/GOES.}
\end{figure}

In the Earth, the subtropical jet stream is driven by the large-scale meridional circulation (see Fig.~\ref{fig:annual_winds}\hyperref[fig:annual_winds]{c}) initiated by the enhanced convective activity which is concentrated in a narrow zonal band near the equator, called the ITCZ (Intertropical Convergence Zone) (cf.~Fig.~\ref{fig:ITCZ}). This axisymmetric circulation produces a nearly angular momentum conserving zonal flow by transferring angular momentum from the deep tropics to the poleward upper part of the Hadley cell, where the subtropical jet maximum is located (Fig.~\ref{fig:annual_winds}\hyperref[fig:annual_winds]{c}). %The Hadley cell has been the subject of theoretical work that has provided the theoretical basis for its understanding \parencite{Schneider-Lindzen-1977,Held-Hou-1980,Lindzen-Hou-1988}.
The polar jet is maintained from the momentum convergence of the turbulent motions, which themselves owe their existence to the very jet they support. %\footnote{If there was full geometrical similarity between Jupiter and the Earth we would expect Jupiter to have 20 eastward zonal jets instead of 15 eastward jets alternating with 15 westward jets, given that the radius of Jupiter is 10 times larger than the radius of the Earth and also given that the Jovian atmosphere is unlikely to support a subtropical type of jet because of the weak solar forcing of the planet. (Jupiter receives $12.5\ \textrm{W}\,\textrm{m}^{-2}$ from the Sun and $5.7\ \textrm{W}\,\textrm{m}^{-2}$ from its interior, primarily from the gravitational contraction of the planet. The Earth receives $340\ \textrm{W}\,\textrm{m}^{-2}$ from the Sun.)}
 It should be clarified that the turbulent motions responsible for the maintenance of the polar jet is the midlatitude turbulence, with typical length scales of $1000\ \textrm{km}$ and time scales of a week. This large-scale atmospheric turbulence is often referred to in the meteorological literature as the macroturbulence \parencite{Held-1999,Schneider-Walker-2006}. While the dynamics of the subtropical jet has been fully elucidated~\parencite{Schneider-Lindzen-1977,Held-Hou-1980,Lindzen-Hou-1988}, the theory for the formation and maintenance of the eddy-driven jets is far from complete. This thesis will present a new theory for the formation and maintenance of eddy-driven jets in planetary turbulence. % This thesis attempts to propose a theory for the formation and maintenance of jets in turbulent anisotropic fluids.

%The polar jet is from the large-scale turbulence that in turn exists because of the polar jet. Turbulent eddy motions systematically converge momentum to the location of the jet maintaining it this way against dissipation. By turbulence here we refer to the turbulence associated with the cyclone scale\footnote{However turbulent motions in a molecular level may also act upgradiently as well, as it is in the case in turbulent flows in pipes where turbulence organizes to form and maintain the large roll/streak circulation (see for example \textcolor{red}{\textbf{(cite who? is there a work showing that eddies feed the roll/streak prior to S3T?)}}\textcite{Farrell-Ioannou-2012,Constantinou-etal-Madrid-2014}}, which is of the order of $1000\ \textrm{km}$.

%from the large-scale turbulence that in turn exists because of the polar jet. By turbulence here we refer to the turbulence associated with the cyclone scale\footnote{However turbulent motions in a molecular level may also act upgradiently as well, as it is in the case in turbulent flows in pipes where turbulence organizes to form and maintain the large roll/streak circulation (see for example \textcolor{red}{\textbf{(cite who? is there a work showing that eddies feed the roll/streak prior to S3T?)}}\textcite{Farrell-Ioannou-2012,Constantinou-etal-Madrid-2014}}, which is of the order of $1000\ \textrm{km}$. Unlike its subtropical counterpart, the theory of the formation and maintenance of the eddy-driven jets is not complete and this thesis provides a contribution in the direction of building such a theoretical understanding.

The idea that smaller scale turbulence transfers momentum and maintains larger scale flows, thus fluxing momentum upgradiently, is provocative and has been called in the literature, equally provocatively, a ``negative viscosity'' phenomenon \parencite{Starr-1953}. This idea, expressed in this manner, seems to violate the natural entropic tendency of physical systems towards states of greater disorder, which is consistent with the usual downgradient action of diffusion: for example, as \textcite{Reynolds-1883} has shown, high density ink is spread and homogenized by turbulent flow. Interestingly, as we will discuss shortly, it has been shown that the fine-grain maximum entropy states in barotropic flows correspond to macrostates with large-scale jets and vortices \parencite{Miller-1990,Robert-Sommeria-1991,Bouchet-Venaile-2012}.

\begin{figure}
\centering
\hspace{3mm}\includegraphics[width=.65\textwidth]{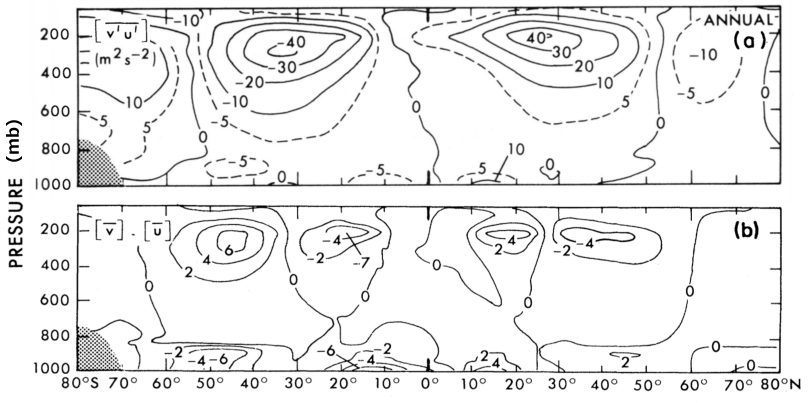}
\caption{\label{fig:anual_eddyfluxes} Zonal mean cross section of zonal mean of northward flux of zonal momentum (cf.~\eqref{eq:Jy}) by (a) the eddies, $\[\overline{u'v'}\]$ ($\textrm{m}^2\,\textrm{s}^{-2}$) and (b) the mean meridional circulation, $\[\overline{u}\,\overline{v}\]$ ($\textrm{m}^2\,\textrm{s}^{-2}$). Brackets denote time average for a period of a year. %Notice that both eddy and mean meridional momentum fluxes have a structure so that they reinforcing the polar and subtropical jets.
It is clear that the dominant contribution comes from the eddies. Note the eddy momentum flux convergence occurs in the location of the polar jet while mean meridional momentum flux converges at the location of the subtropical jet. (Taken from \textcite{Peixoto-Oort-1984}.)}
\end{figure}

\textcite{Jeffreys-1926-winds} was the first to propose that large-scale atmospheric circulation is eddy-driven. Until then people were trying to obtain understanding of the general circulation of the atmosphere neglecting the non-axisymmetric dynamics of the flow as well as the effect of the mean quadratic eddy stresses (the divergence of the Reynolds stresses) on the mean axisymmetric circulation. Jeffreys demonstrated that an exclusively axisymmetric point of view is inadequate by analyzing the zonally averaged momentum balance of a whole column of air located at the midlatitudes with westerly (or eastward) winds at the ground. Since the momentum of the whole column (about $10^5\ \textrm{N}\,\textrm{s}$ per $\textrm{m}^{2}$) is lost at the ground at the rate of $0.1\ \textrm{N}\,\textrm{m}^{-2}$ (this is the surface drag), it can be shown that it cannot be replenished by the flux of momentum by the observed mean axisymmetric circulations, and thus he concluded that the surface westerlies had to be maintained by the horizontal convergence of momentum from the non-axisymmetric motions (the eddy motions), i.e., he argued that the horizontal momentum divergence of the non-axisymmetric motions must be responsible for maintaining the mean momentum of the column and also for the westerlies at the ground.\footnote{At the time of Jeffreys the existence of the polar jet was not known and in his~\citeyear{Jeffreys-1926-winds} paper there is no mention of upper-level jets. Evidence of strong upper-level winds was obtained from kite measurements first in Japan in the 1920's, but the existence of the polar jet as a global feature of the climatology was established from aircraft measurements during the Second World War. The term ``jet stream'' was coined by Rossby in 1942 \parencite{Lewis-2003}.} The paper of \citeauthor{Jeffreys-1926-winds} introduced the idea that the eddy motions in the atmosphere (the cyclones) should not be viewed as unstable perturbations to the axisymmetric mean circulation but rather as an integral component for the maintenance of the very axisymmetric circulation that gives rise to them (for a historical discussion cf.~\textcite{Lorenz-67}). Jeffreys also stressed for the first time that the horizontal eddy barotropic dynamics are responsible for the maintenance of the large-scale structure. This point of view was further advanced and given a theoretical basis by \textcite{StaffMembers-1947,Rossby-1947}. In the meanwhile the detailed upper air observations, that became available after the Second World War, gave solid observational support to the proposition that the Earth's polar jet is supported by barotropic upgradient momentum fluxes. Such modern measurements of the momentum flux compiled by~\textcite{Peixoto-Oort-1984} are shown in Fig.~\ref{fig:anual_eddyfluxes}. In these plots the zonally averaged momentum flux $\overline{u\,v}$ is decomposed as\footnote{$u$ is the zonal velocity with eastward flow being $u>0$, $v$ is meridional velocity with northward flow corresponding to $v>0$ and overbar denotes zonal average.} 
\be
\overline{ u\,v} = \overline{u}\,\overline{v} + \overline{u'v'}\ ,\label{eq:Jy}
\ee 
into the momentum flux due to the axisymmetric motions, $\overline{u}\,\overline{v}$, and the mean momentum flux due to the eddies (the motions that deviate from axisymmetry), $\overline{u'v'}$. It can be seen in Fig.~\ref{fig:anual_eddyfluxes} that the eddy momentum fluxes are larger than the momentum fluxes from the meridional circulations and dominate at the latitudes of the polar jet. There is a large eddy momentum flux convergence around 55$\deg$N/S and at $300\ \textrm{mb}$, precisely at the location of the polar jet, and at this location the momentum fluxes from the axisymmetric motions is negligible, showing that the polar jet is maintained by eddy momentum flux convergence. The second region of convergence occurs at 30$\deg$N/S where the subtropical jet is located (cf.~Fig.~\ref{fig:anual_eddyfluxes}\hyperref[fig:anual_eddyfluxes]{b}). The dominant flux convergence in this case is from the axisymmetric motions indicating that the subtropical jet is a result of an axisymmetric circulation.

\begin{figure}
\centering
\includegraphics[width=3.5in]{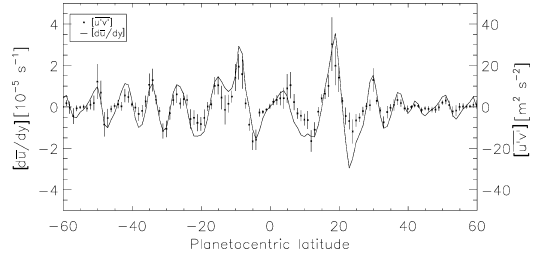}
\caption{\label{fig:Jupiter_uv_Uy} The structure of the eddy momentum flux, $[\overline{u'v'}]$ (dots), and the zonal mean flow shear, $[\df \overline{u} /\df y]$ (solid), as a function of latitude. $[\overline{u'v'}]$ error bars correspond to 2 standard deviations from the mean. There is a distinct positive correlation between the two curves. (Taken from \textcite{Salyk-etal-2006}.)}
\end{figure}

The Jovian jets are eddy-driven, like the Earth's polar jets. This was confirmed through systematic and repeated analysis of the turbulent velocity fields at cloud level \parencite{Ingersoll-etal-1981,Ingersoll-90,Salyk-etal-2006,Galperin-etal-2014}. Ingersoll and coworkers demonstrated the upgradient action of the eddy momentum fluxes by plotting the eddy momentum flux, $\overline{u'v'}$, together with the shear of the mean flow, $\df\overline{u}/\df y$, as a function of latitude. They noted that these two quantities are positively correlated and to a great degree of accuracy satisfy the linear law,
\be
\overline{u'v'} = \kappa \frac{\df\overline{u}}{\df y}\ ,\label{eq:antidiffusion}
\ee
with $\kappa \approx10^6\ \textrm{m}^2\,\textrm{s}^{-1}$, as can be seen in Fig.~\ref{fig:Jupiter_uv_Uy}. That $\kappa>0$ implies the remarkable fact that the momentum fluxes on Jupiter act anti-diffusi\-vely, since the rate of change of zonal momentum (disregarding dissipation) obeys
\be
\frac{\partial \overline{u}}{\partial t} = -\frac{\partial}{\partial y}\(\,\overline{u'v'}\,\) = -\kappa\,\frac{\df^2\overline{u}}{\df y^2}\ ,
\ee
which is a diffusion equation with negative diffusion coefficient.

\section{Theories for jet formation and current understanding}

Since atmospheric motions on Earth are confined in a thin shell $10\,000\ \textrm{km}$ in the horizontal and $10\ \textrm{km}$ in the vertical (the mean depth of the troposphere, where most of the mass of the atmosphere is located) the motions are quasi-horizontal and one would expect that barotropic dynamics, which involve only the height-averaged flow fields, would be sufficient to describe the dynamics of the Earth's atmosphere. However, this is erroneous: the vertical shear of the mean zonal wind (i.e. the derivative of the zonal wind with height), which is associated with the temperature contrast between the Equator and the Pole, gives rise to powerful baroclinic growth processes that produce the cyclones, which are the ``atoms'' of atmospheric turbulence responsible for the transport of heat and momentum in the atmosphere.

Interestingly, the cyclones grow first through the baroclinic process of drawing potential energy from the mean flow and transporting in this way heat to the poles, and then they assume a nearly barotropic structure (i.e. height independent). The barotropic dynamics responsible for the redistribution of momentum in the upper troposphere and the formation of jets is also referred to as the ``barotropic governor'', because this mechanism of barotropic exchange that forms the jets is the very mechanism that alleviates the instability of the atmospheric flow and maintains the atmosphere in a state of baroclinic neutrality \parencite{Ioannou-Lindzen-1986,James-1987,Lindzen-1993,Roe-Lindzen-1996}. This duality in the behavior of the baroclinically growing disturbances simplifies the study of jet formation in baroclinic atmospheres. It allows us to consider that the atmospheric dynamics fall into two manifolds: the slow barotropic manifold that controls the formation and evolution of jets in the upper troposphere, and the faster manifold of baroclinic processes that provides the necessary excitation of the barotropic manifold to maintain it in a turbulent state. A confirmation of this point of view has been given by \textcite{DelSole-01a}, who by considering that the upper troposphere is governed by barotropic dynamics excited by the baroclinic activity from lower levels demonstrated that the structure of the momentum fluxes responsible for maintaining the upper level jets could be accurately captured. As a result, in this thesis we will adopt the traditional view and study the formation of jets and other large-scale structure both in the Earth and in Jupiter within the context of barotropic dynamics. 

This barotropic two-dimensional framework has been adopted by most researchers that investigated jet formation in Jupiter and the outer planets starting with \textcite{Williams-78} and more recently with \textcite{Nozawa-and-Yoden-97} and \textcite{Huang-Robinson-98}. Other authors investigated the formation of jets on Jupiter in the primitive equation extension of the quasi-geostrophic barotropic dynamics by modeling the Jovian atmosphere as a shallow-water fluid; but also in these studies jet formation proceeded as in the purely two-dimensional barotropic models \parencite{Cho-Polvani-1996pof,Cho-Polvani-1996,Scott-Polvani-2007}. That these dynamics can produce jet formation has been also demonstrated experimentally by \textcite{Read-etal-2004} in the Coriolis rotating tank in Grenoble and by \textcite{Espa-etal-2010} in Rome. We adopt the simplest framework and study jet formation in Jupiter and the outer planets in the context of barotropic dynamics that are maintained in a turbulent state by external excitation. The excitation represents the introduction of vorticity at cloud level from convective motions induced by the heating source in the interior planet. The typical scale of vorticity injection is $1000\ \textrm{km}$ \parencite{Little-etal-1999,Gierasch-etal-2000}.

\vspace{1em}

We now review the main theories that have been advanced for understanding the formation of jets in turbulence. These theories can be distinguished as those that arise from turbulence theory and are generally phenomenological, and those that consider that the flow perturbations are coherent, like a wave, and study the interaction of this coherent eddy field with the mean flow. The latter theories will be referred to as wave--mean flow interaction theories and are generally more deductive.

\subsection{Turbulence theories\label{sec:turb_theor}}

Jet formation in turbulence theory is viewed as a consequence of a cascade of energy from small scales to large scales. This type of cascade is called ``inverse'' and is the opposite of direct cascades that characteristically operate at small scales in homogeneous isotropic 3D turbulence transferring energy from large scales to small scales where it is dissipated. That turbulence confined on a plane, like the barotropic turbulence that we will study, supports an inverse cascade in energy was first proposed by \textcite{Fjortoft-1953} who argued that this was consequence of the two dimensionality of the flow which implies in the inviscid limit the conservation of the total kinetic energy of the flow, $E=\int \frac1{2}|\uv|^2\,\df A$, as in 3D, but also the conservation of the vorticity of every particle in the flow, which leads to an infinite set of integral invariants. As a result, on the $x\textrm{-}y$ plane the material conservation of the vorticity $\z\equiv(\nablav\times\uv)_z=\partial_x v-\partial_y u$, implies that all integrals over the whole area of the fluid of the form $\int F(\z)\,\df A$, with $F$ any integrable function, are conserved. Enstrophy, defined as $Z=\int \frac1{2}\z^2\,\df A$, is the invariant that is usually considered out of this hierarchy of conserved quantities and \citeauthor{Fjortoft-1953} considered the constraint imposed on the spectral evolution of the flow by the simultaneous conservation of energy and enstrophy. Expressing the 2D incompressible flow field through a streamfunction $\psi$ as $\uv=(u,v)=(-\partial_y\psi,\partial_x\psi)$, implies that $\z=(\partial^2_{xx}+\partial^2_{yy})\psi$, and expanding the streamfunction in Fourier as, $\psi = (2\pi)^{-2}\int \df^2\kv\;\hat{\psi}_\kv\,e^{\i \kv\cdot\xv}$, we have that energy and enstrophy are respectively given as $E = (2\pi)^{-3} \int \df^2\kv\;|\kv|^2 |\hat{\psi}_\kv|^2$ and $Z = (2\pi)^{-3} \int \df^2\kv\;|\kv|^4 |\hat{\psi}_\kv|^2$. This means that the energy and enstrophy spectral densities that correspond to wavenumber $\kv$, $\mathcal{E}(\kv)$ and $\mathcal{Z}(\kv) $ respectively, are related through $\mathcal{Z}(\kv) = k^2 \mathcal{E}(\kv)$, where $k=|\kv|$. \citeauthor{Fjortoft-1953} stated that conservation of energy and enstrophy in 2D flows constrains the energy exchanges between scales in such a manner so that if enstrophy moves to smaller scales energy must move to larger scales.

\begin{figure}
\centering
\includegraphics[width=5.2in,trim = 50mm 0mm 50mm 0mm, clip]{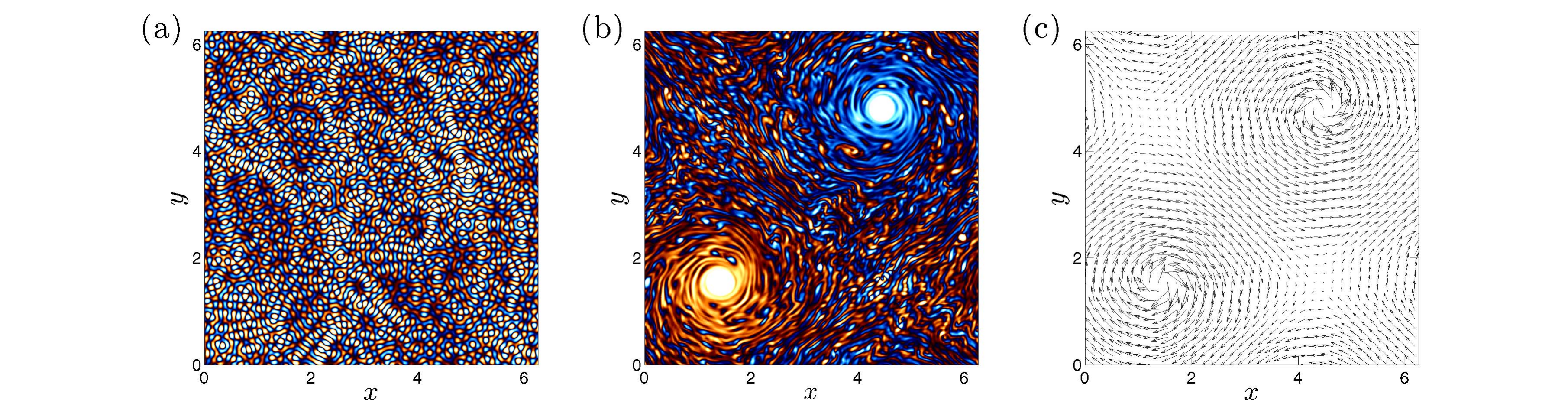}
\caption{\label{fig:NL_forc-diss_beta0} Evolution of the vorticity field in forced--dissipative flow on a doubly periodic domain, $2\pi\times2\pi$. Energy is being injected in the flow at rate $\varepsilon_f$ and at length scale $2\pi/k_f$ with $k_f=40$, and dissipation is done at small scales with 8th-order hyperdiffusion with coefficient $\nu=7\times 10^{-35}$. (a) Typical structure of the forcing field. (b) The vorticity field at time $t/\tau=1265$, with $\tau=(\varepsilon_f k_f^2)^{-1/3}$. Vorticity patches tend to merge creating large-scale vortices which contain most of the energy in the flow. As vortices are advected by the flow characteristic vortex filaments are created transferring enstrophy to smaller scales. (c) Velocity field at time $t/\tau=1265$. For the  simulation a pseudospectral code was used at a $256\times256$ resolution with an exponential filter acting on wavenumbers $k>\frac2{3}k_{\textrm{max}}$, where $k_{\textrm{max}}$ is the maximum resolved scale.}
\end{figure}

However, these energy exchanges in the unforced, inviscid limit are reversible in time and as a result no systematic direction of energy or enstrophy flow can be deduced from these arguments \parencite{Salmon-1998,Tung-Orlando-2003-2DQG}. In irreversible forced--dis\-sipative systems, it can be argued that energy should move to large scales and enstrophy to small scales, as \citeauthor{Fjortoft-1953} envisioned, namely from the scale the energy or enstrophy is being injected towards the scale that each is dissipated. \textcite{Kraichnan-1967} provided such a refinement of \citeauthor{Fjortoft-1953}'s argument, which was further refined by \textcite{Eyink-1996}, by showing that energy and enstrophy conservation imply an inverse energy cascade since if energy and enstrophy are injected at a scale $k_f$ energy must be dissipated at a larger scale $k_r<k_f$ and enstrophy at a smaller scale, $k_v>k_f$.\footnote{\citeauthor{Kraichnan-1967}'s argument: Assume that energy and enstrophy are injected at a scale $k_f$ at rates $\varepsilon_f$ and $\eta_f=k_f^2\,\varepsilon_f$, and that they are being dissipated at two distinct scales: a larger scale $k_r$ (with $k_r<k_f$) at rates $\varepsilon_r$ and $\eta_r$, and a smaller scale $k_v>k_f$ at rates $\varepsilon_v$ and $\eta_v$ and that there is almost no dissipation for $k_r<k<k_v$. At a statistical steady state we expect from conservation of energy and enstrophy that, $\varepsilon_f = \varepsilon_r + \varepsilon_v$ and $\eta_f = \eta_r + \eta_v$, from which we obtain
\be
\frac{\varepsilon_v}{\varepsilon_r} = \frac{1-(k_r/k_f)^2}{(k_v/k_f)^2-1} \ \ ,\ \ \frac{\eta_v}{\eta_r} = \(\frac{k_v}{k_f}\)^2 \(\frac{k_f}{k_r}\)^2\frac{1-(k_r/k_f)^2}{(k_v/k_f)^2-1} \ .\label{eq:Kraichnan_arg}
\ee
% first that if such a steady state is to be maintained, then injection must be done in the intermediate scale, $k_r<k_f<k_v$, since if energy/enstrophy injection was happening at a scale $k_f<k_r<k_v$ then both $\varepsilon_v/\varepsilon_r<0$ and $\eta_v/\eta_r<0$ implying that either in $k_r$ or $k_v$ energy and enstrophy is being injected instead of being dissipated, which is contrary to the initial assumption (contradiction is also achieved if $k_r<k_v<k_f$). Moreover, as the flow approaches the zero viscosity limit then small scale dissipation occurs at even finer (smaller) scales,
Note that because all $\varepsilon$ and $\eta$ are positive such a steady state is possible only if $k_f$ satisfies $k_r < k_f < k_v$, as it was assumed from the start	. Moreover, if $k_v\gg k_f$ then $\varepsilon_v/\varepsilon_r\ll 1$, % and $\eta_v/\eta_r\to (k_f^2/k_a^2)-1$, 
which means that that most of energy flux occurs at large scales $k_r$. Also, if $k_r\ll k_f$ then $\eta_v/\eta_r\ll1$ % and $\varepsilon_a/\varepsilon_r\to (k_a/k_f)^2-1$, 
implying that there is no enstrophy flux at large scale and all the enstrophy moves to small scale $k_v$.} The physical mechanism that decreases the mean scale of the enstrophy of the flow is the stretching of the vorticity as the flow evolves. This is shown in a simulation of forced--dissipative 2D turbulence in Fig.~\ref{fig:NL_forc-diss_beta0}. Vortex filaments form increasing the vorticity gradients and transferring enstrophy to smaller scales and, as has been argued by~\citeauthor{Fjortoft-1953}, energy to larger scales in the form of large vortices.

\begin{figure}
\centering
\begin{minipage}[c]{.95\textwidth}\vspace{0pt}
\includegraphics[height=1.68in]{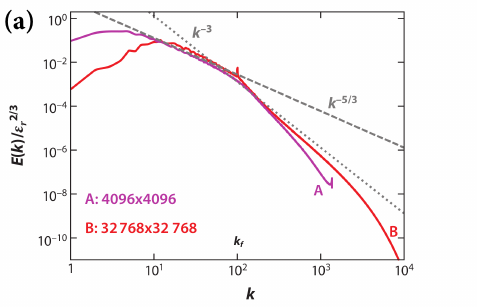}\hspace{1em}
\hspace{1em}\includegraphics[height=1.68in]{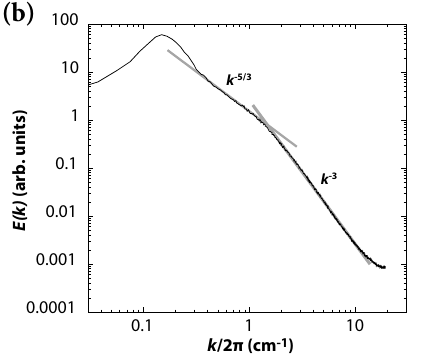}
\end{minipage}\\
%\begin{minipage}[t]{.47\textwidth}\vspace{15pt}
%\hspace{1em}\includegraphics[height=1.68in]{Rutgers-1998-expFig2.pdf}
%\end{minipage}
\begin{minipage}[t]{.47\textwidth}\vspace{0pt}
\includegraphics[width=2.3in]{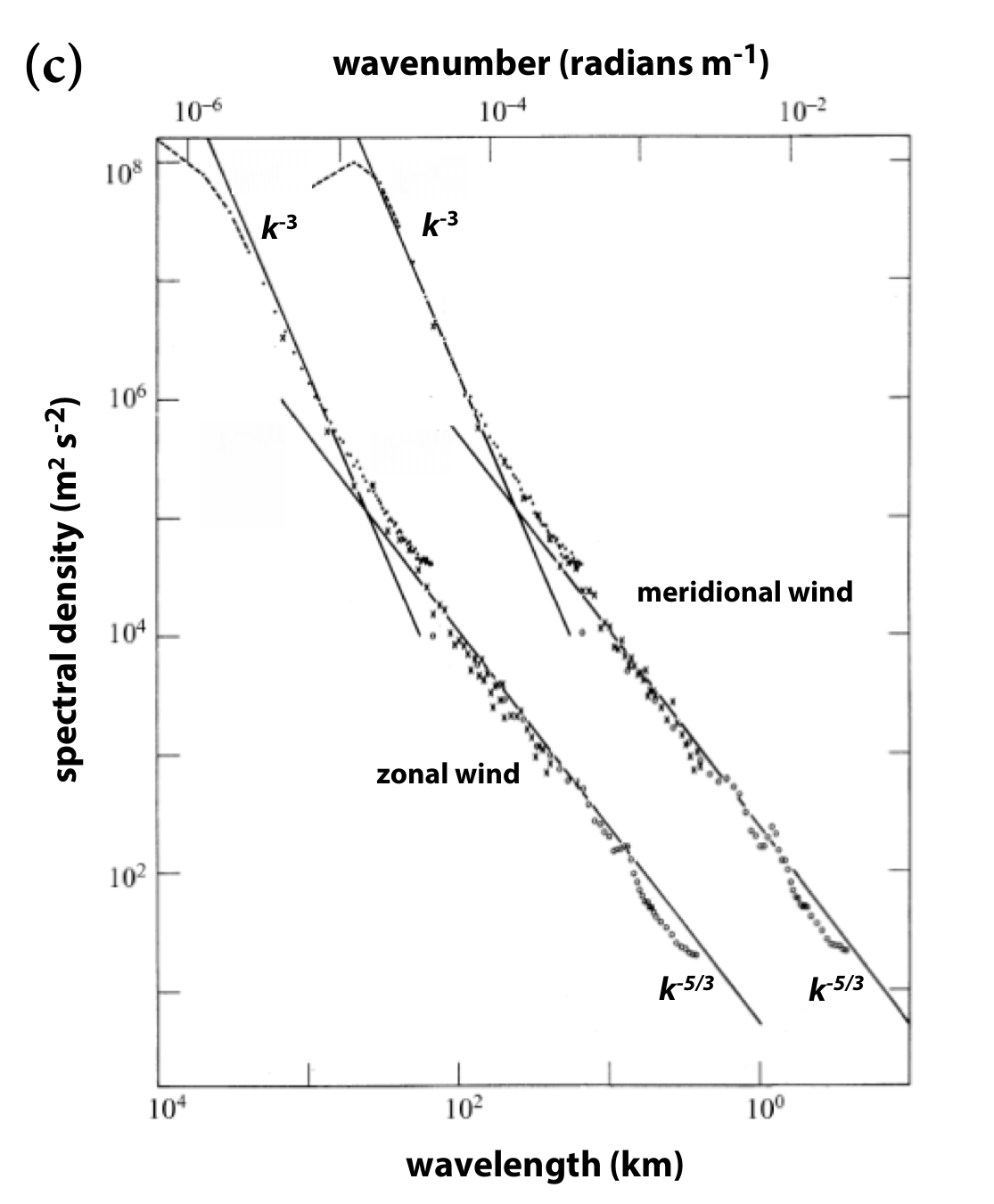}
\end{minipage}%
\caption{\label{fig:2Dcascade} (a) Spectral energy density $E(k)$ vs total wavenumber obtained by numerical simulation of forced--dissipative isotropic two dimensional turbulence, for different resolutions: (A) $4096\times4096$ and (B) $32\,768\times32\,768$. Also shown are the $k^{-5/3}$ (dashed) and $k^{-3}$ (dotted) slopes. (After \textcite{Boffetta-Musacchio-2010}.) %(b) Experimental spectral energy density of a thin layer of electrolyte fluid stochastically forced electromagnetically. Shown here is the eddy kinetic energy spectrum together with the $k^{-5/3}$ slope (after \cite{Xia-etal-2011}).
(b) Experimental spectral energy density at steady state of flowing soap films in which turbulence was excited by grids (arrays of cylinders) lining the walls of the flow channel, showing clear evidence of the dual inverse energy and forward enstrophy cascade. The $k^{-5/3}$ and $k^{-3}$ slopes are also shown. (After \textcite{Rutgers-1998}.) (c) Variance power spectra of winds near the tropopause from commercial aircraft data (by NASA GASP). The spectrum for meridional wind is shifted one decade to the right so it is visible. (After \textcite{Nastrom-Gage-1985}.)}
\end{figure}

\textcite{Kraichnan-1967,Leith-1968,Batchelor-1969} (ofter abbreviated as KBL) suggested that conservation of energy and enstrophy in 2D turbulence results in the formation of two distinct inertial ranges: a range in which energy is transferred upwards to larger scales and a range in which enstrophy is transferred to smaller scales. Using Kolmogorov type non-dimensional arguments, which assume that these ranges are homogeneous, isotropic and self-similar, they showed that the energy density in the energy transferring inertial range should follow the power law $\mathcal{E}(k) = C \varepsilon_r^{2/3} k^{-5/3}$, where $\varepsilon_r$ is the energy transfer rate and $C$ a dimensionless universal constant, while the energy density spectrum in the enstrophy inertial subrange follows the power law $\mathcal{E}(k) = C' \eta_v^{2/3} k^{-3}$, where $\eta_v$ is the enstrophy transfer rate and $C'$ a different dimensionless constant. Evidence of this scalings has been verified in numerical simulations, as shown in Fig.~\ref{fig:2Dcascade}\hyperref[fig:2Dcascade]{a}. Experiments with flowing soap films in which turbulence is excited by grids (arrays of cylinders) lining the walls of the flow channel, provide a physical occurrence of forced--dissipative 2D turbulence and confirm the predictions of KBL for the two inertial ranges, as seen in Fig.~\ref{fig:2Dcascade}\hyperref[fig:2Dcascade]{b}.

All these arguments however, are based on homogeneous and isotropic 2D turbulence. It may be the case that large-scale processes in the atmosphere, including jet formation, are essentially two dimensional, but overall the atmosphere is neither isotropic nor homogeneous. Anisotropy in the atmosphere is induced by the Earth's rotation which distinguishes the zonal from the meridional direction and also by the temperature difference between the Equator and the Pole, while homogeneity is broken by the large-scale jets. Therefore one should use the classical KBL 2D arguments with caution when trying to explain atmospheric motions. Also, the observed atmospheric spectrum of the winds in the upper troposphere, contrary to what the classical 2D picture would expect, shows the $k^{-5/3}$ dependance on the short-wave side of the spectrum at scales ranging from $600\ \textrm{km}$ down to $2\ \textrm{km}$, at the so called ``mesoscales'', as seen in Fig.~\ref{fig:2Dcascade}\hyperref[fig:2Dcascade]{c}. Coincidentally, the direct energy cascade that is typically found at homogeneous isotropic 3D turbulence and is responsible for transferring energy to the smaller scales where is dissipated, also presents a $k^{-5/3}$ dependance. However, the vertical extend of the troposphere limits 3D effects in the atmosphere to appear only at most at scales of $1\textrm{-}2\ \textrm{km}$ and therefore the classical dimensional arguments that are offered to account for the $k^{-5/3}$ inertial range scaling cannot apply to the mesoscales. A concrete explanation for the atmospheric spectrum is an open and challenging problem, which will not be addressed in this thesis. It is interesting to note that if the atmosphere is represented crudely as a two-layer fluid, which is one step of an approximation higher than the one adopted in this thesis, the atmospheric spectrum of \textcite{Nastrom-Gage-1985} is obtained \parencite{Tung-Orlando-2003}.

There is an additional problem with the predictions of the KBL theory in planetary turbulence. While the existence of the inverse energy cascade could provide an explanation for the emergence of the observed large-scale flows in planetary turbulence, the theory predicts that the inverse cascade will lead to the formation of a large-scale condensate, as large as the geometry allows, as shown in Figs.~\ref{fig:NL_forc-diss_beta0}\hyperref[fig:NL_forc-diss_beta0]{b,c}. The structures that emerge in planetary turbulence however are usually not at the largest scale of the flow and moreover they have a very particular structure (see for example Fig.~\ref{fig:JupiterJets}\hyperref[fig:JupiterJets]{c}). As a result if we are to provide a theory for the emergence and maintenance of the large-scale structure in planetary turbulence, we have to go beyond the classical KBL theory and consider the implications of anisotropy and even inhomogeneity in 2D turbulence.

\subsection{Rossby waves}

\begin{figure}
\centering
\includegraphics[width=2.2in]{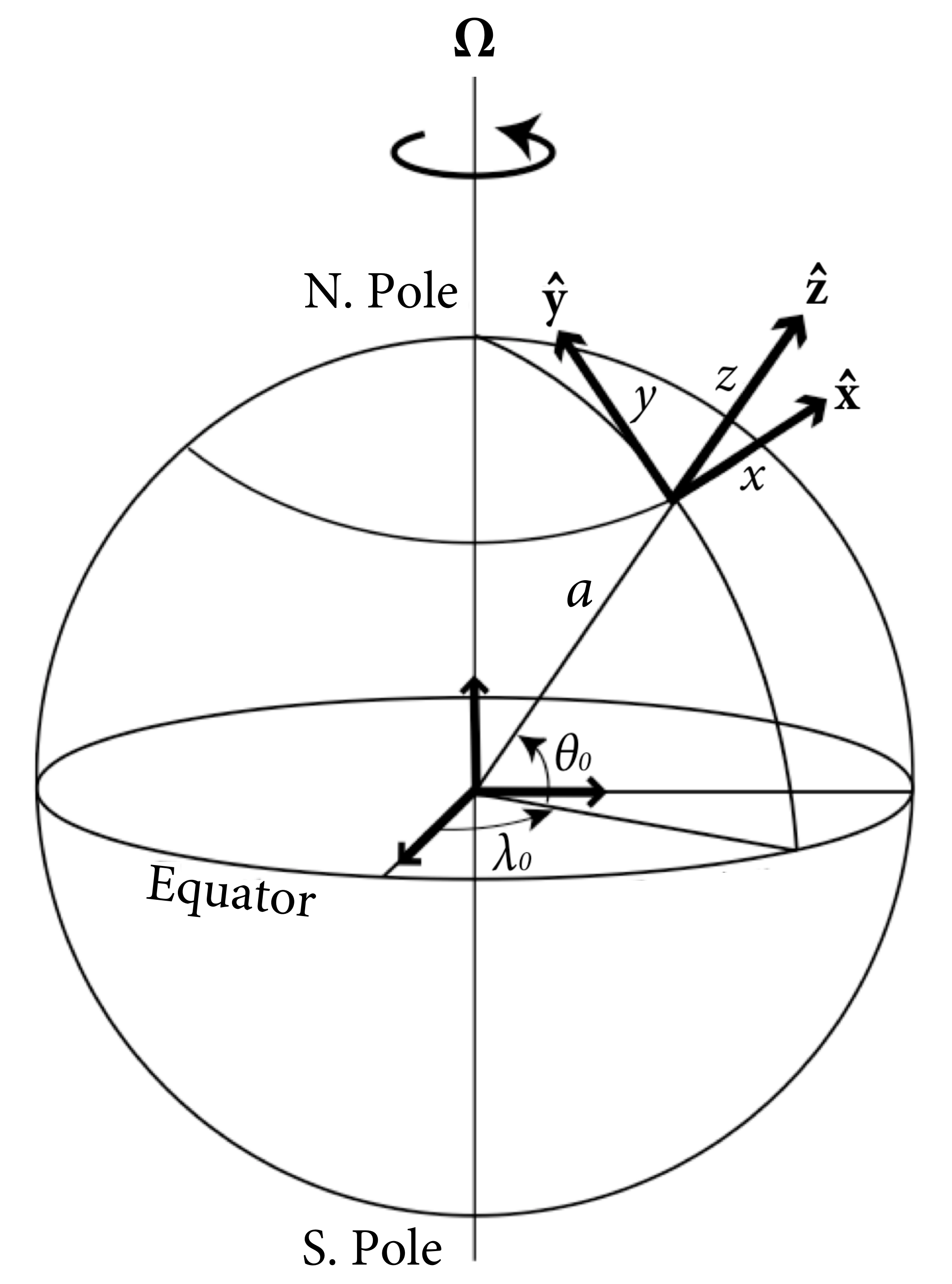}
\caption{\label{fig:bplane} A $\b$-plane at latitude $\theta_0$ approximates a zonal belt of the spherical surface centered at $\theta_0$ with a plane. The coordinate $x$ in that plane corresponds to the zonal direction on the sphere (circles of constant latitude $\theta$) and the direction $y$ is taken to correspond to the meridional direction on the sphere (circles of constant longitude $\lambda$).}
\end{figure}

Flows at rotating planetary atmospheres are anisotropic due to the preferential direction imposed by the rotation. This has important implications to planetary motions because it leads to a new class of exact solutions of the equations of motions that were discovered by Rossby and are referred to as Rossby waves \parencite{Rossby-1939}. Consider a fluid on the nearly spherical surface of a rotating planet at angular velocity $\Omv$, where the magnitude $\Om=|\Omv|$ is the rate of rotation of the planet (for the Earth $\Om=7.27\times10^{-5}\ \textrm{rad}\,\textrm{s}^{-1}=2\pi\ (24\textrm{h})^{-1}$, for Jupiter $\Om=1.76\times10^{-4}\ \textrm{rad}\,\textrm{s}^{-1}\approx 2\pi\ (10\textrm{h})^{-1}$). The velocities of the fluid as measured in an inertial frame of reference and as measured in a frame of reference co-rotating with the planet are related by $\uv_I = \uv_R + \Omv\times\rv$, where subscripts $I$ and $R$ denote quantities measured in the inertial and rotating frame respectively. That the velocity of the flow lies predominantly on the plane tangential to surface of the sphere implies that the vorticity in the inertial frame, $\nablav\times\uv_I \approx \z_I\,\zhat$, is normal to the surface of the sphere ($\zhat$ being the direction normal to the surface of the planet as shown in Fig.~\ref{fig:bplane}). The magnitude of the vorticity, $\z_I$, is equal to $\z_I = \z_R + f$, where $\z_R=(\nablav\times\uv_R)\cdot\zhat$ is the relative vorticity of the fluid and $f\equiv(2\Omv)\cdot\zhat$, is the planetary vorticity, which is also referred to as the Coriolis parameter being the coefficient that multiplies the velocity in the Coriolis force in the equations of motion in a rotating frame. Material conservation of vorticity on this planetary surface implies that $\z_I$ is conserved, i.e.,
\be
\frac{\Df\z_I}{\Df t} = \frac{\Df}{\Df t} (\z_R+f) = \(\bit\partial_t + \uv_R\cdot\nablav\)(\z_R+f)= 0\ ,\label{eq:Dfz_I}
\ee
where $\Df/\Df t\equiv\partial_t + \uv\cdot\nablav$ is the material derivative that determines the time rate of change of flow properties as they move with the fluid, which is frame independent when it acts on scalars. %\footnote{Consider a scalar field, $\phi$. The time rate of change οf $\phi$ moving with the fluid at a fixed point $\rv$ is given as the sum of the change due to variation of $\phi$, $\partial_t\phi$, and the change due to the advection of $\phi$ by the flow field, $\uv_I\cdot\nablav\phi$ and is denoted as $\Df\phi/\Df t\equiv(\partial_t +\uv\cdot\nablav)\phi$. Because the rotational frame moves with velocity $\Omv\times\rv$ relative to the inertial frame, the time rate of $\phi$ is given as $(\partial_t\phi)_I =(\partial_t\phi)_R-(\Omv\times\rv)\cdot\nablav\phi$, while the change due to advection is: $\uv_I\cdot\nablav\phi = (\uv_R+\Omv\times\rv)\cdot\nablav\phi$. Adding the two we get that $(\Df\phi/\Df t)_I=(\Df\phi/\Df t)_R$.}
From hereafter we will drop subscripts $R$ for simplicity and, as it is usually done, all velocities will be considered to be relative to the rotating frame of reference.

The spherical shape of the planet implies that $f =2 \Om \sin\theta$ and consequently $\uv\cdot\nablav f = \b v$, with $v$ the poleward velocity, $\b = (2\Om\cos\theta) /a$, $a$ the radius of the planet and $\theta$ the planet's latitude (cf. Fig.~\ref{fig:bplane}). This $\beta$-term in the equations of motions introduces a new class of anisotropic exact nonlinear solutions, discovered by \textcite{Rossby-1939}, that will be of principal importance in this thesis. Rossby further introduced the $\b$-plane approximation that greatly simplifies the equations of motions. Instead of solving for the barotropic dynamics~\eqref{eq:Dfz_I} on the surface of a sphere we approximate the domain as a planar surface tangent to the surface of the sphere at latitude $\theta_0$ rotating at the constant rate $\Om\sin\theta_0$. The anisotropy due to the sphericity is then simply introduced by keeping only the $\b$-term in the equations of motion with $\b=(2\Om\cos\theta_0) /a$ (at latitude $45\deg$, $\b=1.61\times10^{-11}\ \textrm{m}^{-1}\,\textrm{s}^{-1}$ on the Earth and $\b=3.56\times10^{-12}\ \textrm{m}^{-1}\,\textrm{s}^{-1}$ on Jupiter). Following the seminal work of Rossby and the multitude of theoretical work that followed that employ the $\b$-plane approximation in the study of midlatitude planetary dynamics, we will also adopt in this thesis the $\b$-plane approximation for studying the formation of jets and other large-scale structure in planetary turbulence.

\begin{figure}
\centering
\includegraphics[width=3.5in]{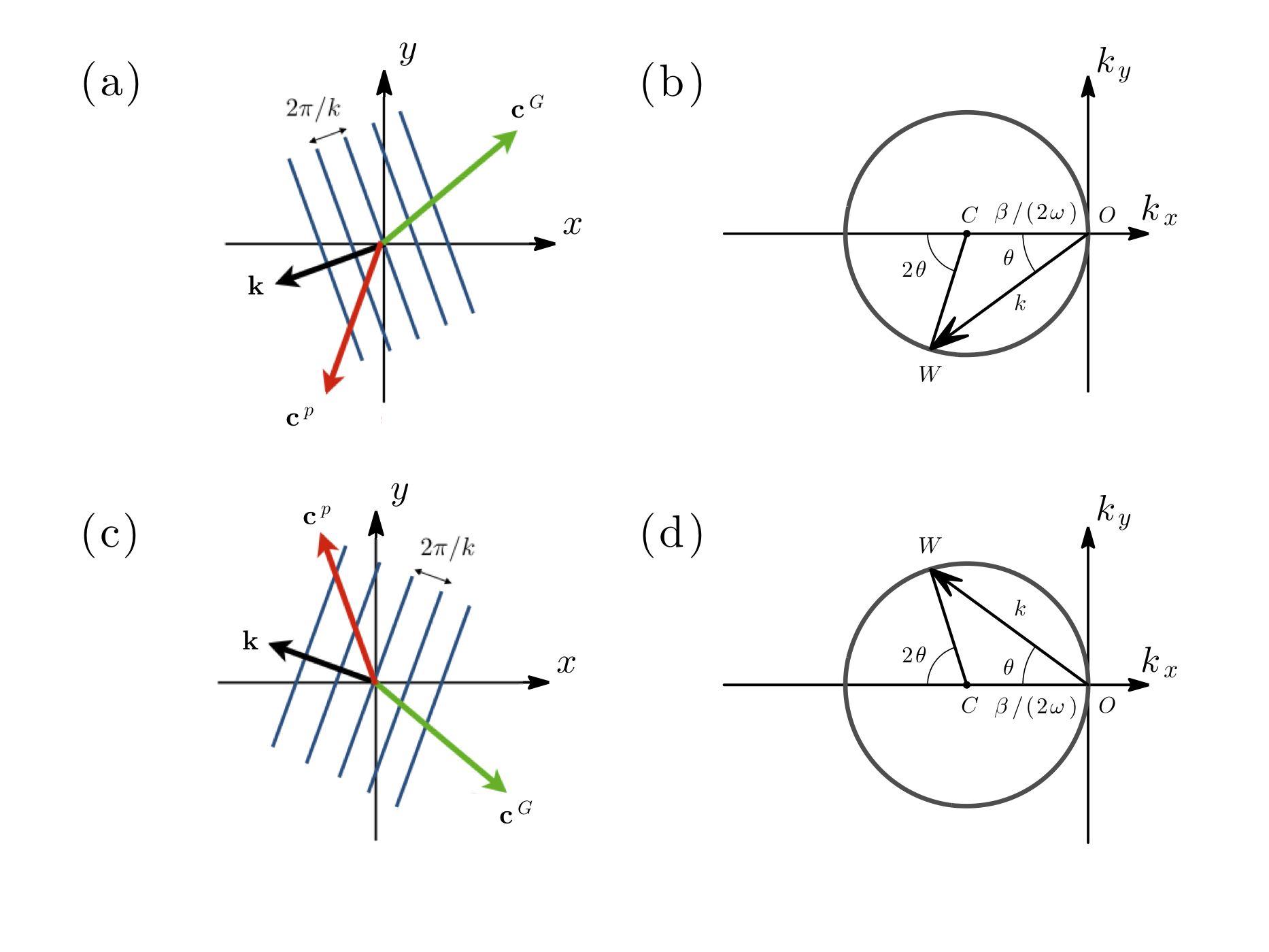}
\caption{\label{fig:LongHigg} (a),(c) Lines of constant phase of a Rossby wave with wavevector $\kv$ subtending an angle $\theta=\arctan(|k_y|/|k_x|)$ with the horizontal. (b),(d) The locus of wavectors $\kv$ that correspond to a fixed value of $\om>0$ is the circle shown with center at $C$ and radius $\b/(2\om)$. Expressing~\eqref{eq:def_omr_cl} as $\om=(\b/k)\cos\theta$ we find that the phase velocity is $\cv^p=(\b/k^2) \,(1,\cot\theta)$, while the group velocity is $\cv^G =\nablav_{\kv}\om=-(\b/k^2)\(\cos(2\theta),\sin(2\theta)\)$, in the direction $\protect \overrightarrow{WC}$. The $y$-component of the group velocity of the Rossby waves is always directed in the opposite direction from the $y$-component of the phase velocity (phase and group velocity directions are drawn in (a) and (c)).}
\end{figure}

We present briefly the basic features of the Rossby wave solutions which will be a central to this thesis. Expressing the two dimensional velocity in terms of a streamfunction as $\uv=\zhat\times\nablav\psi$, the barotropic vorticity equation \eqref{eq:Dfz_I} takes the form
\be
\partial_t \Del\psi + J\(\psi,\Del\psi\) + \b\partial_x\psi= 0\ ,\label{eq:Dfz_I_psi}
\ee
where $J(g,h)\equiv (\partial_x g)(\partial_y h)-(\partial_y g)(\partial_x h)$ is the Jacobian of functions $g$ and $h$ and $\Del g=(\partial^2_{xx}+\partial^2_{yy})g$ the horizontal Laplacian. Equation~\eqref{eq:Dfz_I_psi} reveals that the $\b$-term supports the wave solutions $\psi=A\,e^{\i(\kv\cdot\xv-\om t)}$, with wavevector $\kv=(k_x,k_y)$ and the frequency satisfying the anisotropic dispersion relation
\be
\om(\kv) = -\frac{\b k_x}{k_x^2+k_y^2} \ ,\label{eq:def_omr_cl}
\ee
which is described geometrically in the manner of \textcite{Longuet-Higgins-1964} in Fig.~\ref{fig:LongHigg}. Being aniso\-tropic, wavevectors $\kv$ that correspond to frequency $\om$ do not lie on a circle centered at the origin. Figure~\ref{fig:LongHigg} demonstrates the important property that Rossby wave packets have $y$-phase velocities opposite to their
$y$-group velocities.\footnote{For example in Fig.~\ref{fig:jet_local_homog}\hyperref[fig:jet_local_homog]{a}, which will be discussed later, the phase lines in the region $y>3.6$ are such so that the group velocity is directed towards the center of the channel. Also the phase lines in the region $y<2.6$ are configured so that group velocity is also directed towards the center.}

%Moreover, calculating the group velocity $\cv^G = \nablav_{\kv}\om$ we find that its meridional component, $c^G_y$ is always in the opposite direction of the meridional phase velocity, $\om/k_y$.

Remarkably, because these monochromatic Rossby waves satisfy $J(\psi, \Delta \psi) = 0$, they are also nonlinear solutions of~\eqref{eq:Dfz_I_psi}. Moreover, stationary Rossby waves with $k_x=0$ correspond to sinusoidal zonal jets with streamfunction $\psi=A\,e^{\i k_y y}$ or more generally any zonal flow with $\psi = \int A(k_y) e^{i k_y y}\,\df k_y$ is also nonlinear solution of~\eqref{eq:Dfz_I_psi} (other nonlinearly valid Rossby wave solutions are presented in Appendix~\ref{app:MI}). %\footnote{All streamfunctions of the form $\psi=g(y)$, with $\uv=-(\partial_y g) \hat{\mathbf{x}}$, satisfy $J(\psi,\Del\psi)=0$ implying that jets are both exact nonlinear solutions of the inviscid equation~\eqref{eq:Dfz_I_psi} and also Rossby waves with $k_x=0$.}
We will demonstrate in this thesis that the emergence of large-scale features in turbulent flow in the form of jets and other Rossby waves (or zonons) can be traced to the property that exact nonlinear solutions can serve as good repositories into which the eddy energy may ``condensate''.

\vspace{5em}

\subsection{Anisotropic turbulence on a beta-plane}

\begin{figure}
\centering
\includegraphics[height=2.5in]{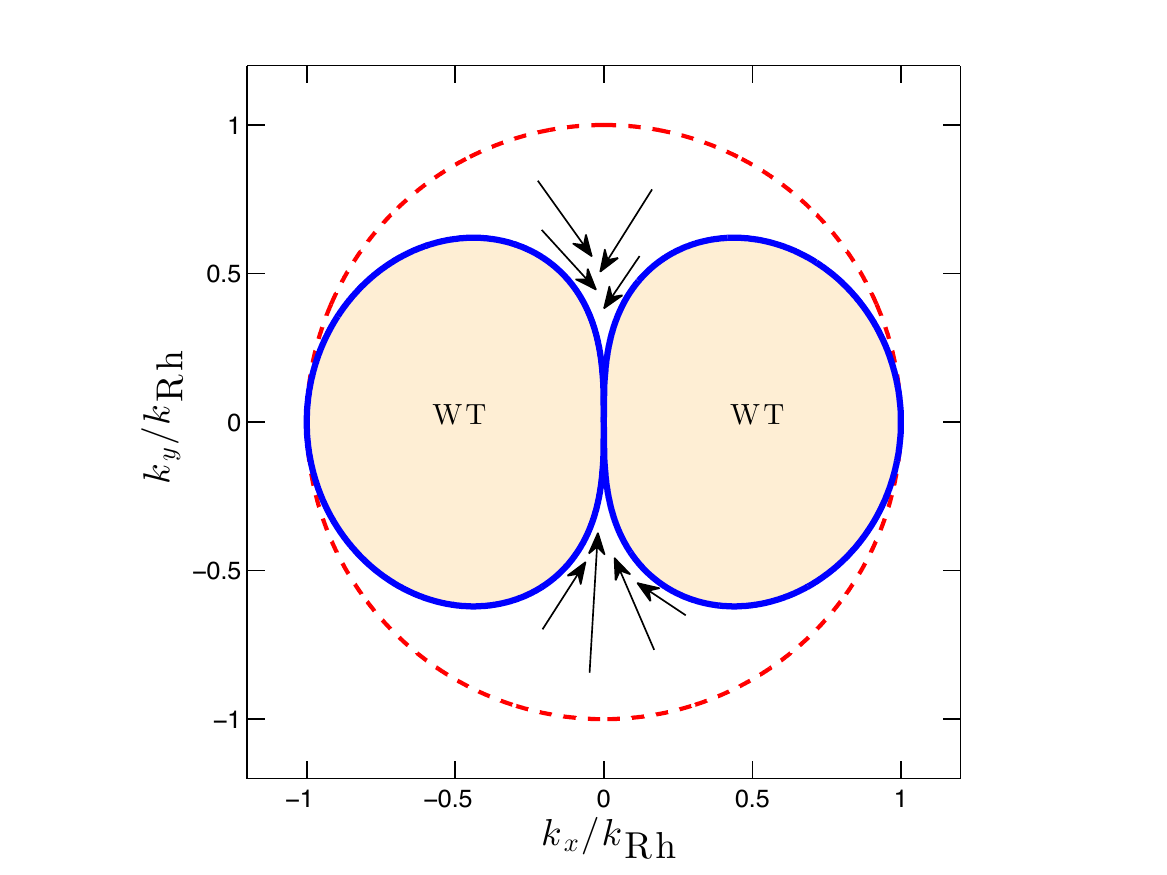}
\caption{\label{fig:vallis_dumbbell} The interior of the dumbbell or lazy eight (shaded) is the region in which Rossby waves are prevalent. According to~\textcite{Rhines-1975} the excitation of the Rossby waves halts and anisotropizes the cascade, energy proceeds as shown towards the $k_x=0$ axis.}
\end{figure}

The presence of Rossby waves affects the structure of turbulence in the degree that the $\b$-term dominates the advection term, $J(\psi, \Delta \psi)$, in the vorticity equation~\eqref{eq:Dfz_I_psi}. The ratio of these terms is $|\b\, \partial_x \psi| \big/ |J(\psi, \Delta \psi)| =\omega_{\textrm{Rossby}}/\omega_{\textrm{turb}}= \Ocal\(\bit \b / (k^2\mathcal{U}_{\textrm{rms}} )\)$, where $\omega_{\textrm{Rossby}}= \Ocal(\b/k)$ is the Rossby wave frequency and $\omega_{\textrm{turb}} = k\, \mathcal{U}_{\textrm{rms}}$ is the inverse of the shear time associated with nonlinear advection, with $\,\mathcal{U}_{\textrm{rms}}$ the root-mean-square velocity of the flow at scale. It is reasonable to expect that when $\omega_{\textrm{Rossby}} \ll \omega_{\textrm{turb}}$ the $\b$-term barely affects the turbulent dynamics. Since $\omega_{\textrm{Rossby}}$ increases while $\omega_{\textrm{turb}}$ decreases as $k$ decreases, we expect only the large scales to be influenced by $\b$. The wavenumber $k_{\textrm{Rh}}=\sqrt{\b/\mathcal{U}_{\textrm{rms}}}$ at which the shear time scale is equal to the Rossby wave period separates, according to Rhines, the wavenumber space in two regions: a region of wave turbulence in which coherent Rossby wave motions are manifest in the flow and nonlinearly interact as waves and a region in which wave motions are not discernible, the flow is considered incoherent and the nonlinear interactions are no longer constrained to be among waves. The scale that separates these two regions is being referred to as the ``Rhines's scale'' and the locus of the wavenumbers that satisfy the requirement $\om_{\textrm{Rossby}}=\om_{\textrm{turb}}$ is the popular and iconic dumbbell shape of \textcite{Vallis-Maltrud-93}, shown in Fig.~\ref{fig:vallis_dumbbell}, and which will be  encountered in this thesis under a different exegesis. \textcite{Rhines-1975} argued that for scales larger than the Rhines's scale ($k<k_{\textrm{Rh}}$) the selectivity imposed to the nonlinear interactions by wave turbulence, which allows interactions only among waves, retards the inverse energy cascade and anisotropizes it. \citeauthor{Rhines-1975} in this way explained that in anisotropic $\b$-plane turbulence the inverse cascade should not be expected to proceed to the largest scale available and is halted at $k_{\textrm{Rh}}$. This led to the general prediction that the large-scale structure in $\b$-plane turbulence should have a characteristic length scale of the order of $1/k_{\textrm{Rh}}$.

\begin{figure}
\centering
\includegraphics[width=4in]{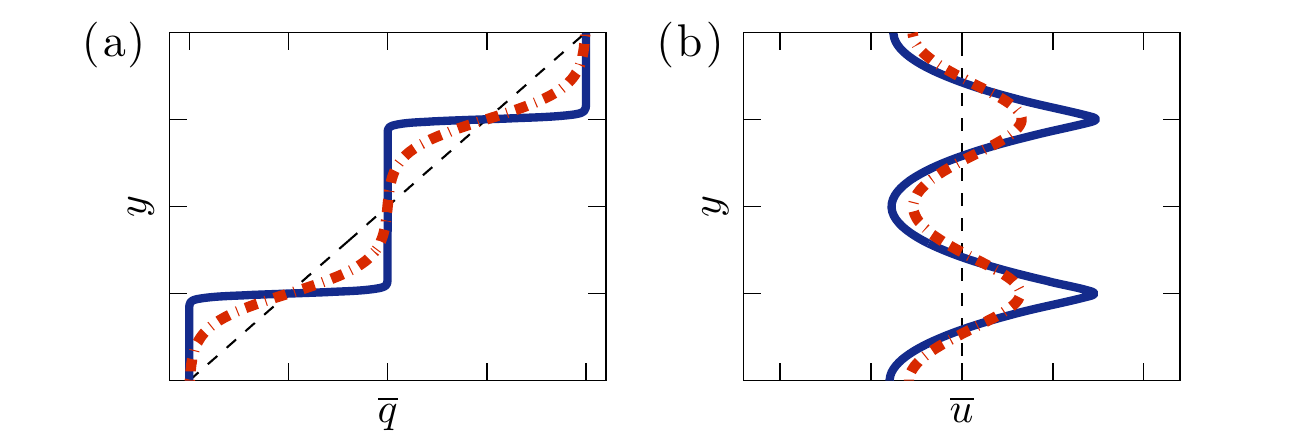}
\caption{\label{fig:staircases} Jet formation through homogenization of PV, $q=\z+f$. Shown are (a) zonal mean PV profiles, $\overline{q}=f_0+\b y-\df \overline{u}/\df y$, and (b) the corresponding zonal mean flows, $\overline{u}$. A linear PV distribution (dashed) is associated with no zonal mean flow. Westward jets are associated with regions of homogenized PV. These jets become stronger as the homogenization increases (compare the jets associated with the dashed-dot PV partially homogenized distribution with the staircase profile). Note that PV mixing theory implies that PV gradient should be everywhere non-negative.}
%\end{figure}
%
\vspace{1em}
%
%\begin{figure}
\centering
\includegraphics[height=2.5in,trim = 0mm 0mm 1mm 0mm, clip]{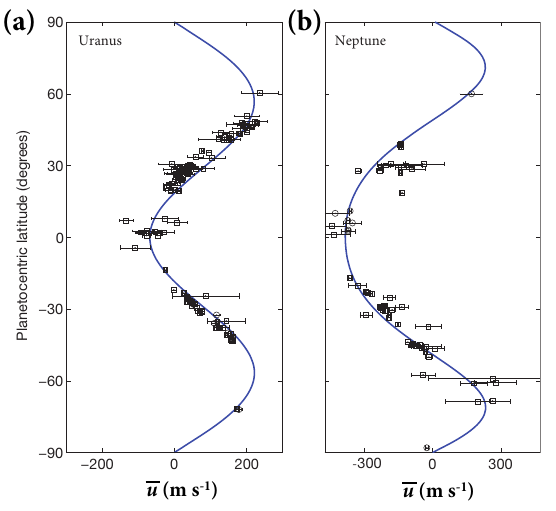}
\includegraphics[height=2.55in]{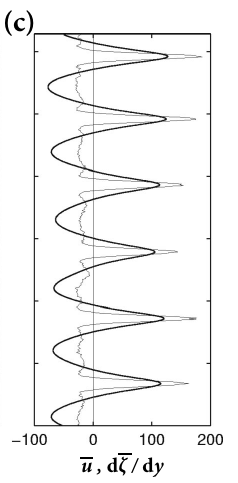}
\caption{\label{fig:neptune_uranus_danilov} (a),(b): Observed zonally averaged zonal winds on (a) Uranus and (b) Neptune at cloud level. Data taken from \emph{Voyager 2} (circles) and from \emph{Hubble Space Telescope} measurements (squares). Solid lines are empirical fits on the data. The equatorial retrograde jets in both planets (up to approximately 25$\deg$N/S for Uranus and 35$\deg$N/S for Neptune) can be approximated very well by a parabola. (Taken from~\textcite{Kaspi-etal-2013}.) (c) Numerical simulation of forced--dissipative turbulence on a $\b$-plane with large-scale dissipation in the form of second order hypofriction, $-\kappa\,\Del^{-2}$. Shown are the zonal mean vorticity gradient, $\df\overline{\z}/\df y$, (thin line) averaged over 100 time units after the simulation has reached steady state and also zonal velocity, $\overline{u}$, (thick line) multiplied with 100 to fit the same axes. For the simulation a pseudospectal code at $512\times512$ resolution is used and turbulence is maintained against dissipation by energy injection in the form of isotropic excitation at wavenumber $k_f$. The parameters of the simulation are: $\b k_f/\kappa=49$ and $\varepsilon_f\, k_f^8 /\kappa^3=7.7\times10^4$. (Taken from~\textcite{Danilov-04}.) }
\end{figure}

Still unanswered remains how zonal jets finally emerge. Broadly speaking three different approaches attempt to answer this question within Rhines's phenomenology. \textcite{Rhines-1975,Vallis-Maltrud-93,Chekhlov-etal-1996,Smith-Waleffe-1999} argue that the cascade proceeds through local interactions up to the dumbbell where the cascade process is halted and the upscale flow of energy is directed to move tangentially along the dumbbell (in the direction of the arrows in Fig.~\ref{fig:vallis_dumbbell}) towards the bottleneck at $k_x=0$, thereby forming jets. McIntyre, Dritschel, Scott and collaborators \parencite{Baldwin-etal-2007,Dritschel-McIntyre-2008,Dritschel-Scott-2011,Scott-Dritschel-2012} argue that jet formation is the inevitable and universal result of irreversible mixing of the potential vorticity (PV) of the flow, $q = \z+f_0+\b y$, that occurs in turbulent flows, which tends to wipe out the large-scale PV gradients producing a flow that approximately satisfies $\nablav q = 0$ or the large-scale flow satisfies $\partial^2_{xx} v = \partial^2_{xy} u$ and $\partial^2_{yy} u- \partial^2_{xy}v = \beta$. They further postulate that because the zonal direction, $x$, is a homogeneous direction the resulting well mixed large-scale flows must be independent of $x$ and consequently irreversible mixing in the presence of $\b$ produces only mean zonal jets $\overline{u}(y)$ at large scale with a parabolic profile satisfying $\df^2 \overline{u} / \df y^2 = \b$ (as for example the central jet of Fig.~\ref{fig:staircases}\hyperref[fig:staircases]{b}). Their argument has however a further twist: there is an important feedback between Rossby waves and turbulence that occurs in the boundary of the dumbbell. When waves are strongly present the turbulent mixing is inhibited, while when the turbulence is strongly nonlinear turbulent interactions iron out the PV gradient. In the presence of a zonal jet, the effective Rossby restoring force (``Rossby elasticity'') is not $\b$ but rather,
\be
%\b_{\textrm{eff}}=\df\overline{q}/\df y=\b-\df^2\overline{u}/\df y^2\ .\label{eq:beta_eff}
\b_{\textrm{eff}}=\frac{\df \overline{q}}{\df y}=\b-\frac{\df^2 \overline{u}}{\df y^2}\ ,\label{eq:beta_eff}
\ee
which means that in regions of eastward jet maxima Rossby wave excitation is reinforced since $\b_{\textrm{eff}}>\b$ and $\om_{\textrm{Rossby}}$ increases inhibiting mixing, while at westward jet maxima the excitation of Rossby waves is not comparatively favored, since $\b_{\textrm{eff}}<\b$, and this leads to increased mixing of PV that reduces further the PV gradient. These two effects result in inhomogeneous PV mixing in the anisotropic $y$ direction producing a staircase PV profile, as in Fig.~\ref{fig:staircases}\hyperref[fig:staircases]{a}, with regions in which the PV gradients have been severely reduced and the flow is retrograde and parabolic and regions in which the potential vorticity gradient is very large with very sharp prograde jets, as shown in Fig.~\ref{fig:staircases}\hyperref[fig:staircases]{b}. It is remarked by McIntyre that the same interaction mechanism was proposed by \textcite{Phillips-1972,Phillips-77} in order to account for the widespread occurrence of a succession of layers of uniform stratification in the stratified ocean.

\begin{figure}
\centering
\includegraphics[height=1.8in,trim = 14mm 0mm 18mm 0mm, clip]{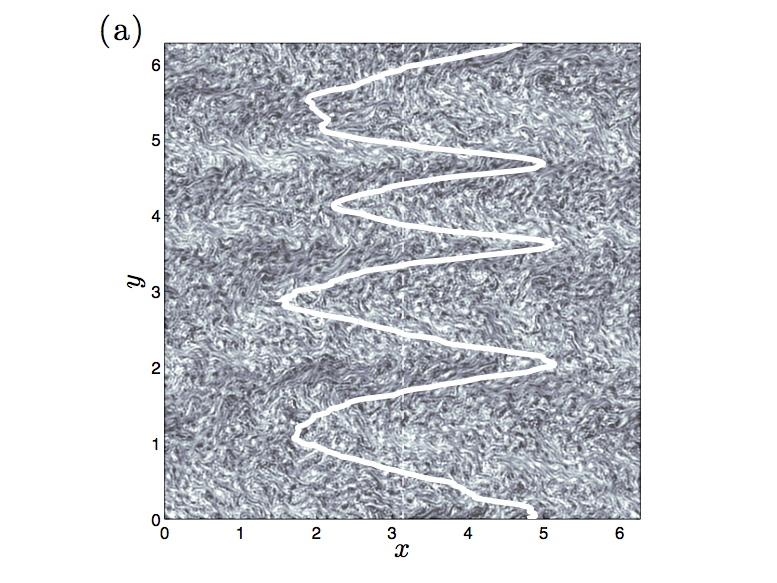}
\includegraphics[height=1.8in]{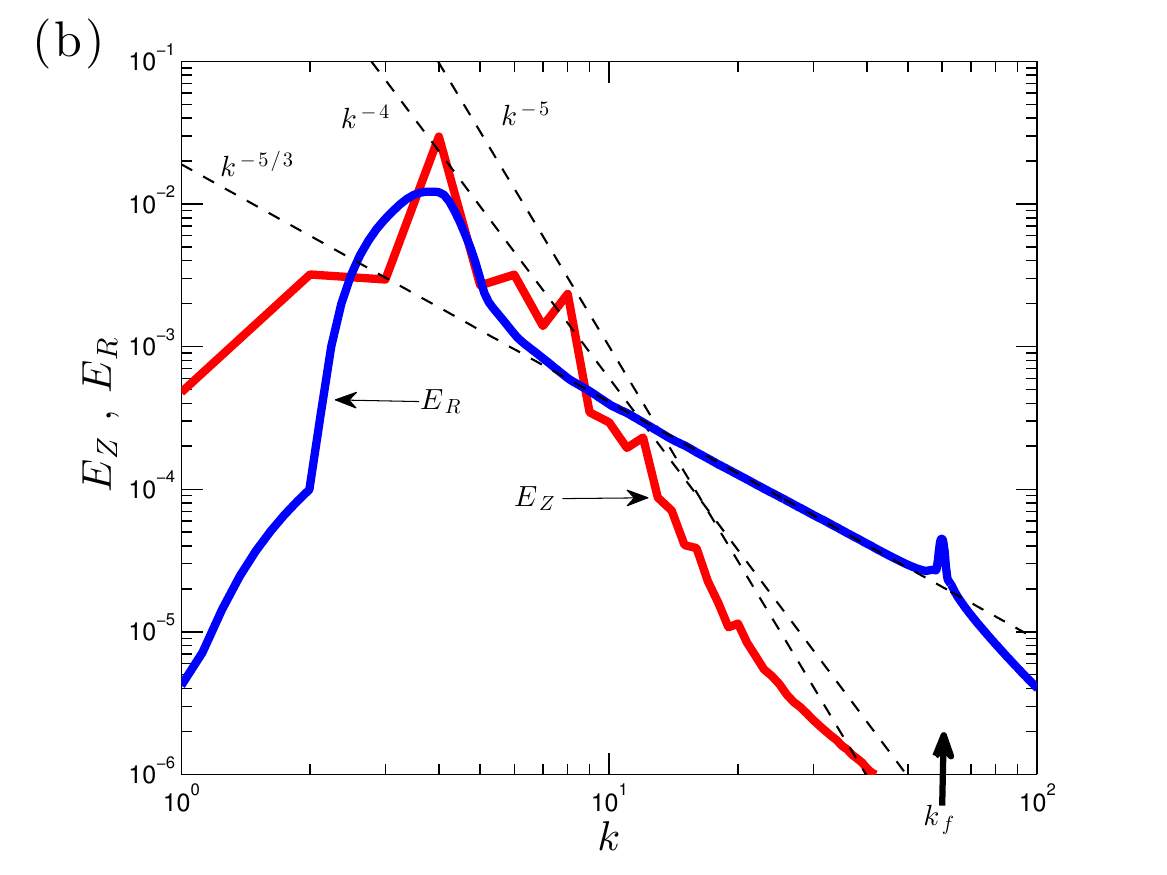}
\caption{\label{fig:nikos_NL_EQL} (a) A snapshot of the vorticity field of forced--dissipative turbulence on a $\b$-plane with the associated instantaneous zonal mean zonal flow, $\overline{u}$, also indicated. The zonal mean flow contains 45\% of the total energy of the flow. (b) The energy spectrum of the zonal flow, $E_Z$ (red), and the spectrum of the remaining energy, $E_R$ (blue). The turbulence is forced at the scale $k_f$ and $E_R$ develops the universal $k^{-5/3}$ spectrum while the large scales develop a sharper $k^{-4}$-$k^{-5}$ spectrum as expected from the near discontinuity of the prograde jets shown in (a). A pseudospectal code was used at $512\times512$ resolution with $\b/(k_f r)=42$ and $\varepsilon_f\, k_f^2 /r^3=2.3\times10^7$, with $r$ the coefficient of linear damping. Also plotted are the $k^{-4}$, $k^{-5}$ and $k^{-5/3}$ slopes (dashed). (Taken from~\textcite{Bakas-Ioannou-2015-book}.)}
\end{figure}

The observation that the maintained jets in planetary $\beta$-plane turbulence are prograde jets joined with parabolic wind profiles is astute. It gives the shape of the $24\deg$N jet on Jupiter, shown in Fig.~\ref{fig:JupiterJets}\hyperref[fig:JupiterJets]{c}, and of the equatorial retrograde jets on Neptune and Uranus (see Fig.~\ref{fig:neptune_uranus_danilov}\hyperref[fig:neptune_uranus_danilov]{a,b}). In numerical steady state simulations of an almost inviscid turbulent flow on a doubly periodic $\b$-plane channel \textcite{Danilov-04} were able to produce the mean flow shown in Fig.~\ref{fig:neptune_uranus_danilov}\hyperref[fig:neptune_uranus_danilov]{c}, which almost exactly conforms to the above specification. With this successful prediction it is very tempting to cease further effort and accept that the phenomenon of jet formation has in essence been resolved by the above inhomogeneous mixing arguments. However, the above arguments are phenomenological, qualitative, descriptive and they do not comprise a deductive theory that proceeds from the equations of motion. In this thesis we will present a predictive and quantitative theory that proceeds directly from the equations of motion that can account for the observations. Moreover, this theory leads to different conclusions about the role of the physical processes that lead to jet formation. For example, it will be shown that the tendency for jet emergence occurs even in the absence of $\b$ (in fact $\b$ may actually retard the tendency for the emergence of jets) and that $\beta$ is required in order to obtain steady state and hydrodynamically stable equilibrated flows, which requires that the PV gradient, $\df \overline{q}/\df y=\beta - \df^2 \overline{u} / \df y^2$, be of one sign (here positive) and as result in the retrograde parts of the flow, where $\df^2 \overline{u} / \df y^2 >0$, the equilibrated flow must become parabolic satisfying $\beta = \df^2 \overline{u} / \df y^2$, while the prograde jets can become infinitely sharp with no constraint other than mean diffusive dissipation. The tendency towards discontinuity of the derivative of the mean flow implies that the turbulent spectra at large scales have a $k^{-4}$ power law behavior, which is not far from the observed $k^{-4}$-$k^{-5}$ spectrum, shown in Fig.~\ref{fig:nikos_NL_EQL}. Note also that the demand that the mean flow does not violate the Rayleigh-Kuo criterion for barotropic instability also sets the scale of the jets to be nothing else but the Rhines's scale ($k_{\textrm{Rh}}=\sqrt{\b/U}$) given that the maximum of the mean flow velocity $U$, rather than r.m.s.~turbulent velocity, must at most satisfy the condition $k^2 U = \b$. More basically, this scale should be expected to emerge irrespectively of mechanism because it is the only length scale that can be formed from the mean flow $U$ (units $\textrm{L}\,\textrm{T}^{-1}$) and $\b$ (units $\textrm{L}^{-1}\,\textrm{T}^{-1}$).

For scales within the dumbbell the $\b$-term dominates over the advection term and the flow is well approximated by a sea of weakly interacting Rossby waves. By transforming~\eqref{eq:Dfz_I_psi} in Fourier space it can be shown that the nonlinear Jacobian is transformed into a convolution with the property that Fourier component $\pv$ of the flow interacts with Fourier component $\qv$ to produce Fourier component $\kv$ only if the wavevectors  form a triangle, i.e., satisfy $\pv+\qv = \kv$. In the wave regime however, the interacting Fourier components must produce a wave motion and significant response is obtained only when the wavenumbers satisfy additionally the resonant condition\footnote{Consider a streamfunction $\psi=\psi_\pv + \psi_\qv$, which is a sum of two monochromatic Rossby waves $\psi_\pv=A\,e^{\i[\pv\cdot\xv-\om(\pv) t]}$ and  $\psi_\qv=A\,e^{\i[\qv\cdot\xv-\om(\qv) t]}$. The advection term in \eqref{eq:Dfz_I_psi} is then given as:
\bdm
J(\psi,\Del\psi) = -A^2 (p^2-q^2) (\pv\times\qv)\cdot\zhat\,e^{\i\left\{ (\pv+\qv)\cdot\xv - [\om(\pv) +\om(\qv) ]t\right\}\ .} 
\edm
In the special case for which $\om(\pv) +\om(\qv) =\om(\pv+\qv)$ the r.h.s. is proportional to a third Rossby wave $\psi_{\pv+\qv}$ which can then resonantly grow.}:
\be
\om(\kv) = \om(\pv) + \om(\qv)\ .\label{eq:resonace}
\ee
The turbulence that results  from these  resonant interactions among waves is referred to as ``weakly nonlinear turbulence'' or ``wave turbulence'' (WT) \parencite{Zakharov-1965,Zakharov-1992,Hasselmann-1966,Hasselmann-1967}. Balk discovered that Rossby waves in the WT regime do not only conserve  energy and enstrophy  in the inviscid limit, but also new independent invariant. Balk with Zakharov and Nazarenko have demonstrated that this additional invariant, which they named ``zonostrophy'', is responsible for the anisotropization of the cascades and leads to the emergence of zonal jets \parencite{Balk-1991,Balk-etal-1991,Nazarenko-09,Balk-etal-2009}.

Finally, a  theory  of very different character has been advanced for the emergence of zonal jets which is based on the property  that  basic flows consisting of  infinitely coherent monochromatic Rossby waves are hydrodynamically unstable to zonal jets \parencite{Lorenz-1972,Gill-1974}. This instability is called a ``modulational instability'' (MI) because of its similarity with the Benjamin-Feir instability of surface gravity waves~\parencite{Benjamin-1967,Benjamin-Feir-1967,Yuen-Lake-1980,Zakharov-Ostrovsky-2009} and has recently resurfaced in relation to zonal jet formation in planetary turbulence and also in drift-wave turbulence in plasmas, both of which are governed by the Charney-Hasegawa-Mima equation, which is formally equivalent to the barotropic vorticity equation with finite radius of deformation \parencite{Connaughton-etal-2010}. The MI theory for the emergence of jets departs considerably from the cascade theories of jet formation. It does not require that jets emerge through a sequence of local interactions in wavenumber space transferring energy upscale, but rather jet emerge from the instability of the primary Rossby wave to a zonal jet perturbation, an interaction that involves a non-local interaction in wavenumber space, i.e., the primary Rossby wave with wavenumber $\kv=(k_x,k_y)$ gives energy to the spectrally removed unstable zonal jet $\pv=(0,p_y)$, which is nothing else but a zero frequency Rossby wave. In this thesis we will investigate the relation of the MI theory for the emergence of jets with the statistical theory that will be studied in this thesis. We will show that MI is a special case of the more general instability that occurs in the theory we will present (cf. chapter~\ref{ch:MI}).

\subsection{Statistical approaches for large-scale structure formation}

Turbulent flows involve enormous complexity and a large number of degrees of freedom so it is tempting to describe turbulent flows by statistical methods reducing in this way its complexity, in a similar way thermodynamics dramatically reduce the complexity of the gas molecules movements in a box while still fully describe the gas macrostate. However, turbulent flows are usually far from equilibrium and the application of equilibrium statistical mechanics might not be possible. Interestingly, due to the enstrophy conservation in 2D flows an equilibrium statistical mechanical description of the flow in the inviscid limit is feasible.\footnote{In 3D flows, there can be energy dissipation due to vortex stretching even in the inviscid limit, a phenomenon called ``anomalous dissipation'' \parencite{Onsager-1949,Kaneda-etal-2003}, not allowing 3D flows to reach equilibrium state.} The statistical mechanics of a set of inviscid point vortices in 2D flow goes back to \textcite{Onsager-1949} (cf.~\textcite{Eyink-Sreenivasan-2006}) and the first statistical mechanical formulation of 2D unforced inviscid flows was developed by \textcite{Miller-1990} and \textcite{Robert-Sommeria-1991} (the MSR theory).\footnote{For a historic review of statistical mechanical methods in turbulence refer to~\textcite{Bouchet-Venaille-2014-book}.} %Since in that limit the vorticity evolves into finer and finer structures (cf.~Fig.~\ref{fig:NL_forc-diss_beta0}\hyperref[fig:NL_forc-diss_beta0]{b}), MSR described this small-scale vorticity structures statistically. 
Their theory postulates that the emergent equilibrium structures will be the ones that maximize entropy while conserving energy, enstrophy and all the hierarchy of invariants in 2D. \textcite{Bouchet-Sommeria-2002} extended the MSR theory to  quasi-geostrophic barotropic turbulence and showed that the most probable structures are zonal jets or large-scale vortices (for a review see \textcite{Bouchet-Venaile-2012}).

However, planetary flows are both strongly forced and dissipated and therefore out of equilibrium. A non-equilibrium statistical approach is more suitable  for the description of the statistical dynamics of the turbulent state. Such a non-equilibrium statistical mechanical  theory has been advanced by~\textcite{Farrell-Ioannou-2003-structural} and will be discussed in this thesis.

%. The equilibrium statistical mechanics applicability is expected to be limited. The advance of non-equilibrium statistical methods seems therefore more adequate, and this method will be followed in this thesis. Surprisingly, it turns out that inhomogeneous turbulence, as is turbulent flows in the presence of jets or large-scale vortices, is much more amenable to statistical analysis compared to homogeneous turbulence. In this thesis we will develop a non-equilibrium statistical theory for the description of a turbulent flow which will be based on the dynamical interaction of waves or eddies with the large-scale mean flows.

\subsection{Wave--mean flow interaction theories}

It has been known for a long time that waves in a material medium can interact with the medium and form mean flows. The acoustic streaming experiments of Rayleigh demonstrate this phenomenon (cf. \textcite{Rayleigh-1896,Lighthill-1978}). In acoustic streaming strong jet-like winds are generated by powerful ultrasound sources. Acoustic streaming results from the divergence of Reynolds stresses induced by the acoustic waves as they dissipate. As with acoustic streaming, jets in planetary atmospheres can emerge from Rossby wave streaming in the presence of dissipation. This is the basis of the wave--mean flow interaction theories for the emergence of jets in the atmospheres. For this mechanism to work the region of excitation of the waves and the region of dissipation of the waves should be separated. In this case prograde flows emerge in the excitation region while retrograde flows emerge in the regions of dissipation.\footnote{Because of momentum conservation the integrated mean flow acceleration induced by the waves integrates to zero.} In the Earth the source of the equivalent barotropic planetary waves in the upper troposphere, following \textcite{Kuo-1951,Hoskins-1983,Held-Hoskins-1985} and as discussed earlier, is the baroclinic activity in the lower troposphere. The equivalent barotropic Rossby waves are radiated way to the North and South of the source region where they eventually dissipate maintaining the upper-level polar jets (for a model of this see \textcite{DelSole-01a}). The presence of dissipation the emergence of large-scale mean flows at steady state since in this case the wave--mean flow non-interaction theorem \parencite{Eliassen-Palm-1961,Charney-Drazin-1961,Andrews-McIntyre-1976-I,Boyd-1976-thesis,Boyd-1976} does not hold. (The dissipation of the flow is mainly due to Ekman spin-down and additionally to breaking of the waves at the ``equatorial surf zone'' at the critical layers in the Equator-wards flank of the midlatitude jet.)

\begin{figure}
\centering
\includegraphics[width=3.2in]{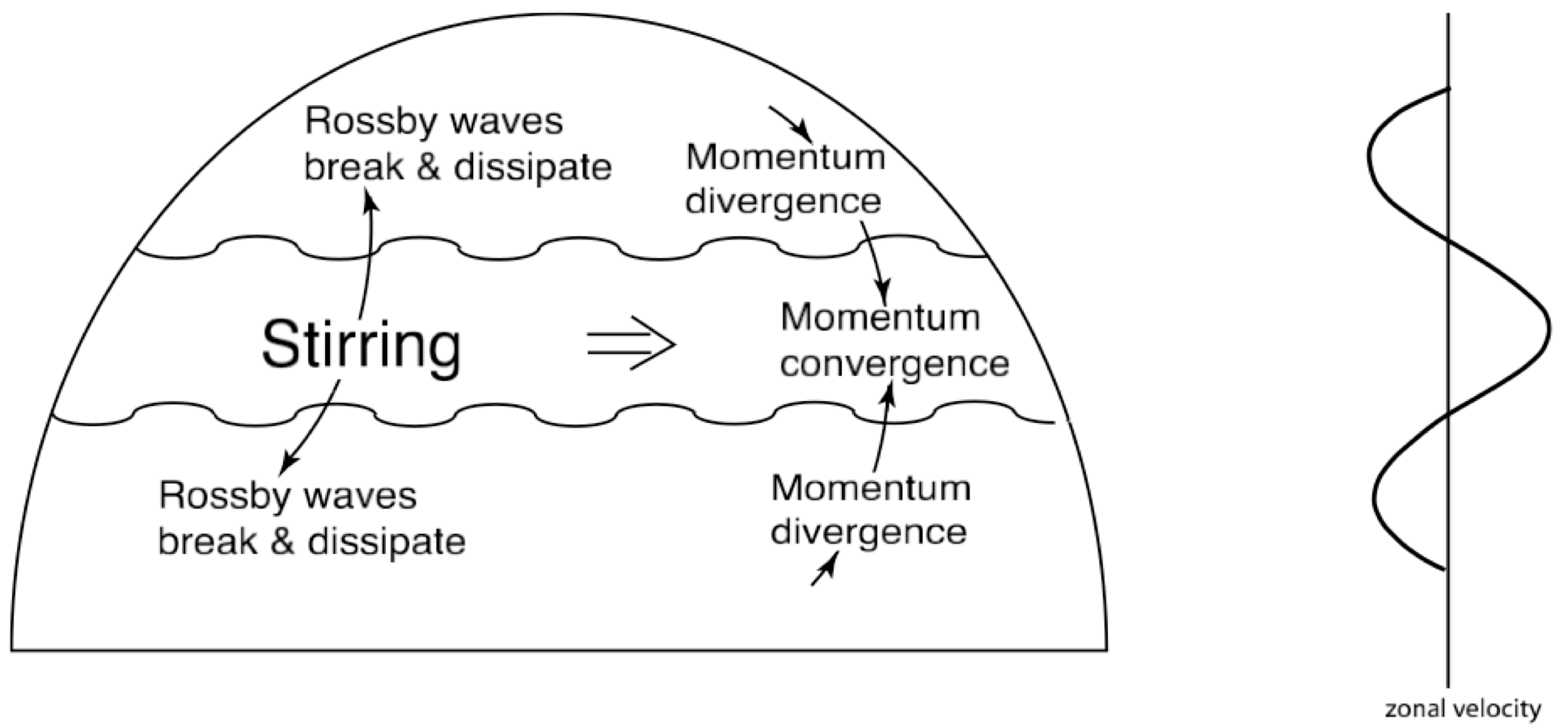}
\caption{\label{fig:rossby_wave_jetgen} Schematic explaining the emergence of the upper-level eddy-driven tropospheric jet in the Earth's atmosphere according to \textcite{Kuo-1951,Held-Hoskins-1985}. Equivalently barotropic waves are excited by baroclinic processes at the baroclinically active latitudes and as they propagate way from the region converge prograde momentum into the region giving rise to westerly flows. The waves are dissipated far from the region of excitation forming easterly flows at these latitudes.}
\end{figure}

\begin{figure}
\centering
\includegraphics[height=1.8in]{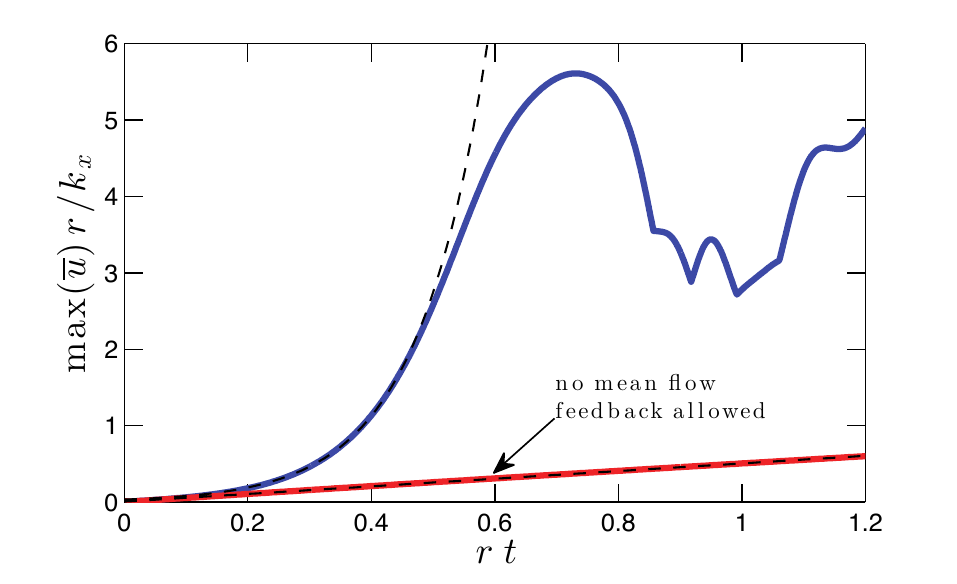}
\caption{\label{fig:S3TvsHeld} Zonal jet emergence under localized statistically stationary forcing. The fluid has initially no mean flow. The red curve shows the linear jet amplitude growth that results in the mean from momentum convergence into the excitation region according to the classical wave--mean flow interaction theory, in which the modification of the eddy structure by the emerging mean flow is neglected. The blue curve shows the ensemble mean jet amplitude predicted to emerge by the statistical dynamical theory described in this thesis. The theory predicts exponential jet growth induced by the active eddy--mean flow interaction (dashed), followed by equilibration of the instability. The stochastic forcing consists of an ensemble of temporally delta-correlated waves with zonal wavenumber $k_x = 8$ with Gaussian structure in $y$, i.e., $e^{-y^2/d^2}$, with $d=0.8$. Other parameters are $\b =1.4$ and $r=0.1$.}
\end{figure}

To demonstrate this mechanism consider a Rossby wave source region. Rossby waves that radiate away from this region  converge wave momentum into the excitation region, i.e., $\partial_y (\overline{u' v'}) <0$, inducing a positive mean flow acceleration in this region. This is because Rossby waves with positive group velocities propagating to the North (cf.~Fig.~\ref{fig:LongHigg}\hyperref[fig:LongHigg]{a}) have $\overline{u'v'}<0$, and Rossby waves with negative group velocities propagating to the South (cf.~Fig.~\ref{fig:LongHigg}\hyperref[fig:LongHigg]{c}) have positive $\overline{u'v'} >0$. As a result $\partial _y (\overline{u'v'})<0$ in the stirring region and because the zonal mean flow is governed by $\partial_t \overline{u} = -\partial_y (\overline{u'v'})$, a positive mean flow acceleration occurs in the stirring region. In the regions of dissipation momentum divergence leads to the emergence of negative flows, as shown in Fig.~\ref{fig:rossby_wave_jetgen}. If the wave excitation is statistically stationary the momentum convergence, $-\partial_y (\overline{u'v'})$, will also be statistically stationary and the eastward mean flow will grow in the mean linearly at a rate proportional to the energy input power, $\varepsilon_f$, as shown in Fig.~\ref{fig:S3TvsHeld}.\footnote{This explanation can be found in \textcite{Thompson-1971,Thompson-1980} who proposed that this Rossby wave radiation mechanism is responsible for the emergence of strong eastward currents in the oceans and also conducted a laboratory experiment demonstrating the process \parencite{McEwan-etal-1980}.} In conclusion: this wave--mean flow mechanism for the emergence of flows predicts linear mean growth of the jet in the regions of stirring and requires a localized forcing region and a propagation mechanism (here guaranteed because of the positive PV gradient due to the existence of~$\b$) in order for the waves to dissipate away from the source region. As a corollary: if the forcing is distributed homogeneously and the dissipation coefficients are constant (with no preferred dissipation regions) then this mechanism cannot induce any mean flows.

\begin{figure}
\centering
\includegraphics[height=1.8in]{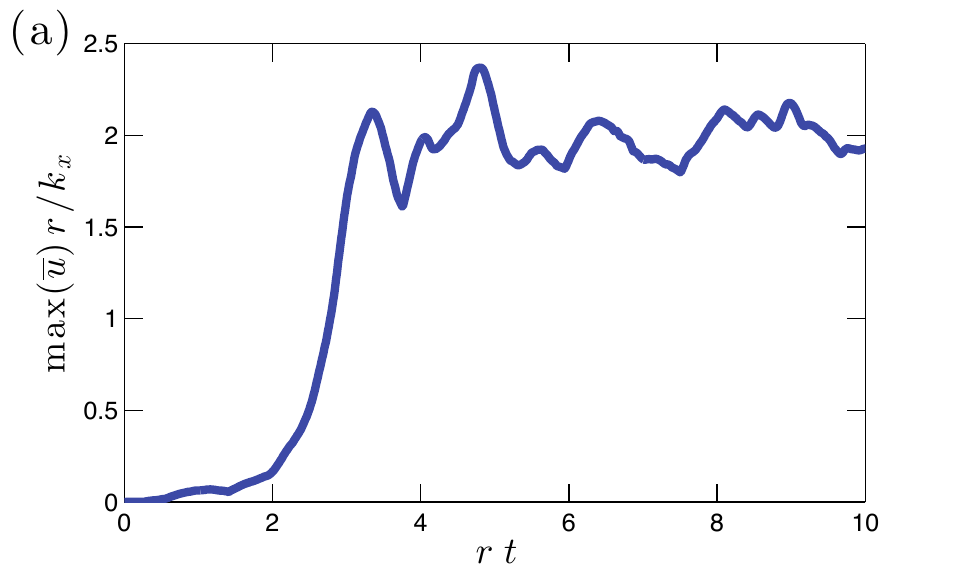}
\includegraphics[height=1.8in]{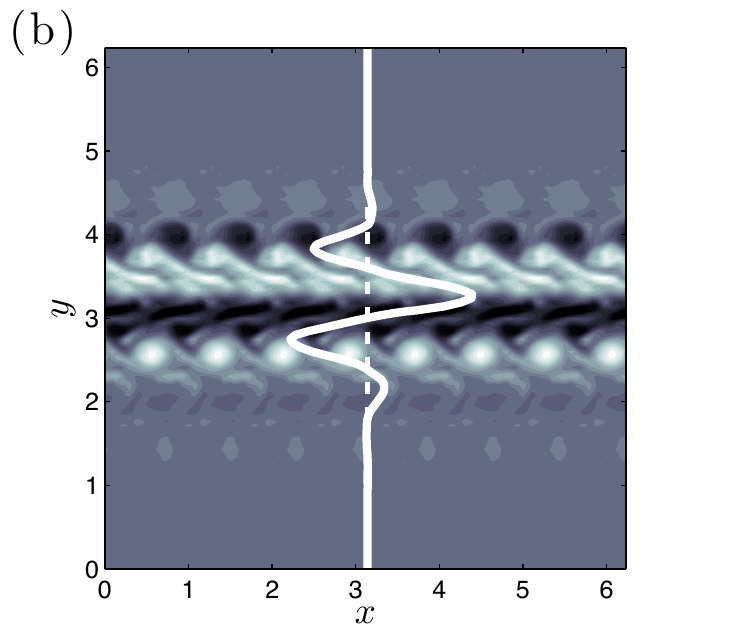}
\vspace{-.5em}\caption{\label{fig:NL_local_Kx8} Reflection of the statistical dynamical results shown in Fig. \ref{fig:S3TvsHeld} obtained by integrating the nonlinear barotropic equation~\eqref{eq:Dfz_I_psi} under a single realization of the forcing used in Fig.~\ref{fig:S3TvsHeld}. (a): the time development of the jet amplitude. (b): a snapshot of the vorticity field. The realization reflects the predictions of the statistical theory that initially the jet amplitude grows exponentially.}
%\end{figure}
%
%
%
%
\vspace{1em}
%\begin{figure}
\centering
\includegraphics[width=5.2in]{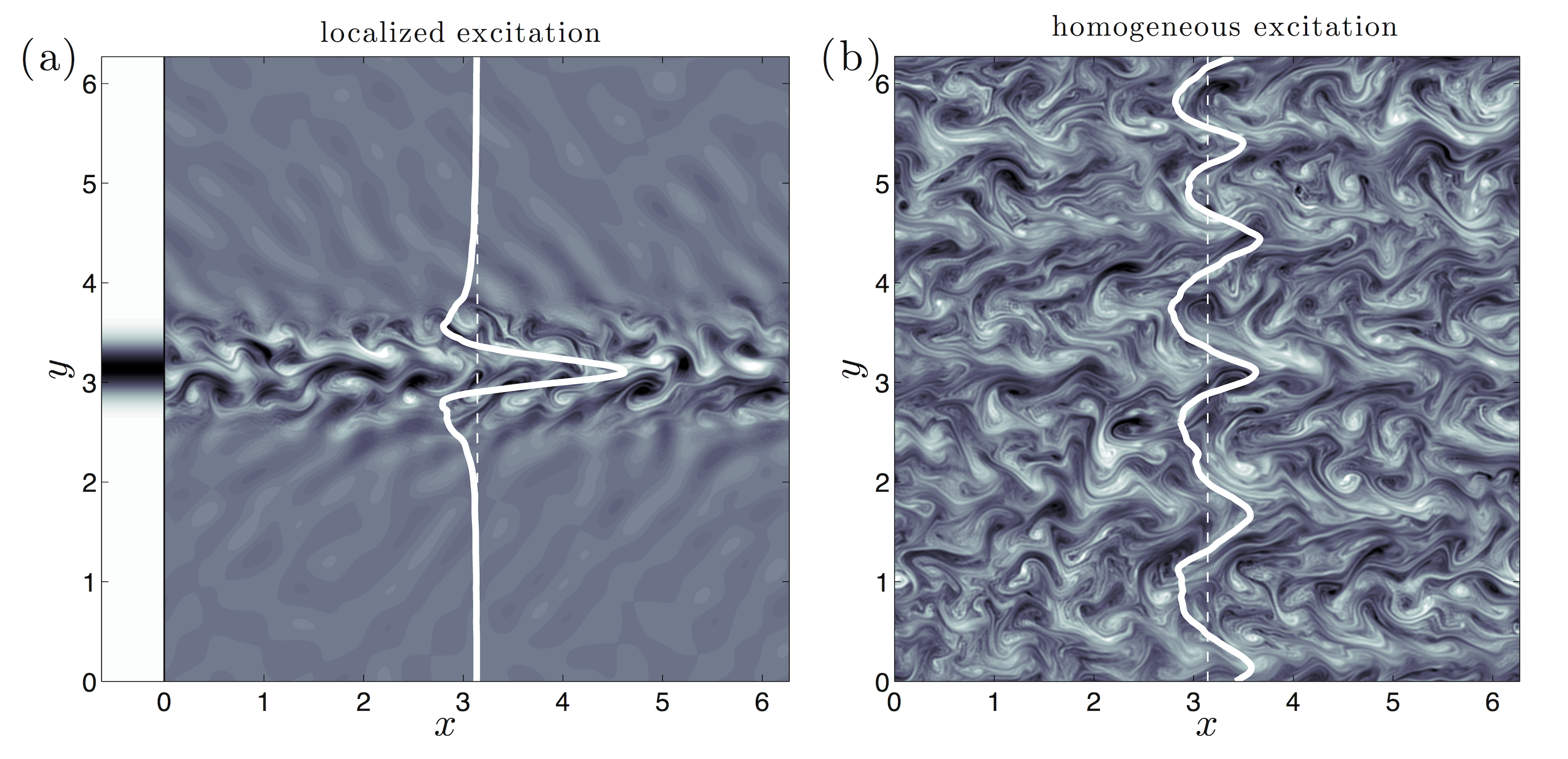}%change to jet_local_homog.eps
\vspace{-.5em}\caption{\label{fig:jet_local_homog} (a) Instantaneous snapshot of the vorticity field resulting from localized stirring in the region centered at $y=\pi$, indicated on the left of the panel, together with a snapshot of the structure of the zonal mean velocity (thick white line). This case differs from that of Fig.~\ref{fig:NL_local_Kx8} in the spectrum of the excitation. Here the excitation is obtained by convolving a homogeneous excitation with isotropic spectrum centered at total wavenumber $k_f= 15$ with a Gaussian localizing the excitation only in the $y$ direction to the region around $y=\pi$. The dependence of the large-scale flows that emerge on the spectrum of the excitation will be addressed in this thesis. The jets in this case form from the momentum convergence into the excitation region resulting from the Rossby wave propagation and from the active feedback of the jet on the waves that leads to the intensification of the preexisting jet. (b) Snapshot of the zonal mean flow and of the vorticity field when the excitation is spatially homogeneous. While in this case  classical wave-mean flow theory predicts that no mean flow should emerge, the statistical dynamics discussed in this thesis predict that a bifurcation occurs as the energy input, $\varepsilon_f$, increases or the dissipation coefficient, $r$, decreases. In the specific case, for parameter values $\varepsilon_f k^2_f / r^3 < 3.3\times 10^3$ the turbulent flow remains homogeneous with no jets, while for $\varepsilon_f k^2_f / r^3 > 3.3\times 10^3 $ the turbulent flow transitions to an inhomogeneous state with large-scale jets. The nonlinear integration shown here at $\varepsilon_f\, k_f^2 /r^3=5\times10^4$ demonstrates that the predictions of the statistical dynamics are reflected in individual realizations of the flow. It will be demonstrated in this thesis that the predictions of the statistical dynamical theory are reflected also for parameters near the critical. Other parameters: $\b/(k_f r)=70$ and the numerical integration was performed at resolution $512\times512$ with a pseudospectral code.}
\end{figure}

The above theory assumes that the dominant and most relevant mechanism for jet emergence is the momentum convergence in the stirring region by the propagating Rossby waves. It is assumed that as the jet emerges its influence on turbulence (which is not all waves) is negligible and as a result it can, at first, be neglected (of second order if the jet is infinitesimal). However, we will show in this thesis and demonstrate immediately with an example, that the influence of the emerging jet on turbulence (which is neglected in the above theory) is the important and dominant mechanism for the emergence and maintenance of jets. Actually, it is dominant even for jets of infinitesimal amplitude. This active feedback of the mean flow on the turbulence results in a new type of instability that leads to exponential growth of the amplitude of the jet with the result that the amplitudes diverge exponentially from the linear growth predicted by classical wave--mean flow theory. It is important to note that this instability is an instability of the statistical dynamics of the turbulent flow. We will present in this thesis a second-order cumulant approximation to the full statistical dynamics of the turbulent flow that reveals this instability of the interaction between large-scale structure and turbulence. To demonstrate the implications of the statistical dynamical formulation of the wave--mean flow dynamics we plot in Fig.~\ref{fig:S3TvsHeld} the jet amplitude evolution predicted by the second-order closure theory discussed in this thesis under the same forcing. The jet grows initially exponentially and then equilibrates, indicating that the second-order closure incorporates also the dynamics of equilibration. The instability manifests only in the ideal dynamics of the statistical state of the flow and is only partially reflected in individual simulations. For example, in Fig.~\ref{fig:NL_local_Kx8}\hyperref[fig:NL_local_Kx8]{a} we show the development of the jet in a sample integration of the nonlinear equations of motion under a realization of the excitation that was used in Fig.~\ref{fig:S3TvsHeld} and in Fig.~\ref{fig:NL_local_Kx8}\hyperref[fig:NL_local_Kx8]{b} a snapshot of the vorticity field and of the latitudinal structure of the jet that emerges. The nonlinear simulation confirms that the growth is faster than linear at first, but this sample integration can neither establish that there is an underlying instability nor make analytic predictions  as what jet structure is expected to emerge at first or the parameter range that leads to jet emergence. Although this instability is revealed only when the dynamics of the statistical state of the turbulent flow are examined, the predictions of the statistical theory are reflected in sample simulations of the nonlinear dynamics. Moreover, this instability of interaction that leads to the emergence of jets does not even require that the forcing be localized. Jets may emerge even if the forcing is homogeneous, contrary to the predictions of classical wave--mean flow theory. In Fig.~\ref{fig:jet_local_homog} we demonstrate in sample nonlinear simulations the emergence of robust jet structure both under spatially inhomogeneous forcing (Fig.~\ref{fig:jet_local_homog}\hyperref[fig:jet_local_homog]{a}) and, most importantly, under homogeneous forcing (Fig.~\ref{fig:jet_local_homog}\hyperref[fig:jet_local_homog]{b}).

In this thesis we will use a non-equilibrium statistical theory to address formation and maintenance  of jets and large-scale structures in turbulence. The proposed theory differs greatly from current theories that involve turbulent cascades and it has its basis in wave--mean flow interaction theories which consider that the most important interaction is the non-local in wavenumber space interaction between large-scale flows and the smaller scale eddies. Systematic investigation of the energy and enstrophy transfers among spectral components in numerical simulations has revealed that indeed the upgradient energy transfer from the small scales to the large-scale flow is mainly due to the highly non-local interactions in wavenumber space with a clear scale separation between them \parencite{Shepherd-1987,Huang-Robinson-98}. In this thesis we will demonstrate that not only local wavenumber interactions are not the main contributors to large-scale structure formation but moreover, they are not even required.

% !TEX root = ../thesis.tex

%\begin{savequote}[75mm] 
%This is some random quote to start off the chapter.
%\qauthor{Firstname lastname} 
%\end{savequote}

\chapter{Formulation of the S3T statistical state dynamics of turbulent flows on a $\b$-plane\label{ch:formulation}}

The formation and maintenance of zonal jets in planetary atmospheres is essentially governed by barotropic processes. The simplest setting in which we can study planetary barotropic processes is a planar flow on a rotating $\beta$-plane which conserves the absolute vorticity of the flow in the absence of dissipation. Turbulence on a $\b$-plane does not self-sustain and a turbulent state must be externally forced in order to be maintained against dissipation. This forcing may model processes absent from the 2D barotropic dynamics, such as energy injected by baroclinic instabilities or turbulent convection. Because of the erratic and unpredictable nature of these vorticity sources in planetary turbulence, the forcing is modeled as a white-noise process in time given that the fluctuations of the forcing have a short autocorrelation time compared to the time scales of the barotropic dynamics. We also assume that the forcing is spatially homogeneous and if the turbulent flow becomes inhomogeneous this should be attributed to the dynamics.  

In the following sections we formulate the quasi-linear approximation of the nonlinear stochastically forced barotropic vorticity equations and derive the equations of the S3T statistical dynamics of the turbulent flow on a barotropic $\b$-plane. S3T is an acronym for Stochastic Structural Stability Theory, which was initially abbreviated as SSST. The reason for this acronym will become apparent in this chapter.

\section{Formulation of the S3T dynamics on a $\b$-plane}

Consider a non-divergent, barotropic flow on a infinite $\b$-plane with planetary vorticity gradient, $\bv = (0,\b)$. % In most applications the flow is in {\color{red}a doubly periodic channel of size $L_x\times L_y$}.
 The velocity field being non-divergent can be  expressed in terms of a streamfunction, $\psi$, as $\uv=\hat{\mathbf{z}}\times\nablav\psi$ which implies $(u,v)=(-\partial_y \psi, \partial_x \psi)$  ($\hat{\mathbf{z}}$ is the unit vector normal to the $\b$-plane, see Fig.~\ref{fig:bplane}). The vorticity of the fluid is $\nablav\times\uv = \z\, \hat{\mathbf{z}}$ with $\z=\partial_x v-\partial_y u=\Del\psi$, and $\Delta \equiv \nablav\cdot\nablav=\partial^2_{xx} + \partial^2_{yy}$ the two-dimensional Laplacian. In the presence of stochastic forcing and dissipation, the potential vorticity, $q=\z+f_0+\bv\cdot\xv$,  which here is simply the absolute vorticity, evolves as:
\begin{align}
\frac{\Df q}{\Df t} = \partial_t \z + J\(\psi,\z+ \betav \cdot\mathbf{x} \) = -r\,\z +\sqrt{\e}\,\xi\ ,\label{eq:nl}
\end{align}
where $\Df/\Df t\equiv \partial_t+\uv\cdot\nablav$ is the material derivative along the fluid flow. The advection term, $(\uv\cdot\nablav)q$, is alternatively expressed as $J(\psi,q)$ where  $J(g,h)\equiv (\partial_x g)(\partial_y h)-(\partial_y g)(\partial_x h)$ is the Jacobian of functions $g$ and $h$. The flow is dissipated with linear damping at a rate $r$, which typically models Ekman drag in planetary atmospheres. Turbulence is maintained by the external stochastic forcing $\sqrt{\e}\xi(\xv,t)$. We assume that $\sqrt{\e}\xi$ is a homogeneous random stirring and we model this excitation as temporally delta-correlated Gaussian process with zero mean, i.e., $\<\xi(\xv,t)\>=0$, and with spatial correlation prescribed by $Q$,
\be
\<\bit \xi(\xv_a,t)\xi(\xv_b,t')\> = Q(\xv_a-\xv_b)\,\d(t-t')\ .
\ee
The brackets denote the ensemble average over forcing realizations. (For details regarding the stochastic excitation refer to Appendix~\ref{app:forcing}.) We render~\eqref{eq:nl} non-dimensional using as a time scale the dissipation time scale $1/r$ and as a length scale the typical length scale of the stochastic excitation, $L_f$. With this non-dimensiona\-lization~\eqref{eq:nl} becomes an equation for the variables
\begin{subequations}\label{eq:nondim}\be
 \zeta^*=\frac{\z}{r}\ ,\ \ \psi^*=\frac{\psi}{r L_f^2}\ ,\ \ \xi^*=\frac{\xi}{r^{1/2}L_f^{-1}}\ ,
\ee
and parameters
\be
\b^*=\frac{\b}{ rL_f^{-1}}\ ,\ \ \e^*=\frac{\e}{r^3 L_f^2}\ ,\ \ r^*=1\ .\label{eq:nondim_param}
\ee\end{subequations}
From here on we will work with the non-dimensional equation and drop the asterisks. Typical values for the non-dimensional parameters are $\b^*=15$, $\varepsilon^*=1200$ for the Earth's atmosphere, $\b^*=3$, $\varepsilon^*=1600$ for the ocean and $\b^*=450$, $\varepsilon^*=4\times10^7$ for Jupiter, based on the parameter values of the table~\ref{tab:pla_values}.

%\begin{table}
%\footnotesize\caption{Relevant parameters in geophysical flows.}
%\label{tab:pla_values}\centering
%%\begin{tabular}{ p{3 cm} p{1.8 cm} p{1.8 cm} p{1.8 cm} p{1.8 cm} p{1.3 cm} p{1.5 cm}}
%\begin{tabular}{ l c c c c c c }
% &  $1/k_f$~[km]   &   $1/r$~[days]  & $\b~[10^{-11}\textrm{m}^{-1}\,\textrm{s}^{-1}]$ & $\e~[\textrm{m}^{-2}\,\textrm{s}^{-3}]$ & $\b^*$ & $\e^*$\\
%	\hline
%Earth's atmosphere & 1000 & 10 & $1.6$ & $3\times 10^{-4}$ & 15 & 190\\
%\hline
%Earth's ocean & 20 & 1000 & $1.6$ & $10^{-9}$ & 40 & \textcolor{red}{2500}\\
%\hline
%Jovian atmosphere & 1000 & 5800 & $0.25$ & $0.5\times 10^{-5}$ & 125 & $6.25\times10^8$\\
%\end{tabular}
%\end{table}

\begin{table}[h]
\footnotesize\caption{Typical parameter values for geophysical flows. The typical forcing length scale is taken as the deformation radius in each geophysical setting.} %The energy input rate is calculated assuming that the observed root mean square velocity fluctuations, $\mathcal{U}_{\textrm{rms}}$, satisfy $\mathcal{U}_{\textrm{rms}}^2=\varepsilon/r$.}
\label{tab:pla_values}
\begin{center}
{\fontsize{7}{7}\selectfont
\begin{tabular}{l c c c c c c c}
\topline
 &   \begin{tabular}{@{}c@{}}$1/k_f$\\$[\textrm{km}]$\end{tabular}    &    \begin{tabular}{@{}c@{}}$1/r$\\$[\textrm{day}(=24\textrm{h})]$\end{tabular}   &    \begin{tabular}{@{}c@{}}$\mathcal{U}_{\textrm{rms}}$\\$[\textrm{m}\,\textrm{s}^{-1}]$\end{tabular}   & \begin{tabular}{@{}c@{}}$\b$\\$[10^{-11}\textrm{m}^{-1}\,\textrm{s}^{-1}]$\end{tabular} & \begin{tabular}{@{}c@{}}$\varepsilon$\\$[\textrm{m}^{-2}\,\textrm{s}^{-3}]$\end{tabular}   & $\b^*$ & $\varepsilon^*$\\
\midline
Earth's atmosphere & 1000 & 10 & $15$ & $1.6$ & $2\times 10^{-3}$ & 15 & 1300\\
Earth's ocean & 20 & 100 & $0.1$ & $1.6$& $10^{-9}$ & 3 & 1600\\
Jovian atmosphere & 1000 & 1500 & $50$ & $0.35$& $0.5\times 10^{-5}$ & 450 & $4\times10^7$\\
\botline
\end{tabular}}
\end{center}
\end{table}

The first step in constructing the S3T dynamical system is to decompose the vorticity flow field into an averaged or mean field, $Z = \tav{\z}$, and deviations from the mean vorticity, $\z' = \z-Z$, which is referred to as eddy vorticity. The averaging operator $\Tcal$ determines the type of mean field we want to study. We employ two types of averaging operators: i) an average over the zonal $x$ direction, i.e., $\tav{\phi}= L_x^{-1} \int_0^{L_x} \df x\; \phi(\xv,t)$ and ii) a Reynolds average in which $\tav{\phi}$ produces a coarse-grained field which is obtained by averaging over  an intermediate time scale or length scale which is larger than the time scale or length scale of the turbulent motions but smaller than the time scale or length scale of the coarse-grained field. In the first interpretation of the averaging operator the mean flows are zonal jets while in the second they may be either zonal jets or slowly moving traveling waves.

With this decomposition, the barotropic vorticity equation~\eqref{eq:nl} is equivalently rewritten as a system for the joint evolution of the mean and the eddy vorticity:
\begin{subequations}\begin{align}
\partial_t Z &+ J \(\Psi, Z+  {\bm\beta}\cdot\mathbf{x} \) = -\tav{ J \(\psi', \z'\)}-Z \ ,\label{eq:enl_mean}\\
\partial_t \z' &=\underbrace{- J\(\psi',Z+ {\bm\beta}\cdot\mathbf{x} \) -J\(\Psi,\z'\) -\z'}_{\Acal(\Uv)\,\z'}+
\underbrace{\tav{  J \(\psi', \z'\) }-J \(\psi', \z'\)}_{f_{\textrm{nl}}}+\sqrt{\e}\,\xi \ ,\label{eq:enl_pert}
\end{align}\label{eq:enl}\end{subequations}
where $\Psi=\tav{ \psi }$ is the mean streamfunction and $\psi'$ is the eddy streamfunction.
Equations~\eqref{eq:enl} are referred to as the NL system. The stochastic excitation is assumed to have zero mean, $\tav{\xi}=0$, and consequently the mean equations are unforced. %\footnote{\color{red} Stochastic forcing $\xi$ satisfies $\tav{\xi}=0$ by construction. For example if $\Tcal$ is a zonal average then $\tav{\xi}=0$ since $\xi$ is homogeneous in $x$, or if $\Tcal$ is an average over an intermediate time scale then $\tav{\xi}=0$ since it varies much more rapidly than any other motion in the flow. There is no ergodic assumption at this point that $\tav{\xi}=\<\xi\>=0$.}
The first  term  on the r.h.s. of~\eqref{eq:enl_pert}, $\Acal(\Uv)\z'$, represents advection of eddy vorticity by the mean flow and is a bilinear functional of the eddies and the mean flow, while the second nonlinear term, $f_{\textrm{nl}}\equiv\tav{  J \(\psi', \z'\) }-J \(\psi', \z'\)$, represents advection of the eddy field by itself. The operator $\Acal(\Uv)$, which governs the linear dynamics of the eddy field if the mean flow $\Uv=\zhat\times\nablav\Psi$ is prescribed, can be written as:
\be
\Acal(\Uv)\equiv -\Uv\cdot\nablav + \[\bit(\Del\Uv)\cdot\nablav+\zhat\cdot\( \bv\times\nablav\)\]\Del^{-1} -1\ .\label{eq:def_Acal}
\ee

We also make the ergodic assumption that the $\Tcal$ average of a flow field (i.e. the zonal average or the Reynolds average over the intermediate time or length scale) is equal to the ensemble average over the forcing realizations, i.e., $\Tcal\[\phi(\xv,t)\]=\<\phi(\xv,t)\>$,
where the brackets denote the ensemble average. The identification of the ensemble average
with an averaging operation is crucial for the realization of the statistical quantities in a single planet and the validity of the ergodic assumption is established by experiment. 

%is chosen appropriately according to which type of mean flow we want to consider. If for example we interested in the description of zonal jets (as in~Fig.~\ref{fig:NL3cases}\hyperref[fig:NL3cases]{c}) then $\Tcal$ may by chosen as a zonal average over a latitude circle or over the zonal direction, $x$, in the $\b$-plane, while if we want to describe large-scale traveling waves (as in Fig.~\ref{fig:NL3cases}\hyperref[fig:NL3cases]{b}) then a zonal average is not appropriate and $\Tcal$ 

%\vspace{1em}

\begin{figure}
\centering
\includegraphics[height=1.28in]{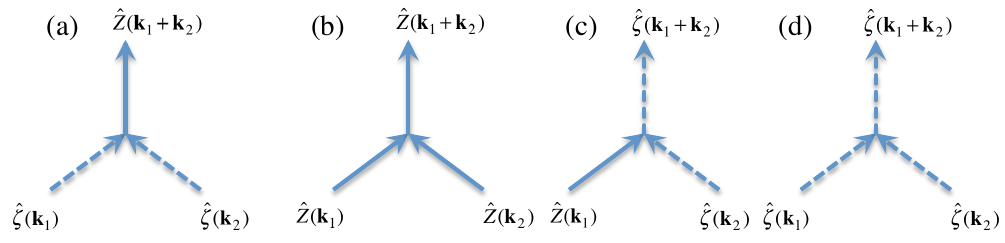}
\caption{\label{fig:triads-gen} Schematic of all possible wavenumber triad interactions in the NL and QL systems.
(a) Two eddies with wavenumber vectors	 $\kv_1$ and $\kv_2$ combine to form a mean flow with wavenumber $\kv_1+\kv_2$. This interaction is in both NL and QL (it is term $\tav{ J \(\psi', \z'\) }$ in~\eqref{eq:enl_mean} and~\eqref{eq:eql_mean}).
(b) A wavenumber $\kv_1$ mean flow interacts with a $\kv_2$ mean flow to produce a mean flow with wavenumber $\kv_1+\kv_2$. This interaction is in both NL and QL (it is term $J(\Psi,Z)$ in~\eqref{eq:enl_mean} and~\eqref{eq:eql_mean}).
(c) A wavenumber $\kv_1$ mean flow  interacts with a wavenumber $\kv_2$ eddy to produce a $\kv_1+\kv_2$ eddy. This interaction is in both NL and QL (it is term $\Acal(\Uv)\z'$ in~\eqref{eq:enl_pert} and~\eqref{eq:eql_pert}).
(d) An wavenumber $\kv_1$ eddy interacts with a $\kv_2$ eddy to produce a $\kv_1+\kv_2$ eddy.
This interaction is included in  NL  (term $f_{\textrm{nl}}$ in~\eqref{eq:enl_pert}) but neglected
in QL.}
\end{figure}

In order to obtain a closed statistical description of the turbulent flow we restrict the nonlinearity in the NL equations by neglecting the eddy--eddy term, $f_{\textrm{nl}}$, in~\eqref{eq:enl_pert} or parametrize it as stochastic excitation. We obtain in this way the quasi-linear (QL) approximation to the NL system~\eqref{eq:enl}: 
\begin{subequations}\begin{align}
\partial_t Z &+ J \(\Psi, Z+  {\bm\beta}\cdot\mathbf{x} \) = -\tav{ J \(\psi', \z'\) }-Z \ ,\label{eq:eql_mean}\\
\partial_t \z' &=\Acal(\Uv)\,\z'+\sqrt{\e}\,\xi \ .\label{eq:eql_pert}
\end{align}\label{eq:eql}\end{subequations}
A schematic comparing the nonlinear interactions operating in NL and QL system is shown in Fig.~\ref{fig:triads-gen}. In NL the term$f_{\textrm{nl}}$, which is neglected in QL, neither injects nor dissipates energy (see Appendix~\ref{app:forcing}) and therefore the QL system, in the absence of forcing and dissipation, has the same invariants as the NL system, namely it conserves both energy and enstrophy.

% Note that~\eqref{eq:eql} is an ensemble of quasi-linear equations, in which the eddies evolve according to~\eqref{eq:eql_pert} for each forcing realization and the mean flow is driven by the average of the vorticity flux divergence over the different forcing realizations. For a stochastic forcing with gaussian statistics,~\eqref{eq:eql} admits a second order closure of its statistics.

%\vspace{1em}

The QL system has the attribute that its statistical dynamics close at second order. To obtain the statistical  dynamics of the quasi-linear system~\eqref{eq:eql} we use the ergodic assumption to identify $Z=\<\,\z\,\>$ and the second cumulant of the vorticity between points $\xv_a$ and $\xv_b$, 
\be
C(\xv_a,\xv_b,t)\equiv\<\z'(\xv_a,t)\z'(\xv_b,t)\bit\>\ ,\label{eq:def_C}
\ee
with $\tav{\z'(\xv_a,t)\,\z'(\xv_b,t)}$. Then, the average $\tav{J(\psi',\z')}=\<\,J(\psi',\z')\,\>$ can be expressed as a linear functional of $C_{ab}(t)\equiv C(\xv_a,\xv_b,t)$. To show that we use the incompressibility condition to rewrite $J(\psi',\z')= \nablav\cdot \[(\hat{\mathbf{z}}\times\nablav\psi')\, \z'\]$, and proceed as follows:
\begin{align}
\Tcal\left\{\bit  \nablav\cdot \[(\hat{\mathbf{z}}\times\nablav\psi')\, \z'\]\right\}&= \nablav\cdot \tav{ (\hat{\mathbf{z}}\times\nablav\psi')\, \z' }= \nablav\cdot \< (\hat{\mathbf{z}}\times\nablav\psi')\, \z'\>\nonumber\\%=-\nablav\cdot \< \uv'\,\z' \>= -\nablav\cdot\<\frac1{2}(\uv'_a \z'_b+\uv'_b \z'_a)\>_{\xv_a=\xv_b}\nonumber\\
%  & = -\partial_x\,\[- \frac1{2}\(\Del^{-1}_{a}\partial_{y_a}\!+\!\Del^{-1}_{b}\partial_{y_b}\) C\]_{\xv_a=\xv_b} \hspace{-.8em}- \partial_y\,\[ \frac1{2}\(\Del^{-1}_{a}\partial_{x_a}\!+\!\Del^{-1}_{b}\partial_{x_b}\) C\]_{\xv_a=\xv_b} \\
  & = \nablav\cdot\<\frac1{2}\hat{\mathbf{z}}\times\(\nablav_a \psi'_a\, \z'_b+\nablav_b\psi'_b\, \z'_a\)\>_{\xv_a=\xv_b}\nonumber\\
   &  =\nablav\cdot\[\frac1{2}\hat{\mathbf{z}}\times\(\nablav_a \Del_{a}^{-1}+\nablav_b \Del_{b}^{-1}\) C_{ab}\]_{\xv_a=\xv_b}\ .\label{eq:def_Rcal}
\end{align}
%through
%\begin{align}
%\Rcal( C ) &\equiv -\partial_x\,\[- \frac1{2}\(\Del^{-1}_a\partial_{y_a}\!+\!\Del^{-1}_b\partial_{y_b}\) C\]_{\xv_a=\xv_b} \hspace{-.8em}- \partial_y\,\[ \frac1{2}\(\Del^{-1}_a\partial_{x_a}\!+\!\Del^{-1}_b\partial_{x_b}\) C\]_{\xv_a=\xv_b} \nonumber\\
%&= -\<\,J(\psi',\z')\,\> \ .\label{eq:defR} %\nonumber\\&-\tav{J(\psi',\z')}&=-\<\,J(\psi',\z')\,\>\nonumber\\
%\end{align}
%

The subscript $a$ or $b$ in functions denotes hereafter the value of the function at $\xv_a$ or $\xv_b$, i.e. $\z'_a\equiv\z'(\xv_a,t)$, the subscript $a$ or $b$ in operators denotes the action of the operators only on the variables $\xv_a$ or $\xv_b$ respectively, and the subscript $\xv_a=\xv_b$  denotes that any expression depending on the two variables $\xv_a$ and $\xv_b$ should be evaluated at $\xv_a=\xv_b=\xv$.\footnote{For example: $\partial_y\,\left\{\bit\partial_{x_b}\[\z'(\xv_a,t)\z'(\xv_b,t) \]\right\}_{\xv_a=\xv_b} = \partial_y\[\bit \z'(\xv_a,t)\,\partial_{x_b} \z'(\xv_b,t)\]_{\xv_a=\xv_b} = \partial^2_{yx}\[ \z'(\xv,t)^2/2\bit\]$.} The operator $\Delta^{-1}$ is the inverse of the Laplacian which has been rendered unique by incorporating the boundary conditions. Equation~\eqref{eq:def_Rcal} shows that $\Tcal[ J(\psi',\z') ]$ is a linear functional of $C$. We denote the linear functional given in~\eqref{eq:def_Rcal} by $\Rcal$ and set
\be
\Rcal(C)\equiv-\Tcal[ J(\psi',\z') ]\ .\label{eq:def_Rcalv2}
\ee 
The first cumulant, $Z$, of the flow field therefore evolves according to:
\begin{align}
\partial_t Z &+ J \(\Psi, Z+  {\bm\beta}\cdot\mathbf{x} \) = \Rcal( C )-\,Z\ .
\end{align}
To obtain the evolution equation of $C$ we take the time derivative of~\eqref{eq:def_C} and obtain:\footnote{In writing~\eqref{eq:s3t_C_xizeta} we adopt the  Stratonovich interpretation for stochastic differential equations. However, because the stochastic forcing in our case is additive, both Stratonovich and It\^o interpretations lead to the exact same results (cf.~ Appendix~\ref{app:forcing}).}
\begin{align}
\partial_t C_{ab} & = \[\bit\Acal_a(\Uv) + \Acal_b(\Uv)\]C_{ab} +\sqrt{\e}\,\<\xi_a\z'_b+ \z'_a\xi_b\bit\>\ ,\label{eq:s3t_C_xizeta}
\end{align}
where $\Acal_a(\Uv)$ indicates that the coefficients of the operator $\Acal(\Uv)$ are evaluated at $\xv_a$ and that the differential operator act only on the variable $\xv_a$ of $C(\xv_a,\xv_b,t)$. (Similarly for $\Acal_b(\Uv)$.)

It can be shown (see Appendix~\ref{sec:proof_xizeta}) that for temporally delta-correlated stochastic forcing term $\<\xi_a\z'_b+ \z'_a\xi_b\bit\>$ is independent of the state of the system and exactly equal to $\sqrt{\e}\,Q(\xv_a-\xv_b)\equiv\sqrt{\e}\,Q_{ab}$.\footnote{The dependence of the spatial covariance of the forcing on the difference coordinate $\xv_a-\xv_b$ indicates that the stochastic forcing is spatially homogeneous (see Appendix~\ref{app:forcing}).} The rate of energy injection is thus independent of the state of the system and is prescribed by the spatial forcing covariance $Q$ and the amplitude factor $\e$. The same is true for the NL system. In both systems the energy injection rate is $\e \[(2\pi)^{-2}\int \df^2 \kv\;  \hat{Q}(\kv)/(2k^2)\]$, where $\hat{Q}(\kv)$ is the Fourier transform of $Q$,
\be
\hat{Q}(\kv) = \int\df^2(\xv_a-\xv_b)\; Q(\xv_a-\xv_b)\,e^{-\i \kv\cdot(\xv_a-\xv_b)}~,\label{eq:def_Qhat}
\ee
with $\kv=(k_x,k_y)$. Because $Q$ is a homogeneous covariance its Fourier transform is real and non-negative, i.e., $\hat{Q}(\kv) \ge 0$, for all wavenumbers $\kv$. The quantity $\hat{Q}/k^2$, where $k\equiv|\kv|$, determines the energy spectrum of the forcing. We normalize $Q$ so that
\bdm
\int \frac{\df^2 \kv}{(2\pi)^2}  \frac{\hat{Q}(\kv)}{2k^2}=1\ ,
\edm
and the energy injection rate per unit area is $\e$. For details see Appendix~\ref{app:forcing}. 

The joint evolution of the first two cumulants of the flow field, $Z$ and $C$, define the S3T statistical state dynamics of the turbulent flow which is governed by the autonomous system of deterministic equations:\begin{subequations}
\begin{align}
\partial_t Z &+ J \(\Psi, Z+  {\bm\beta}\cdot\mathbf{x} \) = \Rcal( C )-\,Z\ ,\label{eq:s3t_mean}\\
\partial_t C_{ab} & = \[\bit\Acal_a(\Uv) + \Acal_b(\Uv)\]C_{ab} +\e\,Q_{ab}\ .\label{eq:s3t_C}
\end{align}\label{eq:s3t}\end{subequations}

The S3T system~\eqref{eq:s3t} corresponds to a second-order closure of the full statistical dynamics of the turbulent flow. This closure became possible because of the adoption of the quasi-linear approximation. If the quasi-linear approximation were not made, then the evolution of the second cumulant, $C$, would also involve terms of the form $\<f_{\textrm{nl}}(\xv_a,t)\z'(\xv_b,t) \>$, which are related to the third cumulant and as a result the equations for the first two cumulants would not close. Neglecting or parametrizing the eddy--eddy terms in~\eqref{eq:enl_pert} by a state independent Gaussian stochastic process leads to a closed set of equations for the evolution of the first two cumulants of a Gaussian approximation of the statistical state dynamics of the turbulent flow. This approximation is also referred to as ``CE2''. Note that  higher order truncations of the cumulant equations is problematic. \textcite{Marcinkiewicz-1939} has shown that  truncations of the cumulant equations at order $n_0>2$, obtained by setting all $n$-th order cumulants for $n>n_0$ equal to zero, produce non positive probability density functions (pdf). Therefore the only physically realizable cumulant truncation is at second order.

% 
%\item
%The solutions of the S3T system remain bounded for all times.
%%\item
%%When the mean flow component of the S3T state variable, $Z$, does not depend on $x$ then the mean flow has a zonal jet structure, otherwise it has non-zonal structure. Non-zonal structures typically propagate westwards.
%\end{enumerate}

\section{Formulation of the S3T dynamics of zonal mean states}\label{sec:s3tz}

The most common mean flows that appear in planetary turbulence are zonal jets. In order to address the statistical dynamics of zonal jets in turbulence we may choose the averaging operator $\Tcal$ to be directly the average over the zonal direction, $x$, i.e.,
\be
\tav{\phi} = \frac1{L_x}\int_0^{L_x}\phi(x',y,t)\,\df x' = \Phi(y,t)\ .\label{eq:zonal_edd_dec}
\ee
The zonal average of a flow field is also denoted with an overbar, for example $\overline{\phi(\xv,t)}\equiv$ ${L_x}^{-1}\int_0^{L_x}\phi(x',y,t)\,\df x'$. This zonal S3T closure simplifies significantly the QL and S3T systems and it is the easiest to interpret because the separation between mean and eddy is unequivocal.

With the zonal average the mean flow vorticity $Z$ is related to the zonal flow mean flow, $U$, through $Z=-\partial_y U$, and because non-divergence implies $\partial_y V=0$, without any loss of generality, we can assume that $V=0$. Since $\bv=(0,\b)$ the advection of the mean potential vorticity flow, $Z+\bv\cdot\xv$, by the mean flow field, $\Uv$, vanishes, i.e., $J \(\Psi, Z+  {\bm\beta}\cdot\mathbf{x} \)=0$, and the zonal average vorticity flux divergence simplifies to:
\be
\overline{J(\psi',\z')} = \overline{ \nablav\cdot\(\uv'\,\z'\)} = \partial_y\(\overline{ v'\z'}\)\ .
\ee

 With these simplifications the NL system takes the form:
\begin{subequations}\begin{align}
\partial_t U &  = \overline{ v'\z' } - U\ ,\label{eq:nlz_mean}\\
\partial_t \z' &=\Acal_{\textrm{z}}(U)\,\z' + \underbrace{\partial_y\(\overline{v'\z'}\)-\nablav\cdot \(\uv'\,\z'\)}_{f_{\textrm{nl,z}}}+\sqrt{\e}\,\xi \ ,\label{eq:nlz_pert}
\end{align}\label{eq:nlz}\end{subequations}
while the QL system becomes
\begin{subequations}\begin{align}
\partial_t U &  = \overline{ v'\z' } - U\ ,\label{eq:qlz_mean}\\
\partial_t \z' &=\Acal_{\textrm{z}}(U)\,\z'+\sqrt{\e}\,\xi \ ,\label{eq:qlz_pert}
\end{align}\label{eq:qlz}\end{subequations}
where in both~\eqref{eq:nlz} and~\eqref{eq:qlz} operator $\Acal_{\textrm{z}}$ is
\be
\Acal_{\textrm{z}}(U) = -U\partial_x - \(\bit\b- \partial^2_{yy}U\)\partial_x\Del^{-1} -1\ .
\ee
(Roman subscript $\textrm{z}$ denotes that the zonal mean--eddy decomposition was used.)

\begin{figure}
\centering
\includegraphics[height=1.28in]{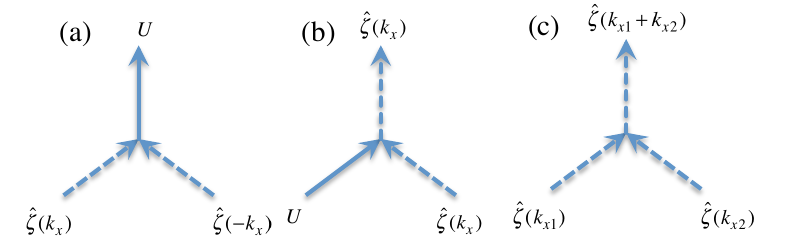}
\caption{\label{fig:triads-zon} Schematic of the triad interactions among the zonal or $x$-wavenumber fields in the NL system~\eqref{eq:enl}. The wavenumbers in this figure
refer only to the zonal or $x$-wavenumber components of the full wavenumber vector $\kv$.
(a) Two eddies with wavenumbers $k_x$ and $-k_x$ interact to form a zonal mean flow, $U$ (with $k_x=0$). This interaction is in both NL and QL (it is term $\overline{v'\z'}$ in~\eqref{eq:nlz_mean} and~\eqref{eq:qlz_mean}).
(b) A $k_x$ wavenumber eddy interacts with mean flow $U$ ($k_x=0$) to produce another eddy with zonal wavenumber $k_x$. This interaction is in both NL and QL (it is term $\Acal_{\textrm{z}}(U)\,\z'$ in~\eqref{eq:nlz_pert} and~\eqref{eq:qlz_pert}).
(c) A $k_{x1}$ wavenumber interacts with a $k_{x2}$ wavenumber eddy to produce a $k_{x1}+k_{x2}$ wavenumber eddy. This interaction is included in NL (term $f_{\textrm{nl,z}}$ in~\eqref{eq:nlz_pert}) but neglected in QL.}
\end{figure}

The three types of nonlinear triad interactions that can occur between the mean quantities and the eddies  are shown in Fig.~\ref{fig:triads-zon}. In the QL approximation we neglect $f_{\textrm{nl,z}}$ or parameterize it as stochastic noise. It should be noted that \textcite{Bouchet-etal-2013} have established that in this zonal mean--eddy decomposition the QL approximation becomes exact in the limit of $\e^*=\e/(r^3 L_f^2)\to\infty$ (cf.~\eqref{eq:nondim_param}).

%\begin{subequations}\begin{align}
%\partial_t(-\partial_y U) &+ \b_x U  = -\partial_y\(\overline{ v'\z' }\) - (-\partial_y U) \ \label{eq:qlz_mean}\\
%\partial_t \z' &=\Acal_{\textrm{z}}(U)\,\z'+\sqrt{\e}\,\xi \ \label{eq:qlz_pert}
%\end{align}\label{eq:qlz}\end{subequations}
%with
%\be
%\Acal_{\textrm{z}}(U) = -U\partial_x + \[\bit (\partial^2_{yy}U)\partial_x+ \(\b_x\partial_y-\b_y\partial_x\)\]\Del^{-1} -1\ 
%\ee

Because $\Acal_{\textrm{z}}(U)$ is invariant under the translation $x\to x+\alpha$ for any constant $\alpha$, the eddy vorticity equation is homogeneous in $x$ and therefore the eddy vorticity covariance will always be homogeneous in $x$ and consequently of the form: 
\be
C(\xv_a,\xv_b,t) = C(x_a-x_b,y_a,y_b,t)\ .\label{eq:C_zonal}
\ee
The zonal homogeneity of $C$ allows us to simplify the flux divergence to:
\be
\Rcal_{\textrm{z}}( C ) = - \partial_y\,\[ \frac1{2}\(\Del^{-1}_a\partial_{x_a}\!+\!\Del^{-1}_b\partial_{x_b}\) C_{ab}\]_{\xv_a=\xv_b}\ ,\label{eq:defRzonal} 
\ee
and the S3T system takes the form
\begin{subequations}
\begin{align}
\partial_t U & = \[ \frac1{2}(\Del^{-1}_a\partial_{x_a}\!+\!\Del^{-1}_b\partial_{x_b}) C_{ab}\]_{\xv_a=\xv_b}- U\ ,\label{eq:s3tz_mean}\\
\partial_t C_{ab} & = \[\bit\Acal_{\textrm{z},a}(U) + \Acal_{\textrm{z},b}(U)\]C_{ab} +\e\,Q_{ab}\ .\label{eq:s3tz_C}
\end{align}\label{eq:s3tz}
\end{subequations}
This system will be denoted as S3Tz (for S3T-zonal).

Solutions of~\eqref{eq:s3tz} are also solutions of the generalized S3T system~\eqref{eq:s3t}, i.e., a solution $\(\bit U(y,t),C(x_a-x_b,y_a,y_b,t)\)$ that satisfies~\eqref{eq:s3tz} also satisfies~\eqref{eq:s3t} as well; the converse however is not true. S3Tz system~\eqref{eq:s3tz} has tremendous advantage in numerical simulations over the generalized S3T system~\eqref{eq:s3t} because its state variables have significantly fewer degrees of freedom. The method of numerical integration of the stochastic NL and QL and of the deterministic S3T equations is discussed in Appendix~\ref{app:numerical_method}.

%The fact that the covariance, $C$, is homogeneous in $x$ gives a tremendous advantage in the numerical integration of the S3T system, since it reduces significantly its degrees of freedom. More details on how NL, QL and S3T systems are integrated numerically can be found in Appendix~\ref{app:numerical_method}.

%\begin{subequations}
%\begin{align}
%\partial_t (-\partial_y U) & +\b_x U= -\partial_y\[ \frac1{2}(\Del^{-1}_a\partial_{x_a}\!+\!\Del^{-1}_b\partial_{x_b}) C\]_{\xv_a=\xv_b}-(-\partial_y U)\ \label{eq:s3tz_mean}\\
%\partial_t C & = \[\bit\Acal_{\textrm{z},a}(U) + \Acal_{\textrm{z},b}(U)\]C +\e\,Q\ \label{eq:s3tz_C}
%\end{align}\label{eq:s3tz}
%\end{subequations}

\section{S3T statistical equilibria and their stability}

S3T systems~\eqref{eq:s3t} and~\eqref{eq:s3tz} are autonomous and may admit equilibrium (fixed point) solutions $\(\bit Z^e(\xv),C^e(\xv_a,\xv_b)\)$. These equilibria are statistical equilibria of the turbulent flow and consist of the mean flow vorticity $Z^e(\xv)$ and an eddy field with covariance $C^e(\xv_a,\xv_b)$.

%A different type of solutions that the S3T system may posses are periodic orbits (limit circles), $\(\bit Z^e(\xv,t),C^e(\xv_a,\xv_b,t)\)$. These time-depended solutions are found with the mean flow in the form of coherent waves traveling westward. They can be rendered stationary by the Galilean transformation $\xv\to\xv_0=\xv +(V\hat{\mathbf{x}})\,t$, where $V$ is the velocity with which these solutions travel westward. In the new coordinates $\(\bit Z^e(\xv,t),C^e(\xv_a,\xv_b,t)\)\to\(\bit Z^e(\xv_0),C^e(\xv_{0,a},\xv_{0,b})\)$.

Remarkably, both S3T systems~\eqref{eq:s3t} and~\eqref{eq:s3tz} admit the stationary homogeneous equilibrium
\be
 Z^e=0\ ,\ \ C^e=\frac{\e}{2} Q\ ,\label{eq:hom_equil}
 \ee
 for all values of $\e$ and $\bv = (\b_x,\b_y)$ with components $\b_x$ and $\b_y$, under the condition that the forcing covariance is homogeneous. This statistical equilibrium has no mean flow and a homogeneous eddy field. To confirm this note that $\Acal^e\equiv\Acal(\Uv^e=0)=-1+\zhat\cdot(\bv\times\nablav)\Del^{-1}$, and
\begin{align}
(\Acal^e_a+\Acal^e_b) C^e &  =\left\{\bit -2+\zhat\cdot\[\bv\times\(\nablav_a\Del_a^{-1}+\nablav_b\Del_b^{-1}\) \]\right\}\frac{\e}{2}Q\nonumber\\
%&  =\left\{\bit -2-\bv\cdot\[\zhat\times\(\nablav_a\Del_a^{-1}+\nablav_b\Del_b^{-1}\) \]\right\}\frac{\e}{2}Q\nonumber\\
&=-\e Q + \frac{\e}{2} \zhat\cdot\[\bv\times\(\nablav_a\Del_a^{-1}+\nablav_b\Del_b^{-1}\) \]\int\frac{\df^2\kv}{(2\pi)^2}\; \hat{Q}(\kv) e^{\i \kv\cdot(\xv_a-\xv_b)}\nonumber\\
&=-\e Q - \frac{\e}{2} \int\frac{\df^2\kv}{(2\pi)^2}\;  \zhat\cdot\[\bv\times\(\frac{\i\kv}{-k^2}+\frac{-\i \kv}{-k^2}\)\]  \hat{Q}(\kv) e^{\i \kv\cdot(\xv_a-\xv_b)}\nonumber\\
&=-\e Q\ ,\label{eq:homog_proof_eq1}
\end{align}
showing  that $C^e$ of~\eqref{eq:hom_equil} satisfies~\eqref{eq:s3t_C}. Further, from~\eqref{eq:def_Rcalv2},
\begin{align}
\Rcal(C^e)  %&  =-\frac{\e}{4}\nablav\cdot\[\bit\hat{\mathbf{z}}\times\(\nablav_a \Del_{a}^{-1}+\nablav_b \Del_{b}^{-1}\) Q\]_{\xv_a=\xv_b}\nonumber\\
&=-\frac{\e}{4}\nablav\cdot\[\hat{\mathbf{z}}\times\(\nablav_a \Del_{a}^{-1}+\nablav_b \Del_{b}^{-1}\) \int\frac{\df^2\kv}{(2\pi)^2}\; \hat{Q}(\kv) e^{\i \kv\cdot(\xv_a-\xv_b)}\]_{\xv_a=\xv_b}\nonumber\\
&  =-\frac{\e}{4}\nablav\cdot\[\hat{\mathbf{z}}\times\int\frac{\df^2\kv}{(2\pi)^2}\;\(\frac{\i\kv}{-k^2}+\frac{-\i\kv}{-k^2}\) \hat{Q}(\kv) e^{\i \kv\cdot(\xv_a-\xv_b)}\]_{\xv_a=\xv_b} = 0\ ,\label{eq:homog_proof_eq2}
\end{align}
which in turn confirms that~\eqref{eq:s3t_mean} is also satisfied. While the homogeneous state~\eqref{eq:hom_equil} is always an equilibrium of the S3T system it may only be an approximate equilibrium of the full hierarchy of cumulant equations. However, we show in Appendix~\ref{app:MI} that for the case of isotropic delta function ring forcing, i.e., for $\hat{Q}(\kv) = 4\pi k_f^2\, \d(k-k_f)$, this homogeneous statistical equilibrium is also an equilibrium of the full hierarchy of cumulants.

% it may only be approximately valid in when the full nonlinearity is retained as will be discussed in chapter~\ref{ch:NLvsS3Tjas}. 

%\vspace{1em}

The stability of any S3T equilibrium solution $(Z^e,C^e)$ is addressed by considering small perturbations $(\d Z,\d C)$ about this equilibrium and performing an eigenanalysis of the  linearized S3T equations about this equilibrium:
\begin{subequations}\begin{align}
\partial_t \,\d Z & = \Acal^e\,\d Z + \Rcal( \d C )\ ,\label{eq:s3t_pert_dZgen}\\
\partial_t \,\d C_{ab} & = \(\bit\Acal^e_a + \Acal^e_b \)\d C_{ab} +\(\bit\d\Acal_a + \d\Acal_b\)C^e_{ab}\ ,\label{eq:s3t_pert_dCgen}
\end{align}\label{eq:s3t_dZdCgen}\end{subequations}
where $\Acal^e\equiv\Acal(\Uv^e)$ and  $\d\Acal \equiv \Acal(\Uv^e+\d\Uv)-\Acal^e$.

%\vspace{1em}

When the equilibrium is unstable the statistics of the flow bifurcate to a new state. So the stability of an S3T equilibrium implies the structural stability of the turbulent flow, while the marginally stable S3T equilibria identify the critical states at which the turbulent flow becomes structurally unstable and transitions to a new statistical state. 

S3T stability involves the stability of the statistics of the turbulent state and is fundamentally different from the hydrodynamic stability of a mean state. It can be shown that if the S3T
equations admit the equilibrium $(Z^e,C^e)$ then by necessity the associated mean state is hydrodynamically stable (cf.~Appendix~\ref{app:s3t-equil-prop}). However, the hydrodynamic stability of a mean state does not imply the S3T stability. Most notable example is the homogeneous equilibrium with no mean flow~\eqref{eq:hom_equil}. The state of zero mean flow is clearly hydrodynamically stable but it will be shown that at a critical parameter $\e$ the homogeneous equilibrium becomes S3T unstable and the turbulent flow reorganizes to an inhomogeneous state.

\section{Bibliographical Note}

The S3T theory was introduced by \textcite{Farrell-Ioannou-2003-structural}. The continuous formulation of the theory was developed by \textcite{Srinivasan-Young-2012}. The cumulant interpretation was discussed by \textcite{Marston-etal-2008} who refer to it as CE2 (see also \textcite{Marston-2012}). The cumulant representation of the statistical dynamics of the flow were developed by \textcite{Hopf-1952}. The statistical stability of the homogeneous state in S3T (or CE2) and the subsequent formation of zonal jets is investigated in barotropic flows by~\textcite{Farrell-Ioannou-2007-structure,Bakas-Ioannou-2011,Srinivasan-Young-2012,Parker-Krommes-2014-generation}. Earlier, \textcite{Carnevale-Martin-1982} using field theoretic techniques arrived at the same equations for the statistical description of fluctuations about a homogeneous state but the relevance for the emergence of zonal jets was not discussed. The statistical stability of inhomogeneous states in S3T is investigated by \textcite{Farrell-Ioannou-2003-structural,Parker-Krommes-2014-generation}. Statistical state dynamics with higher order cumulant truncations are discussed by \textcite{Marston-2012,Marston-etal-2014}. The generalized coarse-grained mean flow interpretation of S3T that allows non-zonal solutions was introduced by~\textcite{Bernstein-Farrell-2010} in an investigation of the phenomenon of blocking in a two-layer baroclinic atmosphere and was studied recently for barotropic flows by \textcite{Bakas-Ioannou-2013-prl,Bakas-Ioannou-2014-jfm}. %A different but related numerical implementation of the coarse graining averaged equations is discussed by \textcite{Sapsis-Majda-2013,Grooms-Majda-2014}.

% !TEX root = ../thesis.tex

%\begin{savequote}[75mm] 
%This is some random quote to start off the chapter.
%\qauthor{Firstname lastname} 
%\end{savequote}

\chapter{Emergence of coherent structures out of homogeneous turbulence through S3T instability}
\label{ch:st3hom}

%{\color{red}
%
%S3T instability and formation of coherent structures occurs if a seed mean flow organizes the turbulent eddies so that the eddy flux divergence $\Rcal( \d C )$ reinforces it producing in this way a positive feedback. The eddy-mean flow dynamics of this feedback underlying the structure forming instability of the homogeneous turbulent equilibrium will be thoroughly investigated in the following chapter.
%
%}
%
%
%In this chapter we discus thoroughly the S3T instability of the homogeneous turbulent equilibrium and the mechanisms by which turbulence self-organizes in order to reinforce the mean flow. The main points of this chapter are included in~\upmax\textcite{Bakas-etal-2014}\dnmax.

\section{S3T instability of homogeneous turbulent equilibrium}

We have seen in the previous chapter that for spatially homogeneous forcing there is always a homogeneous equilibrium of the S3T system~\eqref{eq:s3t}. This equilibrium is given by
\be
Z^e=0\ \ ,\ \ C^e(\xv_a-\xv_b)=\frac{\varepsilon}{2} Q(\xv_a-\xv_b)\ .\label{eq:ZeCe}
\ee

We want to determine the statistical stability of this equilibrium as a function of the parameters available in the problem. These parameters are the non-dimensional $\e$ and $\b$ defined in~\eqref{eq:nondim_param} and also the spectrum of $Q$. We examine cases in which the spectrum of $Q$ is isotropic and cases in which it is anisotropic. The stability of this equilibrium is determined by the linearized S3T perturbation equations~\eqref{eq:s3t_dZdCgen} about the homogeneous equilibrium~\eqref{eq:ZeCe}, which take the form:
\begin{subequations}\begin{align}
\partial_t \,\d Z &= \Acal^e\,\d Z+ \Rcal( \d C )\ ,\label{eq:s3t_pert_dZ}\\
\partial_t \,\d C_{ab} & = \(\bit\Acal^e_a + \Acal^e_b \)\d C_{ab} +\(\bit\d\Acal_a + \d\Acal_b\)C_{ab}^e\ ,\label{eq:s3t_pert_dC}
\end{align}\label{eq:s3t_dZdC}\end{subequations}
where $(\d Z,\d C)$ are the perturbation mean flow and perturbation covariance, $\Acal^e=\hat{\mathbf{z}}\cdot\( {\bm\beta}\times{\bm\nabla}\)\Del^{-1} -1$ and  $\d\Acal = -\d\Uv\cdot{\bm\nabla} + \[\bit(\Del\,\d\Uv)\cdot{\bm\nabla}\]\Del^{-1} $.

\vspace{1em}

The purpose of this chapter is to examine the stability of the homogeneous equilibrium. We derive an analytic expression for the eigenvalues of~\eqref{eq:s3t_dZdC} and show that there is always a critical energy input rate $\e=\e_c$ that renders~\eqref{eq:ZeCe} unstable. When the equilibrium is unstable a mean flow in the form of the most unstable mean flow eigenfunction grows, initially at the rate predicted by the eigenvalue, and the turbulent flow will eventually reorganize to an inhomogeneous state. We study the dependance of $\e_c$ on non-dimensional $\b$ for isotropic and anisotropic forcing spectra and also determine which type of mean flow (zonal jets or non-zonal flows) is the most unstable.

\section{Eigenanalysis of the homogeneous equilibrium}

We proceed now with the stability analysis of~\eqref{eq:ZeCe}. %Since the equilibrium solution is stationary its stability reduces to the eigenanalysis of~\eqref{eq:s3t_dZdC}. %\footnote{If $(Z^e,C^e)$ was not stationary but a time-periodic solution of~\eqref{eq:s3t} with period $T$ then its stability would require calculation of the eigenvalues of the propagator of~\eqref{eq:s3t_dZdC} for time $T$. Such examples will be considered in chapter~\ref{ch:S3Tnonhom}.}
Consider eigenfunctions of the form $\(\dZ,\dC\)e^{s t}$. The eigenvalue $s$ and $\dZ(\xv)$ and $\dC(\xv_a,\xv_b)$ satisfy the eigenvalue problem
\begin{subequations}\begin{align}
s \,\dZ &=\Acal^e\,\dZ+ \Rcal( \dC )\ ,\label{eq:s3t_pert_dZ_s}\\
s \,\dC_{ab} & = \(\bit\Acal^e_a + \Acal^e_b \)\dC_{ab} +\(\bit\d\tilde{\Acal}_a + \d\tilde{\Acal}_b\)C^e_{ab}\ .\label{eq:s3t_pert_dC_s}
\end{align}\label{eq:s3t_dZdC_s}\end{subequations}
The eigenfunctions can be assumed in the form 
\begin{subequations}\begin{align}
\dZ_\nv(\xv) &= e^{\i\nv \cdot\xv}\ ,\label{eq:dZtilde}\\
\dC_\nv(\xv_a,\xv_b) &=\tilde{C}_\nv^{(\textrm{h})}\(\bit\xv_a-\xv_b\)\,e^{\i\nv\cdot(\xv_a+\xv_b)/2}\ ,\label{eq:dCtilde}
\end{align}\label{eq:eigendZdC}\end{subequations}
with $\nv=(n_x,n_y)$ the wavevector of the eigenfunction (see Appendix~\ref{app:S3Tgr}). The mean flow component of the eigenfunction~\eqref{eq:dZtilde} is a zonal jet when $n_x=0$ and a non-zonal flow, a plane wave, when $n_x\ne0$.% (cf. Fig.~\ref{fig:einx}). 

%\begin{figure}
%\centering
%\includegraphics[width=4in]{einx.pdf}
%\caption{\label{fig:einx}  (a) Constant phase lines of a non-zonal mean flow perturbation eigenfunction $\dZ=e^{\i\,\nv\cdot\xv}$ with wavevector $\nv$ forming angle $\varphi =30\deg$ with the meridional. (b) Constant phase lines of a zonal jet mean flow eigenfunction with $n_x=0$. For zonal jet perturbation the wavevector $\nv$ is parallel to the meridional. The arrow in each plot shows the direction of the wavevector $\nv$.}
%\end{figure}
%

Note that the mean flow eigenfunction $\dZ_\nv$ is also an eigenfunction of $\Acal^e$ with eigenvalue $-(\i\om_\nv+1)$,
\be
\Acal^e\,\dZ = -(\i\om_\nv+1)\,\dZ\ ,\label{eq:AedZ_eigen}
\ee
where $\om_{\nv}$ is the Rossby frequency 
\be
\om_\nv \equiv \frac{\hat{\mathbf{z}}\cdot(\bv\times\nv)}{n^2}\ ,\label{eq:def_omRossby}
\ee
with $n=|\nv|$ and as a result~\eqref{eq:s3t_pert_dZ_s} can be written as $(\s+1) \dZ = \Rcal(\dC)$ with $\s\equiv s+ \i \om_\nv$. Because the Reynolds stress associated with the perturbation covariance, $\Rcal(\dC)$, is proportional to $\e$ (as $C^e$ is proportional to $\e$) it can be written as 
\be
\Rcal(\dC) = \e\,f(\s) \,\dZ\ ,\label{eq:sens}
\ee
where $f$ is the Reynolds stress feedback or eddy feedback, and $\s$ satisfies the dispersion relation
\begin{align}
\s +1  &=   \varepsilon \, f(\s)~.\label{eq:sigma_nz}
\end{align}
We remind the reader that the 1 in the l.h.s. of~\eqref{eq:sigma_nz} is  the rate of dissipation and therefore the homogeneous state is unstable when $\real(\s)>0$ or $\e\real\[\bit\f(\s)\]>1$, i.e., the mean flow acceleration by the Reynolds stress feedback exceeds the decay due to dissipation. The term $f(\s)$ measures the feedback on the mean flow $\dZ$ by the eddy perturbation field after being distorted by the mean flow $\dZ$. When $\real\[f(\s)\]>0$ the feedback on the mean flow by the eddy perturbation field has the tendency to reinforce the existing mean flow and the vorticity fluxes due to the eddies are upgradient. It is necessary for instability to have upgradient vorticity fluxes but it is not sufficient, because they have to overcome the dissipation. In Appendix~\autoref{app:S3Tgr} we show that the function $f(\s)$ is
(cf.~Appendix~\autoref{app:S3Tgr}, eq.~\eqref{eq:RdC_efdZ}):
\be
f(\s) = \int \frac{\df^2 \kv}{(2\pi)^2}\;\frac{|\mathbf{k}\times{\mathbf{n}}|^2\, (k^2_s-k^2) (k^2-n^2) }{k^4 k^2_s n^2  \[ (\s+2) +\i\( \om_{\kv+\nv}- \om_{\nv} - \om_{\kv} \)\bit \] }\frac{\hat{Q}(\kv)}{2}\ , \label{eq:f_factor}
\ee
with $\kv_s=\kv+\nv$ and $k_s=|\kv_s|$ and the dispersion relation for the stability of the homogeneous equilibrium is 
\be
\s+1 =  \e\int \frac{\df^2 \kv}{(2\pi)^2}\;\frac{|\mathbf{k}\times{\mathbf{n}}|^2\, (k^2_s-k^2) (k^2-n^2) }{k^4 k^2_s n^2  \[ (\s+2) +\i\( \om_{\kv+\nv}- \om_{\nv} - \om_{\kv} \)\bit \] }\frac{\hat{Q}(\kv)}{2} \ .\label{eq:s3tgr_z_nz}
\ee

{}

%\textcolor{red}{If in the rotated frame we have that:
%\be
%\dvz= -\frac{\df^2}{\df {y'}^2}\dU(y')\ 
%\ee
%in the unrotated frame this translated to:
%\begin{align}
%\Rcal(\dC)&= -(\hat{\mathbf{n}}\cdot\nablav)(\hat{\mathbf{n}}\cdot\nablav\dZ(x,y))\nonumber\\
%&= -\sum_{i,j}  \hat{n}_i\hat{n}_j\,\partial^2_{x_i x_j} \dZ(x,y)\ 
%\end{align}
%}

\begin{figure}
\centering
\includegraphics[width=5in]{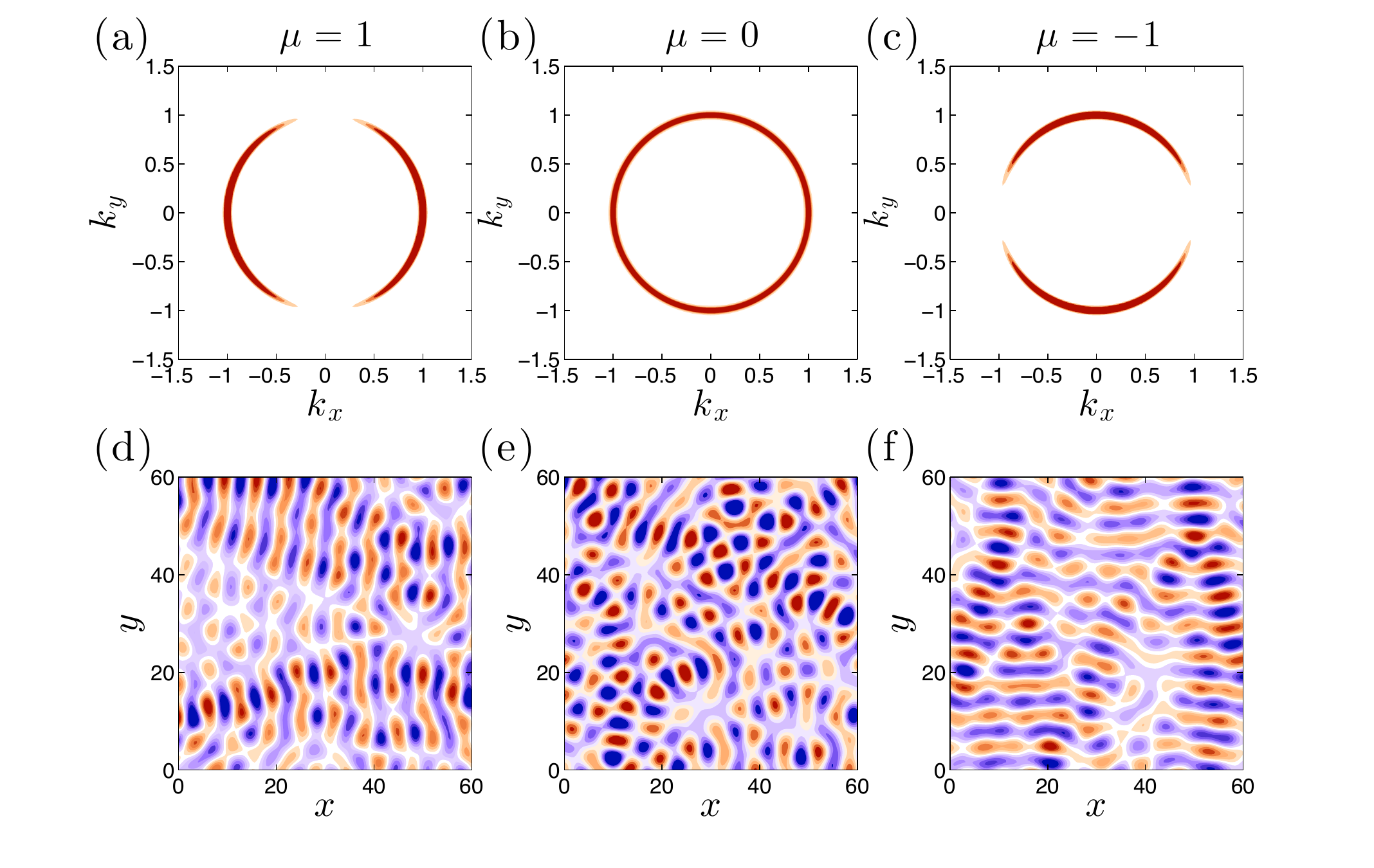}
\caption{\label{fig:RF}  Top panels: the forcing covariance spectrum, $\hat{Q}(\kv)= 4\pi\,\d(k-1)\,\[1+\mu \cos(2\gamma)\]$,  for (a) $\mu=1$, (b) $\mu=0$ and (c) $\mu=-1$ (the support of the delta function is represented as a thin  ring). Bottom panels:  contours of the vorticity field  induced by a realization of the stochastic forcing for (d) $\mu=1$, (e) $\mu=0$ and (f) $\mu=-1$.}
\end{figure}

We investigate the stability of the homogeneous equilibrium under stochastic forcing with spectrum
\be
\hat{Q}(\kv) = 4\pi\,\Gcal(\gamma)\,\d(k-1)\ ,\label{eq:Qhat_deltaG}
\ee
with $\gamma=\arctan{(k_y/k_x)}$ and
\be
\Gcal (\gamma) = 1 + \mu \cos (2 \gamma)\ .\label{eq:defG}
\ee
This forcing excites an eddy field at total wavenumber $k_f=1$ (in dimensional units $k_f=1/L_f$) and the parameter $\mu$ measures the anisotropy of the forcing. Parameter $\mu$ takes values $|\mu|\le1$ so that $\Gcal(\gamma)\ge0$ for all $\gamma$. For $\mu=0$ the forcing is isotropic (see Fig.~\ref{fig:RF}\hyperref[fig:RF]{b,e}). For $\mu>0$ the stochastic forcing is anisotropic (see Fig.~\ref{fig:RF}\hyperref[fig:RF]{a}) favoring structures aligned with the meridional axis (i.e. with $k_y=0$), as shown in Fig.~\ref{fig:RF}\hyperref[fig:RF]{d}, while $\mu<0$ (see Fig.~\ref{fig:RF}\hyperref[fig:RF]{c}) favors structures aligned with the zonal axis (i.e. with $k_x=0$), as shown in Fig.~\ref{fig:RF}\hyperref[fig:RF]{f}. In Jupiter because the excitation models vorticity input by turbulent convection we expect excitation to be of the $\mu=0$ type, while in the Earth because the excitation models injection of vorticity due to baroclinic processes we expect excitation closer to $\mu=1$.

We determine the critical energy input rate $\ecz$ that renders the homogeneous equilibrium unstable to zonal jet perturbations and the critical energy input rate $\ecnz$ that renders the homogeneous equilibrium unstable to non-zonal perturbations. $\ecz$ is the minimum $\e$ for which $\real(\s)=0$ for an eigenfunction with wavevector $\nv=(0,n_y)$ and $\ecnz$ is the minimum $\e$ for which $\real(\s)=0$ for an eigenfunction with wavevector $\nv=(n_x,n_y)$ and $n_x\ne0$. When $\e>\min{(\ecz,\ecnz)}\equiv\e_c$ the homogeneous equilibrium is unstable and the structure that first emerges is zonal or non-zonal according to whether the minimum $\e$ is $\ecz$ or $\ecnz$. 

\begin{figure}
\centering
\includegraphics[width=4.25in]{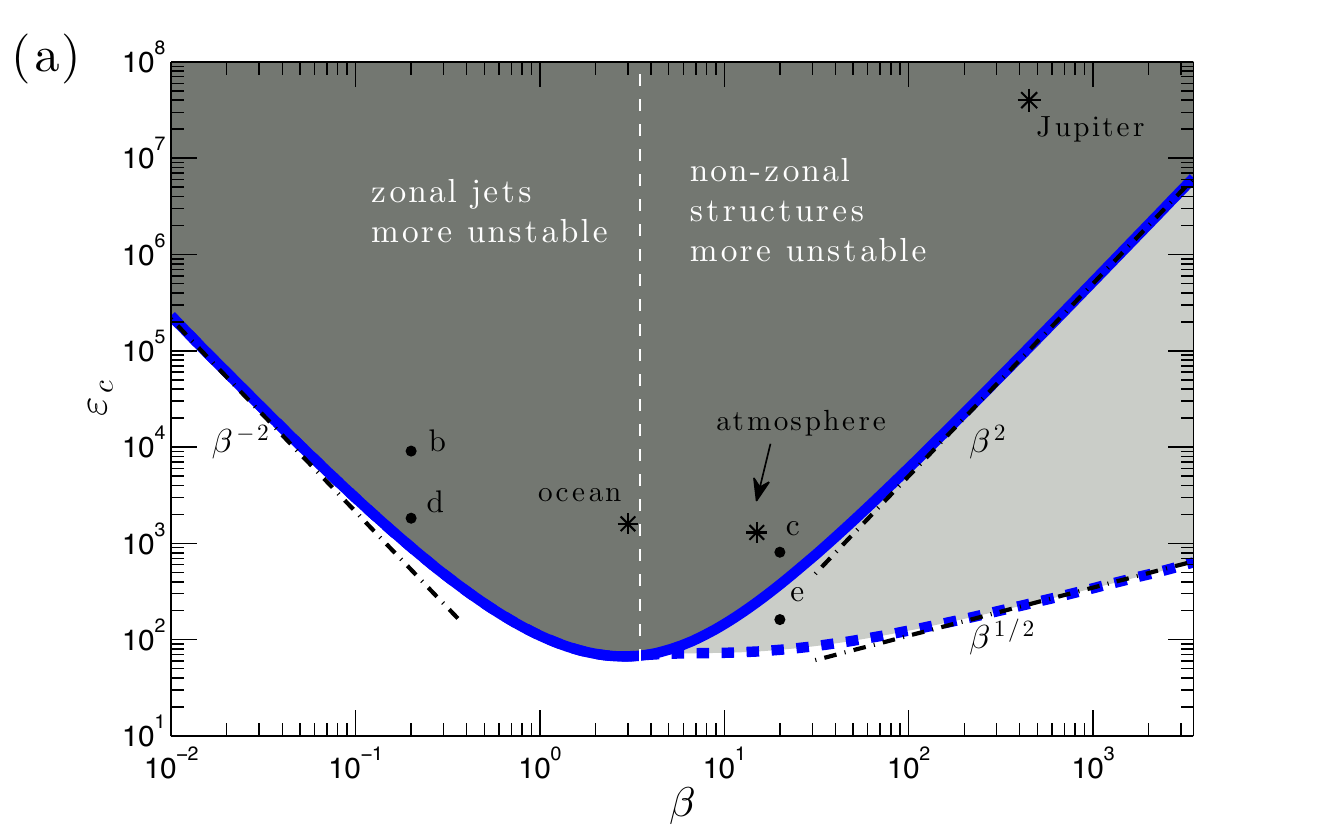}\\
\includegraphics[width=3.5in]{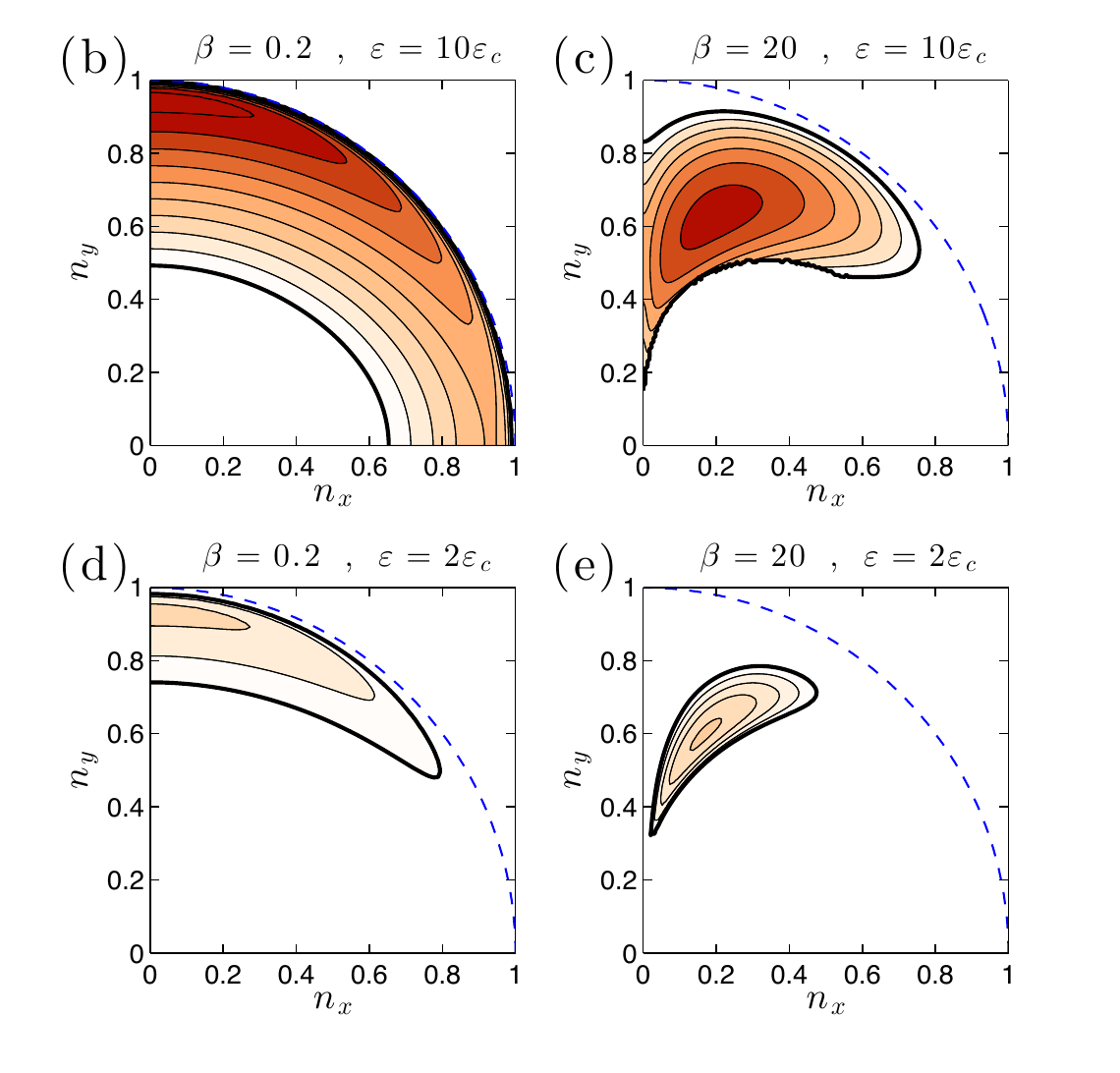}
\caption{\label{fig:s3tgr_isotropic} (a) The critical energy input rates $\ecz$ (solid) and $\ecnz$ (dashed) that render the homogeneous equilibrium unstable to zonal jet perturbations  ($\nv=(0,n_y)$) and non-zonal perturbations ($\nv=(n_x,n_y)$) respectively, as a function of  $\b$ for isotropic forcing covariance spectra ($\mu=0$, Fig.~\ref{fig:RF}\hyperref[fig:RF]{b}). Also shown are the slopes $\b^{-2}$, $\b^2$ and $\b^{1/2}$ (dash-dot). Typical values of the Earth's atmosphere and ocean and Jupiter's atmosphere (found in table~\ref{tab:pla_values}) are marked with stars. For $\b<3.5$ zonal jet eigenfunctions grow the most whereas for $\b>3.5$ non-zonal structures grow the most. In the light shaded region only non-zonal coherent structures are unstable, while in the dark shaded region both zonal jets and non-zonal coherent structures are unstable.  (b)-(e) S3T growth rates, $\s_r$, as a function of the eigenfunction wavevector $\nv=(n_x,n_y)$ for the four cases marked in (a). The thick line corresponds to the $\s_r=0$ contour. For (b), (d), (e) the contour interval is 0.15 while in (c) the contour interval is 0.5. The dashed line marks $n=1$ and corresponds to the forcing scale.}
\end{figure}

The critical energy input rates, $\ecz$ and $\ecnz$ as a function of $\b$ for isotropic forcing ($\mu=0$) is shown in Fig.~\ref{fig:s3tgr_isotropic}\hyperref[fig:s3tgr_isotropic]{a}. For $\b<3.5$ the structures that become first unstable are zonal jets ($\ecz<\ecnz$) and for supercritical energy input rates always zonal jets are more unstable than non-zonal perturbations. For $\b>3.5$ non-zonal structures become first unstable and for a range of energy input rates $\ecnz<\e<\ecz$ only them are unstable. For $\e>\ecz$ zonal jets become unstable but with less growth rates compared to non-zonal structures. For $\b<3.5$ zonal jet eigenfunctions grow the most whereas for $\b>3.5$ non-zonal structures grow the most. In the light shaded region only non-zonal coherent structures are unstable, while in the dark shaded region both zonal jets and non-zonal coherent structures are unstable. Growth rates, $\s_r$, as a function of the eigenfunction wavevector $\nv=(n_x,n_y)$ for 4 different choices of $\b$ and $\e$ are shown in Figs.~\ref{fig:s3tgr_isotropic}\hyperref[fig:s3tgr_isotropic]{b-e}.

\begin{figure}
\centering
%\hspace{.05in}\includegraphics[width=3.05in]{book_emin_stars_a.pdf}\\
%\vspace{-.1in}
\includegraphics[width=4.25in]{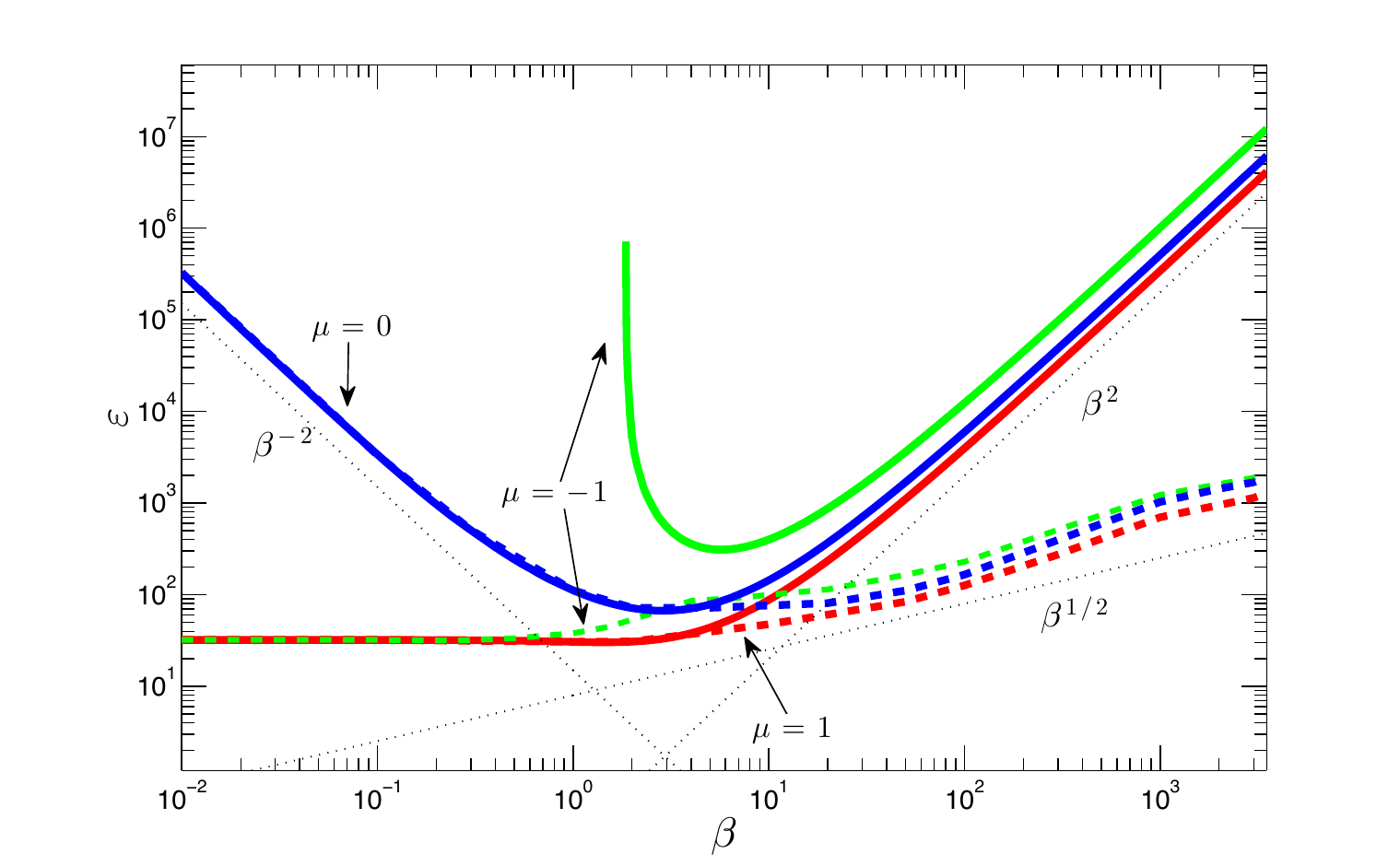}
\caption{\label{fig:ec_RFg_z_nz}  The critical energy input rates $\ecz$ (solid) and $\ecnz$ (dashed) that render the homogeneous equilibrium unstable to zonal jet perturbations  ($\nv=(0,n_y)$) and non-zonal perturbations ($\nv=(n_x,n_y)$) respectively as a function of  $\b$. Shown are the  $\ecz$ and $\ecnz$ for the three forcing covariance spectra seen in Fig.~\ref{fig:RF}. Also shown are the slopes $\b^{-2}$, $\b^2$ and $\b^{1/2}$ (dotted line). For $\b\gtrsim4$ the equilibrium first becomes unstable to non-zonal perturbations regardless of $\mu$. For $\mu=-1$ zonal jet perturbations are unstable only for $\b\gtrsim 1.8$.}
\end{figure}

The $\ecz$ and $\ecnz$ for both isotropic as well as anisotropic forcing are shown in Fig.~\ref{fig:ec_RFg_z_nz}. This figure shows that the homogeneous equilibrium becomes unstable for all values of $\b>0$. The homogeneous equilibrium becomes also unstable even for $\b=0$, unless the excitation is exactly isotropic (cf.~Appendix~\ref{appsec:eigenvalue}). This is an important result because it shows that the  dynamics that lead to the initial emergence of large-scale structure does not require the presence of $\b$. We show in the next sections that for isotropic forcing both $\ecz$ and $\ecnz$ increase as $\b^{-2}$ as $\b\rightarrow 0$, but for anisotropic forcing  $\e_c=32/|\mu| + \Ocal(\b^2)$ for small $\b$. The homogeneous equilibrium is rendered unstable with the least $\e$ in the range $1\lesssim\b\lesssim10$. For $\b \gtrsim 4$ the equilibrium becomes first unstable to non-zonal perturbations. As $\b$ increases the homogeneous equilibrium becomes more stable and larger $\e$ is required to destabilize it. It is shown that zonal jet emergence requires $\ecz  \sim\b^2$ as $\b\to\infty$, which means that the effective feedback on the mean flow falls as $\real[f(\s)]\sim \b^{-2}$ as $\b\to\infty$, but for the emergence of non-zonal structure $\ecnz  \sim\b^{1/2}$ as $\b\to \infty$ ($\real[f(\s)]\sim \b^{-1/2}$) because of the occurrence  of fortuitous resonances that are explained in the next sections. The asymptotic behavior of  $\ecz$ and $\ecnz$ for large $\b$ is independent of the forcing spectrum. 

%in the expressions~\eqref{eq:fr}-\eqref{eq:defF}. The wavevector $\nv=(n\sin\varphi,n\cos\varphi)$ forms angle $\varphi$ with the planetary vorticity gradient vector, $\bv=(0,\b)$, which is in the direction of the meridional. All spectral components of the forcing prescribed by~\eqref{eq:Qhat_deltaG} have total non-dimensional wavenumber $k=1$ and are taken as $\kv=(\cos\thet,\sin\thet)$, where $\thet$ is the complementary angle between wavevectors $\kv$ and $\nv$. The angle $\phi=\arctan(k_y/k_x)$ is $\phi=\thet-\phi$.

\section{Eddy--mean flow dynamics underlying the S3T instability of homogeneous turbulent equilibrium}

%The equilibrium~\eqref{eq:ZeCe} is unstable when the growthrate $\s_r>0$. From~\eqref{eq:sigma_nz} we have that:\begin{subequations}
%\begin{align}
%\s_r &= -1 + \varepsilon\,\real[f(\s)]\ ,\\
%\s_i &=  \varepsilon\,\imag[f(\s)]\ ,
%\end{align}\end{subequations}
%so instability occurs only if the real part of the eddy feedback factor $\real(f)$ is positive. If this condition is satisfied, the induced vorticity fluxes are upgradient and the homogeneous equilibrium becomes unstable for $\varepsilon > \varepsilon_c\equiv 1/\real(f)$.

%a function of parameters for excitation amplitudes at the stability boundary $\e=\e_c$. Therefore we set $\s_r = 0$ and $\s_i=0$ (or equivalently $s_i=-\om_\nv$).\footnote{While the phase speed of the marginally unstable non-zonal structures almost matches the corresponding Rossby phase speed for $\b\gg 1$ it overestimates the Rossby phase speed by almost by a factor of 2 when $\b \sim \Ocal(1)$ or smaller. However, at these values of $\b$ we have found that the results presented in this work are not sensitive to the value of the frequency.}  

\begin{figure}
\centering
\includegraphics[width=3in,trim = 10mm 10mm 10mm 5mm, clip]{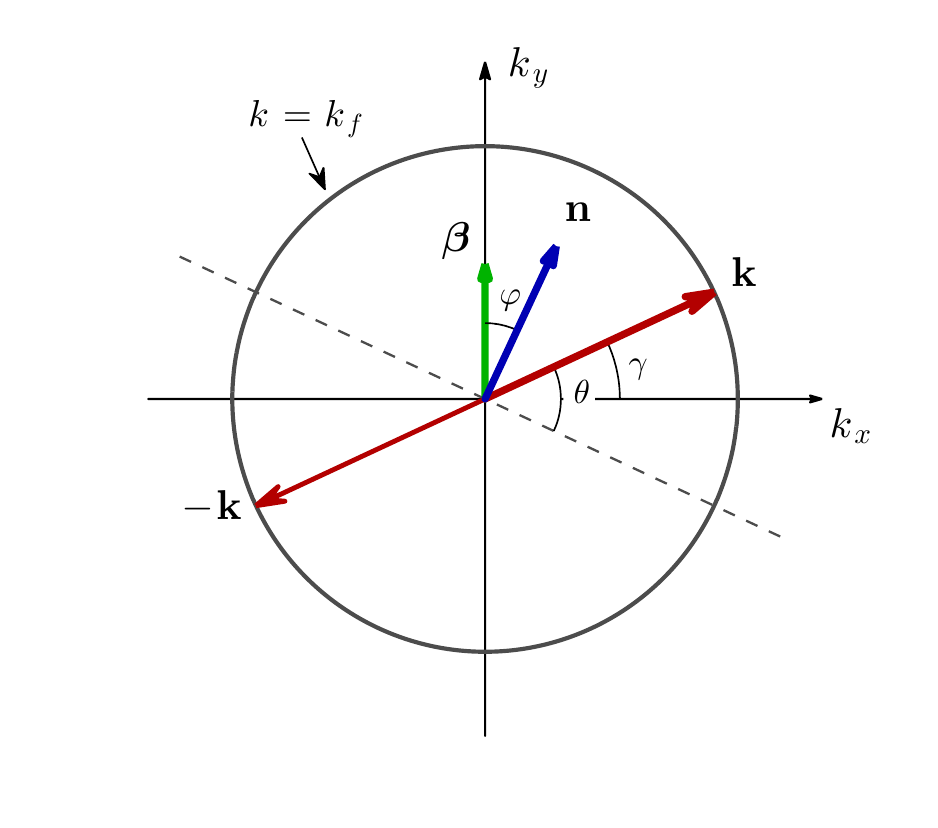}
\caption{\label{fig:einx_angles}  A non-zonal plane wave perturbation with   wavevector $\nv$
at an angle  $\varphi$ to the northward direction (the direction of $\bv$)
becomes  a zonal perturbation when the coordinate frame is rotated clockwise by  $\varphi$.  Under this rotation the components of the wavevector  $\kv=(\cos\gamma,\sin\gamma)$ are transformed
to $\kv=(\cos\thet,\sin\thet)$, with $\thet=\gamma+\varphi$.
$\Fcal(n,\thet)$ in~\eqref{eq:fr} is the mean momentum flux convergence from plane wave  perturbations that arise from excitations with wavevectors $\kv$ and $-\kv$.}
\end{figure}

In order to analyze the dynamics underlying the S3T instability we study the behavior of $\real(f)$ at the critical $\e$ at which the eigenfunction with wavevector $\nv$ becomes neutral and set $\s_r=0$ and $\s_i=0$ in~\eqref{eq:f_factor} and \eqref{eq:s3tgr_z_nz}.\footnote{That $\s_i=0$ or equivalently $s_i=-\om_\nv$ for all wavevectors $\nv$ at the stability boundary is an approximation but it can be shown that is a valid approximation.} We denote the eddy feedback on the mean flow perturbation with wavenumber $\nv$ in this approximation as $\fr\equiv\real(f(0))$. The eddy feedback for this delta function forcing~\eqref{eq:Qhat_deltaG} can be written as
\begin{align}
\fr &=\int\limits_{0}^{\pi}  \Fcal (\thet, \nv)\,\df\thet\ ,\label{eq:fr}
\end{align} 
where $\Fcal(\thet,\nv)$ is the contribution to $\fr$ from the individual forcing components of $Q$ corresponding to wavenumbers $\kv$ and $-\kv$. For the narrow ring forcing~\eqref{eq:Qhat_deltaG} all forcing components have $k=1$ and are only characterized by angle $\theta$, that is subtended measured from the lines of constant phase of the eigenfunction $\nv$ (see Fig.~\ref{fig:einx_angles}). %%It can be shown that for the spectrum~\eqref{eq:Qhat_deltaG} the eddy feedback $\fr$ may be written as:
%where $\Fcal(\thet,\nv)$ is the contribution to $\fr$  from the individual Fourier components of the forcing spectrum. 
%These components of the forcing spectrum can be also interpreted as waves excited by the stochastic forcing.
%For~\eqref{eq:Qhat_deltaG} these waves all have total wavenumber $k=1$ and are characterized only by an angle $\thet$. We will take the angle $\thet$ to be the angle formed by their phase lines and wavevector $\nv$. 
We also write $\nv=(n\sin\varphi,n\cos\varphi)$ so that zonal jet eigenfunctions correspond to $\varphi=0\deg$, while non-zonal eigenfunctions to $\varphi\ne0\deg$. The angle $\gamma=\arctan(k_y/k_x)$ is given as $\gamma=\thet-\varphi$. The relation between angles $\thet$, $\varphi$ and $\gamma$ is shown in Fig.~\ref{fig:einx_angles}. We can isolate the dependence of this eddy feedback on $\b$ by writing as $\Fcal(\thet,\nv) =\bit F(\thet,\nv)+F(180\deg+\thet,\nv)$ with
\be
F(\thet,\nv)=\frac{\Ncal\,\Dcal_0}{ \Dcal_0^2 + \b^2\,\Db^2 }\ ,\label{eq:defF}
\ee
where, as shown in Appendix~\ref{appsec:fr_theta_phi_n}, functions $\Ncal$, $\Dcal_0$ and $\Db$ do not depend on $\b$. $\Fcal$ measures the feedback on the mean flow from two monochromatic excitations with wavenumbers $\kv$ and $-\kv$ (see Fig.~\ref{fig:einx_angles}). We wish to determine the $\thet$ that produce positive feedback to eigenfunction $\nv$ and contribute to the instability of $\nv$.

In the following sections we determine the contribution of the various waves to the eddy feedback and identify the angles $\thet$ that produces the most significant contribution to this  feedback. We also calculate the eddy feedback $\fr$ as a function of the total mean flow wavenumber $n$ for $0 \le \varphi \le 90\deg$. We limit our discussion to the emergence of mean flows with $n<1$, i.e., with scale larger than the scale of the forcing. (Remember that all wavenumbers are non-dimensionalized with the forcing wavenumber $k_f$.) In section \ref{sec:iso} the analysis is mostly focused to isotropic forcing ($\Gcal=1$) while the effect of  anisotropy is discussed in section~\ref{sec:anisotr}.

\section{\label{sec:iso}Eddy--mean flow dynamics leading to formation of zonal and
non-zonal structures for isotropic forcing}

\subsection{Induced vorticity fluxes when $\b\ll1$\label{subsec:b0}}

\begin{figure}
\centering\includegraphics[width=3.8in]{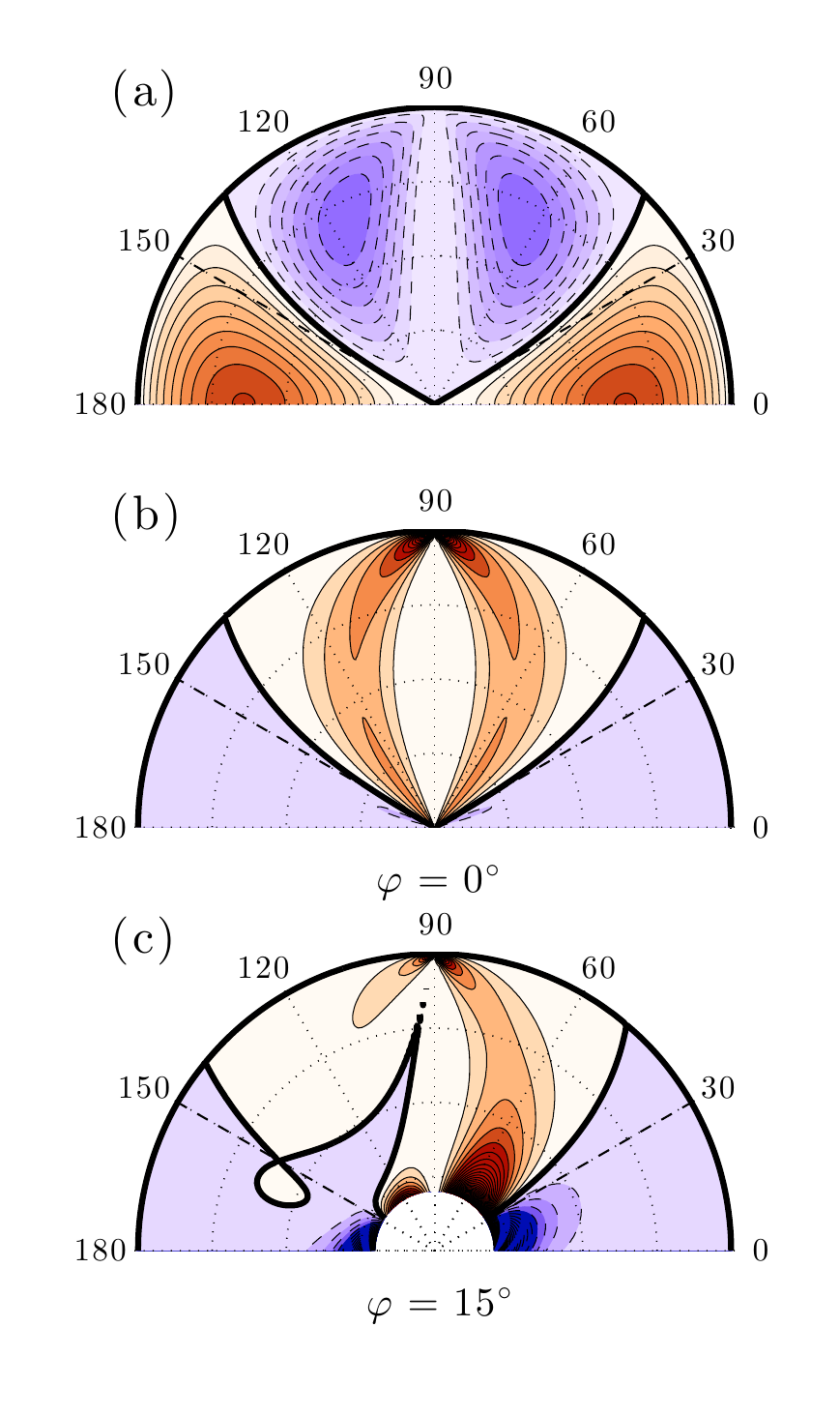}
\vspace{-1em}
\caption{\label{fig:smallb_F} (a) Contours of $\Fcal_0(\thet,n)$ in a $(\thet,n)$ polar plot ($n$ radial and $\thet$ azimuthal). This figure shows the magnitude and sign of the vorticity flux  induced by  waves with phase lines oriented at an angle $\thet$ to the $y$ axis in the presence of an infinitesimal mean flow perturbation of total wavenumber $n$ when $\b=0$. The contour interval is $3\times 10^{-3}$ and note that $\Fcal_0(\thet,n)$ is independent of $\varphi$. (b) Contours of the normalized $\Fcal_2(\thet,n)/n^4$ show the $\mathcal{O}(\b^2)$ correction to  $\Fcal_0(\thet,n)$  for the case of zonal jet perturbations ($\varphi=0\deg$). The contour interval is 0.02.  (c) Same as (b) but for non-zonal perturbations with $\varphi=15\deg$. The contour interval is 0.04. In all panels the forcing is isotropic ($\mu=0$), solid (dashed) lines indicate contours with positive (negative) values, the thick line  is the zero contour, the radial grid interval is $\Delta n=0.25$ and the $30\deg$ wedge is marked (dashed-dot). In panels (a) and (b) the zero contour is the curve $4\sin^2\thet = 1+n^2$ (see~Appendix~\ref{app:fR}).}
\end{figure}

We expand the integrand $\Fcal$ of~\eqref{eq:fr} in powers of
$\b$:
\be
\Fcal=\Fcal_{0} + \b^2\, \Fcal_2 + \Ocal(\b^4)\ ,
\ee
with $\Fcal_2 = \frac1{2}\left . \bit \partial^2_{\b\b} \Fcal \right|_{\b=0}$. The leading order term, $\Fcal_{0}$, is the contribution of each wave
with wavevector $\kv=(\cos \thet,\sin \thet)$ to the eddy feedback
 in the absence of $\b$ and is shown in
Fig.~\ref{fig:smallb_F}\hyperref[fig:smallb_F]{a}. For $\b=0$, the
dynamics are rotationally symmetric and for isotropic forcing $\fr$ is
independent of $\varphi$. Therefore all zonal and non-zonal
eigenfunctions with the same total wavenumber, $n$, grow at the same rate.
Upgradient fluxes ($\Fcal_0>0$) to a mean flow with wavenumber $n$ are
induced by waves with phase lines inclined at angles satisfying
$4\sin^2\thet < 1+n^2$ (cf.~Appendix~\ref{app:fR}).
This implies that all waves with $|\thet| < 30 \deg$ necessarily produce
upgradient vorticity fluxes to any mean flow with wavenumber $n<1$,
while waves with $30\deg<|\thet|<45\deg$ produce upgradient fluxes for any mean
flow with large enough wavenumber (cf.~Fig.~\ref{fig:smallb_F}\hyperref[fig:smallb_F]{a}). The eddy--mean flow dynamics was investigated in the limit of $n\ll1$ by~\citet{Bakas-Ioannou-2013-jas}. 
It was shown that the vorticity fluxes can be calculated from time averaging the fluxes over the life
cycle of an ensemble of localized stochastically forced wavepackets
initially located at different latitudes. For $n\ll 1$, the wavepackets evolve
in the region of their excitation under the influence of the
infinitesimal local shear of $\delta U$ and are rapidly dissipated
before they shear over. As a result, their effect on the mean flow is
dictated by the instantaneous (with respect to the shear time scale)
change in their momentum fluxes. Any pair of wavepackets having a
central wavevector with phase lines forming angles $|\thet|<30\deg$ with
 the $y$ axis surrender instantaneously momentum to the mean flow and
reinforce it, whereas pairs with $|\thet|>30\deg$ gain instantaneously
momentum from the mean flow and oppose jet formation. Therefore,
anisotropic forcing that injects significant power into Fourier
components with $|\thet|<30\deg$ (such as the forcing from baroclinic
instability that primarily excites Fourier components with $\thet=0\deg$)
produces robustly upgradient fluxes that asymptotically behave
anti-diffusively. That is, for a sinusoidal mean flow perturbation $\dU = \sin{( n y
)}$ we have $\int_0^\pi\Fcal_0\,\df\thet = \kappa n^2 $ with
$\kappa$ positive and proportional to the anisotropy factor $\mu$ 
(cf.~Appendix~\ref{app:fR}).

\begin{figure}
\centering\includegraphics[width=3in]{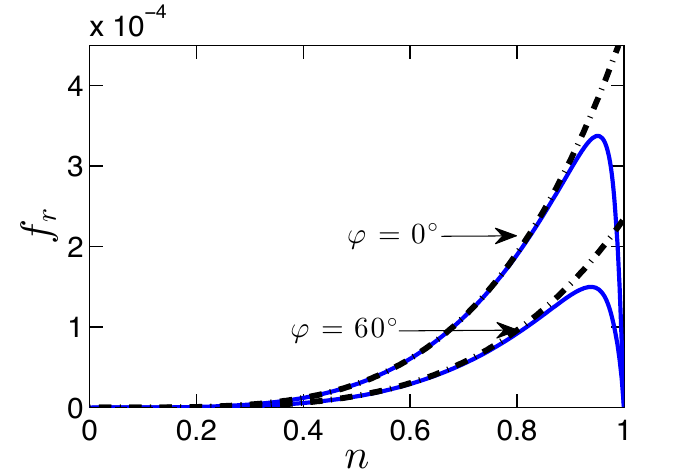}
\caption{\label{fig:vq_n_phi0_phi60_b0p1} Eddy feedback factor $\fr$ as a function
of $n$ for $\b=0.1$ and isotropic forcing. Note that the fluxes are upgradient (i.e. $\fr>0$) for all mean flow wavenumbers $n$. Shown is $\fr$ for $\varphi=0\deg$ and $\varphi=60\deg$ (solid lines), as well as the asymptotic expression~\eqref{eq:fr_bll1_mu0} (dash-dot) derived for the feedback factor in the limit $\b\ll 1$ and $\b/n\ll 1$.}
\end{figure}

For isotropic forcing the net vorticity flux produced by shearing of
the perturbations vanishes, i.e., $\int_0^\pi\Fcal_0\,\df\thet=0$, given
that the upgradient fluxes produced by waves with $|\thet|<30\deg$
exactly balance the downgradient fluxes produced by the waves with
$|\thet|>30\deg$. However, a net vorticity flux feedback is produced
and asymptotically behaves as a negative fourth-order hyperdiffusion
with coefficient $\Ocal(\b^2)$ for $\b\ll 1$ (cf.~\eqref{eq:fr_bll1_mu0} and \citet{Bakas-Ioannou-2013-jas}). In Appendix~\ref{app:fR} it is shown that the feedback factor $\fr$ for isotropic forcing in the limit $\b\ll1$ with
$\b/n\ll1$ is:
\begin{align} \fr &= \b^2 \frac{n^4}{64}
\[\bit2+\cos(2\varphi)\]+\Ocal(\b^4)\ ,\label{eq:fr_bll1_mu0}
\end{align}
which is accurate even up to $n\approx 1$, as shown in
Fig.~\ref{fig:vq_n_phi0_phi60_b0p1}. In order to understand the contribution of $\beta$ to the vorticity flux feedback, we plot $\Fcal_2 / n^4 $ for a zonal (Fig.~\ref{fig:smallb_F}\hyperref[fig:smallb_F]{b}) and a non-zonal perturbation (Fig.~\ref{fig:smallb_F}\hyperref[fig:smallb_F]{c}) as a function of the mean flow wavenumber $n$ and wave angle $\thet$. We choose to scale $\Fcal_2$ by $n^4$ because in~\eqref{eq:fr_bll1_mu0} $f_r$
increases as $n^4$. Consider first the case of a zonal jet. It can be
seen that at every point, $\Fcal_2$ has the opposite sign to $\Fcal_0$,
implying that $\b$ tempers both the upgradient (for roughly $|\thet|<30\deg$)
and the downgradient (for $|\thet|>30\deg$) fluxes of $\Fcal_0$. However, in
the sector $|\thet|>30\deg$ the values of $\Fcal_2$ are much larger than
in the sector $|\thet|<30\deg$ and  the net  fluxes
integrated over all angles  are upgradient, as in~\eqref{eq:fr_bll1_mu0} for  the isotropic case.

The asymptotic analysis of \textcite{Bakas-Ioannou-2013-jas}, which is
formally valid for $n\ll1$, offers understanding of the dynamics that
lead to the inequality $\Fcal_2 \Fcal_0<0$ and to the positive net
contribution of $\Fcal_2$, i.e., to $ \int_0^\pi\Fcal_2 \,\df \thet>0$.
Any pair of wavepackets with wavevectors at angles $|\thet|>30\deg$
instantaneously gain momentum from the mean flow as described above
(i.e. $\Fcal_0<0$ for $|\thet|>30\deg$), but their group velocity is
also increased (decreased) while propagating northward (southward). This 
occurs due to the fact that shearing changes their meridional wavenumber 
and consequently their group velocity. The instantaneous
change in the momentum fluxes resulting from this speed up (slowing down) 
of the wavepackets is positive in the region of excitation leading to
upgradient fluxes ($\Fcal_2>0$). The opposite happens for pairs with
$|\thet|<30\deg$ (cf.~Fig.~3 of~\citet{Bakas-Ioannou-2013-jas}), however the
downgradient fluxes produced are smaller than the upgradient fluxes, leading to a net positive contribution when integrated over all angles. Figure~\ref{fig:smallb_F}\hyperref[fig:smallb_F]{b}, shows that this result is valid for larger mean flow wavenumbers as well.

Consider now the case of a non-zonal perturbation (Fig.~\ref{fig:smallb_F}\hyperref[fig:smallb_F]{c}). We  observe that the angles for which the waves have significant positive or negative
contributions to the vorticity flux feedback are roughly the same as in
the case of zonal jets. In addition, the vorticity flux feedback factor
decreases with the angle $\varphi$ of the non-zonal
 perturbations (cf.~\eqref{eq:fr_bll1_mu0}). As a result, zonal jet perturbations
always produce larger vorticity fluxes compared to non-zonal
perturbations and are therefore the most unstable in the limit $\b\ll1$.
Additionally, these results show that for $\b\ll1$, the mechanism for
structural instability of the non-zonal structures is the same as the
mechanism for zonal jet formation, which is shearing of the eddies by
the infinitesimal mean flow.

\subsection{Induced vorticity fluxes when $\b\gg1$ \label{sec:largeb_limit}}

When $\b\gg1$ by inspecting~\eqref{eq:defF} we expect that $\fr$ should fall as $\b^{-2}$. This indeed is the case, as we will demonstrate, for zonal jet perturbations. However, non-zonal perturbations may render $\Db=0$ and in that case, as we will show, the eddy feedback $\fr$ again falls, but as $\b^{-1}$ or for some special non-zonal perturbations even as $\b^{-1/2}$.

\begin{figure}
\centering\includegraphics[width=3.5in]{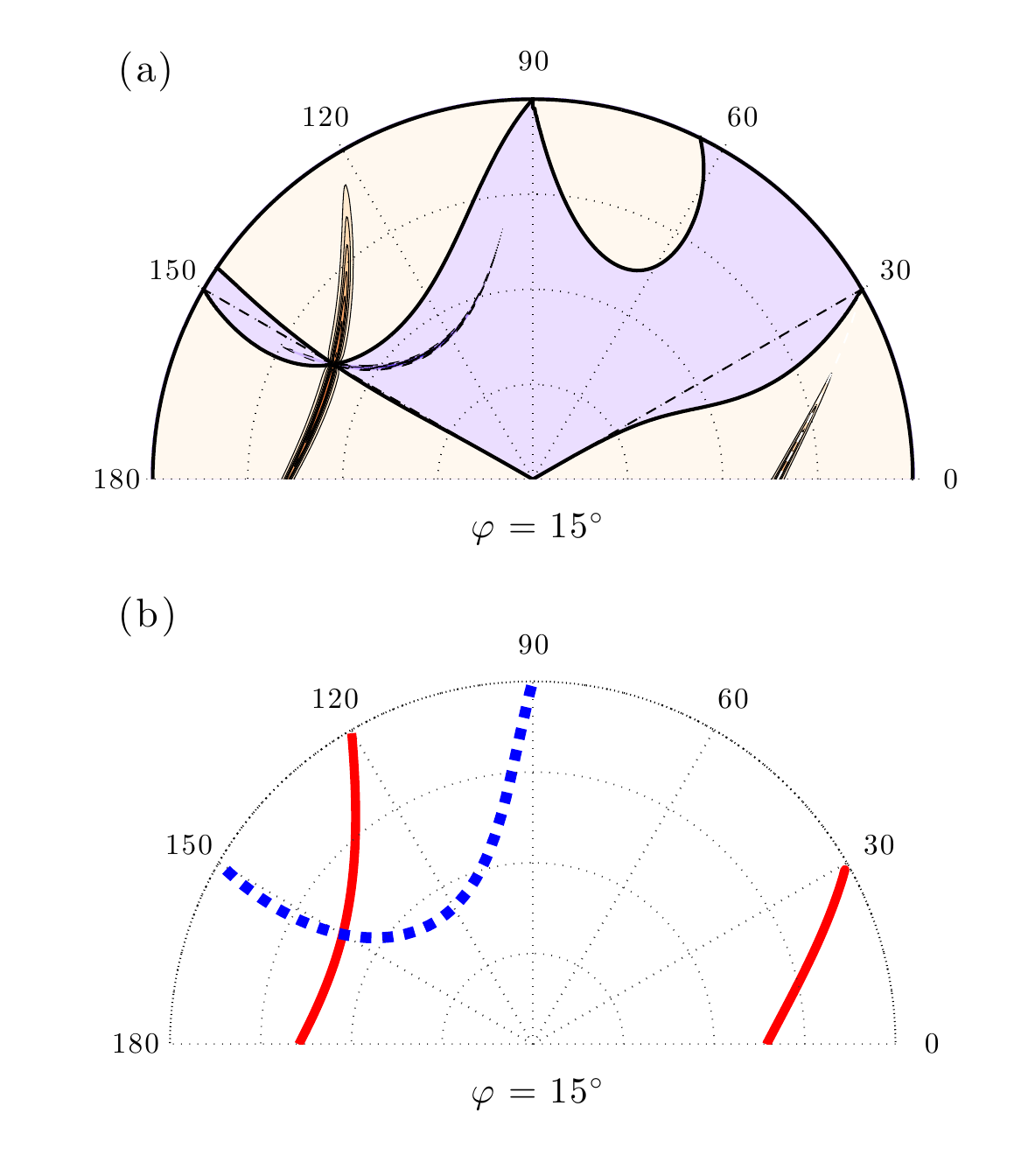}
\caption{\label{fig:Fr_Db} (a) Contours of $\Fcal(\thet,n)$ in a $(\thet,n)$ polar plot ($n$ radial and $\thet$ azimuthal) for isotropic forcing and $\b=200$. This panel shows the vorticity fluxes induced by waves with phase lines oriented at an angle $\thet$ in the presence of a non-zonal perturbation with  mean flow wavenumber $n$ and $\varphi=15\deg$. Solid (dashed) lines indicate contours with positive (negative) values, the contour interval is $2.5\times10^{-3}$ and the thick line is the zero contour. (b) Locus of the roots of  $\Db(\thet,n)$ on the $(\thet,n)$ plane for non-zonal perturbations with $\varphi=15\deg$. The roots correspond to resonant interaction  between waves with phase lines oriented at an angle $\thet$ and non-zonal perturbations with mean flow wavenumber $n$. Thick solid (dashed) lines indicate whether the vorticity fluxes produced by the resonant waves are upgradient (downgradient). The radial grid interval in both panels is $\Delta n=0.25$.}
\end{figure}

Consider first the emergence of non-zonal structures in the limit
$\b\gg1$. The contribution of each Fourier component of the forcing to
the vorticity flux feedback $\Fcal$ for the case of non-zonal structures
at $\b = 200$ is shown in Fig.~\ref{fig:Fr_Db}\hyperref[fig:Fr_Db]{a}.
In contrast to the cases with $\b\ll1$ (or $\b=\Ocal(1)$, discussed in
section~\ref{subsec:O1}), there is only a small band of
Fourier components that contribute significantly to the vorticity flux
feedback, as indicated with the narrow tongues in
Fig.~\ref{fig:Fr_Db}\hyperref[fig:Fr_Db]{a}. The reason for this
selectivity in the response is that for $\b\gg1$ the components that
produce appreciable fluxes, as seen from~\eqref{eq:defF}, are
concentrated on the $(\thet,n)$ curves that satisfy $\Db=0$ (shown in
Fig.~\ref{fig:Fr_Db}\hyperref[fig:Fr_Db]{b}) or equivalently for the
$(\thet,n)$ that satisfy the resonant condition
$\om_\kv+\om_\nv=\om_{\kv+\nv}$ (cf.~\eqref{eq:D2zero}). This is the
resonant condition satisfied when a Rossby wave with wavevector $\kv$
and frequency $\om_\kv$ forms a resonant triad with the non-zonal
structure with wavevector $\nv$ and frequency $\om_\nv$.  We concentrate
our analysis to these ``resonant contributions''  because  they  dominate the
 eddy feedback of non-zonal perturbations for $\b \gg 1$.

\begin{figure}
\centering\includegraphics[width=.97\textwidth,trim = 15mm 0mm 15mm 0mm, clip]{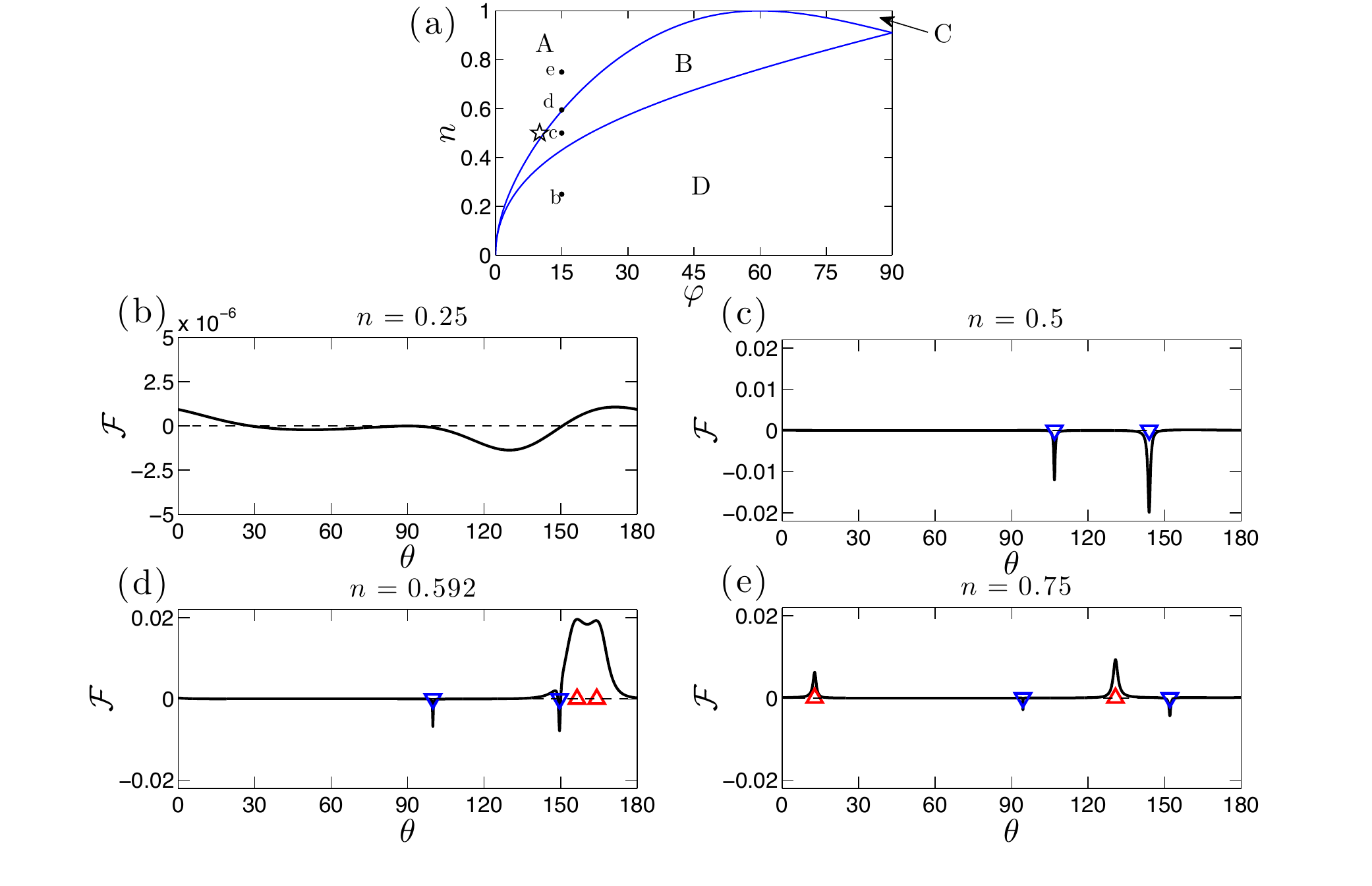}
\caption{\label{fig:R_Fcal} (a) The curves separating the regions in the $(n,\varphi)$ plane for which $\Db$ has no roots (region~D), 2 roots (region~B) and four roots (regions A and C). Waves with $\thet$ corresponding to two out of the four roots of $\Db$ found in region~A produce upgradient fluxes. (b)-(d) The vorticity fluxes $\FcalR$ as a function of the angle $\thet$ subtended by the phase lines of the waves and the $y$ axis in the presence of a non-zonal perturbation with $\varphi=15\deg$ at $\b=200$. The mean flow wavenumber is (b) $n=0.25$ (in region~D), (c) $n=0.5$ (in region~B), (d) $n=0.592$ (in region~A) and (e) $n=0.75$ (in region~A). The resonant angles (i.e. the roots of $\Db$) are marked by upper (lower) triangles when the waves induce upgradient (downgradient) fluxes. Note that the scale in (b) is much smaller.}
\end{figure}

Resonant triads do not occur for all mean flow perturbations $\nv$. For
$(n, \varphi)$ in region~D of Fig.~\ref{fig:R_Fcal}\hyperref[fig:R_Fcal]{a}, $\Db$ has no roots and
therefore there are no Fourier components with
$\kv=(\cos\thet,\sin\thet)$ that form a resonant triad with the mean
flow perturbation $\nv$ and the eddy feedback is determined
by the sum over the non-resonant contributions as illustrated in
Fig.~\ref{fig:R_Fcal}\hyperref[fig:R_Fcal]{b}. In region~B of
Fig.~\ref{fig:R_Fcal}\hyperref[fig:R_Fcal]{a}, there are only two
resonant angles $\thet$. The resonant and non-resonant contribution for
a typical case in region~B is shown in
Fig.~\ref{fig:R_Fcal}\hyperref[fig:R_Fcal]{c}. Note that it is the
resonant contributions that determine the eddy feedback.
However, they produce a negative eddy feedback (a
downgradient tendency), which is stabilizing, a result that holds for
all $(n,\varphi)$ in region~B. In regions A and C, there exist four
resonant angles $\thet$ which dominate the vorticity flux. In C all
resonant contributions are stabilizing and therefore C is also a stable
region. In region~A, which at most extends to $\varphi=60\deg$ (cf.~Appendix~\ref{app:fR}),
two of the four resonances give positive contributions to $\fr$
(cf.~Figs.~\ref{fig:R_Fcal}\hyperref[fig:R_Fcal]{d,e}). Therefore only
for $(n,\varphi)$ in region~A, does a destabilizing eddy feedback occur. The largest destabilizing feedback occurs when the positively contributing resonances are near coalescence (i.e.~as in Fig.~\ref{fig:R_Fcal}\hyperref[fig:R_Fcal]{d}), which occurs for $(n,\varphi)$ close to the curve separating regions A and B. The reason is that when the
resonances are apart, as in Figs.~\ref{fig:R_Fcal}\hyperref[fig:R_Fcal]{c,e}, the 
significant contributions come from near-resonant waves with angles within a
band of $\Ocal(1/\b)$ around the resonant angles and the integrated resonant
contributions to the vorticity flux are $\Ocal(1/\b)$. However, when the
resonances are near coalescence, as for the case shown in Fig.~\ref{fig:R_Fcal}\hyperref[fig:R_Fcal]{d}, the band of near-resonant waves contributing significantly increases
as the integrand assumes a double humped shape and, as shown in Appendix~\ref{app:fR}, the
destabilizing vorticity flux feedback becomes $\Ocal(1/\sqrt{\b})$. Note that as $\b\to\infty$, the width over which we have significant contributions
diminishes and therefore $\fr\to0$ unless an infinite amount of energy is injected
exactly at the resonant angles (as is assumed in modulational instability studies).

\begin{figure}
\centering\includegraphics[width=3in]{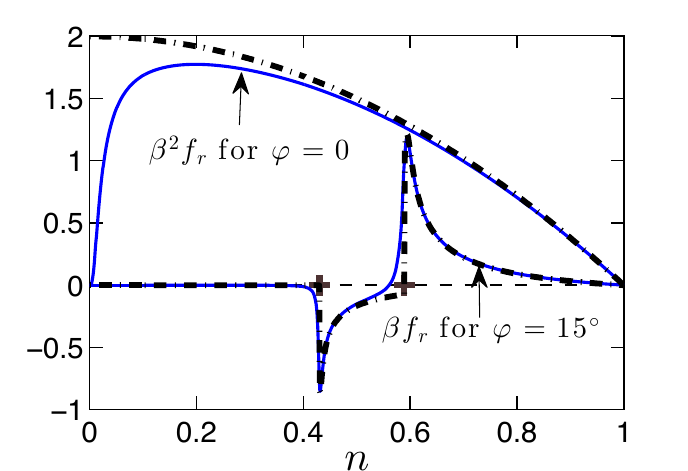}
\caption{\label{fig:fr_n_b200} Eddy feedback $\fr$ as a function of $n$ for $\b=200$. Positive (negative) values correspond to upgradient (downgradient) fluxes. Shown is $\fr$ for $\varphi=0\deg$ (multiplied by $\beta^2$) and for $\varphi=15\deg$ (multiplied by $\beta$). Also the asymptotic expressions~\eqref{eq:dvzR_largeb} for $\varphi =0\deg$ and~\eqref{eq:frR} for $\varphi=15\deg$ are shown (dash-dot). The crosses mark the mean flow wavenumbers $n=0.43$ and $n=0.59$ that separate regions A, B and D in Fig.~\ref{fig:R_Fcal}\hyperref[fig:R_Fcal]{a} for $\varphi=15\deg$.}
\end{figure}

It can be shown (cf.~Appendix~\ref{app:fR}) that the resonant contribution for $\b\gg 1$ asymptotically approaches
\be
\fr^{(\textrm{R})} = \frac{1}{\sqrt{\b}}\sum_{j=1}^{N_r}\frac{\pi \, \Ncal_j\,\eta_j}{ 2\,\Dcal_{0,j}^{1/2}|\rho_j|^{1/2}}\ ,\label{eq:frR}
\ee
where the subscript $j$ indicates the value of the functions at
 the $j$-th out of the $N_r$ roots of $\Db$ and $\rho =
\partial^2_{\thet\thet}\Db$. The values $\Ncal_j$, $\Dcal_{0,j}$,
$\rho_j$ are all $\Ocal(1)$, whereas $\eta_j$ is always positive and the
only quantity that has dependence on $\b$. It is $\Ocal(1)$ only for $(n,\varphi)$
just above the separating boundaries of regions A and B and regions B and D in
Fig.~\ref{fig:R_Fcal}\hyperref[fig:R_Fcal]{a} yielding
$\fr^{(\textrm{R})}\sim 1/\sqrt{\b}$ and is $\Ocal(1/\sqrt{\b})$
elsewhere yielding $\fr^{(\textrm{R})}\sim 1/\b$, as also qualitatively
described above. The sign of the $j$-th resonant contribution to the
total eddy feedback depends only on the sign of $\Ncal_j$. For $(n,\varphi)$ just above the
boundary separating regions~B and~D, $\Ncal_j<0$ and $\fr$ attains its
minimum value, which corresponds to the largest stabilizing tendency.
This is illustrated in Fig.~\ref{fig:fr_n_b200}, showing the eddy feedback $\fr$ as a function of $n$.
 For $(n,\varphi)$ just above the boundary separating regions A and B, coalescence of the two positive contributing resonances occurs and $\fr$ attains its maximum value,
which corresponds to the largest destabilizing tendency. For small mean
flow wavenumbers $n$ (corresponding to region~D) the eddy feedback is
negative and $\Ocal(\b^{-2})$ due to the absence of resonant
contributions.

%\footnote{{\color{Red}Note that as can be seen from~\eqref{eq:defF} and
%discussed in \hyperref[sec:asymptotics]{Appendix~\ref{app:fR}}, the effect of the
%resonances starts appearing for $\thet$ satisfying $|\Db|\lesssim
%1/\sqrt{\b}$. As a result, the boundaries separating resonant from
%non-resonant regions in Fig.~\ref{fig:R_Fcal}\hyperref[fig:R_Fcal]{a}
%that are derived for the $\b\to+\infty$ limit, have a finite width
%(proportional to $1/\sqrt{\b}$) for finite values of $\beta$, as shown
%in Fig~\ref{fig:fr_n_b200}.%. For example: $\fr$ for $\varphi=15\deg$
%and $\beta=200$ attains positive values for $0.56<n<0.59$ which are
%mean flow wavenumbers in region~B or $\fr$ attains negative values of
%$\Ocal(1/\b)$ for $0.41<n<0.43$ which are in region~D (cf.
%Fig~\ref{fig:fr_n_b200}).}}. Just above the boundary separating regions
%B and D we have the maximum negative $\fr$, while just above the
%boundary of A and B we have the maximum positive $\fr$.

\begin{figure}
\centering\includegraphics[width=3.5in]{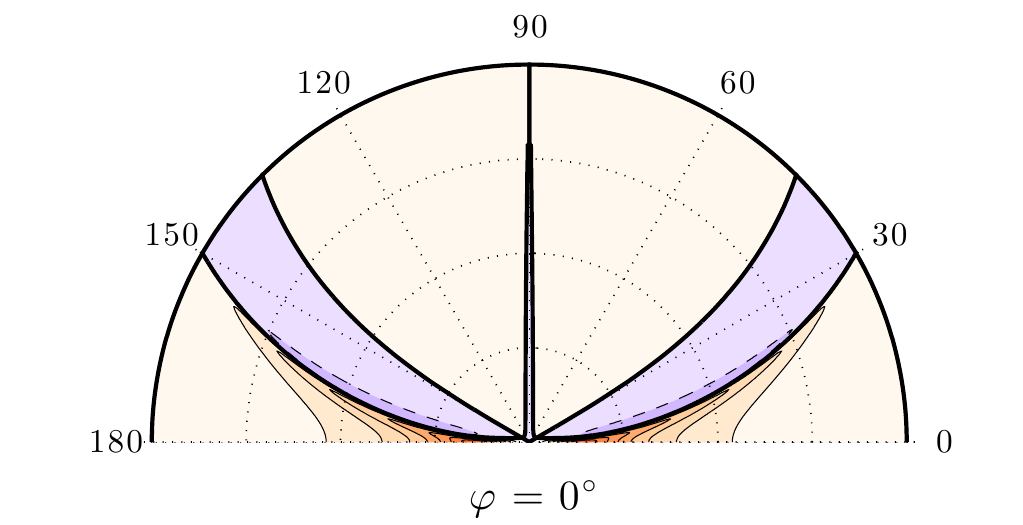}
\caption{\label{fig:F_z_b200} Contours of $\FcalR(\thet,n)$ in a $(\thet,n)$ polar plot
($n$ radial and $\thet$ azimuthal) for zonal jet perturbations ($\varphi=0\deg$) and $\b=100$. Solid (dashed) lines indicate contours with positive (negative) values, the contour interval is $2\times10^{-4}$, the thick line is the zero contour and the radial grid interval is $\Delta n=0.25$.}
\end{figure}

An interesting exception to the results discussed above occurs for the
important case of zonal jet perturbations ($\varphi=0\deg$). In that case,
$\Ncal_j=0$ in~\eqref{eq:frR} as the roots of $\Db$ and $\Ncal$ coincide
and the resonant contribution~\eqref{eq:frR} is exactly zero. As shown
in Fig.~\ref{fig:F_z_b200}, positive vorticity flux feedback is obtained from a
broad band of the non-resonant Fourier components with $\gamma=\thet \approx
0\deg$, corresponding to waves with lines of constant phase nearly
aligned with the $y$ axis (remember that for smaller $\b$ the region
that produces destabilizing fluxes extends up to $|\thet| \approx
30\deg$). For large $\b$ the vorticity flux $\fr$ is always
destabilizing for all zonal jet perturbations with $n<1$, as shown by
\eqref{eq:dvzR_largeb} and Fig.~\ref{fig:fr_n_b200}, and the largest
destabilizing vorticity flux, $\f_{r,\textrm{max}}=(2+\mu) \b^{-2}$, is
obtained for jets with the largest allowed scale. The reason for the
weak fluxes and the preference for the emergence of jets of the largest
scale in this limit is understood by noting that the stochastically
forced eddies for $\b\gg1$ propagate with $\Ocal(\b)$ group
velocities. Therefore in contrast to the limit of $\b\ll 1$ in which
they evolve according to their local shear, the forced waves 
respond to the integrated shear of the sinusoidal perturbation over
their large propagation extend, which is very weak.

\begin{figure}
\centering\includegraphics[width=3.2in]{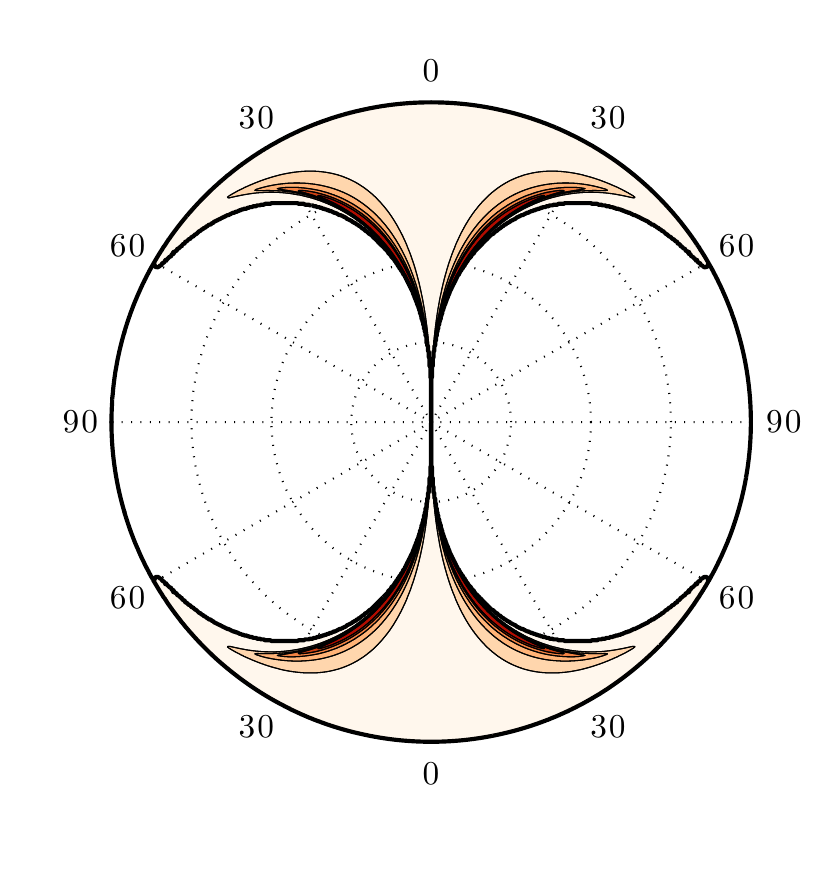}
\caption{\label{fig:dumbbell_b100} Contours of the eddy feedback $\fr$ in a $(\varphi,n)$ polar plot ($n$ radial and $\varphi$ azimuthal) for the case $\b=200$. Shown are contours of positive values, so the white area corresponds to negative values indicating downgradient vorticity fluxes. The contour interval is $10^{-3}$ and the radial grid interval is $\Delta n=0.25$. Note that the feedback factor is always negative (downgradient fluxes) for $\varphi\ge 60\deg$ (cf. Appendix~\ref{app:fR}).}
\end{figure}

To summarize: Although zonal jets and most non-zonal perturbations
induce fluxes that decay as $1/\b^2$ for large $\b$, resonant
and near resonant interactions arrest the decay rate of certain non-zonal
perturbations by a factor of $\Ocal(\b^{3/2})$ leading to fluxes that decay as
$1/\sqrt{\b}$. This makes the non-zonal perturbations to be the most S3T
unstable perturbations for $\b \gg 1$. Also in contrast to $\b\ll1$
when $\fr$ is positive for all $n$ and $\varphi$
(cf.~Fig.~\ref{fig:vq_n_phi0_phi60_b0p1}), the vorticity flux feedback
is negative for $(n,\varphi)$ in regions~B and~D of
Fig.~\ref{fig:R_Fcal}\hyperref[fig:R_Fcal]{a}. As a result, the mean
flows that produce negative fluxes and are by necessity S3T stable are
interestingly in the interior of the dumbbell shown in
Fig.~\ref{fig:dumbbell_b100}, illustrating $\fr$ in a polar $(n, \varphi)$ plot. The largest
destabilizing fluxes occur in the narrow region adjacent to the outer
boundaries of the dumbbell shape, which demarcates the boundary
separating regions A and B. Because of the selectivity of the resonances
these results do not depend on the forcing anisotropy, as we will see in the
next section.

\begin{figure}
\centering\includegraphics[width=5in]{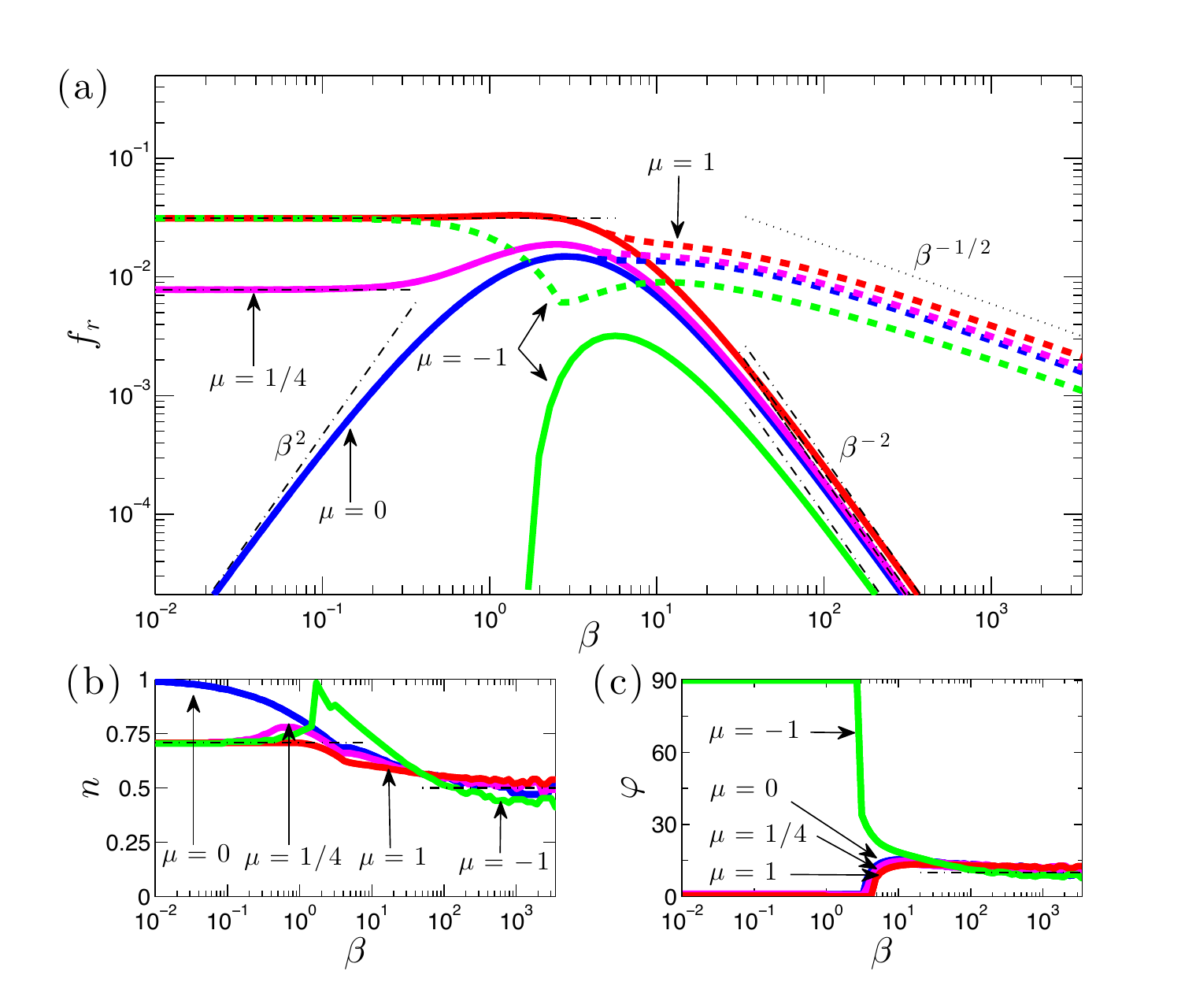}
\caption{\label{fig:frmax} The maximum value of $\fr$ over all wavenumbers $n$ for zonal jets (solid), and the maximum value of $\fr$ over all wavenumbers $n$ and angles $\varphi\ne0\deg$ for non-zonal perturbations (dashed) as a function of the planetary vorticity, ${\beta}$ for the three forcing covariance spectra seen in Fig.~\ref{fig:RF} and for $\mu=1/4$. Also shown are the asymptotic expressions~\eqref{eq:frmax_smallb_mu}, \eqref{eq:frmax_smallb_mu0} and~\eqref{eq:sr_largeb_max} (dash-dot) and the $\b^{-1/2}$ slope (dotted). For $\mu=-1$ zonal jet perturbations are stable for $\b<1.67$. (b) The mean flow wavenumber $n$ and (c) the angle $\varphi$ for which the maximum value of $\fr$ (shown in (a)) is attained. The asymptotes $n=1/\sqrt{2}$ (for $\b\ll1$) and $n=0.5$ (for $\b\gg1$) are shown in (b) (dash-dot) as well as the asymptote $\varphi=10\deg$ (for $\b\gg1$) is also shown in (c) (dash-dot).}
\end{figure}

\begin{figure}
\centering\includegraphics[width=.9\textwidth,trim = 30mm 1mm 30mm 1mm, clip]{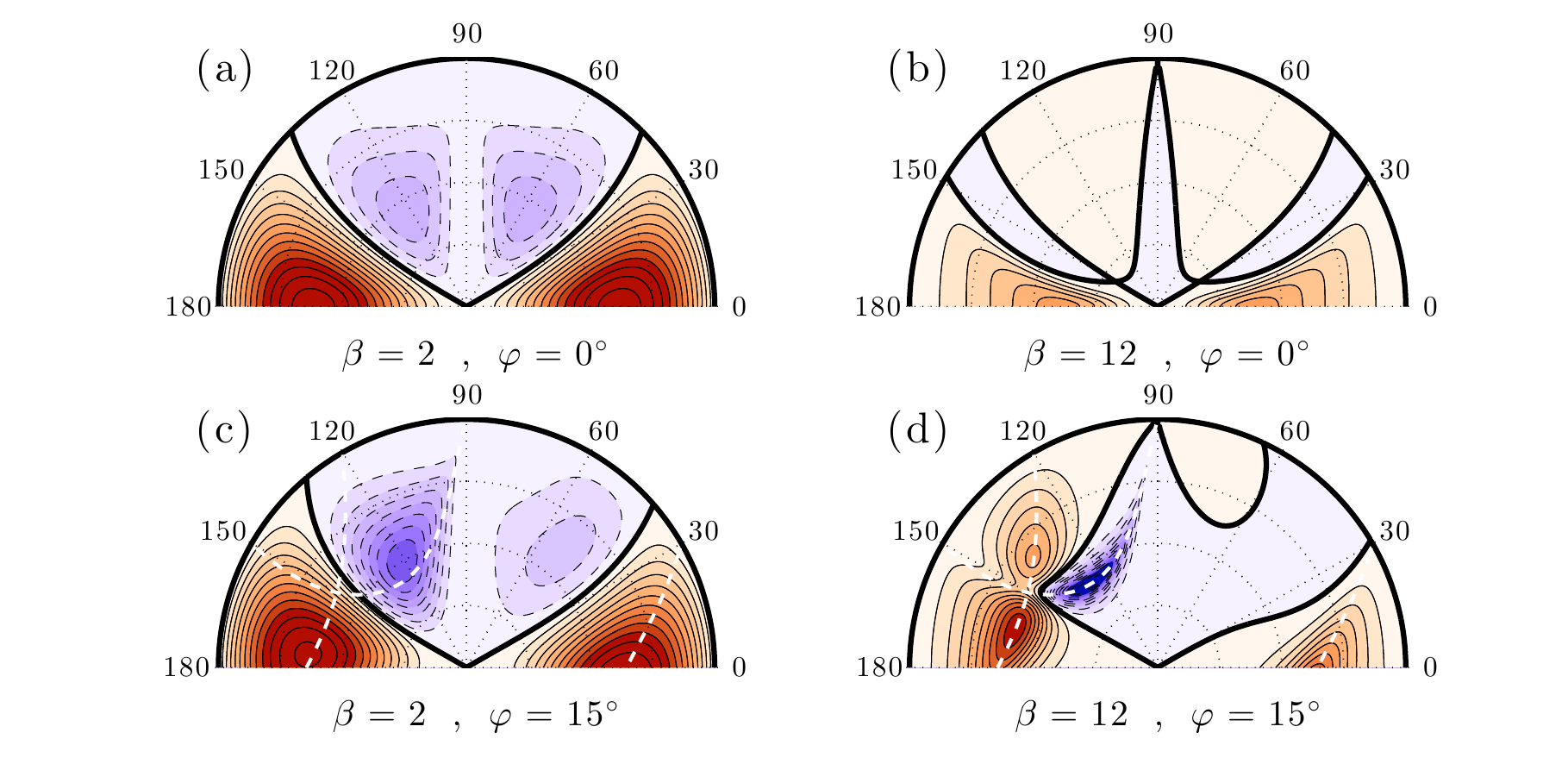}
\caption{\label{fig:F_order1} Contours of the $\Fcal(\thet,n)$ in a $(\thet,n)$ polar plot ($n$ radial and $\thet$ azimuthal). Shown is (a) $\Fcal$ for a zonal jet perturbation ($\varphi=0\deg$) and (c) a non-zonal perturbation with $\varphi=15\deg$ when $\b=2$. Panels (b) and (d) are the same as (a) and (c) for the case $\b=12$. In all panels, solid (dashed) lines indicate contours with positive (negative) values, the contour interval is $2\times10^{-3}$, the thick lines indicate the zero contour and the radial grid interval is $\Delta n=0.25$. White dashed lines in (c), (d) correspond to the locus of the roots of $\Db(\thet,n)$ on the $(\thet,n)$ plane.}
\end{figure}

\subsection{Induced vorticity fluxes for $\b\sim\Ocal(1)$}
\label{subsec:O1}

We have seen that in the singular case of isotropic forcing the only process available for the
emergence of mean flows is the fourth-order anti-diffusive vorticity
feedback induced by the variation of the group velocity of the forced
eddies due to the mean flow shear. For $\b\ll1$, the waves interact with
the local shear producing fluxes proportional to $\b^2\, \df ^4 \delta U
/ \df y^4 $. As $\b$ increases this growth is reduced since the waves
interact with an effective integral shear within their propagation
extent which is weak and eventually, as we have seen in the previous
section, for $\b\gg1$ the fluxes decay as $\b^{-2}$. Therefore, the
fluxes attain their maximum at an
intermediate value of $\b$. This occurs for $\b\approx 3.5$, as can be
seen in Fig.~\ref{fig:frmax}\hyperref[fig:frmax]{a} where the maximum
$\fr$ over all $(n,\varphi)$ is shown. It is demonstrated in the
next section that this intermediate $\b$ maximizes the S3T instability
for all forcing spectra.

While the eddy--mean flow interaction of both zonal and non-zonal
perturbations is dominated by the same dynamics when $\b\ll1$, for
$\b\gg1$ the eddy--non-zonal flow interaction is dominated by
resonances which do not occur for zonal  flow perturbations. The resonant interactions lead to the possibility of arrested decay of the eddy feedback at the rates of $\b^{-1/2}$ and $\b^{-1}$,
instead of the $\b^{-2}$ decay in the absence of resonances. The
vorticity flux attains its maximum at an intermediate value
$\b\sim\mathcal{O}(1)$ for non-zonal mean flows as well, which is
nonetheless large enough for the resonant contributions to reinforce
the contribution from the shearing mechanism. Figure \ref{fig:F_order1}
shows the contribution to the eddy feedback induced by the
various wave components that are excited for two values of $\b$ ($\b=2$
and $\b=12$) in the case of zonal jets ($\varphi=0\deg$) and non-zonal
perturbations ($\varphi=15\deg$). As $\b$ increases, the resonant
contributions start playing an important role for non-zonal
perturbations as there is enhanced contribution to the eddy 
feedback in the vicinity of the $\Db=0$ curves, indicated by the white
dashed lines. These resonant contributions enhance the vorticity fluxes
relative to the fluxes obtained for zonal jets and render the non-zonal
structures more unstable compared to zonal jets when $\b \gtrsim
3.5$~\citep{Bakas-Ioannou-2014-jfm}.

\section{Effect of anisotropic forcing on S3T instability\label{sec:anisotr}}

In this section we investigate the effect of the anisotropy of the
excitation on the S3T instability. The maximum vorticity flux feedback
$\fr$ for three cases of anisotropy ($\mu=\pm1$ and $\mu=1/4$) and for
isotropic forcing ($\mu=0$) is shown in Fig.~\ref{fig:frmax}\hyperref[fig:frmax]{a}. For $\b\gg1$, the main contribution to $\fr$ for zonal jet perturbations, comes from forced
waves with nearly meridional constant phase lines (angles near
$\thet=\gamma=0\deg$, cf.~Fig.~\ref{fig:F_z_b200}). Therefore, the eddy feedback
 $\fr$, attains larger (smaller) values for a stochastic
forcing that injects more (less) power in waves with angles near
$\gamma=0\deg$, that is for positive (negative) anisotropicity factor
$\mu$ (cf.~Fig.~\ref{fig:RF}). The maximum value of $\fr$ over
all wavenumbers $n$ depends in this case linearly on $\mu$ (cf.~Appendix~\ref{app:fR}),
\be
\f_{r,\textrm{max}} = (2+\mu) \b^{-2} +
\Ocal(\b^{-4})\ .
\ee
For non-zonal perturbations, the main contribution comes from forced
waves satisfying the resonant condition $\om_\kv+\om_\nv=\om_{\kv+\nv}$
and $\fr$ depends only on the sum of the resonant contributions. The
sign of $\Ncal_j$ that determines whether the resonant contribution is
positive or negative (cf.~\eqref{eq:frR}), depends only on the sign of
$\sin\thet_j+n/2$ and not on the anisotropicity factor $\mu$
(cf.~\eqref{eq:defNcal}). The anisotropicity affects only
the magnitude of $\Ncal_j$. For any $0<\varphi<90\deg$ it is found that the
resonances giving positive contribution occur at angles $\thet_j$ for which
$|\gamma_j|=|\thet_j-\varphi|<45\deg$. A stochastic excitation, which injects more
power near $\gamma=0\deg$ ($\mu>0$) gives larger positive resonant
contributions and therefore $\fr$ increases with $\mu$. However, the effect on the
maximum vorticity feedback is weak, as the spectral selectivity of the
resonances renders the characteristics of the most unstable non-zonal
structure independent of the spectrum of the forcing. That is, the
$(n,\varphi)$ that correspond to the maximum $\fr$ asymptotes to
$n\approx 0.5$, $\varphi\approx 10\deg$ (marked with star in
Fig.~\ref{fig:R_Fcal}\hyperref[fig:R_Fcal]{a}) as $\b\to\infty$, a
result that is very weakly dependent on $\mu$ (cf.~Figs.~\ref{fig:frmax}\hyperref[fig:frmax]{b,c}).

For $\b\ll1$, the characteristics of the S3T instability are dependent on the anisotropy of the
stochastic forcing. The eddy feedback is at leading order proportional to $\mu$:
\be
\fr =\frac1{8} \mu\,n^2 \left(1-n^2\right) \cos (2 \varphi ) + \Ocal(\b^2)\ .
\ee 
This shows that there can be upgradient vorticity fluxes leading to S3T instability for $\b = 0$ as long as $\mu\cos{(2\varphi)} >0$. For $\mu>0$, the maximum $\fr= \mu/32$ is achieved by zonal jets ($\varphi=0\deg$), while for $\mu<0$ any non-zonal perturbation with $\varphi>45\deg$ can grow, with the maximum $\fr=|\mu| / 32$ achieved for $\varphi=90\deg$ when the non-zonal perturbations assume the form of jets in the $y$ direction (meridional jets) (cf. Fig.~\ref{fig:frmax}\hyperref[fig:frmax]{c}).

It is worth noting that \citet{Srinivasan-Young-2014} also find that that the eddy momentum fluxes  are proportional to $\mu$ when  a constant shear  flow is stochastically forced with power spectrum~\eqref{eq:Qhat_deltaG}. This result is intriguing as the two studies address two different physical regimes. This chapter treats the limit appropriate for emerging structures in which the shear time is far larger than the dissipation time-scale  with the fluxes  determined by the instantaneous response of the eddies on the shear. \citet{Srinivasan-Young-2014} study the opposite limit in which the mean flow shear is finite and the shear time is much shorter than the dissipation time-scale with the fluxes determined by the integrated influence of the shear on the eddies over their whole life cycle, which may include complex effects such as reflection and absorption at critical levels.

In summary:
\begin{enumerate}
\item[a.] The S3T instability of the homogeneous state is a monotonically increasing
function of $\mu$ for all $\b$
\item[b.] The forced waves that contribute most to the instability are structures with small $\gamma$, i.e., waves with phase lines nearly aligned with the $y$ axis, as Fig.~\ref{fig:RF}\hyperref[fig:RF]{a}.
\item[c.] The anisotropy of the excitation affects prominently the S3T stability of the homogeneous state
only for $\b\lesssim 3.5$.
\end{enumerate}

%Positive values of $\Fcal_0$ indicate that the stochastically forced waves with phase lines inclined at angle $\thet$ with respect to the wavevector $\nv$ (cf. Fig.~\ref{fig:einx_angles}) induce upgradient vorticity fluxes to a mean flow with wavenumber $n$ when $\b=0$. 

\section{Bibliographical note}
This chapter is an excerpt from the paper by \upmax\textcite{Bakas-etal-2014}\dnmax. The S3T instability of the homogeneous turbulent equilibrium was studied by \textcite{Farrell-Ioannou-2007-structure}. Analytical results for zonal jet perturbations for $\b=0$ and finite doubly periodic domains were obtained by \textcite{Bakas-Ioannou-2011}. Results for infinite $\b$-planes and for any $\b$ were obtained by \textcite{Srinivasan-Young-2012}. The forcing spectrum used in this chapter was introduced by \textcite{Srinivasan-Young-2014}. The dispersion relation for the stability of non-zonal perturbations was derived by \textcite{Bakas-Ioannou-2013-prl}. A physical interpretation of the S3T instability of the homogeneous equilibrium to zonal jet perturbations for small $\b$ and $n$ is discussed in \citet{Bakas-Ioannou-2013-jas}.

%{\color{Green}
%\section{large $\b$}
%
%The non-zonal dispersion relation has on the l.h.s. in the denominator the term:
%\be
%\i\b K^2\[ 2k_+\(k_+ m+\l_+ n\) -m (K^2+K^2_s)/2 \bit\]\ 
%\ee
%Before shifting $k\to k-m/2$, $\l\to\l-n/2$ we had:
%\begin{align}
%\i\b\[ 2k \(k m+\l n\)  -m (K_-^2+K_+^2)/2 \bit\]=\i\b K^2\( k_- K_+^2-k_+ K_-^2 \bit\)\ 
%\end{align}
%
%If we assume that for the large $\b$ the imaginary part of the eigenvalue is that of the Rossby wave, $\s_i = m\b/N^2$, then we have that the terms in the denominator proportional to $\b$ are:
%\begin{align}
%\i\b K^2 \(  \frac{m}{N^2} K^2_+ K^2_-  +  k_- K_+^2-k_+ K_-^2 \) &=\i\b \frac{K^2}{K^2_+ K^2_- } \(  \frac{m}{N^2} - \frac{k_-}{K_-^2} -\frac{k_+}{K_+^2} \)\\
%&= \frac{K^2}{K^2_+ K^2_- } \( -\omega_{N} + \omega_{K_-} + \omega_{K_+} \)
%\end{align}
%}
%
%

\addcontentsline{toc}{chapter}{A note on non-dimensionalization}
% !TEX root = ../thesis.tex

%\begin{savequote}[75mm] 
%This is some random quote to start off the chapter.
%\qauthor{Firstname lastname} 
%\end{savequote}

\chapter*{A note on non-dimensional units used in the following chapters}
\label{sec:nondim}

In the next chapters we will not scale our fields and parameters as described in~\eqref{eq:nondim}. Instead, we will non-dimensionalize everything using typical values that correspond to the Earth's midlatitude atmosphere, that is a length scale of $L=5000\ \textrm{km}$ and a velocity of $U=40\ \textrm{m}\,\textrm{s}^{-1}$. Using this scales the time unit is $T=1.5\ \textrm{day}$ and the Earth's meridional planetary vorticity gradient at the midlatitudes corresponds to the non-dimensional value $\b=10$.

All numerical simulations that will be presented in the following chapters will be implement using periodic boundary conditions on a $\b$-plane with non-dimensional size $L_x=L_y=2\pi$. Therefore non-dimensional wavenumbers assume only integer values.

%The flow is dissipated with linear damping at rate $r$ and hyperviscosity with coefficient $\nu_4$. Periodic boundary conditions are imposed in $x$ and $y$ with periodicity $2 \pi L$.  Distances have been nondimensionalized by $L=5000\;\textrm{km}$ and time by $T=L/U$, where $U=40\;\textrm{m\,s}^{-1}$, so that the time unit is $T=1.5\;\textrm{day}$ and $\beta=10$ corresponds to a midlatitude value. Turbulence is maintained by stochastic forcing with  spatial and temporal structure, $F$, and amplitude $\varepsilon$.

% !TEX root = ../thesis.tex

%\begin{savequote}[75mm] 
%This is some random quote to start off the chapter.
%\qauthor{Firstname lastname} 
%\end{savequote}

\chapter{Relation of the S3T system with the  4MT system of the modulational instability of Rossby waves}
\label{ch:MI}

In this chapter we investigate the relation of the modulational instability (MI) of Rossby waves with the S3T theory regarding the emergence and equilibration of large-scale structures in $\b$-plane turbulence. It was established by~\textcite{Lorenz-1972} and~\textcite{Gill-1974} that Rossby waves are hydrodynamically unstable, and under certain conditions the greatest instability is a zonal jet. This instability is an instability that has been characterized in the literature as a MI because of its similarity with the Benjamin-Feir instability of surface gravity waves~\parencite{Benjamin-1967,Yuen-Lake-1980}. More recently this instability has been proposed to be the mechanism for the formation of zonal jets in barotropic but also baroclinic turbulence~\parencite{Berloff-etal-2009a,Connaughton-etal-2010}, in the sense that at the Rhines's scale the turbulent state is dominated by relatively coherent wave structures that become modulationally unstable and give rise to jets. In this chapter we demonstrate the formal equivalence between the 4MT system, that approximates well the MI of coherent Rossby waves, and the S3T instability of a homogeneous turbulent state that has the power spectrum of the Rossby wave that undergoes MI. This equivalence embeds the MI results into the more general and physical framework of S3T which can address the instability of more general states, like the structural instability of the attractor of a turbulent flow. We also compare the predictions of the 4MT and S3T systems with nonlinear simulations regarding the initiation of the MI and its equilibration. We demonstrate that the 4MT dynamical framework is inadequate for capturing the finite amplitude equilibration of the instabilities.

\section{MI  of a Rossby wave and the 4MT approximation\label{sec:MI_4MT}}

Consider the stability of a Rossby wave with streamfunction $\psi_\pv = A\,\cos ( \pv\cdot\xv-\om_\pv t )$ (and vorticity $\z_\pv = -p^2\psi_\pv$) that satisfies the inviscid barotropic vorticity equation:
\be
 \partial_t \z_\pv + J\(\psi_\pv,\z_\pv+ \betav \cdot\mathbf{x} \) =0\ .\label{eq:barotrop_inv}\ee
The stability of these nonlinear traveling wave solutions, referred to in MI studies as the primary waves, is addressed by perturbing the primary wave , i.e.,  by writing $\z=\z_\pv+\d\z$ and studying the evolution of the perturbation $\d\z$  in the linear approximation,
\be
\partial_t\,\d\z = \Lcal(\z_\pv)\,\d\z\ ,\label{eq:dz_lorenz}
\ee
where $\Lcal(\z_\pv)$ is a time-dependent linear operator. With the change of the frame of reference:\be
\xv_0 \equiv \xv-\frac{(\zhat\times\bv)\,t}{p^2}\ ,\label{eq:x0_trans}% =(x_0,y_0)\equiv (x+\b_y t/p^2,y-\b_x t/p^2)\ ,%\xv_0 =(x_0,y)\equiv \(x-(\om_\pv/k_x) t,y\)\ ,
\ee
the primary wave assumes the  stationary form: $\psi_\pv = A\,\cos ( \pv\cdot\xv_0)$, and the operator $\Lcal$ becomes time-independent but with the spatial periodicity of the primary wave.  The eigensolutions of~\eqref{eq:dz_lorenz} according to Bloch's theorem, are
\be
\d\z_\nv(\xv_0,t) = e^{s_\nv t}   e^{\i \nv\cdot\xv_0}\,g(\xv_0)\ ,\label{eq:dz}
\ee
and each eigenfunction is indexed by a wavevector $\nv$ which satisfies $|\nv|\le|\pv|/2$. The function $g$ is a periodic function with the periodicity of the primary wave $\psi_\pv$ (cf.~Appendix~\ref{app:bloch}, eq.~\eqref{eq:f_xtoxma}), and can be assumed in the form:
\be
g(\xv_0) = \sum_{m=-\infty}^{+\infty} a_m\,e^{\i m\,\pv\cdot\xv_0}\ .\label{eq:ux0}
\ee
In the original coordinates, because:\begin{subequations}
\begin{align}
g(\xv_0)&\to g\(\bit\xv-(\zhat\times\bv)\,t/p^2\)  = \sum_{m=-\infty}^{+\infty} a_m\,e^{\i m\(\pv\cdot\xv-\om_\pv t\)}\ ,\\
 e^{\i \nv\cdot\xv_0}&\to e^{\i \nv\cdot\[\xv-(\zhat\times\bv)\,t/p^2\]}   =  e^{\i (1-n^2/p^2)\om_\nv t }e^{\i \(\nv\cdot\xv-\om_\nv t\)} \ ,
\end{align}\end{subequations}
($\om_\kv$ denotes the frequency of a Rossby wave $\kv$, cf.~\eqref{eq:def_omRossby}), the eigenfunction $\nv$ can be written as
\be
\d\z_\nv(\xv,t) = e^{s_\nv t}    e^{\i (1-n^2/p^2)\om_\nv t }\d\hat{\z}_\nv(\xv,t)\ ,
\ee
with
\begin{align}
\d\hat{\z}_\nv (\xv,t) &=a_0 e^{\i (\nv\cdot\xv-\om_\nv t)}+\sum_{\substack{m=-\infty\\m\ne0}}^{+\infty}a_m\,e^{\i \[\(\nv+m \pv\)\cdot\xv - \(\om_\nv+m\om_\pv \)t\]} \ .\label{eq:eigen_lorenz}
\end{align}
Written in this form the eigenfunction~\eqref{eq:eigen_lorenz} is a superposition of a nonlinear Rossby wave solution of~\eqref{eq:barotrop_inv} (the $a_0$ harmonic) and satellite modes with wavenumbers $\nv\pm m\pv$, $m=1,2,\dots$, that are Rossby wave solutions only when $\om_\nv+m\om_\pv=\om_{\nv+m\pv}$.

\begin{figure}
\centering
\includegraphics[width=5.2in,trim = 10mm 0mm 10mm 0mm, clip]{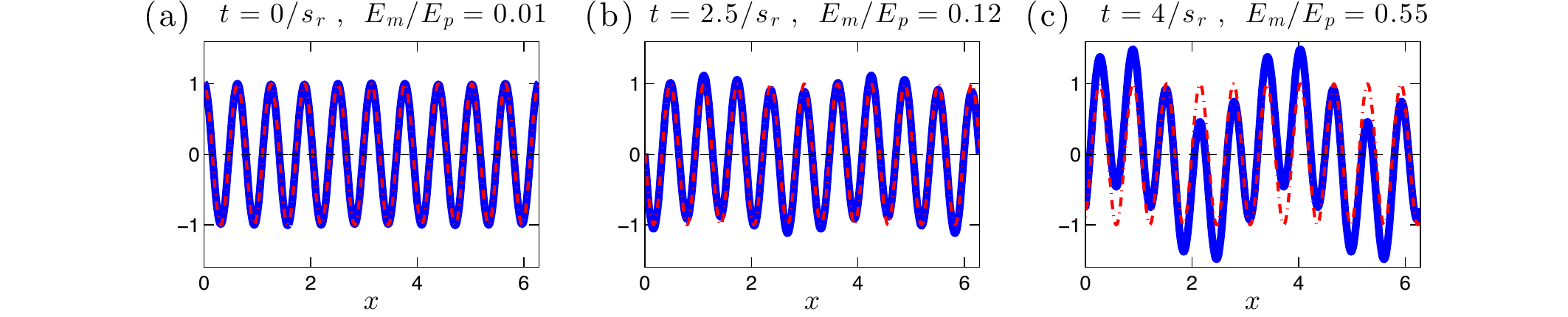}
\caption{\label{fig:whycallMI} Modulational instability of a primary wave $f(x,t)=\cos{(k x-\om_k t)}$ with $k=10$ to a large-scale perturbation $\d f(x,t)$ that grows at rate $s_r$: $\d f(x,t) =0.1\, e^{s_r t}\cos{(n x-\om_n t)}$ with $n=k/5$. Solid lines correspond to $f+\d f$ while dash-dotted lines correspond to the unperturbed $f$. Initially the ``energy'' of large-scale perturbation, $E_m$, is 1\% of the ``energy'' of the finite amplitude of the primary wave, $E_p$. The instability manifests as a modulation to the wave amplitude of the primary wave (cf. panel (b)).}
\end{figure}

By inserting~\eqref{eq:dz}-\eqref{eq:ux0} into~\eqref{eq:dz_lorenz}  an infinite homogeneous linear system for the coefficients $a_m$ is obtained. The eigenvalues $s_\nv$ are obtained from the requirement that this system has non-trivial solutions. This implies that the $s_\nv$ are roots of the associated characteristic polynomial, which is nominally of infinite degree. However, because the physically realizable solutions correspond to the convergent series~\eqref{eq:eigen_lorenz}, the coefficients of physically realizable eigenfunctions will have the property that $a_m\to0$ as  $m\to\pm\infty$; actually $a_m\sim b^m/m!$  for some constant $b$~\parencite{Lorenz-1972}. This enables us to determine accurate approximations of the eigenvalues from finite truncations of this infinite system. One obtains a good approximation of the eigenvalue even if only terms up to $|m|=1$ are kept. This truncation is referred to as the 4 mode truncation, or ``4MT'' system, because only four waves are allowed to interact: the primary wave $\pv$ and perturbation waves $\nv$, $\nv\pm\pv$. It can be also shown from a Fj{\o}rtoft type argument that unstable eigenvalues with $s_r=\real(s)>0$ exist only for $|\nv|<|\pv|$~\parencite{Lorenz-1972}. The instability manifests as a modulation of the amplitude, as shown in  Fig.~\ref{fig:whycallMI}.

% {\color{red}It was found by \textcite{Lorenz-1972} that the coefficients $a_m$ decrease with increasing $|m|$.} Therefore the eigenfunction is dominated by
%$a_{0}e^{\i(\nv\cdot\xv-\om_\nv t)}$, which is a Rossby wave with wavevector $\nv$.
%%If for example $\psi_\pv$ is unstable to a wave $\psi_\nv$ corresponding to a zonal state (i.e. with $n_x=0$) then a zonal jet will grow.
%\textcite{Lorenz-1972} further that you can solve for $\s$ and the coefficients $a_m$ can be obtained if we truncate the sum in r.h.s. of~\eqref{eq:eigen_lorenz}. Truncating up to $|m|=1$  gives very good results compared to truncating at higher $m$.\footnote{especially if $n\ll1$~\parencite{Lorenz-1972}} This truncation is typically referred to as the 4 mode truncation, or ``4MT'' system (the primary wave $\pv$ and perturbation waves $\nv$, $\nv\pm\pv$).
%

\section{Equivalence  of the MI in the 4MT approximation and  the S3T stability of a homogeneous turbulent state}

There is a close relation between the 4MT approximation of the MI and the S3T. \textcite{Parker-Krommes-2014-book} have shown that in the inviscid limit there is a formal equivalence between the modulational instability of the Rossby wave, $\psi_\pv = A\,\cos ( \pv\cdot\xv-\om_\pv t )$, in the 4MT approximation  with the S3T instability of the homogeneous state with eddy vorticity covariance with  the  same power spectrum  as  the Rossby wave, i.e., with $\hat{C}^e(\kv)= (2 \pi)^2 p^4 | A |^2\[\d(\kv-\pv) + \d(\kv+\pv)\]$. The connection is formal because physically the two problems are very different. In MI the stability of a basic state in the form of a coherent Rossby plane wave is studied, while S3T addresses the statistical stability of an incoherent state with equilibrium covariance having the power spectrum of the Rossby wave. In that sense, as noted by \citet{Parker-Krommes-2014-book}, S3T stability analysis embeds the modulational instability results into a more general physical framework. 

We proceed here to show that  this result does not only hold for monochromatic waves but can be generalized to any solution of the barotropic vorticity equation. That is, we show the formal equivalence between the MI of any time-dependent solution of the baro\-tropic equations with stationary power spectrum 
in the dynamical framework of a generalized 4MT with the S3T instability of a homogeneous state with the same power spectrum. The proof can be found in Appendix~\ref{app:MI}. Such a nonlinear solution of the inviscid barotropic vorticity equations is for example a superposition of any number of Rossby waves:
\be
\psi = \sum_{\substack{j=1\\|\pv_j|=p}}^N \underbrace{ A_{j}\cos(\pv_j\cdot\xv - \om_{\pv_j}t) }_{\psi_{\pv_j}}\ ,
\label{eq:C1}
\ee
all with the same total wavenumber\footnote{The vorticity of~\eqref{eq:C1}, $\Delta \psi = - p^2 \psi$, is proportional to $\psi$ and as a result $J(\psi, \Delta \psi)=0$.}, that forms a non-dispersive structure moving westwards (cf.~Appendix~\ref{app:MI}). The MI of~\eqref{eq:C1} can be carried out in a similar manner as described in section~\ref{sec:MI_4MT}. The generalized 4MT system is obtained by keeping the $N$ primary waves, $\psi_{\pv_j}$, the perturbation wave, $e^{\i\nv\cdot\xv_0}$, and the corresponding $2N$ satellite modes, $e^{\i(\nv\pm\pv_j)\cdot\xv_0}$. The eigenvalue relation in this truncation coincides with the S3T eigenvalue relation of the equilibrium covariance with spectral power:\be
\hat{C}^e(\kv) =  (2\pi)^2 p^4 \sum_{\substack{j=1}}^N |A_{j}|^2 \[\d(\kv-\pv_j)+ \d(\kv+\pv_j)\bit\]\ .
\ee 
The MI of a base state in the form of~\eqref{eq:C1} as well as the mechanisms responsible for instability have been studied by \textcite{Lee-Smith-2003}. %nterestingly, for states consisting of a large number of Rossby waves all having the same amplitude, $A_j=A$, they find that there is a transition at an intermediate value of $\b/(A p^3)$ of the structure producing maximum instability from zonal to non-zonal perturbations, in a similar manner as the stability of the isotropic S3T homogeneous turbulent equilibrium (cf.~chapter~\ref{ch:st3hom}, case with $\mu=0$).

{
}

\section{Comparison of MI and S3T predictions with nonlinear simulations for the emergence and equilibration of jets\label{sec:MIcompare}}

\textcite{Connaughton-etal-2010} compared the predictions of the 4MT system with direct numerical simulations and found that the 4MT system captures the initial growth of the instability, but fails to predict the later stages of zonal flow evolution. Contrary to the 4MT system, S3T dynamics capture both the emergence of the large-scale flow instability and also its equilibration. %That the equilibration of the incipient instabilities as well as the characteristics of the equilibrated flow as predicted by nonlinear simulations are captured by the S3T dynamics is demonstrated and discussed in chapter~\ref{ch:NLvsS3Tjas}.
 Here we present an example of jet emergence and equilibration as predicted by the 4MT and S3T systems and compare them with nonlinear simulations of the barotropic vorticity equation.  Details regarding the methods used for performing the numerical simulations can be found in Appendix~\ref{app:numerical_method}.

\begin{figure}
\centering
\includegraphics[width=3.8in]{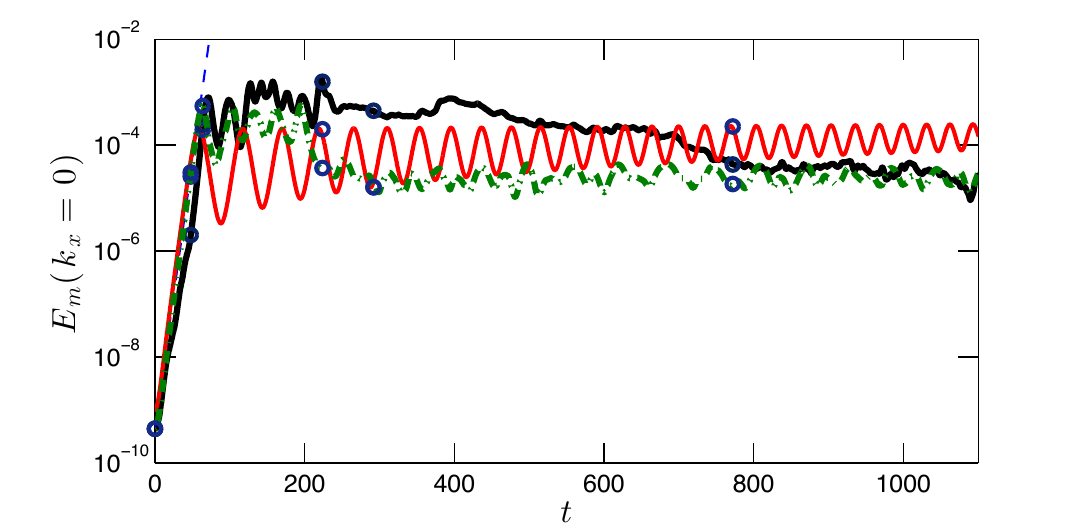}
\caption{\label{fig:Em_NL_NL4MT_decay} Evolution of the zonal energy $E_m(k_x=0)$ in the 4MT system (solid red line), the \NLinv\ system (dash-dotted green line) and the forced--dissipative NL system (solid black line).  Both the \NLinv\ and 4MT systems are initiated with a plane wave with wavenumber $\pv=(7,0)$ and $E_p(t=0)=2\times10^{-3}$ and a zonal jet perturbation with wavenumber $\nv=(0,3)$ and energy $E_m(t=0)=10^{-9}$.
The parameters for the NL are: linear damping coefficient $r=0.01$, stochastic forcing with single harmonics with wavenumber $\pv$ and energy injection rate: $\e=2r E_p(t=0)=4\times10^{-5}$. The predicted growth of the $\nv=(0,3)$ zonal jet perturbation by S3T is shown with the dashed line. Remarkably, the energy $E_m(t)$ of the mean flow grows at the same rate in the unforced and inviscid \NLinv\ and the forced--dissipative NL. Typical snapshots of the streamfunction fields for the three simulations are shown in Fig.~\ref{fig:NL_decay_snapshots} (\NLinv),~Fig.~\ref{fig:NL_stoch_snapshots_psi} (NL) and~Fig.~\ref{fig:NL4MT_decay_snapshots} (4MT), for the times marked with circles. Both the S3T and 4MT predict the initial growth of the mean flow but the 4MT fails to capture the finite amplitude state of the system. In all simulations $\b=4.9$.}
\end{figure}

\begin{figure}[h!]
\centering
\includegraphics[width=4.2in,trim = 15mm 0mm 0mm 0mm, clip]{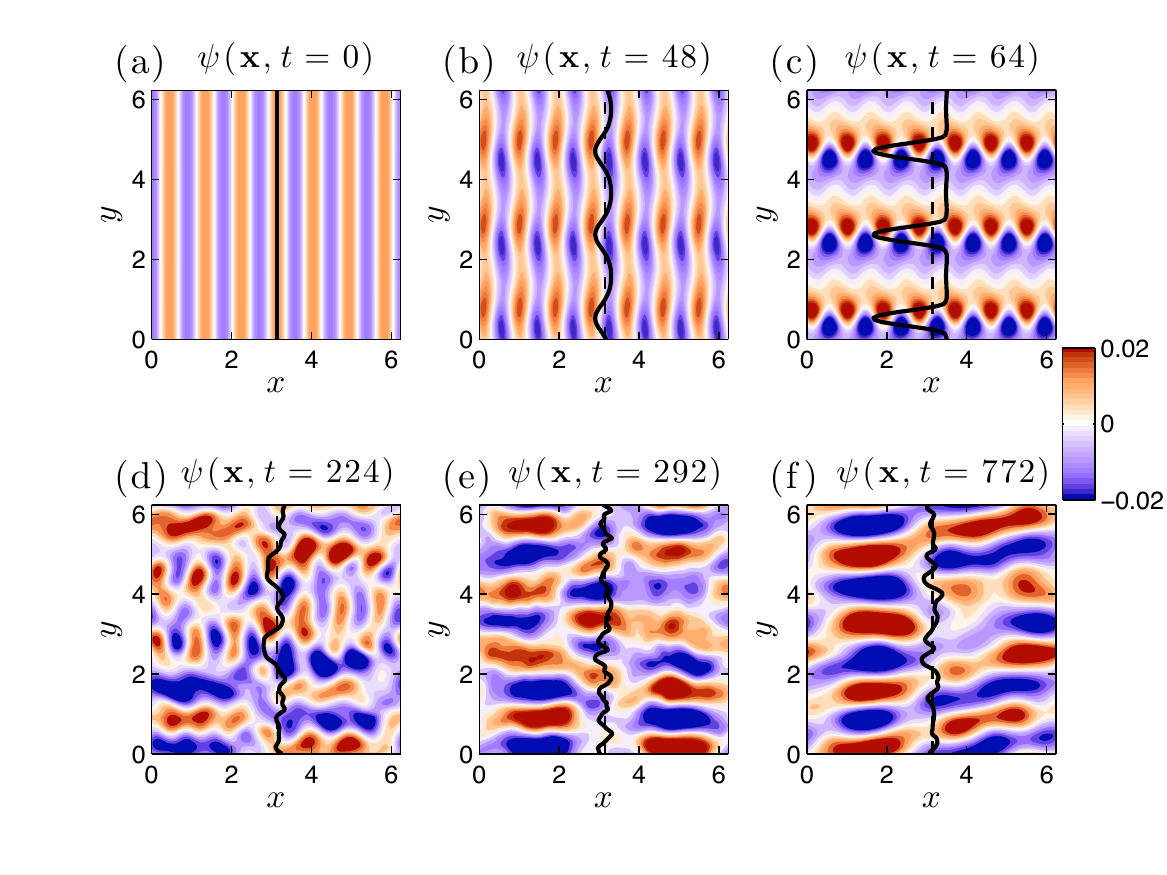}\vspace{-1em}
\caption{\label{fig:NL_decay_snapshots} Snapshots of streamfunction $\psi(\xv,t)$ together with the zonally averaged zonal velocity $U(y,t)$ (thick black line) for the \NLinv\ system at the indicated times with circles in Fig.~\ref{fig:Em_NL_NL4MT_decay}. Initially the zonal mean perturbation $\nv=(0,3)$ grows to finite amplitude (panels (a)-(c)) and at $t\approx200$ the zonal flow reorganizes and becomes a $(1,4)$ westward traveling wave (panels (d)-(f)).}
%\end{figure}
%
%
%\begin{figure}
\centering
\includegraphics[width=4.2in,trim = 15mm 0mm 0mm 0mm, clip]{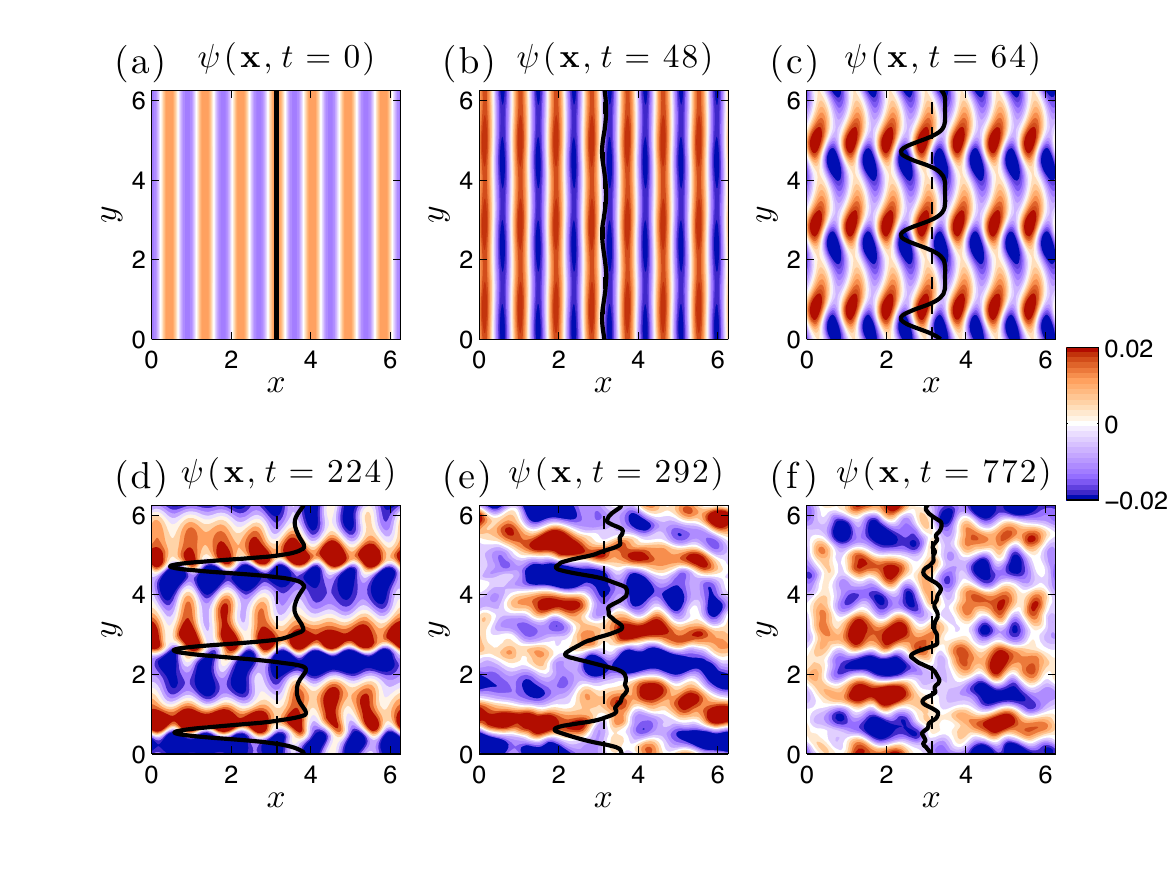}\vspace{-1em}
\caption{\label{fig:NL_stoch_snapshots_psi} Same as~Fig.~\ref{fig:NL_decay_snapshots} but for the stochastically forced--dissipative NL system. Remarkably, the NL system exhibits the same large-scale structure evolution with the \NLinv\ and transitions at approximately the same time to a (1,4) traveling wave structure.}
\end{figure}

\begin{figure}
\centering
\includegraphics[width=4.2in,trim = 15mm 0mm 0mm 0mm, clip]{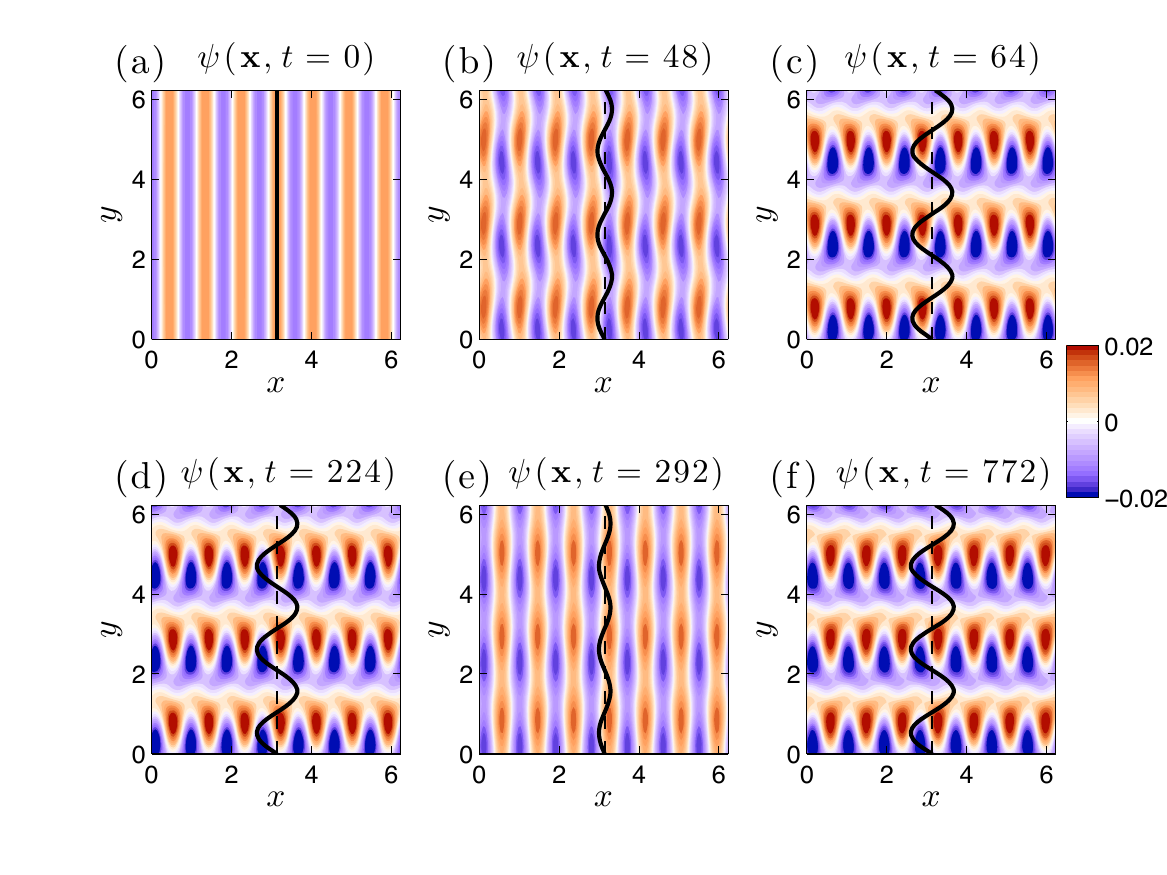}\vspace{-1em}
\caption{\label{fig:NL4MT_decay_snapshots} Snapshots of streamfunction $\psi(\xv,t)$ together with the zonally averaged zonal velocity $U(y,t)$ (thick black line) for the 4MT system at the indicated times with circles in Fig.~\ref{fig:Em_NL_NL4MT_decay}. Initially the zonal mean perturbation $\nv=(0,3)$ grows to finite amplitude but then it alternates between a state with strong zonal mean flow component (i.e. panel (d)) and a state with weak zonal flow component (i.e. panel (e)).}
\end{figure}

We start by performing a simulation of the inviscid and unforced version of the NL system~\eqref{eq:nl}, referred to as \NLinv. We initiate the simulation with a state $\psi(\xv,t=0)$ that consists of a primary wave $\psi_\pv = A \cos{(\pv\cdot\xv)}$ with $\pv=(7,0)$ and energy $E_p(t=0)=2\times10^{-3}$ and a zonal jet perturbation $\psi_\nv = a \cos{(\nv\cdot\xv)}$ with $\nv=(0,3)$ and energy $E_m(t=0)=10^{-9} = 0.5\times10^{-7}E_p(t=0)$. This $\nv=(0,3)$ zonal flow perturbation has been chosen because it is predicted by both S3T or 4MT to be the most unstable large-scale structure. We also perform a 4MT simulation which is initiated with the initial state of the \NLinv.  In the 4MT dynamics we allow only interactions between the Fourier components with wavenumbers $\pm\pv$, $\pm\nv$ and $\pm(\nv\pm\pv)$.  The evolution of the zonal mean flow energy, $E_m$, in the two simulations is plotted in Fig.~\ref{fig:Em_NL_NL4MT_decay}; for comparison we also plot the growth of the energy of the emerging instability as predicted by S3T (or 4MT). Snapshots of the evolution of the flow streamfunction, $\psi$, are shown in Fig.~\ref{fig:NL_decay_snapshots} (\NLinv) and~Fig.~\ref{fig:NL4MT_decay_snapshots} (4MT), for the time instants marked in Fig.~\ref{fig:Em_NL_NL4MT_decay}. Additionally, we perform an integration of the stochastically forced--dissipative NL system~\eqref{eq:nl}, in which the forcing excites structures $\cos{(\pv\cdot\xv+\thet)}$ with $\thet$ a randomly chosen phase. (The spatial covariance of this forcing is $Q(\xv_a-\xv_b)\sim\cos{\[\pv\cdot(\xv_a-\xv_b)\]}$.) For the chosen coefficient of linear damping, $r$, the energy input rate, $\e$, is adjusted so that the steady state equilibrium total energy of the stochastically forced flow is equal to the total energy of the primary wave of \NLinv\ and  4MT, i.e.,  $\e=2rE_p(t=0)$ (cf.~\eqref{eq:Einfinity}). The evolution of the zonal mean flow as well as snapshots of the flow streamfunction are shown in Fig.~\ref{fig:Em_NL_NL4MT_decay} and Fig.~\ref{fig:NL_stoch_snapshots_psi} respectively.

\newpage

\clearpage

Initially the zonal mean flow $E_m(t)$ in the \NLinv\ and NL grows at the  rate predicted by the S3T and 4MT stability analysis. In both the NL and  \NLinv\ the amplitude of the zonal flow reaches a plateau and then the flow reorganizes at $t\approx220$  producing a traveling wave moving westwards with maximal power at wavenumber $(1,4)$ (cf.~Fig.~\ref{fig:NL_decay_snapshots}\hyperref[fig:NL_decay_snapshots]{e,f}). Remarkably, the forced--dissipative NL  undergoes  the same flow reorganization to a traveling wave mean flow at approximately the same time as the \NLinv\ (cf.~Figs~\ref{fig:Em_NL_NL4MT_decay} and~\ref{fig:NL_stoch_snapshots_psi}). The 4MT system however, fails to capture this structural reorganization of the flow. Instead, it oscillates between a state with a strong zonal mean flow component (i.e.~Fig.~\ref{fig:NL4MT_decay_snapshots}\hyperref[fig:NL4MT_decay_snapshots]{d}) and a state with weak zonal mean flow component (i.e.~Fig.~\ref{fig:NL4MT_decay_snapshots}\hyperref[fig:NL4MT_decay_snapshots]{e})
that have no reflection in the \NLinv.

%\begin{figure}
%\centering
%\includegraphics[width=4.2in,trim = 15mm 0mm 0mm 0mm, clip]{NL_stoch_snapshots_psi.pdf}\vspace{-2em}
%\caption{\label{fig:NL_stoch_snapshots_psi} Snapshots of streamfunction $\psi(\xv,t)$ with the structure of the zonal mean velocity (thick line) for the evolution of the forced--dissipative NL system at the time instants marked with circles in Fig.~\ref{fig:Em_NL_NL4MT_decay}.}
%\end{figure}

In order to investigate whether the S3T system is able to  produce  the NL large-scale flow state 
we perform a forced--dissipative S3T time integration of~\eqref{eq:s3t} with  the  parameters of the NL simulation. Snapshots of the large-scale flow streamfunction, $\Psi$,  that emerges in the S3T simulation are plotted in Fig.~\ref{fig:S3T_snapshots_Psi}. First a zonal mean flow emerges, the zonal flow equilibrates producing finite amplitude jets, which then become unstable and give way to
 a traveling wave with structure  similar to  that of the NL simulation (see~Fig.~\ref{fig:S3T_snapshots_Psi}\hyperref[fig:S3T_snapshots_Psi]{e,f}).

\begin{figure}[t]
\centering
\includegraphics[width=3.7in]{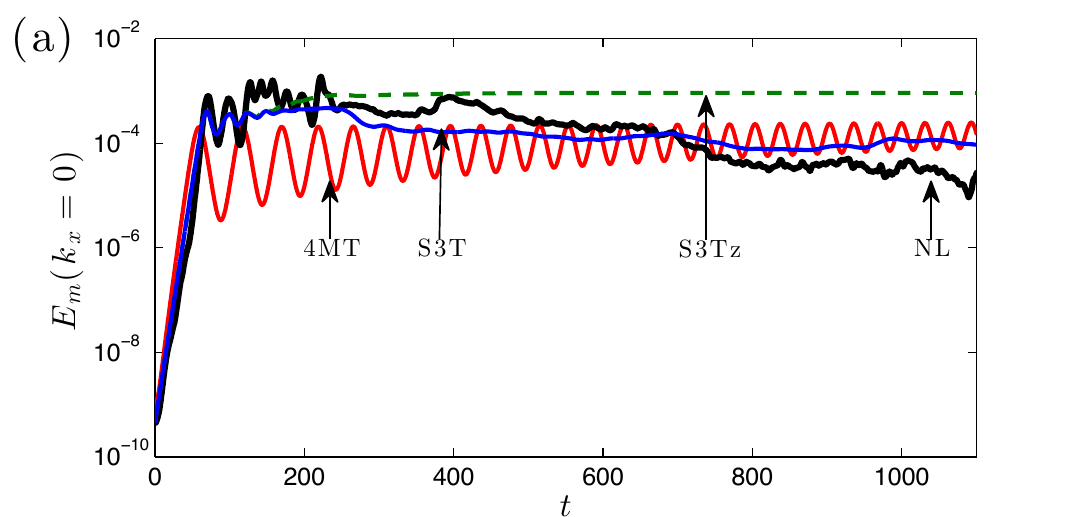}\\\includegraphics[width=3.7in]{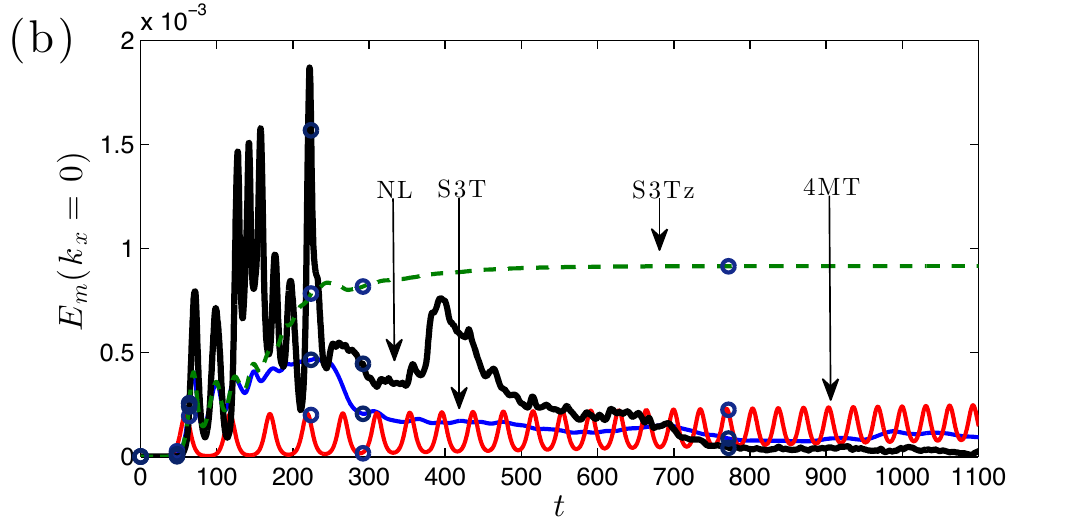}
\caption{\label{fig:Em_NL4MT_decay_NL_S3T_S3Tz_r0p01_n0_3_ab} Evolution of the zonal energy $E_m(k_x=0)$ in logarithmic scale (panel (a)) and in linear scale (panel (b)) in the 4MT system (solid red line), the NL system (black red line), the S3T system (blue solid line) and the S3Tz system (dashed line). It can be seen that the S3T prediction follows closely the NL. The S3Tz system mean flow energy evolution initially coincide with the S3T evolution and at $t\approx150$ the two systems diverge. The S3Tz system equilibrates to a zonal mean flow statistical equilibrium while the S3T system transitions to a traveling wave mean flow with most power at wavenumber $(1,4)$. For all integrations the planetary vorticity gradient is $\b=4.9$. Circles mark the time instants for the snapshots plotted in Figs.~\ref{fig:NL_stoch_snapshots_psi},~\ref{fig:NL4MT_decay_snapshots},~\ref{fig:S3T_snapshots_Psi} and~\ref{fig:S3Tz_snapshots_psi}.}
\end{figure}

%\caption{\label{fig:Em_NL_NL4MT_decay} Evolution of the zonal energy $E_m(k_x=0)$ in the 4MT system (red), the \NLinv\ system (dash-dotted green line) and the forced--dissipative NL system (solid black line).  Both the \NLinv\ and 4MT systems are initiated with a plane wave with wavenumber $\pv=(7,0)$ and $E_p(t=0)=2\times10^{-3}$ and a zonal jet perturbation with wavenumber $\nv=(0,3)$ and energy $E_m(t=0)=10^{-9}$. The parameters for the NL are: linear damping coefficient $r=0.01$, stochastic forcing with single harmonics with wavenumber $\pv$ and energy injection rate: $\e=2r\(E_m(t=0)+E_p(t=0)\)=4\times10^{-5}$. The predicted growth of the $\nv=(0,3)$ zonal jet perturbation by S3T is shown with the dashed line. Remarkably, the energy $E_m(t)$ of the mean flow grows at the same rate in the unforced and inviscid \NLinv\ and the forced--dissipative NL. Typical snapshots of the streamfunction fields for the three simulations are shown in Figs.~\ref{fig:NL_decay_snapshots} (\NLinv),~\ref{fig:NL_stoch_snapshots_psi} (NL) and~Fig.~\ref{fig:NL4MT_decay_snapshots} (4MT), for the times marked with circles. Both the S3T and 4MT predict the initial growth of the mean flow but the 4MT fails to capture the finite amplitude state of the system. In all simulations $\b=4.9$.}

At this point we want to emphasize  that the S3T that succeeded to faithfully produce
the NL flow state was the S3T system~\eqref{eq:s3t} in which the ensemble mean was identified with an average over fast time scales. The simplest S3T system in which the ensemble mean is identified with a zonal mean, that we  denote as S3Tz and obeys equations~\eqref{eq:s3tz}, is able to reproduce the initial instability and equilibration of the zonal jet but is incapable to capture the transition to a non-zonal large-scale flow and instead it equilibrates to a zonal jet state with 3 jets (see  Fig.~\ref{fig:S3Tz_snapshots_psi}). 
 A comparison of the evolution of the zonal mean flow energy for the S3T and S3Tz systems is shown in Fig.~\ref{fig:Em_NL4MT_decay_NL_S3T_S3Tz_r0p01_n0_3_ab}. The S3Tz system mean flow energy evolution initially coincides with the S3T energy evolution and at $t\approx150$ the two energy evolutions diverge: the S3Tz system is attracted to a zonal jet mean flow  while the S3T system transitions to a turbulent state characterized by a traveling wave mean flow with most power at wavenumber $(1,4)$.

%It is also worth investigating whether the simpler S3T system with the zonal mean--eddy decomposition~\eqref{eq:s3tz}, denoted as S3Tz,
% is able to capture the finite amplitude equilibration. The S3Tz system, by construction, only describes mean flows that do not depend on $x$ and therefore cannot describe a non-zonal mean flow state. It is revealed that after the initial growth of the zonal mean flow to finite amplitude the S3Tz system is not able to capture the transition to a wavenumber $(1,4)$ mean flow turbulent state. Contrary however to the 4MT system, the S3Tz system does not show a vacillation between high and low zonal mean flow energy states. Rather, after the initial zonal mean flow growth it equilibrates to an inhomogeneous turbulent zonal jet statistical equilibrium that is characterized by 3 zonal jets 
% 
% (see  Fig.~\ref{fig:S3Tz_snapshots_psi}). A comparison of the evolution of the zonal mean flow energy for the S3T and S3Tz systems is shown in Fig.~\ref{fig:Em_NL4MT_decay_NL_S3T_S3Tz_r0p01_n0_3_ab}. The S3Tz system mean flow energy evolution initially coincide with the S3T energy evolution and at $t\approx150$ the two energy evolutions diverge: the S3Tz system attracts to a zonal jet mean flow statistical equilibrium while the S3T system transitions to a turbulent state characterized by a traveling wave mean flow with most power at wavenumber $(1,4)$.

\begin{figure}
\centering
\includegraphics[width=4.2in,trim = 15mm 0mm 0mm 0mm, clip]{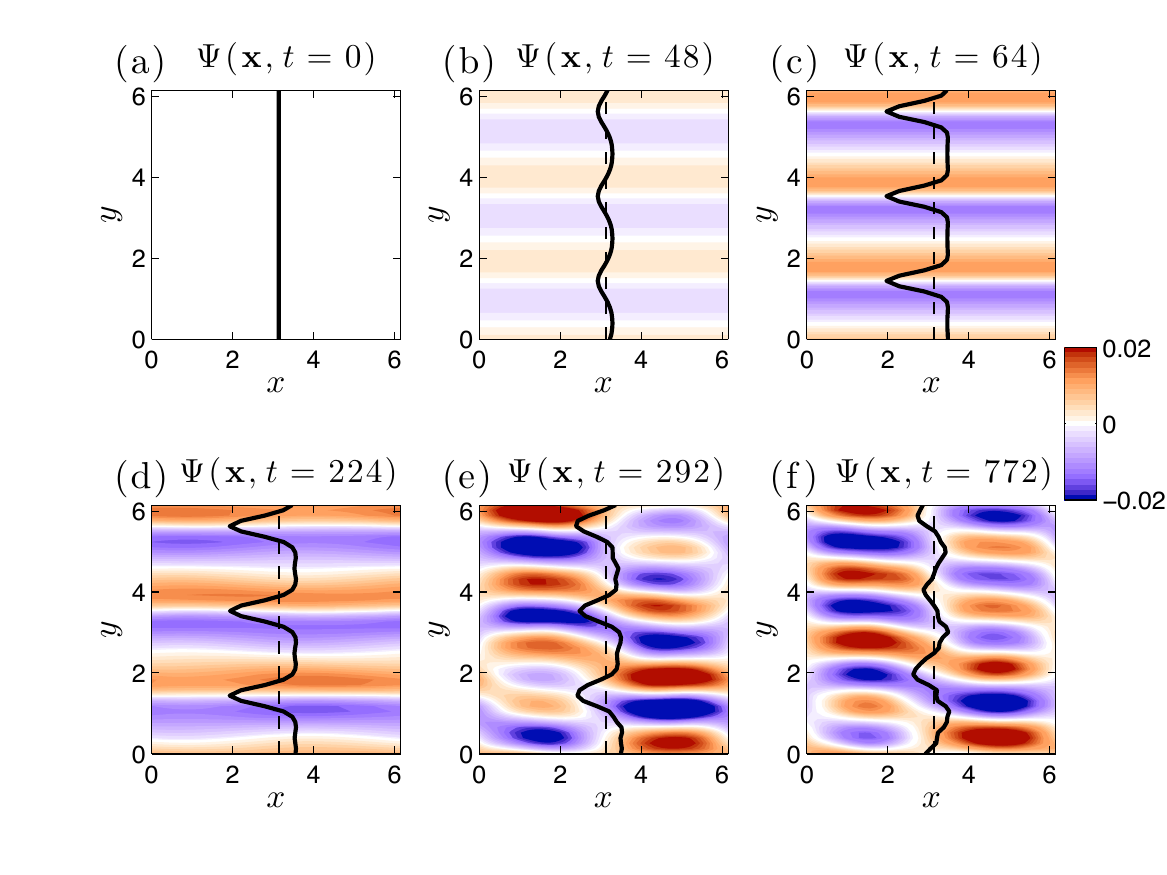}\vspace{-2em}
\caption{\label{fig:S3T_snapshots_Psi} Snapshots of mean flow streamfunction $\Psi(\xv,t)$ together with the zonally averaged zonal velocity $U(y,t)$ (thick black line) for the S3T system at the indicated times with circles in Fig.~\ref{fig:Em_NL4MT_decay_NL_S3T_S3Tz_r0p01_n0_3_ab}\hyperref[fig:Em_NL4MT_decay_NL_S3T_S3Tz_r0p01_n0_3_ab]{b}. Initially a zonal mean flow emerges and equilibrates to finite amplitude (panel (d)). This state however becomes unstable and transitions to a traveling wave with structure similar to that of the NL.}
%\end{figure}
%\begin{figure}
\centering
\includegraphics[width=4.2in,trim = 15mm 0mm 0mm 0mm, clip]{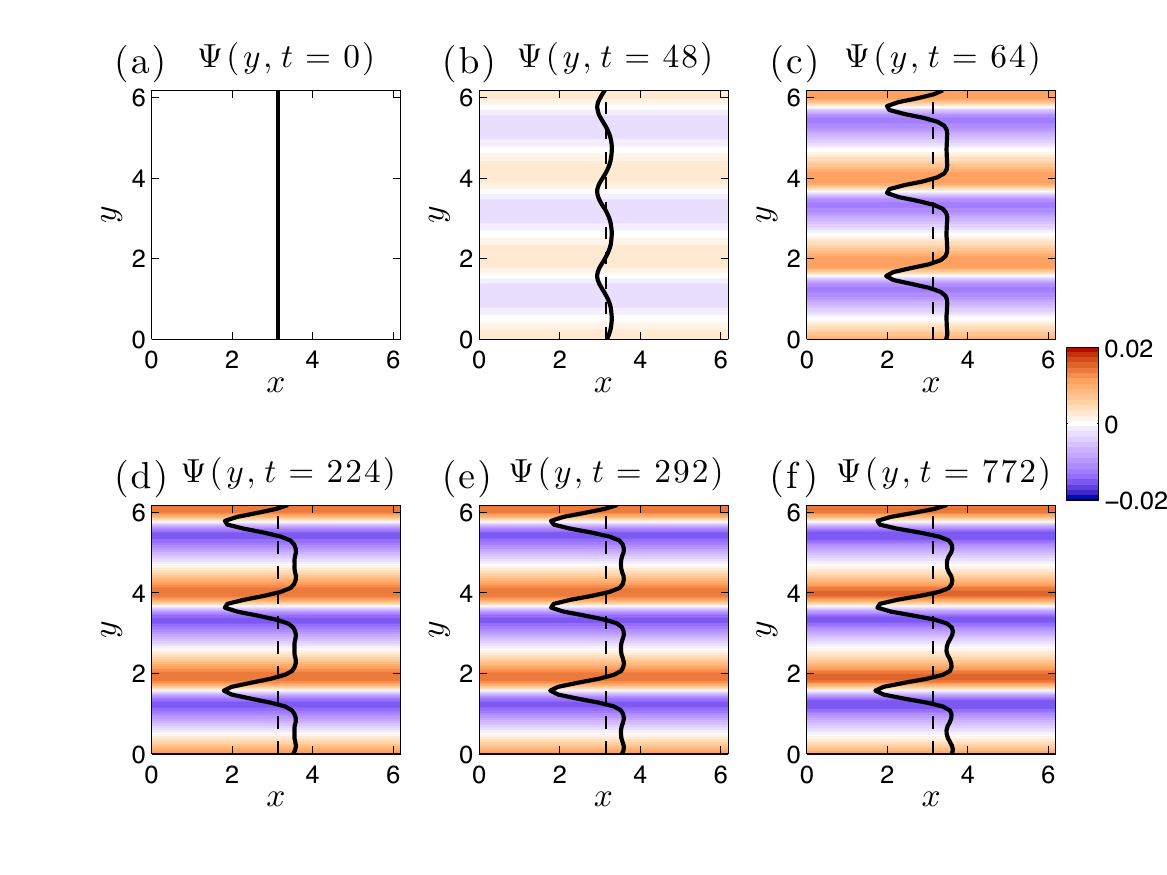}\vspace{-2em}
\caption{\label{fig:S3Tz_snapshots_psi} Snapshots of mean flow streamfunction $\Psi(y,t)$ together with the zonally averaged zonal velocity $U(y,t)$ (thick black line) for the S3Tz system at the indicated times with circles in Fig.~\ref{fig:Em_NL4MT_decay_NL_S3T_S3Tz_r0p01_n0_3_ab}\hyperref[fig:Em_NL4MT_decay_NL_S3T_S3Tz_r0p01_n0_3_ab]{b}. In the S3Tz system the mean flow is by construction  $x$-independent and therefore is able to reproduce the initial instability and equilibration of the zonal jet but is incapable to capture the transition to a non-zonal large-scale flow. Instead the zonal flow initially grows and then equilibrates to a zonal jet state with 3 jets.}
\end{figure}

%\begin{figure}
%\centering
%\includegraphics[width=4.8in,trim = 30mm 0mm 22mm 0mm, clip]{S3T_snapshots_Psi+psipert.pdf}
%\caption{\label{fig:S3T_snapshots_Psi} Snapshots of streamfunction $\Psi(\xv,t)+\psi'(\xv,t)$ for the evolution of the S3T system for the time instants marked with circles in Fig.~\ref{fig:Em_NL_NL4MT_decay}.}
%%\end{figure}
%%
%%
%%\begin{figure}
%\centering
%\includegraphics[width=4.8in,trim = 30mm 0mm 22mm 0mm, clip]{S3Tz_snapshots_Psi+psipert.pdf}\caption{\label{fig:S3Tz_snapshots_psi} Snapshots of streamfunction $\Psi(\xv,t)+\psi'(\xv,t)$ for the evolution of the S3Tz system for the time instants marked with circles in Fig.~\ref{fig:Em_NL_NL4MT_decay}.}
%\end{figure}
%

The stationary statistical equilibrium that the S3Tz system is attracted is also a stationary state of the S3T system and it can be shown that is unstable to non-zonal mean flow perturbations. With the methods described in chapter~\ref{ch:S3Tnonhom} we determine that the most unstable eigenfunction of the large-scale flow corresponds to a traveling wave with wavenumber $(1,4)$ (cf.~section~\ref{sec:jet_to_nonzonal}). This demonstrates that the final state of the NL is an equilibrated secondary instability of the S3T dynamics.

%\section{Bibliographical note}
%
%The modulational instability of Rossby waves was first studied by \textcite{Lorenz-1972} and \textcite{Gill-1974}. 

% !TEX root = ../thesis.tex

%\begin{savequote}[75mm] 
%This is some random quote to start off the chapter.
%\qauthor{Firstname lastname} 
%\end{savequote}

\chapter{Emergence and equilibration of jets in $\b$-plane turbulence as predicted by S3T and its reflection in nonlinear simulations}
\label{ch:NLvsS3Tjas}

\section{Introduction}

Stochastic structural stability theory (S3T) addresses turbulent jet dynamics as a two-way interaction between the mean flow and its consistent field of turbulent eddies~\parencite{Farrell-Ioannou-2003-structural}. The mean flow is supported by its interaction with a broad turbulence spectrum through non-local interactions in wavenumber space. In fact, S3T is a non-equilibrium statistical theory that provides a closure comprising a dynamics for the evolution of the mean flow together with its consistent field of eddies. In S3T the dynamics of the turbulence statistics required by this closure are obtained from a stochastic turbulence model (STM), which provides accurate eddy statistics for the atmosphere at large scale~\parencite{Farrell-Ioannou-1993d, Farrell-Ioannou-1994a,Farrell-Ioannou-1995,Zhang-Held-99}.

 \textcite{Marston-etal-2008} have shown that the S3T system is obtained by truncating the infinite hierarchy of cumulant expansions to second order and they refer to the S3T system as the second order cumulant expansion (CE2). In S3T, jets initially arise as a linear instability of the interaction between an infinitesimal jet perturbation and the associated eddy field and finite amplitude jets result from nonlinear equilibria continuing from these instabilities.
Analysis of this jet formation instability determines the bifurcation structure of the jet 
formation process as a function of parameters. In addition to jet formation bifurcations, S3T predicts jet breakdown bifurcations as well as the structure of the emergent jets, the structure of the finite amplitude equilibrium jets they continue to, and the structure of the turbulence accompanying the jets. Moreover, S3T is a dynamics so it predicts  the time dependent trajectory of the statistical mean turbulent state as it evolves and, remarkably, the mean turbulent state is often predicted by S3T to be time dependent in the sense that the statistical mean state of the turbulence evolves in a manner predicted by the theory \parencite{Farrell-Ioannou-2009-plasmas}.
The formation of zonal jets in planetary turbulence was studied as a bifurcation problem in S3T by \textcite{Farrell-Ioannou-2003-structural, Farrell-Ioannou-2007-structure, Farrell-Ioannou-2008-baroclinic, Farrell-Ioannou-2009-equatorial, Farrell-Ioannou-2009-closure,Bakas-Ioannou-2011, Srinivasan-Young-2012,Parker-Krommes-2014-generation}.
A continuous formulation of S3T developed by \textcite{Srinivasan-Young-2012} has facilitated analysis of the physical processes that give rise to the S3T instability and 
construction of analytic expressions for the growth rates of the S3T instability in homogeneous $\b$-plane turbulence \parencite{Srinivasan-Young-2012,Bakas-Ioannou-2013-jas,Bakas-etal-2014}. Recently, the analogy between the dynamics of pattern formation and zonal jet emergence in the context of S3T was studied by~\textcite{Parker-Krommes-2014-generation}. 

Relating S3T to jet dynamics in fully nonlinear turbulence is facilitated by studying the 
quasi-linear (QL) model which is intermediate between the nonlinear model and S3T. 
The QL approximation to the full nonlinear dynamics (NL) results when eddy--eddy interactions are not explicitly included in the dynamics but are either neglected entirely or replaced with a simple stochastic parameterization, so that no turbulent cascade occurs in the equations for the eddies, while interaction between the eddies and the zonal mean flow is retained fully in the zonal mean equation.
S3T is essentially QL with the additional assumption of an infinite ensemble of eddies
replacing the single realization evolved under QL. Although the dynamics of S3T and QL 
are essentially the same, by making the approximation of an infinite ensemble of eddies,
the S3T equations provide an autonomous and fluctuation-free dynamics of the statistical mean turbulent state, which transforms QL from a simulation of turbulence into a predictive theory of turbulence.

A fundamental attribute of QL/S3T is that the nonlinear eddy--eddy cascade of NL is suppressed in these systems. It follows that agreement in predictions of jet formation and equilibration between NL and QL/S3T provides compelling evidence that cascades are not required for jet formation and theoretical support for observations showing that the turbulent transfers of momentum maintaining finite amplitude jets are non-local in spectral space.

Previous studies demonstrated that unstable jets maintained by mean flow body forcing can be equilibrated using QL dynamics \parencite{Schoeberl-Lindzen-84,DelSole-Farrell-1996,OGorman-Schneider-2007,Marston-etal-2008}. In contrast to these studies, in this work we investigate the spontaneous emergence and equilibration of jets from homogeneous turbulence in the absence of any coherent external forcing at the jet scale. S3T predicts that infinitesimal perturbations with zonal jet form organize homogeneous turbulence to produce systematic up-gradient fluxes giving rise to exponential jet growth and eventually to the establishment of finite amplitude equilibrium jets. Specifically, the S3T equations predict initial formation of jets by the most unstable eigenmode of the linearized S3T dynamics. In agreement with S3T, \textcite{Srinivasan-Young-2012} found that their NL simulations exhibit jet emergence from a homogeneous turbulent state with subsequent establishment of finite amplitude jets, while noting quantitative differences between bifurcation parameter values predicted by S3T and the parameter values for which jets were observed to emerge in NL. \textcite{Tobias-Marston-2013} also investigated the correspondence of CE2 simulations of jet formation with corresponding NL simulations and found that CE2 reproduces the jet structure, although they noted some differences in the second cumulant, and suggested a remedy by inclusion of higher cumulants.

In this chapter we use NL and its QL counterpart together with S3T to examine further the dynamics of emergence and equilibration of jets from turbulence. Qualitative agreement in bifurcation behavior among these systems, which is obtained for all the spatial turbulence forcing distributions studied, confirms that the S3T instability mechanism is responsible for the formation and equilibration of jets. Quantitative agreement is obtained for bifurcation parameters between NL and QL/S3T when account is taken of the modification of the turbulent spectrum that occurs in NL but not in QL/S3T. Remarkably, a primary component of this spectral modification can itself be traced to S3T instability, but of non-zonal rather than of zonal form. We investigate the formation and equilibration of these non-zonal S3T instabilities and the effect these structures have on the equilibrium spectrum of $\b$-plane turbulence. We also investigate circumstances under which non-zonal structures are modified and suppressed by the formation of zonal jets.

A dynamic of potential importance to climate is the possibility of multiple equilibria of the statistical mean turbulent state being supported with the same system parameters \parencite{Farrell-Ioannou-2003-structural,Farrell-Ioannou-2007-structure,Parker-Krommes-2014-generation}. We verify existence of multiple equilibria, predicted by S3T, in our NL simulations. Finally, we show that weak jets result from stochastic excitation by the turbulence of stable S3T modes, which demonstrates the physical reality of the stable S3T modes. Turbulent fluctuation induced excitation of these weak local jets and the weak but zonally extended jets that form at slight supercriticality in the jet instability bifurcation may explain the enigmatic latent jets of \textcite{Berloff-etal-2011}.

Since the emergence and equilibration of jets is addressed throughout this chapter the zonal mean--eddy decomposition for the flow fields~\eqref{eq:zonal_edd_dec} is used. Therefore the NL, QL and S3T dynamics discussed in this chapter use the zonal mean--eddy decomposition which was presented in section~\ref{sec:s3tz}. The NL system is
\begin{subequations}\begin{align}
\partial_t U & = \overline{ v'\z' } - r\,U\ ,\label{eq:NL_U_r}\\
\partial_t \z' &=\Acal_{\textrm{z}}(U)\,\z' + \partial_y\(\overline{v'\z'}\)-\nablav\cdot \(\uv'\,\z'\)+\sqrt{\e}\,\xi \ ,\label{eq:NL_z_r}
\end{align}\end{subequations}
with 
\be
\Acal_{\textrm{z}}(U) = -U\partial_x - \(\bit\b- \partial^2_{yy}U\)\partial_x\Del^{-1} -r\ ,
\ee
the QL system is
\begin{subequations}\begin{align}
\partial_t U & = \overline{ v'\z' } - r\,U\ ,\label{eq:QL_U_r}\\
\partial_t \z' &=\Acal_{\textrm{z}}(U)\,\z'+\sqrt{\e}\,\xi \ ,\label{eq:QL_z_r}
\end{align}\label{eq:QL_r}\end{subequations}
and the S3T system (S3Tz) is
\begin{subequations}
\begin{align}
\partial_t U & = \[ \frac1{2}(\Del^{-1}_a\partial_{x_a}\!+\!\Del^{-1}_b\partial_{x_b}) C_{ab}\]_{\xv_a=\xv_b}-r\, U\ ,\label{eq:S3T_U_r}\\
\partial_t C_{ab} & = \[\bit\Acal_{\textrm{z},a}(U) + \Acal_{\textrm{z},b}(U)\]C_{ab} +\e\,Q_{ab}\ .\label{eq:S3T_C_r}
\end{align}\label{eq:S3T_r}
\end{subequations}

Throughout this chapter where we mention S3T we refer to the zonal mean--eddy decomposition S3Tz system. The S3Tz system does not allow for mean flows with non-zonal structure and therefore S3T instabilities within~\eqref{eq:S3T_r} are only zonal jet perturbations. However, as it is demonstrated in sections~\ref{sec:influence_eddy} and~\ref{sec:nonzonal}, the emergence of non-zonal coherent structures is of great importance in the quantitive predictions of S3Tz for jet emergence and equilibration.

\section{Specification of the stochastic forcing structure\label{sec:forcing_structure}}

Because the S3T instability mechanism that results in jet bifurcation from a homogeneous turbulent state differs for isotropic and non-isotropic turbulence (cf.~chapter~\ref{ch:st3hom}), we consider examples of both isotropic and non-isotropic turbulence forcing. The jet forming instability in the case of homogeneous, non-isotropic forcing arises from the up-gradient fluxes induced by shearing of the turbulence by the infinitesimal perturbation jet, while the up-gradient fluxes for the case of homogeneous isotropic forcing arise from the refraction of the eddies caused by the variation in the potential vorticity gradient induced by the infinitesimal perturbation jet.
 
\begin{figure}[ht]
\centering
\includegraphics[width=.85\textwidth,trim=5mm 5mm 5mm 0mm,clip]{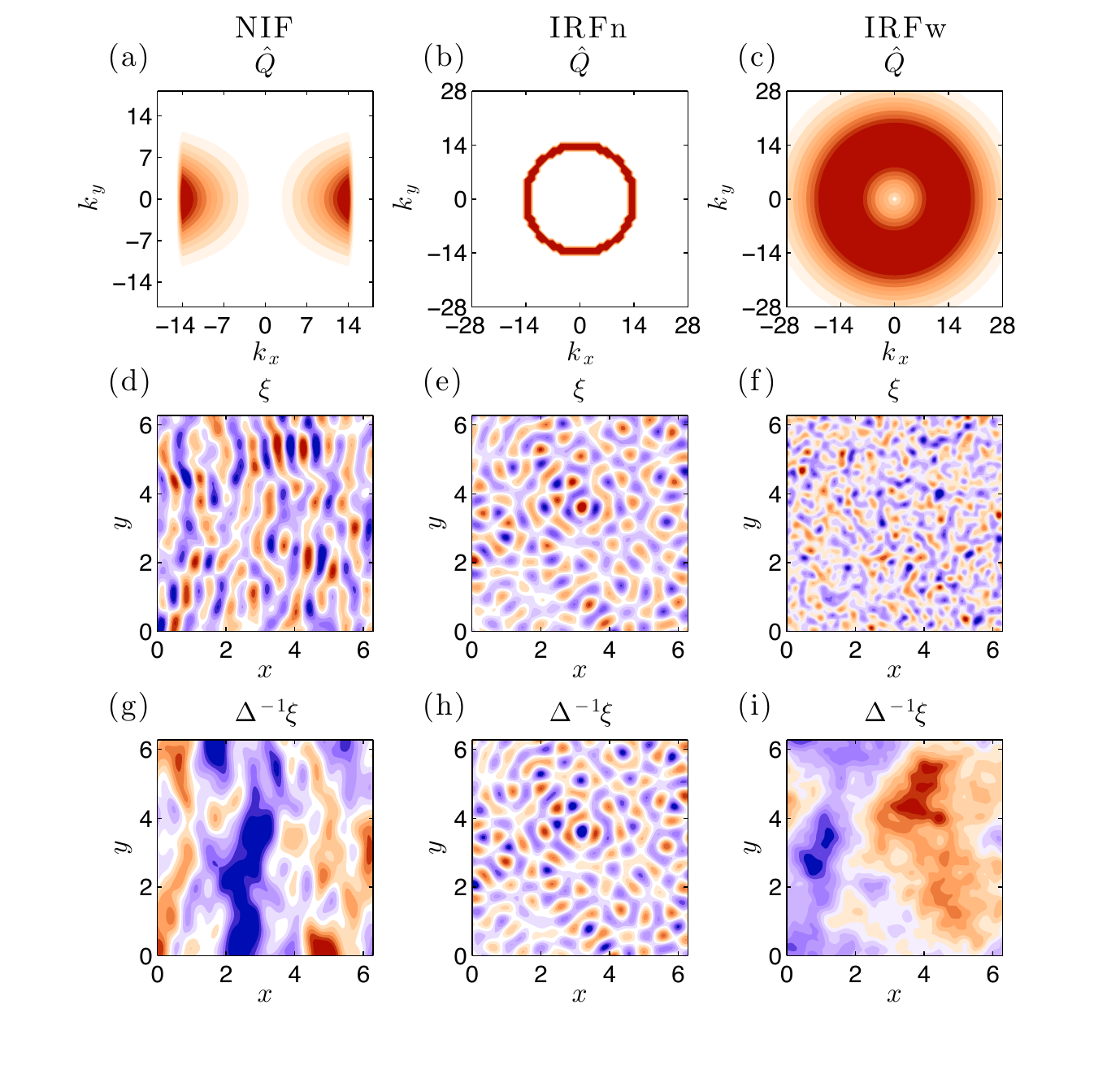}
\caption{ \label{fig:Qkl_Fxy_NIF_IRF} Contour plots of the spatial Fourier coefficients of the forcing vorticity covariances, $\hat{Q}_{\kv}$ (cf.~\eqref{eq:def_Qhat}), used in this study and example realizations of the forcing. 
Panel (a): $\hat{Q}_{\kv}$ for NIF with zonal wavenumbers $k_x=1,\dots,14$ and $d=1/5$. 
Panel (d): $\hat{Q}_{\kv}$ for IRFn at $k_f=14$ and $\d k_f=1$.
Panel (g) $\hat{Q}_{\kv}$ for IRFw at $k_f=14$ and $\d k_f=8/\sqrt{2}$.
In (b), (e) and (h) are shown realizations of these forcings in the vorticity field, 
and in (c), (f) and (i) are shown realizations in the streamfunction field.}
\end{figure}

Three stochastic forcing structures will be used in our investigation of the correspondence among S3T, QL and NL dynamics. The first independently excites a set of zonal wavenumbers. This stochastic forcing is spatially homogeneous but not isotropic and will be denoted as NIF (non-isotropic forcing). %This forcing was first used by \cite{Williams-78} to parametrize excitation of baro\-tropic dynamics by baroclinic instabilities. This forcing was also used by \cite{DelSole-01a} in his study of upper-level tropospheric jet dynamics and in the study of jet formation using S3T dynamics by \cite{Farrell-Ioannou-2003-structural,Farrell-Ioannou-2007-structure} and \cite{Bakas-Ioannou-2011}.
The second forcing, denoted IRFn, is an isotropic narrow ring forcing concentrated near a single total wavenumber, $k_f$. %This forcing structure has been used extensively in studies of $\b$-plane turbulence \citep[cf.][]{Vallis-Maltrud-93} and was also used in the recent study of \cite{Srinivasan-Young-2012}. It was introduced by \cite{Lilly-1969}, in order to isolate the inverse cascade from the forcing in a study of two dimensional turbulence.
The third forcing we use, denoted IRFw, is an isotropic ring forcing in which the forcing is distributed over a wide annular region in wavenumber space around the central total wavenumber. Specification of these stochastic forcing structures are given in Appendix~\ref{appsec:forc_spec_IRF_NIF}. Plots of the corresponding forcing covariance power spectra together with instantaneous realizations both in vorticity and streamfunction for the three types of forcing structures are shown in Fig.~\ref{fig:Qkl_Fxy_NIF_IRF}. Note, that the IRFn ring forcing is peculiar in that it primarily excites vortices of scale $1/k_f$ that are evident in both the vorticity and streamfunction fields, while IRFw produces a streamfunction field dominated by large scale structure similar to the fields excited by the other broadband forcings.

{\color{red}

}

\section{Bifurcations predicted by S3T and their reflection in QL and NL simulations}

We examine the counterpart in NL and QL simulations of the S3T structural instability by comparing the evolution of the domain averaged energy of the zonal flow:
\be
 E_m(t) = \frac1{L_xL_y} \int \frac1{2}U^2\,\df^2\xv\ .\label{eq:Em}
\ee
The amplitude of the zonal flow is measured with the zonal mean flow index (zmf) defined as $\textrm{zmf}=E_m/(E_m+E_p)$, where $E_m$ is  the time average of the domain averaged zonal mean flow, given in~\eqref{eq:Em}, and $E_p$ is the time average of the domain averaged kinetic energy of the eddies,
\be
 E_p(t) = \frac1{L_xL_y} \int \frac1{2}|\uv'|^2\,\df^2\xv\ .
\ee 
 
\begin{figure}
\centering
\includegraphics[width=.98\textwidth,trim=0mm 0mm 2mm 0mm,clip]{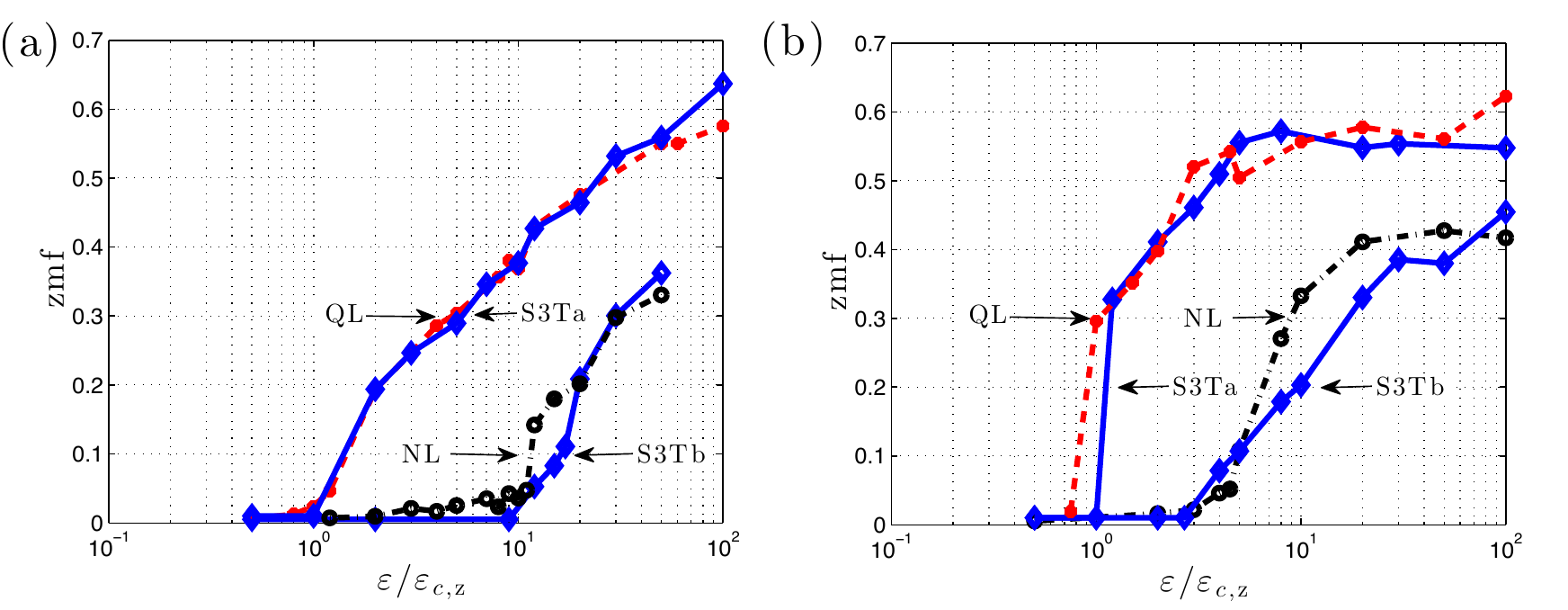}
\caption{\label{fig:bif_r_0p01} Bifurcation structure comparison for jet formation in S3T, QL, and NL. Shown is the zmf index of jet equilibria for (a) NIF and (b) for IRFn forcing as a function of the forcing amplitude $\e/\ecz$ for the NL simulation (dash-dot and circles), the QL simulation (dashed and dots) and the corresponding S3Ta simulation (solid). The bifurcation diagram and the structure of the jet agree in the QL and S3Ta simulation, but the bifurcation in the NL simulations occurs at $\ecz^{(\textrm{NL})}\approx 11\ecz$ for NIF and at at $\ecz^{(\textrm{NL})}\approx 4 \ecz$ for IRFn. Agreement between NL and S3T predictions is obtained if the S3T is forced with the spectrum that reflects the modification of the equilibrium NIF or IRFn spectrum respectively by eddy--eddy interactions (the results of this S3T simulation is indicated as S3Tb, see discussion at section~\ref{sec:influence_eddy}). (For IRFn this spectrum is shown in Fig.~\ref{fig:NLspect_stability_r0p01}\hyperref[fig:NLspect_stability_r0p01]{c}.) This figure shows that the structural stability of jets in NL simulations is captured by the S3T if account is taken of the nonlinear modification of the spectrum. Parameters: $\beta=10$, $r=0.01$.}
\end{figure}

 Zmf is shown as a function of the energy input rate in Fig.~\ref{fig:bif_r_0p01}\hyperref[fig:bif_r_0p01]{a} for NIF forcing and in Fig.~\ref{fig:bif_r_0p01}\hyperref[fig:bif_r_0p01]{b} for IRFn forcing with $r=0.01$. The fundamental qualitative prediction of S3T that jets form as a bifurcation in the strength of the turbulence forcing is verified in these plots. Agreement in the value of the bifurcation parameter is also obtained between S3T and QL while the bifurcation parameter is substantially larger in NL. For example, the NL simulations bifurcate at $\ecz^{(\textrm{NL})}\approx11\ecz$ under NIF forcing  and at $\ecz^{(\textrm{NL})}\approx 4\ecz$ under IRFn forcing. Similar behavior was noted by \textcite{Srinivasan-Young-2012}. The reason for this difference between the NL and S3T bifurcation curves is revelatory of the underlying dynamics of the bifurcation, as we explain in section~\ref{sec:influence_eddy}.
 
\begin{figure}[ht]
\centering
\includegraphics[width=24pc,trim=0mm 7mm 0mm 0mm,clip]{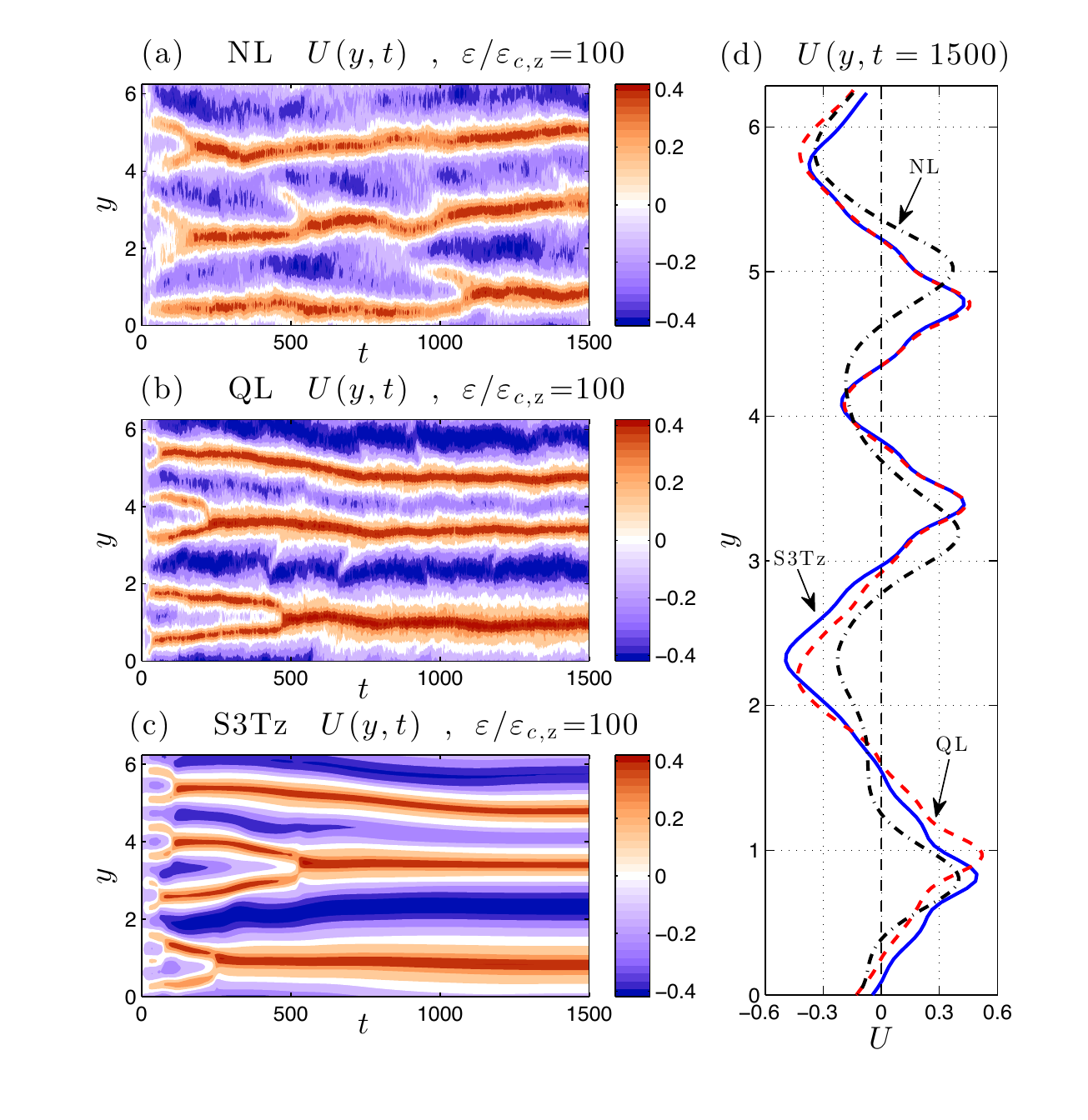}
\caption{Hovm\"oller diagrams of jet emergence in NL, QL and S3T simulations with IRFn forcing at energy input rate $\e= 100\ecz$. Shown is $U(y,t)$ for the NL (panel (a)), QL (panel (b)) and S3T (panel (c)) simulations. Also shown are the equilibrium jets (panel (d)) in the NL (dash-dot), QL (dashed), and S3T (solid) simulations. There is very good agreement between the jet structure in the NL, QL and S3T simulations, despite the difference in the zmf index among them (cf.~Fig.~\ref{fig:bif_r_0p01}\hyperref[fig:bif_r_0p01]{b}). Moreover, in all three simulations similar jet mergers are observed, leading eventually to final equilibrium jets with smaller meridional wavenumber than that of the initial instability. Parameters are $\beta=10$, $r=0.01$.}
\label{fig:hov_IRFh_e100ec_r0p01}
\end{figure}

S3T dynamics not only predicts the emergence of zonal jets as a bifurcation in turbulence forcing, but also predicts the structure of the finite amplitude jets that result from equilibration of the initial jet formation instability. These finite amplitude jets correspond to fixed points of the S3T dynamics. An example for IRFn strongly forced with $\e=100\ecz$ and with damping $r=0.01$ is shown in Fig.~\ref{fig:hov_IRFh_e100ec_r0p01}. This example demonstrates the essential similarity among the jets in NL, QL and S3T simulations. 

\begin{figure}[ht]
\centering
\includegraphics[width=24pc,trim=0mm 5mm 0mm 0mm,clip]{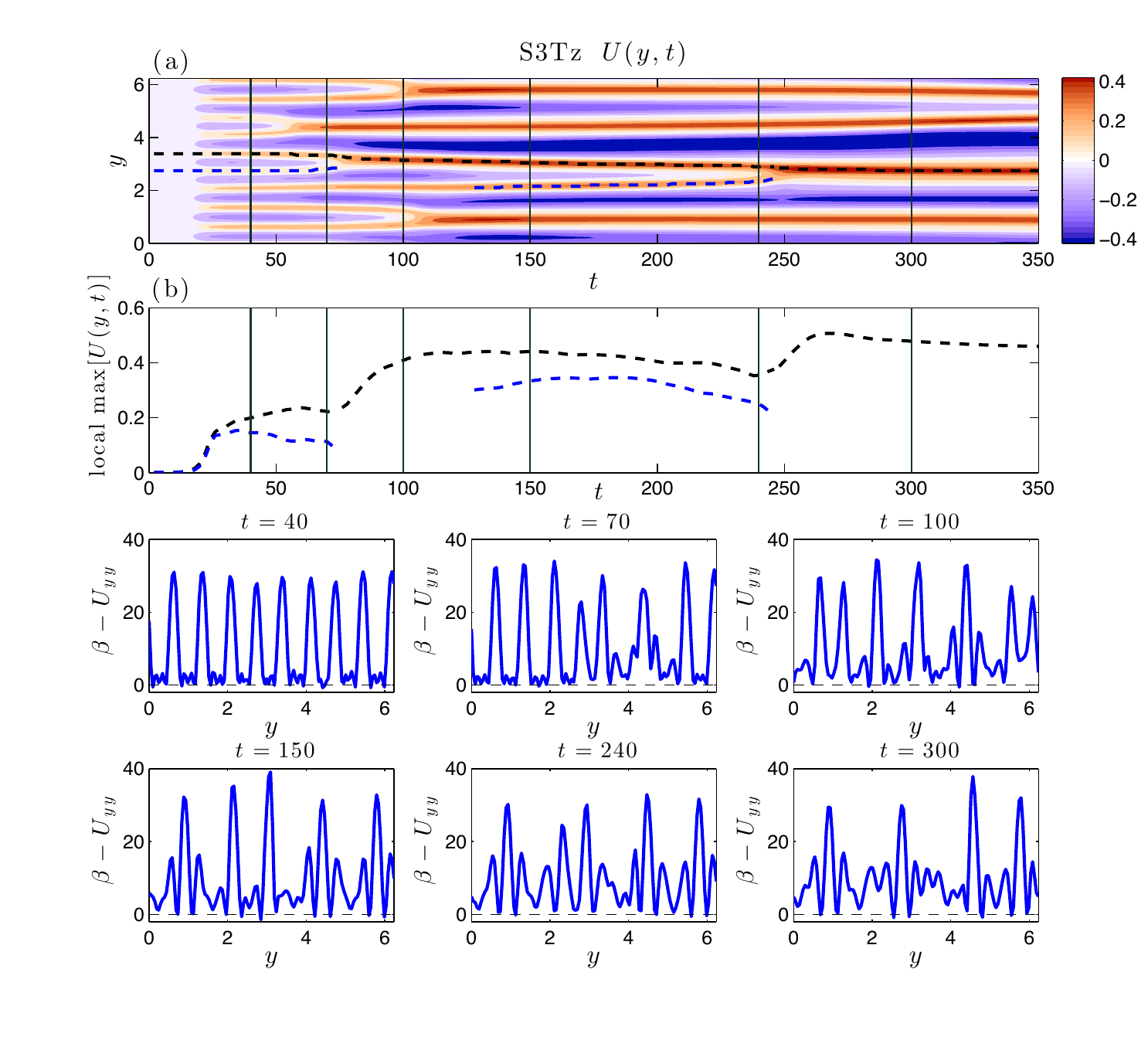}
\vspace{-2mm}
\caption{\label{fig:mergers}(a) Hovm\"oller diagram showing details of the jet mergers for $t\le 350$ in the S3T simulation in Fig.~\ref{fig:hov_IRFh_e100ec_r0p01}. In (b) is shown the amplitude of the jet maxima that appear in (a). Note that only the prograde jets merge. The bottom panels show the mean potential vorticity gradient $\beta-U_{yy}$ as a function of $y$ at the times indicated by vertical lines in (a) and (b). These graphs show that the structure of the jets is configured at each instant to satisfy the Rayleigh-Kuo stability criterion and that jet mergers are the mechanism in S3T for avoiding inflectional instability. Decrease in the amplitude of the jets prior to merger indicates increased downgradient vorticity fluxes as the flow approaches hydrodynamic neutrality.}
\end{figure}

Under strong turbulence forcing the initial S3T jet formation instability typically reaches final equilibrium as a finite amplitude jet at a wavenumber smaller than that of the initial instability. An example is the case of IRFn at $\e=100\ecz$ shown in Fig.~\ref{fig:hov_IRFh_e100ec_r0p01}. In this example, the jets emerge in S3T initially with zonal wavenumber $n_y=10$, in agreement with the prediction of the S3T instability of the homogeneous equilibrium, but eventually equilibrate at wavenumber $n_y=3$ following a series of jet mergers, as seen in the Hovm\"oller diagram. Similar dynamics are evident in the NL and QL simulations. This behavior can be rationalized by noting that if the wavenumber of the jet remains fixed then as jet amplitude continues to increase under strong turbulence forcing violation of the Rayleigh-Kuo stability criterion would necessarily occur. By transitioning to a lower wavenumber the flow is able to forestall this occurrence of inflectional instability. However, detailed analysis of the S3T stability of the finite amplitude equilibria near the point of jet merger reveals that these mergers coincide with the inception of a structural instability associated with eddy--mean flow interaction, which precedes the occurrence of hydrodynamic instability of the jet~\parencite{Farrell-Ioannou-2003-structural,Farrell-Ioannou-2007-structure}.\footnote{Jet mergers occur in the Ginzburg-Landau equations that govern the dynamics of the S3T instability of the homogeneous equilibrium state for parameter values for which the system is close to marginal stability \parencite{Parker-Krommes-2014-generation}. However, these mergers in the Ginzburg-Landau equations are associated with equilibration of the Eckhaus instability, rather than equilibration of the inflectional instability associated with violation of the Rayleigh-Kuo criterion as is the case for mergers of finite amplitude jets (cf.~Fig.~\ref{fig:mergers}). Characteristic of this difference is that in the case of the Ginzburg-Landau equations both the prograde and retrograde jets merge, while in the case of the finite amplitude jets only the prograde jets merge. The same phenomenology as in the Ginzburg-Landau equations occurs in the case of the Cahn-Hilliard equations that govern the dynamics of marginally stable jets in the modulational instability study of \textcite{Manfroi-Young-99}.} The stability of finite amplitude S3T equilibria will be discussed in chapter~\ref{ch:S3Tnonhom}.

\begin{figure}
\centering
\includegraphics[width=24pc,trim=0mm 15mm 0mm 3mm,clip]{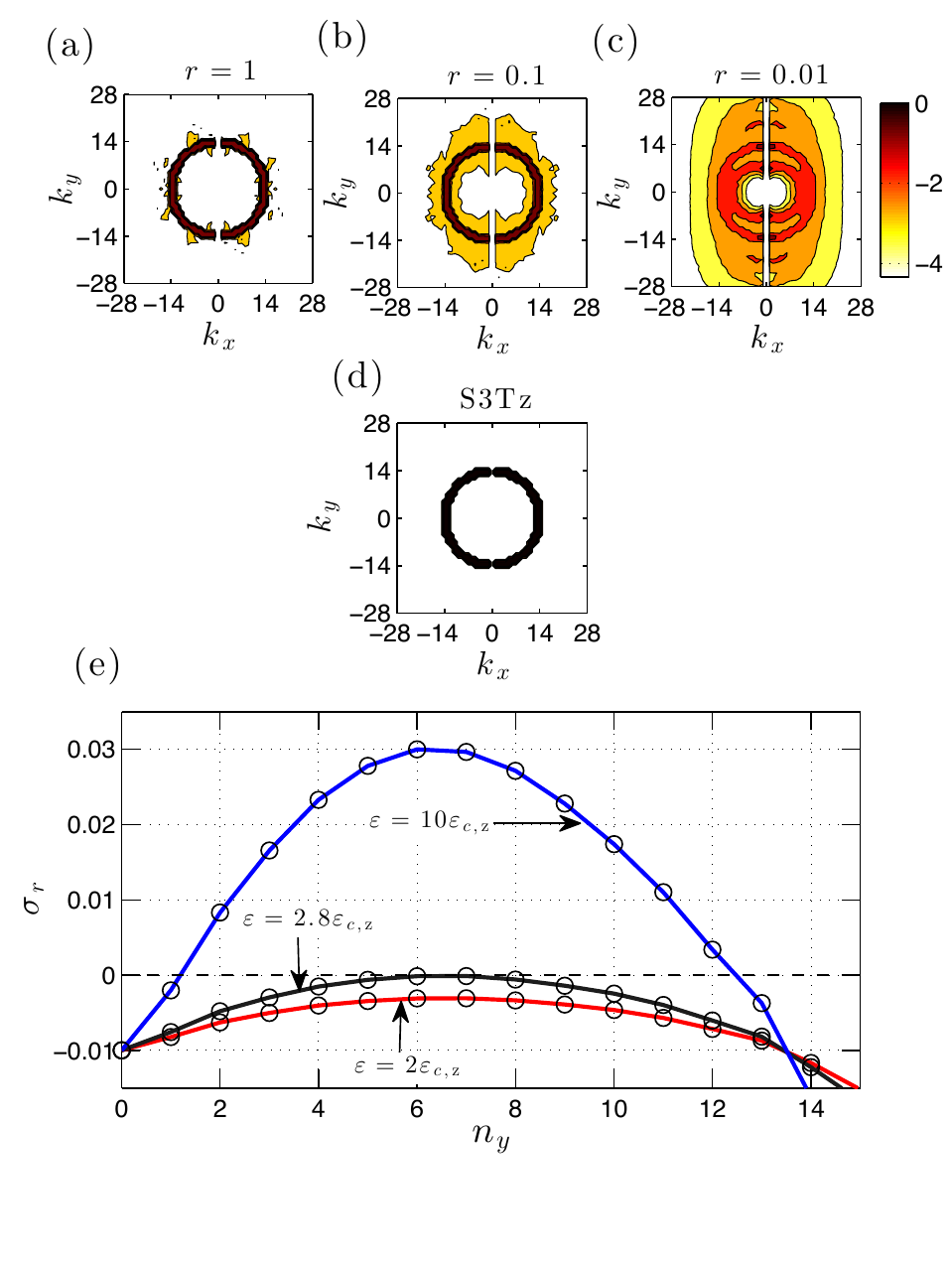}
 \caption{Panels (a)-(d): Equilibrium enstrophy spectrum, $\log{\( \<|\hat{\z}_{\kv}|^2\>\)}$,  of NL simulations, in which eddy--eddy interactions are included and the $k_x=0$ component is excluded, for various damping rates, $r$. The example is for IRFn forcing at $\e=2 \ecz$. Shown are spectra for: (a) $r=1$, (b) $r=0.1$ and (c) $r=0.01$. The critical $\ecz$ is a function of $r$ and is obtained from S3T for each value of $r$. All spectra have been normalized. The equilibrium spectrum of the S3T (identical to QL) is shown in panel (d). This figure shows that for strong damping the spectrum in NL simulations is close to the S3T spectrum while for weak damping the equilibrium spectrum in NL differs substantially from that in S3T. In all cases $\beta=10$. Panel (e): S3T growth rates, $s_r$, as a function of the meridional wavenumber, $n_y$, for the nonlinearly modified spectrum shown in panel (c) ($r=0.01$). Shown are cases for $\e=2\ecz$, $\e=2.8\ecz$ and $\e=10\ecz$. It can be seen that S3T stability analysis forced by this spectrum predicts that jets should emerge at $\e = 2.8 \ecz$ with $n_y=6$. S3T predictions are verified in NL as shown in the bifurcation diagram in Fig.~\ref{fig:bif_r_0p01}\hyperref[fig:bif_r_0p01]{b} (denoted as S3Tb).} 
 \label{fig:NLspect_stability_r0p01} 
\end{figure}

\section{Influence of the turbulence spectrum on the S3T jet formation instability}\label{sec:influence_eddy}

Both QL and S3T dynamics exclude interactions among eddies and include only the non-local interactions between jets, with $k_x=0$, and eddies, with $k_x\ne0$. Therefore, there is no enstrophy or energy cascade in wavenumber space in either QL or S3T dynamics and the homogeneous S3T equilibrium state (cf.~\eqref{eq:hom_equil}) has spectrum, ${\e}\hat{Q}_{\kv} / (2 r)$,which is determined by the spectrum of the forcing ($\hat{Q}_{\kv} $ is the spectral power of the forcing covariance, cf.~Appendix~\ref{app:forcing}).
%, $\tilde{Q}_{\kv}$:
%\begin{equation}
%  \< \bit |\tilde{\z}_{\kv}|^2\> = \frac{\e} {2r} \tilde{Q}_{\kv} ~,
% \label{eq:s3tspectrum}
% \end{equation}
%(cf. Appendix B ).
However, this is not true in NL which includes eddy--eddy interactions producing enstrophy/energy cascades. For example, in NL an isotropic ring forcing is spread as time progresses, becoming concentrated at lower wavenumbers and forming the characteristic dumbbell shape seen in $\b$-plane turbulence simulations (cf.~\textcite{Vallis-Maltrud-93}) and consequently  the homogeneous turbulent state is no longer characterized by the spectrum of the forcing. We can take account of this modification of the spectrum by performing S3T stability on the homogeneous state under the equivalent forcing covariance,
\be
\hat{Q}_{\kv}^\textrm{NL} = \frac{2 r}{\e}\,\<\bit |\hat{\z}_{\kv}|^2\>~,\label{eq:nlspectrum}
\ee
which maintains the observed NL spectrum, $\< \bit |\hat{\z}_{\kv}|^2\> $, in the S3T dynamics. The NL modified eddy vorticity spectrum, $\< \bit |\hat{\z}_{\kv}|^2\> $, is obtained from an ensemble of NL simulations. Plots of $\< \bit |\hat{\z}_{\kv}|^2\>$, under IRFn forcing are shown in Figs.~\ref{fig:NLspect_stability_r0p01}\hyperref[fig:NLspect_stability_r0p01]{a-c} for 
various energy input rates, $\e$, and damping rates, $r$. 
The departure of the NL spectra from the spectra of the QL and S3T equilibria is evident and this 
departure depends on the amplitude of the forcing, $\e$, and the damping, $r$.

\begin{figure}[h!]
\centering
\includegraphics[width=24pc,trim=3mm 7mm 3mm 1mm,clip]{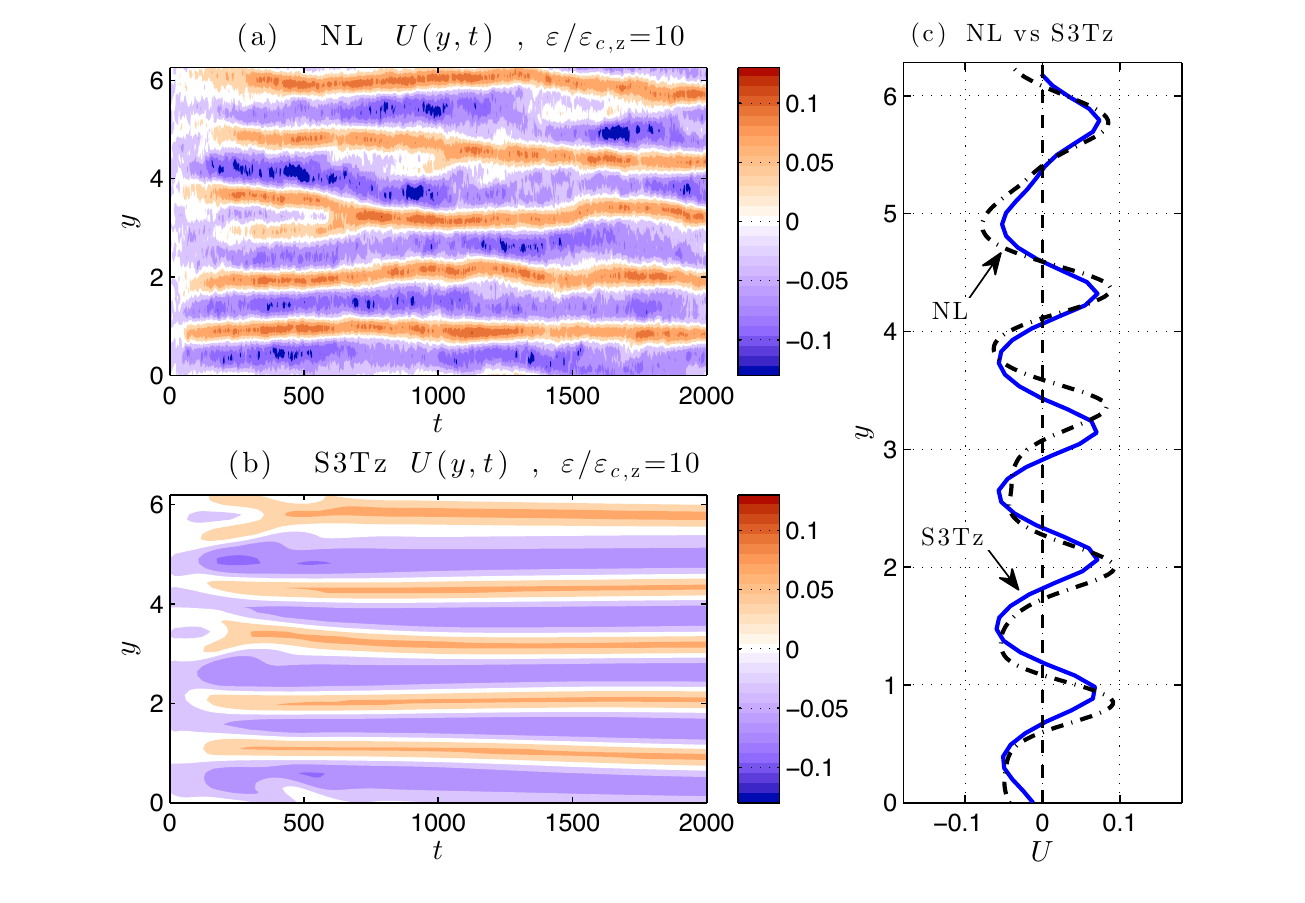}
 \caption{Hovm\"oller diagrams of $U(y,t)$ comparing jet emergence and equilibration in an NL 
 simulation under IRFn forcing (panel (a)) with an S3T simulation under S3Tb forcing (panel (b)). The corresponding time mean jets are shown in panel (c). This figure shows that the S3Tb modification of the forcing spectrum suffices to obtain agreement with NL. Parameters are $\e=10\ecz$, $\beta=10$, $r=0.01$.} 
 \label{fig:hov_IRFh_e10ec_S3Tb} 
\end{figure}

We now demonstrate that while the fundamental qualitative prediction of S3T that jets form as a bifurcation in turbulence forcing and in the absence of turbulent cascades is verified in both QL and NL, a necessary condition for obtaining quantitative agreement between NL and both S3T and QL dynamics is that the equilibrium spectrum used in the S3T and QL dynamics be close to the equilibrium spectrum obtained in NL so that the stability analysis is performed on similar states. 
In the case with IRFn and $r=0.01$, formation of persistent finite amplitude zonal jets occurs in the NL simulations at $\e=2.8 \ecz$ (cf.~Fig.~\ref{fig:bif_r_0p01}\hyperref[fig:bif_r_0p01]{b}). 
In agreement, S3T stability analysis on the NL modified equilibrium IRFn spectrum (denoted S3Tb and shown in Fig.~\ref{fig:NLspect_stability_r0p01}\hyperref[fig:NLspect_stability_r0p01]{c})
predicts instability for $\e \ge 2.8 \ecz$ (cf.~Fig.~\ref{fig:NLspect_stability_r0p01}\hyperref[fig:NLspect_stability_r0p01]{e}).
Moreover, S3T stability analysis with the S3Tb spectrum predicts jet 
formation at $n_y=6$ and in agreement with this prediction jets emerge in NL with $n_y=6$. 
Hovm\"oller diagrams demonstrating similar jet evolution in NL under 
IRFn forcing and in S3T under S3Tb forcing are shown in Fig.~\ref{fig:hov_IRFh_e10ec_S3Tb}. We also note that agreement between NL and S3T in predictions of jet amplitude at large supercriticality is also obtained by using the S3Tb spectrum (cf.~Figs.~\ref{fig:bif_r_0p01}\hyperref[fig:bif_r_0p01]{a} and~\ref{fig:bif_r_0p01}\hyperref[fig:bif_r_0p01]{b}).\footnote{The spectral peaks near the $k_y$ axis do not directly influence the stability of the NL modified spectrum, which is determined by the distorted and broadened ring spectrum. However, while the spectral peaks do not influence the stability directly, they do
influence it indirectly by distorting the incoherent spectrum.}

\begin{figure}
\centering
\includegraphics[width=4.5in,trim=0mm 5mm 0mm 0mm,clip]{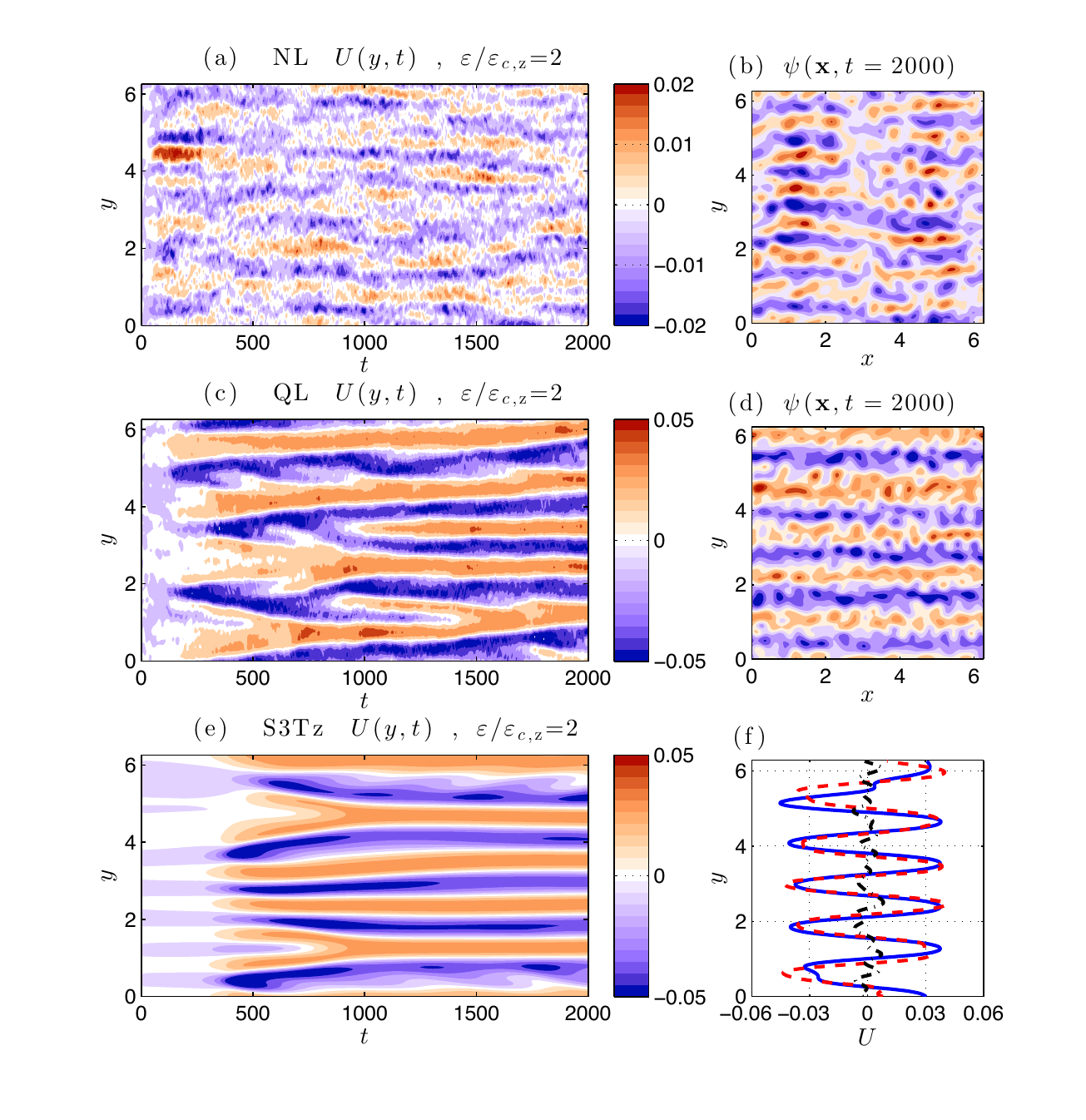}
\caption{\label{fig:hovNLQLS3T_e2_IRFn} Hovm\"oller diagrams of jet emergence in NL, QL and S3T simulations with IRFn forcing at $\e= 2 \ecz$. Shown is $U(y,t)$ for the NL (panel (a)), QL (panel (c)) and S3T (panel (e)) simulations and characteristic snapshots of streamfunction fields at $t=2000$ for the NL and QL simulations (panels (b) and (d)). Notice that in the $U(y,t)$ diagram for NL the color axis is scaled differently. Also shown are the equilibrium jets in the NL (dash-dot), QL (dashed), and S3T (solid) simulation (panel (f)). At $\e=2\ecz$ in the NL simulation no jets emerge but accumulation of energy in non-zonal structures with zonal wavenumber $k_x=1$ and meridional wavenumber $k_y=7$ is discernible. Parameters are $\beta=10$, $r=0.01$.}
\end{figure}

\begin{figure}[h!]
\centering\includegraphics[width=24pc]{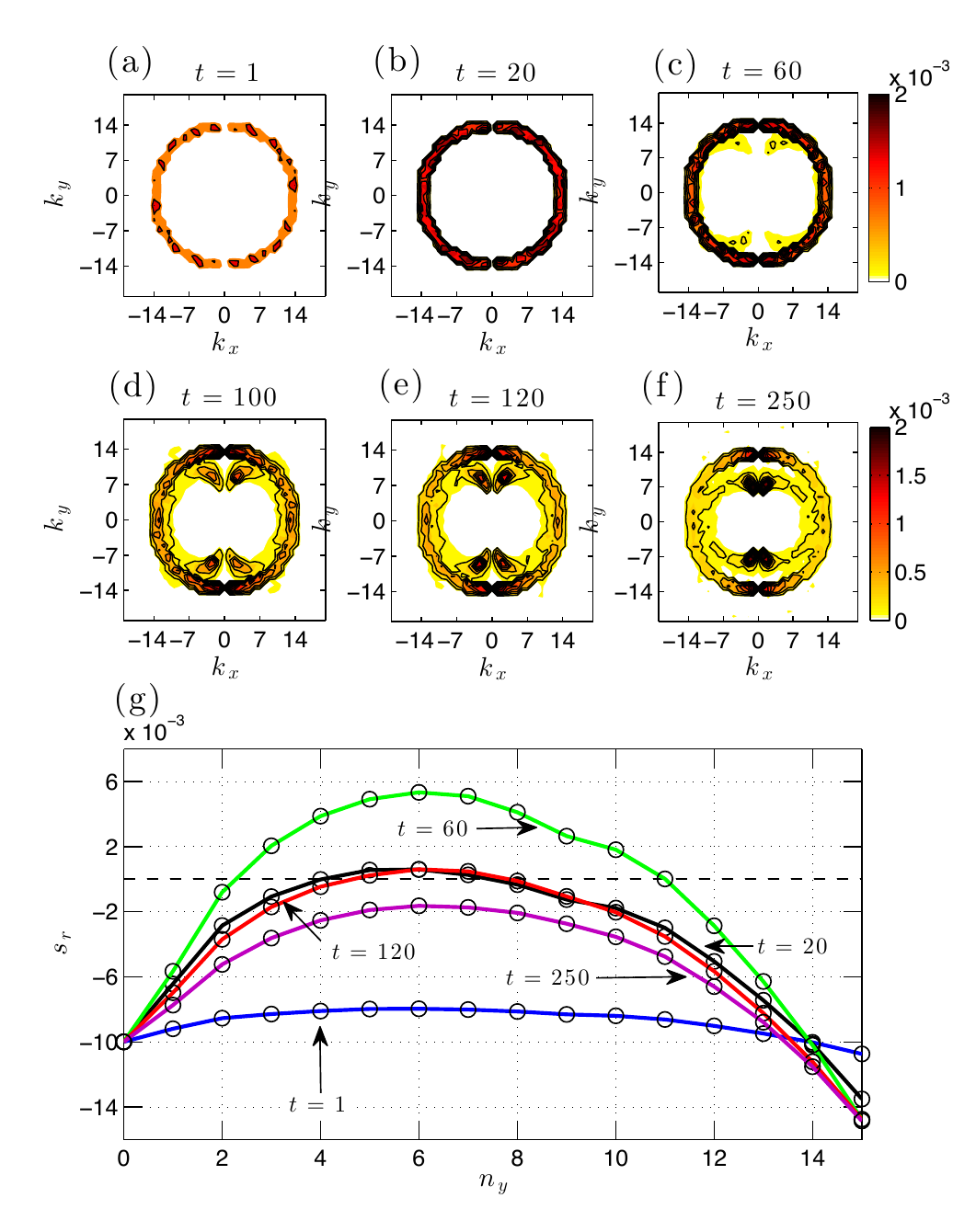}
\caption{Panels (a)-(f): Evolution of the ensemble average enstrophy spectrum, $\<|\hat{\z}_{\kv}|^2\>$, for NL with IRFn forcing at $\e=2\ecz$. Panel (g): Growth rates, $s_r$, as a function of jet meridional wavenumber, $n_y$, predicted by S3T stability analysis performed on the instantaneous spectrum at the times indicated in panels (a)-(f). The evolving spectrum renders the NL simulation S3T unstable at $t\approx20$ and stabilizes it again at $t\approx120$. Parameters are $\beta=10$, $r=0.01$.} 
 \label{fig:NL_spectr_stability}
\end{figure}

This influence of the eddy spectrum on jet dynamics is revealed in the case of IRFn at
energy input rate $\e=2\ecz$, shown in Fig.~\ref{fig:hovNLQLS3T_e2_IRFn}. Although at this energy input rate S3T under IRFn is structurally unstable, no jets emerge in NL. We have shown that agreement in bifurcation structure is obtained between NL and S3T when S3T analysis is performed with the S3Tb spectrum. We now examine the development of the NL spectrum towards S3Tb and demonstrate the close control exerted by this evolving spectrum on S3T stability. The evolving spectrum, shown in Fig.~\ref{fig:NL_spectr_stability}\hyperref[fig:NL_spectr_stability]{a-f}, is obtained using an ensemble of NL simulations, each starting from a state of rest and evolving under a different forcing realization. A sequence of S3T stability analyses performed on this evolving ensemble spectrum is show in Fig.~\ref{fig:NL_spectr_stability}\hyperref[fig:NL_spectr_stability]{g}. The weak NL ensemble spectrum at $t=1$ does not support instability, but by $t=20$ the ensemble spectrum, having assumed the isotropic ring structure of the forcing, becomes S3T unstable.
 This structural instability results in the formation of an incipient $n_y=6$ jet structure which is evident by $t=50$ in the NL simulation shown in Fig.~\ref{fig:hovNLQLS3T_e2_IRFn}.
As the spectrum further evolves, the S3T growth rates decrease and no jet structure is unstable for $t > 120$, and decay rates continue to increase 
until $t=250$  (cf.~Fig.~\ref{fig:NL_spectr_stability}\hyperref[fig:NL_spectr_stability]{g}). 
This example demonstrates the tight control on S3T stability exerted by the spectrum. 
Furthermore, it shows the close association between S3T instability and the emergence of jet structure in NL.

\section{Influence of non-zonal structures predicted by S3T on the turbulence spectrum and on jet dynamics}\label{sec:nonzonal}

\begin{figure}[ht]
\centering\includegraphics[width=4in]{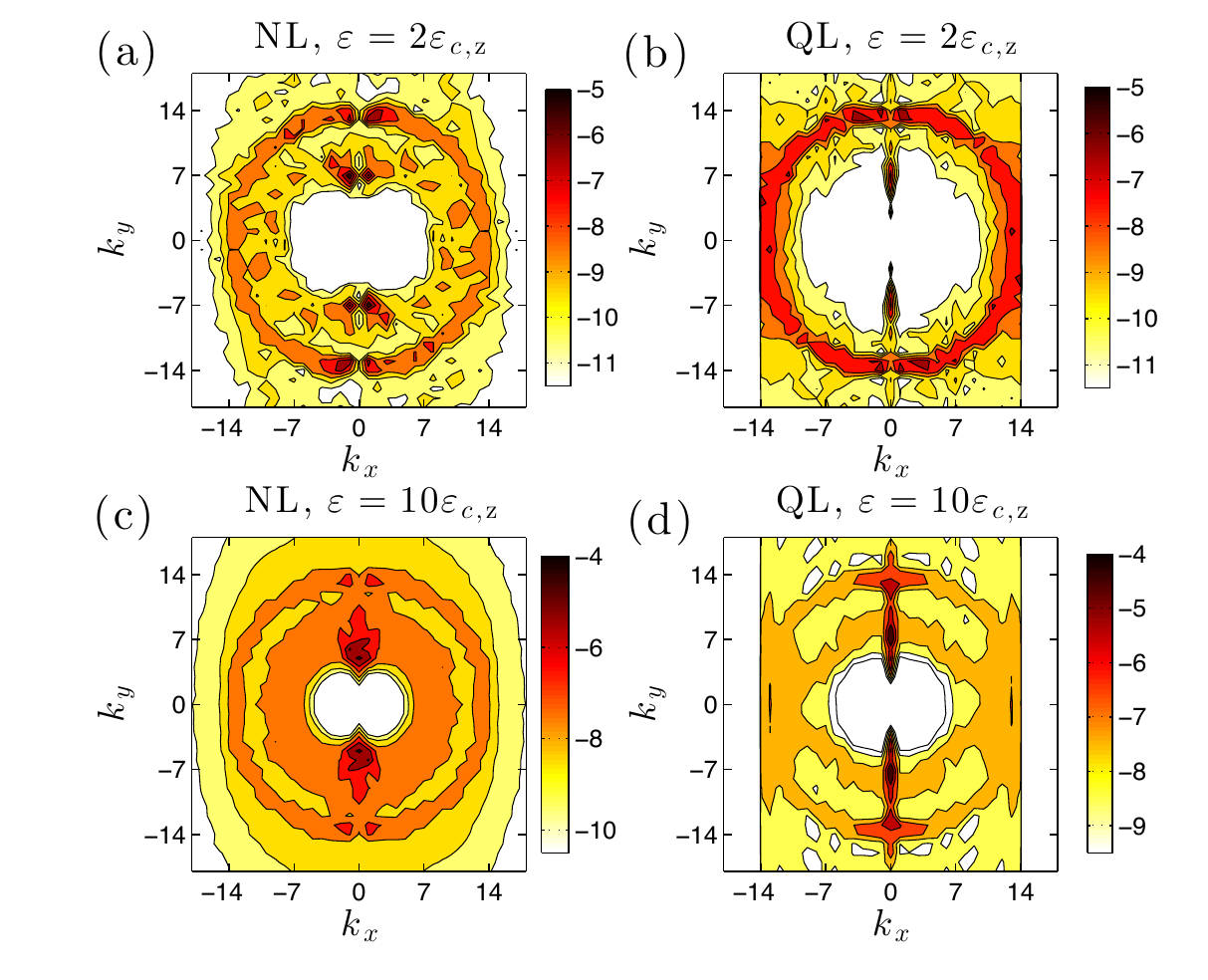}
\caption{\label{fig:spec_NLQL_IRFn_e2e10} The statistical equilibrium enstrophy spectrum, $\log{\( \<|\hat{\z}_{\kv}|^2\>\)}$, for NL and QL simulations under IRFn forcing at $\e=2\ecz$ (panels (a) and (b)) and $\e=10\ecz$ (panels (c) and (d)). For $\e=2\ecz$ the NL simulations do not support zonal jets and energy is seen to accumulate in the non-zonal structure $(|k_x|,|k_y|)=(1,7)$. At $\e=10\ecz$, persistent zonal jets emerge (cf.~Fig.~\ref{fig:hov_IRFh_e10ec_S3Tb}) suppressing the power in the non-zonal structures. Parameters: $\beta=10$, $r=0.01$.}
\end{figure}

\begin{figure}
\centering\includegraphics[width=24pc,trim=7mm 11mm 7mm 1mm,clip]{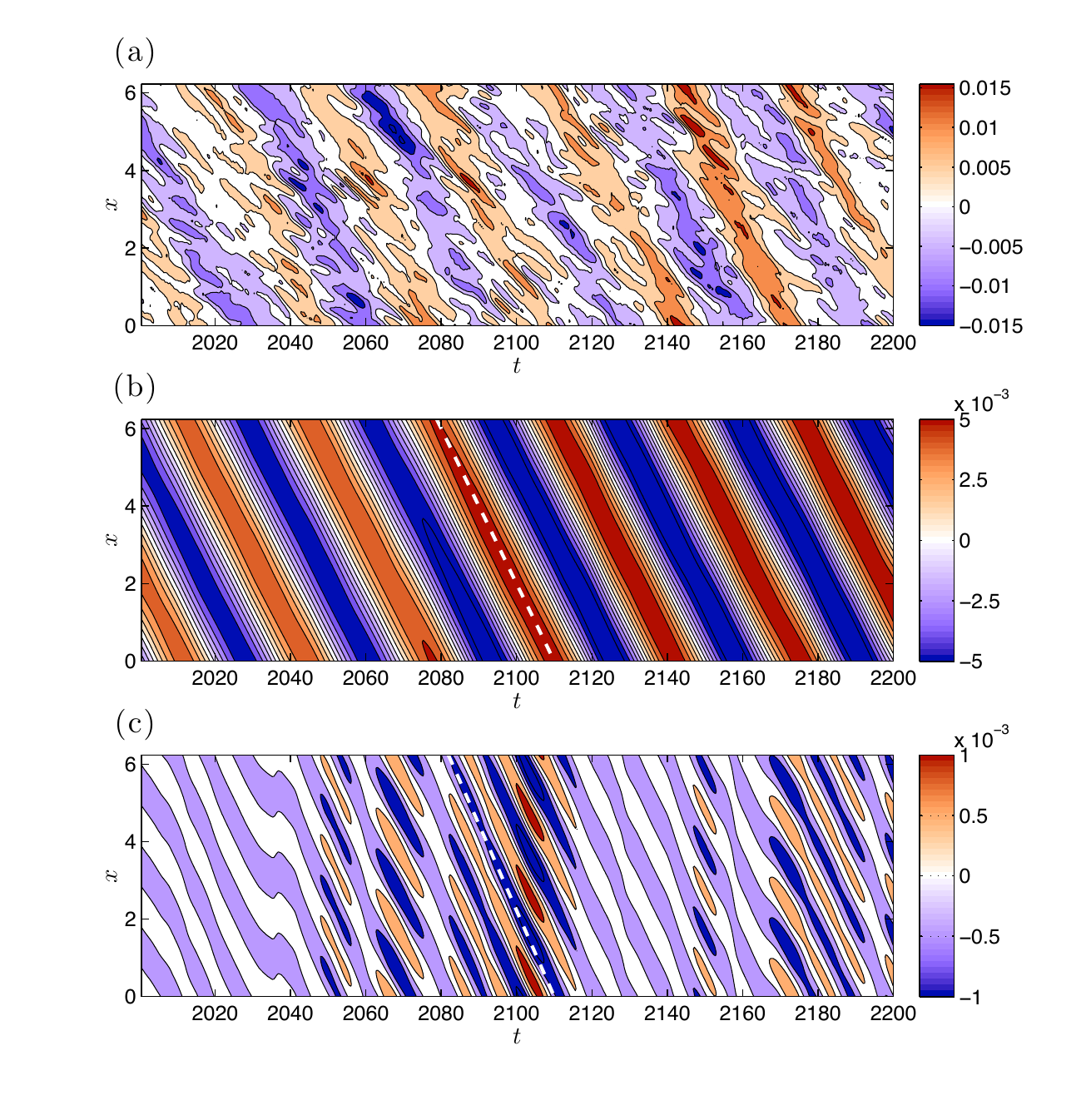}
\caption{\label{fig:nz_hov_IRFh_e2ec_r0p01} Hov\-m\"oller diagrams of the non-zonal structures supported in the NL simulation of Fig.~\ref{fig:hovNLQLS3T_e2_IRFn}. Panel (a): evolution of the total perturbation streamfunction, $\psi(x,y=y_0,t)$, at latitude $y_0=\pi/4$. Panel (b): evolution of the dominant $(|k_x|,|k_y|)=(1,7)$ structure of $\psi(x,y=y_0,t)$ at latitude $y_0=\pi/4$. Almost half of the energy input to the system is captured and dissipated by this mode, which is phase coherent and propagates at the Rossby wave speed indicated by the dashed line. Panel (c): evolution of the $(|k_x|,|k_y|)=(3,6)$ structure at the same latitude. While this structure propagates at the Rossby wave speed it is not phase coherent. Parameters: IRFn forcing at $\e=2\ecz$, $\beta=10$, $r=0.01$.}
\end{figure}

Despite S3T supercriticality, no persistent jets emerge in NL simulations with IRFn forcing in the interval $\ecz<\e<2.8 \ecz$ (cf.~Fig.~\ref{fig:bif_r_0p01}\hyperref[fig:bif_r_0p01]{a}). Comparisons of NL, QL and S3T simulations with IRFn forcing at $\e=2 \ecz $ are shown in Fig.~\ref{fig:hovNLQLS3T_e2_IRFn}. 
Instead of zonal jets, in the NL simulation prominent non-zonal structures are seen to propagate westward at the Rossby wave phase speed. These non-zonal structures are also evident in the concentration of power in the enstrophy spectrum at $(|k_x|,|k_y|)=(1,7)$ (cf.~top panels of Fig.~\ref{fig:spec_NLQL_IRFn_e2e10}).
At this forcing amplitude these structures are essentially linear Rossby waves which, 
if stochastically forced, would be coherent only over the dissipation time scale $1/r$. Coherence on the dissipation time scale is observed in the subdominant part of the spectrum as seen in the case of the $(3,6)$ structure in Fig.~\ref{fig:nz_hov_IRFh_e2ec_r0p01}\hyperref[fig:nz_hov_IRFh_e2ec_r0p01]{c}.
However, the dominant $(1,7)$ structure remains coherent over time periods far exceeding the dissipation time scale (cf.~Hov\-m\"oller diagram Fig.~\ref{fig:nz_hov_IRFh_e2ec_r0p01}\hyperref[fig:nz_hov_IRFh_e2ec_r0p01]{b}).
This case represents a regime in which the flow is dominated by a single non-zonal structure. Both the concentration of power in and the coherence of this structure will be addressed below.

\begin{figure}
\centering\includegraphics[width=24pc]{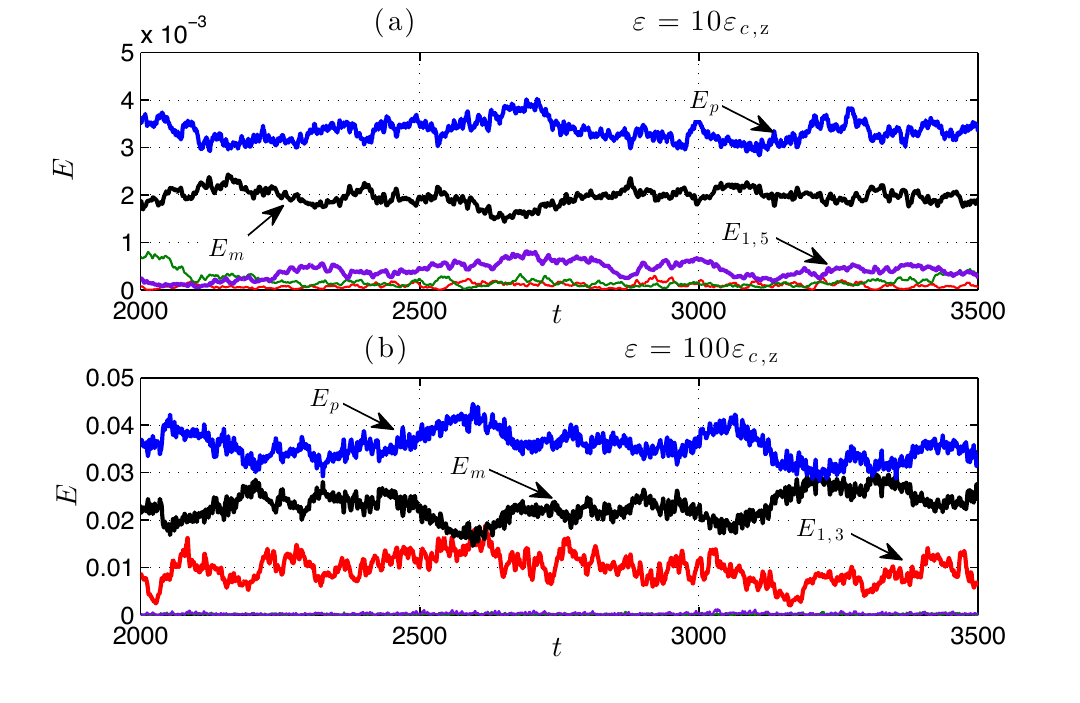}
\caption{Panel (a): Evolution of the mean flow energy, $E_m$, which is concentrated at $(0,6)$, the total eddy energy, $E_p$, and the energy of the (1,5), (1,6) and (1,7) structures for the NL simulation with IRFn forcing at $\e=10\ecz$, shown in Fig.~\ref{fig:hov_IRFh_e10ec_S3Tb}. Panel (b): Evolution of the mean flow energy, $E_m$, the total eddy energy, $E_p$, as well as the energy of the (1,3), (1,5) and (1,6) structures for the NL simulation with IRFn forcing at $\e=100\ecz$, shown in Fig.~\ref{fig:hov_IRFh_e100ec_r0p01}. The mean flow energy is concentrated at $(0,3)$. In both panels the evolution of the energies is shown after statistical steady state has been reached.}
 \label{fig:Em_Eps_f10_f100}
\end{figure}

\begin{figure}
\centering\includegraphics[width=24pc,trim=5mm 1mm 5mm 1mm,clip]{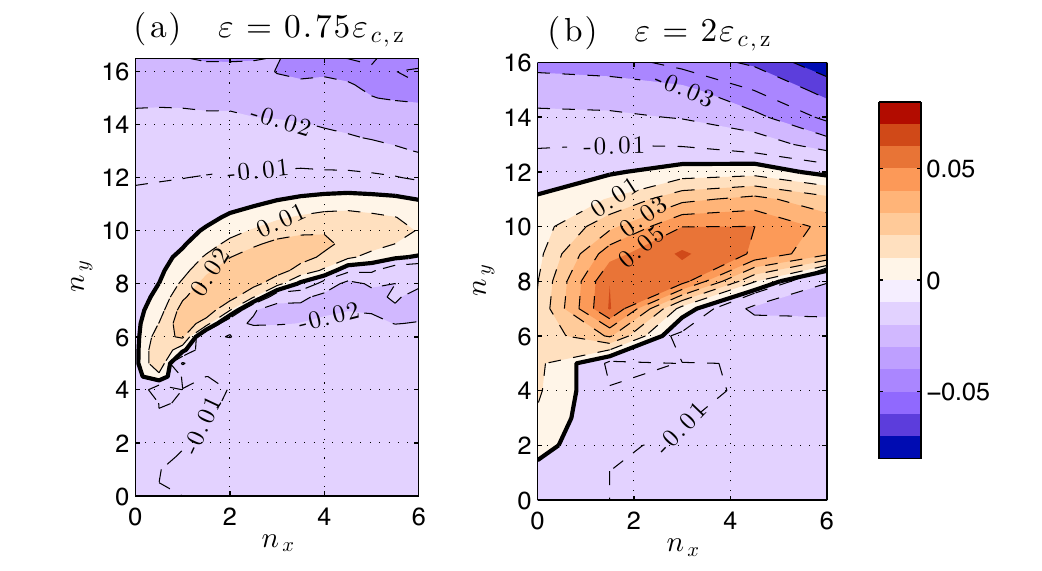}
\caption{Growth rate, $s_r$, of the S3T non-zonal eigenfunction, $e^{\i\nv\cdot\xv}$, as a function of zonal wavenumber $n_x$ and meridional wavenumber, $n_y$ for IRFn at $\e=0.75\ecz$ (panel (a)) and $\e=2\ecz$ (panel (b)). The values at the axis, $(0,n_y)$, give the growth rate of the corresponding jet perturbation. For $\e =0.75 \ecz$ the $n_x=0$ jet eigenfunctions are stable but the non-zonal perturbations are unstable with maximum instability occurring at $\nv=(2,8)$. For $\e=2\ecz$ the $n_x=0$ perturbations are unstable but the non-zonal perturbations are more strongly unstable, with maximum growth at $\nv=(2,8)$ and $\nv=(1,7)$. An NL simulation at $\e=2\ecz$ accumulates energy at $(|k_x|,|k_y|)=(1,7)$ (cf.~Fig.~\ref{fig:spec_NLQL_IRFn_e2e10}) while the vorticity field shows some accumulation at $(|k_x|,|k_y|)=(2,8)$ (cf.~Fig.~\ref{fig:NL_spectr_stability}\hyperref[fig:bif_r_0p1]{f}). The stability boundary ($s_r=0$) is marked with thick solid line. For both panels $\beta=10$ and $r=0.01$.} \label{fig:nz_S3Tgr_IRFh_r_0p01}
\end{figure}

When the forcing is increased to $\e = 10 \ecz$, a $(0,6)$ jet structure emerges, suppresses the non-zonal $(1,7)$ structure, and becomes the dominant structure. A prominent phase coherent non-zonal $(1,5)$ structure propagating with the Rossby wave speed is also present, as shown in Fig.~\ref{fig:Em_Eps_f10_f100}.
 A similar regime of coexisting jets and non-zonal structures is also evident at higher supercriticalities. An example is the case of the equilibrium state at $\e = 100 \ecz$ (cf.~Fig.~\ref{fig:hov_IRFh_e100ec_r0p01}) in which the energy of the flow is shared between the $(0,3)$ jet and the $(1,3)$ structure, as shown in Fig.~\ref{fig:Em_Eps_f10_f100}. At this forcing level the $(1,3)$ structure is not phase coherent, but its phase speed is still given by the Rossby wave speed. At even higher forcing similar non-zonal structures, referred to as zonons, have been reported to coexist with zonal jets while propagating phase incoherently at speeds that differ substantially from the Rossby wave speed \parencite{Sukoriansky-etal-2008}. These cases provide examples of the regime in which jets and non-zonal structures coexist. 
 
In order to study the dynamics of non-zonal structures within the framework of S3T the interpretation of the ensemble mean in the S3T formulation is required: instead of interpreting the ensemble means as zonal means, interpret them rather as Reynolds averages over an intermediate time scale, cf.~\eqref{eq:s3t}. As we have seen in chapter~\ref{ch:st3hom}, the S3T stability of the homogeneous equilibrium state using this interpretation reveals that when the energy input rate reaches the value $\ecz$, which is the S3T stability threshold for the emergence of zonal jets, the state may already be unstable to non-zonal structures (cf.~Fig.~\ref{fig:s3tgr_isotropic}\hyperref[fig:s3tgr_isotropic]{c,e}). This can be also seen in the stability analysis shown in Fig.~\ref{fig:nz_S3Tgr_IRFh_r_0p01}. In agreement with this stability analysis, the spectrum of the NL simulation shows concentration of power in these most S3T unstable wavenumbers (cf.~Fig.~\ref{fig:spec_NLQL_IRFn_e2e10}). 

The dominance and persistence of the structures seen in these NL simulations can be understood from this stability analysis and its extension into the nonlinear regime. Because the stochastic forcing is white in time, the energy injection rate is fixed and state independent and, assuming linear damping at rate $r$ dominates the dissipation, the total flow energy assumes the fixed and state independent mean value $E_m+E_p=\e/ (2 r)$. At finite amplitude the set of S3T unstable structures equilibrate to allocate among themselves most of this energy which results in the dominance of a small subset of these structures. However, we find that in this competition a specific zonal jet structure has primacy so that even if this structure is not the most linearly unstable it emerges as the dominant structure.  

\vspace{1em}

\begin{figure}[b]
\centering
\includegraphics[width=.99\textwidth]{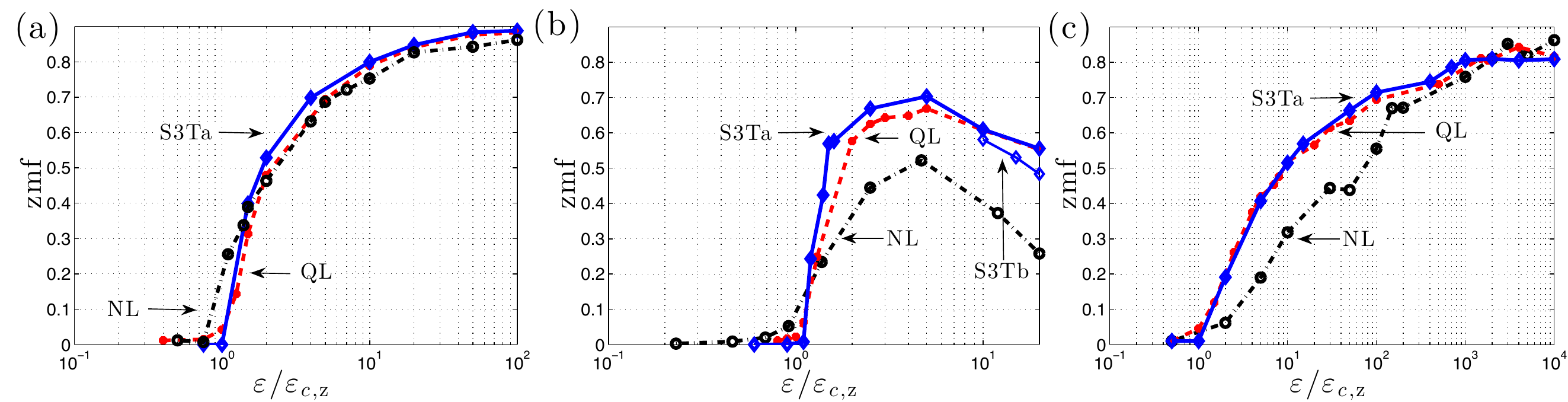}
\caption{\label{fig:bif_r_0p1} Bifurcation structure comparison for jet formation in S3T, QL, and NL with asymmetric damping. Shown is the zmf index of jet equilibria for NIF (panel (a)), IRFn (panel (b)) and IRFw (panel (c)) as a function of the forcing amplitude $\e/\ecz$ for the NL simulation (dash-dot and circles), the QL simulation (dashed and dots) and the corresponding S3Ta simulation (solid and diamonds). Also shown in panel (b) is the zmf that is obtained from S3T simulations forced with the nonlinearly modified S3Tb spectrum (calculated from ensemble NL simulations at $\e=20\ecz$). Parameters are $\beta=10$, $r=0.1$, $\rU=0.01$.}
\end{figure}

An attractive means for exploring the dynamics of the interaction between jets and non-zonal structures is changing the jet damping rate in the mean flow equation~\eqref{eq:NL_U_r} from $r$ to $\rU$ and allowing it to assume values different from the perturbation damping rate, $r$, in~\eqref{eq:NL_z_r}. (The same change of $r$ to $\rU$ is also done in the QL system~\eqref{eq:QL_r} and the S3T system~\eqref{eq:S3T_r}.) In this way we can control the relative stability of jets and non-zonal structures as well as the finite equilibrium amplitude reached by the jet.
This asymmetric damping may be regarded as a model for approximating
jet dynamics in a baroclinic flow in which the upper level jet is lightly damped, while the active baroclinic turbulence generating scales are strongly Ekman damped. This asymmetry in the damping between upper and lower levels contributes  to making jets in baroclinic turbulence generally stronger than jets in barotropic turbulence \parencite{Farrell-Ioannou-2007-structure, Farrell-Ioannou-2008-baroclinic}.
By appropriate choice of $r$ and $\rU$ a regime can be obtained in which the zonal jet instability appears first as $\e$ increases. 
Because once jets are unstable they dominate non-zonal structures, in this regime zonal jets are the dominant coherent structure and S3T analysis based on the zonal interpretation of the ensemble mean produces very good agreement with NL. For example, a comparison of bifurcation structures among S3T, QL and NL under NIF and IRFn forcing using the asymmetric damping $r=0.1$ and  $\rU=0.01$ demonstrates that jets emerge at the same critical value in S3T, QL and NL (cf.~Figs.~\ref{fig:bif_r_0p1}\hyperref[fig:bif_r_0p1]{a} and \ref{fig:bif_r_0p1}\hyperref[fig:bif_r_0p1]{b}). This agreement, which has been obtained by suppression of the non-zonal instability up to $\ecz$, implies that in the simulations with symmetric damping the disagreement in the S3T prediction for the first emergence of jets (cf.~Fig.~\ref{fig:bif_r_0p01}) can be attributed to modification of the background spectrum by 
the prior emergence of the non-zonal structures. Moreover, zonal structures once unstable immediately dominate non-zonal structures assuring that S3T dynamics based on the zonal mean interpretation of the ensemble mean produces accurate results.

\begin{figure}[!h]
\centering\includegraphics[width=20.3pc,trim=0mm 2.5mm 0mm 5mm,clip]{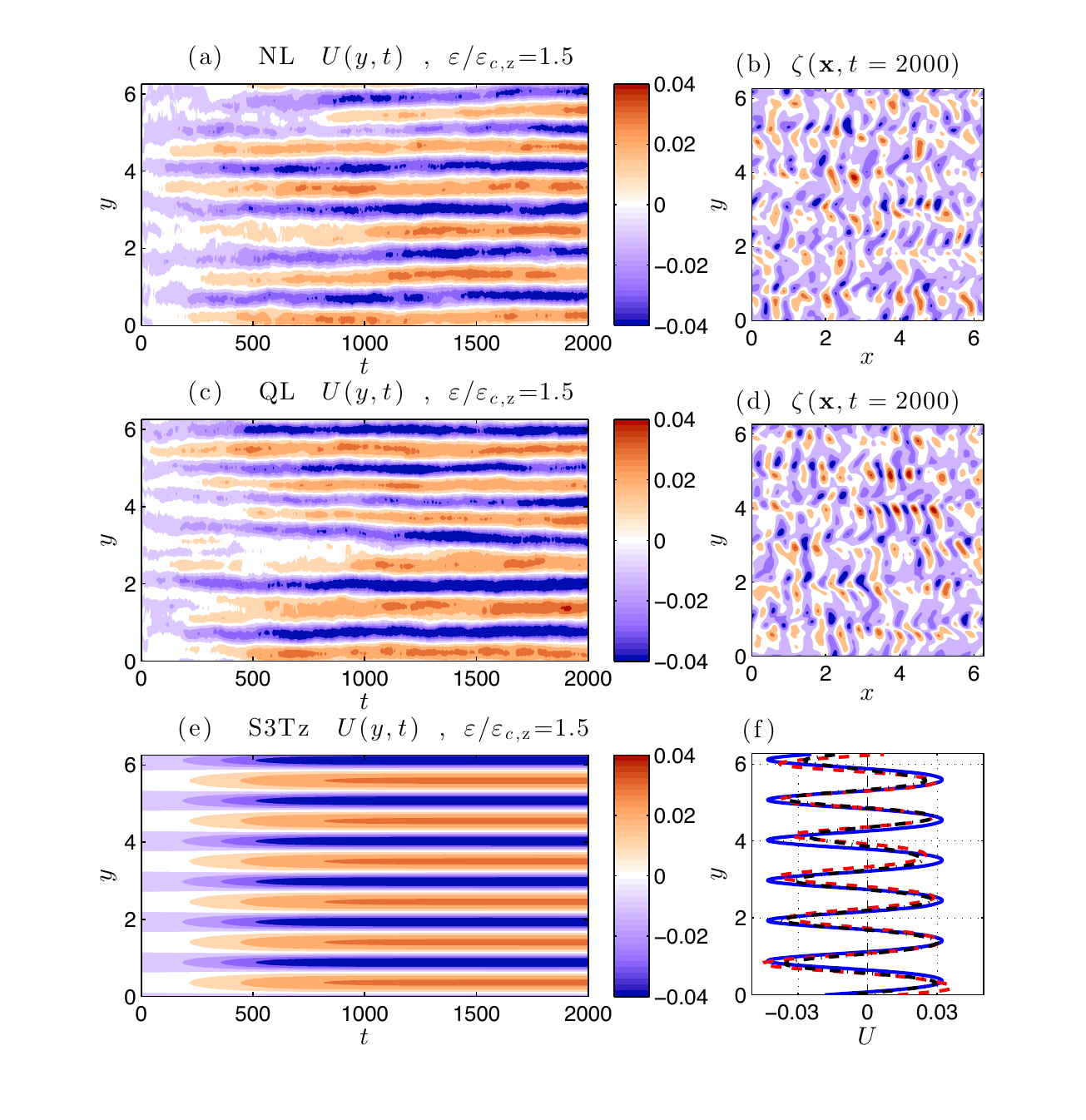}
\vspace{-2em} \caption{\label{fig:hov_NIF_e1p5ec_r0p1_rm0p01} Hovm\"oller diagrams of jet emergence in the NL, QL and S3T simulations for NIF at $\e=1.5\ecz$ with asymmetric damping. Shown is $U(y,t)$ for the NL (panel (a)), QL (panel (c)) and S3T (panel (e)) simulations and also characteristic snapshot of the vorticity fields at $t=2000$ for NL and QL simulations (panels (b) and (d)). Also shown are the equilibrium jets in the NL (dash-dot), QL (dashed), and S3T (solid) simulation (panel (f)). This figure shows that S3T predicts the structure, growth and equilibration of weakly forced jets in both the QL and NL simulations. Parameters are: $\beta=10$, $r=0.1$, $\rU=0.01$.}
%\end{figure}
%
%\begin{figure}
\centering\includegraphics[width=20.3pc,trim=0mm 2.5mm 0mm 0mm,clip]{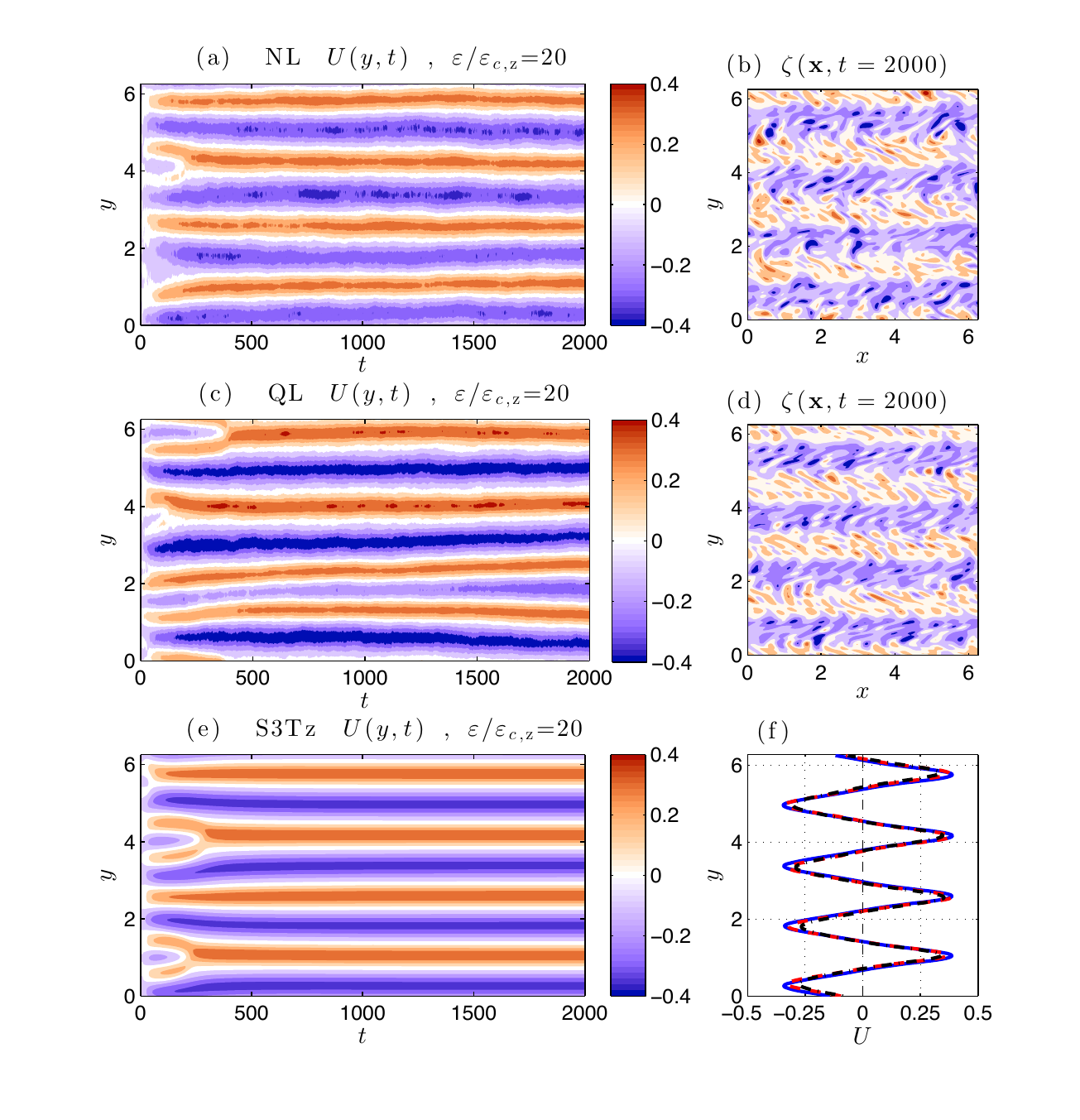}
\vspace{-2em}  \caption{\label{fig:hov_NIF_e20ec_r0p1_rm0p01} Same as Fig.~\ref{fig:hov_NIF_e1p5ec_r0p1_rm0p01} but with forcing at $\e=20\ecz$. While initially jets emerge having the structure of the most unstable S3T mean flow eigenfunction with $n_y=6$, at later times, following a series of mergers they equilibrate to a finite amplitude state with $n_y=4$.}
\end{figure}
%\clearpage

A comparison of the development of jets in S3T, QL, and NL with this asymmetric damping and NIF forcing, shown in Figs.~\ref{fig:hov_NIF_e1p5ec_r0p1_rm0p01} and~\ref{fig:hov_NIF_e20ec_r0p1_rm0p01}, demonstrates the accuracy of the S3T predictions. S3T stability analysis predicts that in this case with NIF forcing maximum instability occurs at $n_y=6$. When these maximally growing eigenfunctions are introduced in the S3T system the jets grow exponentially at first at the predicted rate and then equilibrate. Corresponding simulations with the QL and NL dynamics reveal nearly identical jet growth followed by  finite amplitude equilibration (shown in both Figs.~\ref{fig:hov_NIF_e1p5ec_r0p1_rm0p01} and~\ref{fig:hov_NIF_e20ec_r0p1_rm0p01}). Similar results are obtained with IRFn forcing. This demonstrates that the S3T dynamics comprises both the jet instability mechanism and the mechanism of finite amplitude equilibration.

Although no theoretical prediction of this bifurcation behavior can be made directly from NL or QL, they both reveal the bifurcation structure obtained from the S3T analysis. By suppressing the peripheral complexity of non-zonal structure formation by non-zonal S3T instabilities, these simulations allow construction of a simple model example that provides compelling evidence for identifying jet formation and equilibration in NL with the S3T theoretical framework. Moreover, agreement among the NL, QL and S3T bifurcation diagrams shown in Figs.~\ref{fig:bif_r_0p1}\hyperref[fig:bif_r_0p1]{a} and \ref{fig:bif_r_0p1}\hyperref[fig:bif_r_0p1]{b} provides convincing evidence that turbulent cascades, which are not present in S3T or QL, are not required for jet formation.

While under NIF agreement between NL and S3T equilibrium jet amplitudes extends to all values of $\e$, under IRFn the NL and S3T equilibrium amplitudes diverge at larger values of $\e$ (cf.~Figs.~\ref{fig:bif_r_0p1}\hyperref[fig:bif_r_0p1]{a} and \ref{fig:bif_r_0p1}\hyperref[fig:bif_r_0p1]{b}).
This difference among NL, QL and S3T at large $\e$ cannot be attributed to nonlinear modification of the spectrum, which is accounted for by use of the S3Tb spectrum (cf.~S3Tb response in Fig.~\ref{fig:bif_r_0p1}\hyperref[fig:bif_r_0p1]{b}). Rather, this difference is primarily due to nonlinear eddy--eddy interactions retained in NL that disrupt the up-gradient momentum transfer. This disruption is accentuated by the peculiar efficiency with which the narrow ring forcing, IRFn, gives rise to vortices, as can seen in Fig.~\ref{fig:Qkl_Fxy_NIF_IRF}\hyperref[fig:Qkl_Fxy_NIF_IRF]{d-f}. The more physical distributed forcing structures do not share this property (cf.~Fig.~\ref{fig:Qkl_Fxy_NIF_IRF}).
We verify that the narrow ring IRFn forcing is responsible for depressing NL equilibrium jet strength at high supercriticality by broadening the forcing distribution to assume the form IRFw (cf.~Appendix~\ref{appsec:forc_spec_IRF_NIF} as well as Fig.~\ref{fig:Qkl_Fxy_NIF_IRF} for IRFn--IRFw comparison). Using IRFw while retaining other parameters as in Fig.~\ref{fig:bif_r_0p1}\hyperref[fig:bif_r_0p1]{b}, we obtain agreement between S3T, QL and NL simulations, as is shown in Fig.~\ref{fig:bif_r_0p1}\hyperref[fig:bif_r_0p1]{c}.

\section{Identification of intermittent jets with stable S3T zonal eigenfunctions}

For subcritical forcing S3T predicts a stable homogeneous statistical equilibrium and a set of eigenfunctions that govern the decay of perturbations to this equilibrium. We wish to show that these eigenfunctions are excited in NL by fluctuations in the turbulence and that this excitation gives rise in NL simulations to the formation of intermittent jets with the form of these eigenfunctions.

\begin{figure}[ht]
\centering\includegraphics[width=22pc]{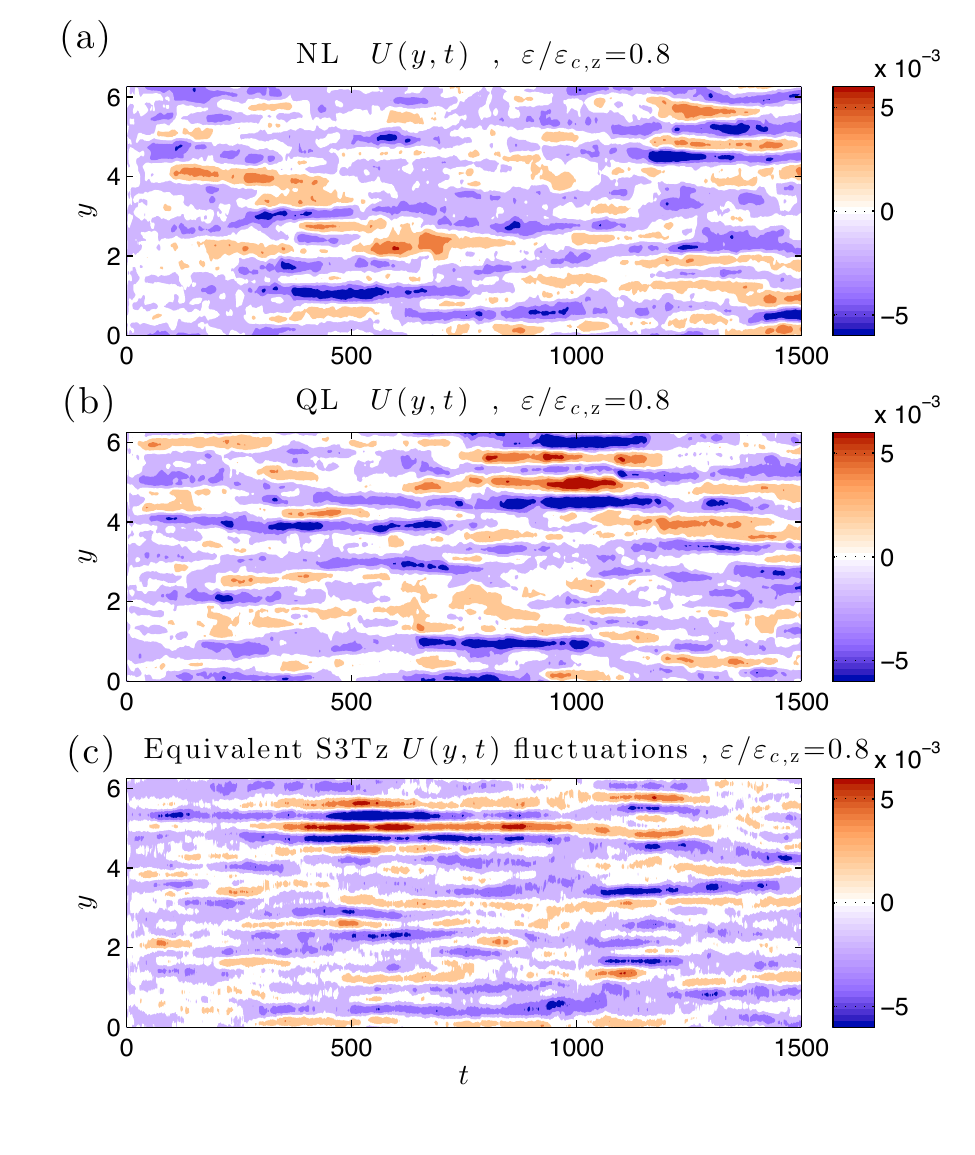}
\vspace{-5mm}\caption{ Hovm\"oller diagrams of intermittent jet structure in NL and QL simulations at subcritical forcing $\e=0.8\ecz$. Shown are $U(y,t)$ for NL (panel (a)) and QL (panel (b)) simulations and the $U(y,t)$ that results from random excitation of the S3T damped modes (panel (c)). These plots were obtained using IRFn forcing with $r=0.1$, $\rU=0.01$. This figure shows that the manifold of S3T damped modes are revealed by being excited in the fluctuating NL and QL simulations. Planetary vorticity gradient: $\beta=10$. } 
 \label{fig:hov_IRFh_e0p8ec_r0p1_rm0p01}
\end{figure}

As an example, consider the simulation with asymmetric damping and IRFn subcritical forcing shown in Fig.~\ref{fig:hov_IRFh_e0p8ec_r0p1_rm0p01}. For these parameters the least damped eigenfunctions are zonal jets and confirmation that the intermittent jets in NL, shown in the top panel of Fig.~\ref{fig:hov_IRFh_e0p8ec_r0p1_rm0p01}, are consistent with turbulence fluctuations exciting
the S3T damped modes is given in the bottom panel of Fig.~\ref{fig:hov_IRFh_e0p8ec_r0p1_rm0p01} where the intermittent jets resulting from stochastic forcing of the S3T modes themselves are shown. 
This diagram was obtained by plotting $U(y,t) = \real{ \[ \sum_{n_y=1}^N \alpha_{n_y}(t) e^{\i n_y y} \]}$, with $\alpha_{n_y}$ independent red noise processes, associated with the damping rates, $|s(n)|$, of the first $N=15$ least damped S3T modes. These $\alpha_{n_y}$ are obtained from the Langevin equation, $\df \alpha_{n_y} \big/ \df t = {s(n_y)}\,\alpha_{n_y} + \xi(t)$, with $\xi(t)$ a $\delta$-correlated complex valued random variable.

The fluctuation-free S3T simulations reveal persistent jet structure only coincident with the inception of the S3T instability, which occurs only for supercritical forcing. However, in  QL and NL simulations fluctuations excite the damped manifold of modes predicted by the S3T analysis to exist at subcritical forcing amplitudes. This observation confirms the reality of the  manifold of S3T stable modes. 
%This emergence of structure at subcritical forcing also provides an explanation for the observation of the so-called latent zonal jets in the oceans~\parencite{Berloff-etal-2009b,Berloff-etal-2011,Cravatte-etal-2012}. 
%This work suggests that these latent jets are fluctuation-forced manifestations of the stable S3T modes. 

In NL and QL simulations these stable modes predicted by S3T are increasingly excited as the critical bifurcation point in parameter space is approached, because their damping rate vanishes at the bifurcation. The associated increase in zonal mean flow energy on approach to the bifurcation point obscures the exact location of the bifurcation point in NL and QL simulations compared to the fluctuation-free S3T simulations for which the bifurcation is exactly coincident with the inception of the S3T instability (i.e. Fig.~\ref{fig:bif_r_0p1}\hyperref[fig:bif_r_0p1]{a},~\ref{fig:bif_r_0p1}\hyperref[fig:bif_r_0p1]{b}~and~\ref{fig:bif_r_0p1}\hyperref[fig:bif_r_0p1]{c}).

\section{Verification in NL of the multiple jet equilibria predicted by S3T}

\begin{figure}[t]
\centering\includegraphics[width=24pc]{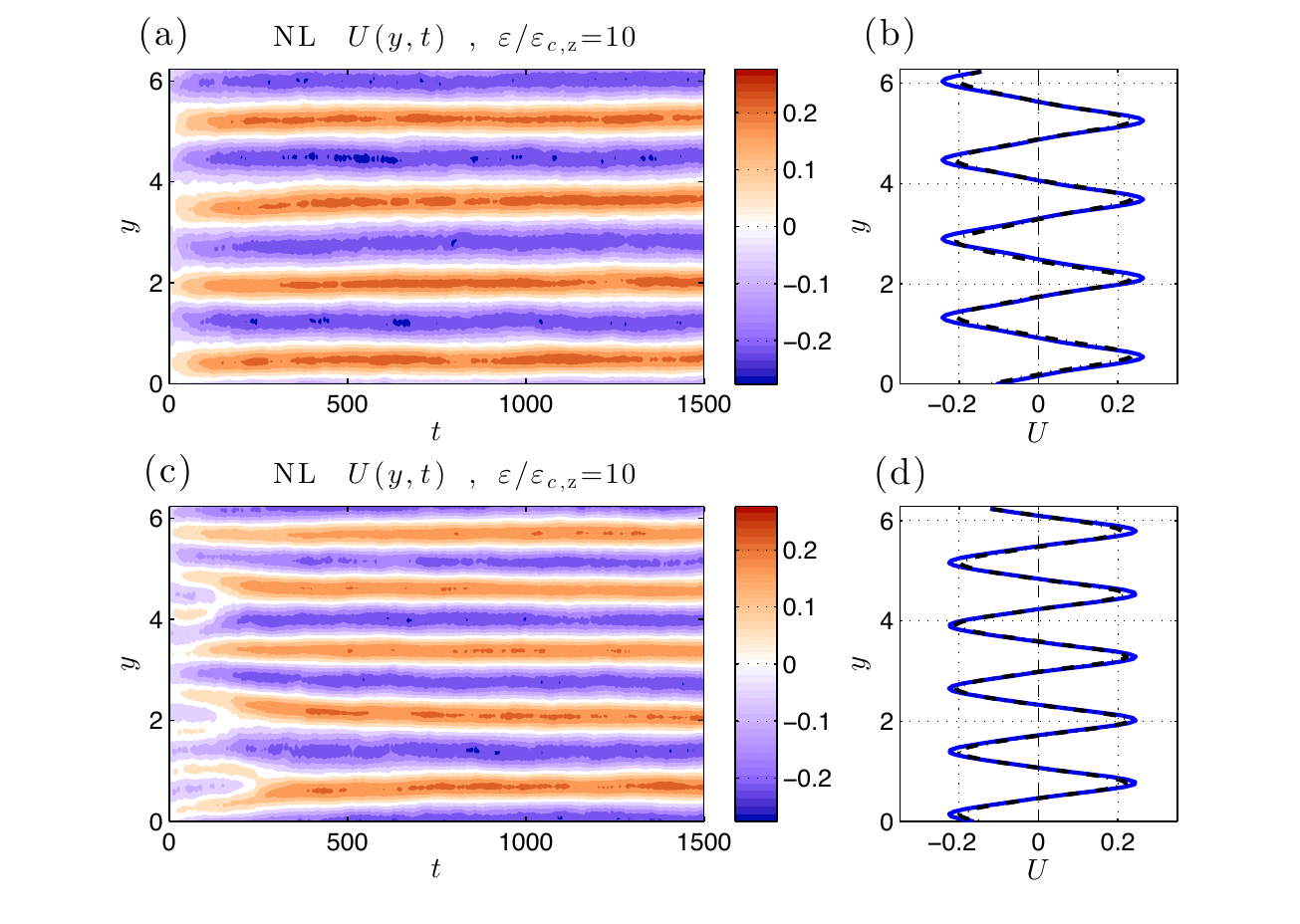}
\vspace{-5mm}\caption{\label{fig:multiple_eq} Realizations in NL simulations of multiple equilibria predicted by S3T. Show are Hovm\"oller diagrams of NL simulations showing the equilibrium with 4 jets (panel (a)) and with 5 jets (panel (c)). Also shown is comparison of the S3T equilibrium jets (solid) with the average jets obtained from the NL simulation (dashed) for the two equilibria (panels (b) and (d)). Parameters: NIF forcing at $\e=10 \ecz$, $r=0.1$, $\rU=0.01$ and $\beta=10$.}
\end{figure}

As is commonly found in nonlinear systems, the finite amplitude equilibria predicted by S3T are not necessarily unique and multiple equilibria can occur for the same parameters.  
S3T provides a theoretical framework for studying these multiple equilibria, their stability and bifurcation structure. An example of two such S3T equilibria are shown in Fig.~\ref{fig:multiple_eq} together with their associated NL simulations. As the parameters change these equilibria may cease to exist or become S3T unstable. 
%This change in the attractor structure manifests as the jet merger phenomenon seen in Hovm\"oller diagrams of jet evolution.
Similar multiple equilibria have been found in S3T studies of barotropic $\b$-plane turbulence \parencite{Farrell-Ioannou-2003-structural,Farrell-Ioannou-2007-structure,Parker-Krommes-2014-generation} and in S3T studies of baroclinic turbulence~\parencite{Farrell-Ioannou-2008-baroclinic,Farrell-Ioannou-2009-closure} and the hypothesis has been advanced that the existence of such multiple jet equilibria may underlie the abrupt transitions found in the record of Earth's climate~\parencite{Farrell-Ioannou-2003-structural, Wunsch-2003}.

\section{Conclusions}

In this chapter predictions of S3T for jet formation and equilibration in barotropic $\b$-plane turbulence were critically compared with results obtained using QL and NL simulations. Throughout this chapter the zonal mean--eddy decomposition (section~\ref{sec:s3tz}) was used for all three NL, QL and S3T systems. The qualitative bifurcation structure predicted by S3T for emergence of zonal jets from a homogeneous turbulent state was confirmed by both the QL and NL simulations. Moreover, the finite amplitude equilibrium jets in NL and QL simulations were found to be as predicted by the fixed point solutions of S3T. Differences in jet formation bifurcation parameter values between NL and QL/S3T were reconciled by taking account of the fact that the spectrum of turbulence is substantially modified in NL. Remarkably, the modification of the spectrum in NL could be traced in large part to emergence of non-zonal structures through S3T instability. When account is taken of the modification of the turbulent spectrum resulting substantially from these non-zonal structures, S3T also provides quantitative agreement with the threshold values for the emergence of jets in NL. The influence of the background eddy spectrum on the S3T dynamics was found to be immediate, in the sense that in spin-up simulations jets 
emerge in accordance with the instability calculated on the temporally developing spectrum. The fact that jets are prominent in observations is consistent with the robust result that when a jet structure emerges it has primacy 
over the non-zonal structures, so that even if the jet eigenfunction is not the most linearly S3T unstable eigenfunction, the jet still emerges at finite amplitude as the dominant structure.

These results confirm that jet emergence and equilibration in barotropic $\b$-plane turbulence results from the cooperative quasi-linear mean flow--eddy instability that is predicted by S3T. These results also establish that turbulent cascades are not required for the formation of zonal jets in $\b$-plane turbulence. Moreover, the physical reality of the manifold of stable modes arising from cooperative interaction between incoherent turbulence and coherent jets, which is predicted by S3T, was verified in this work by relating observations of intermittent jets in NL and QL to stochastic excitation by the turbulence of this manifold of stable S3T modes.

\section{Bibliographical note}
This chapter is an adaptation from the paper by \upmax\textcite{Constantinou-etal-2014}\dnmax. %The S3T instability of the homogeneous turbulent equilibrium was studied by \textcite{Farrell-Ioannou-2007-structure}. Analytical results for zonal jet perturbations for $\b=0$ and finite doubly periodic domains were obtained by \textcite{Bakas-Ioannou-2011}. Results for infinite $\b$-planes and for any $\b$ were obtained by \textcite{Srinivasan-Young-2012}. The forcing spectrum used in this chapter was introduced by \textcite{Srinivasan-Young-2014}. The dispersion relation for the stability of non-zonal perturbations was derived by \textcite{Bakas-Ioannou-2013-prl}. A physical interpretation of the S3T instability of the homogeneous equilibrium for small $\b$ and $n$ is discussed in \textcite{Bakas-Ioannou-2013-jas}.
The NIF forcing used in this chapter was first used by \textcite{Williams-78} in order to to parametrize excitation of baro\-tropic dynamics by baroclinic instabilities. It was also used by \textcite{DelSole-01a} in his study of upper-level tropospheric jet dynamics and in the study of jet formation using S3T dynamics by \textcite{Farrell-Ioannou-2003-structural,Farrell-Ioannou-2007-structure} and \textcite{Bakas-Ioannou-2011}. 
The isotropic narrow ring forcing, IRFn, has been used extensively in studies of $\b$-plane turbulence \citep[cf.][]{Vallis-Maltrud-93} and was also used in the recent study of \textcite{Srinivasan-Young-2012}. It was introduced by \textcite{Lilly-1969}, in order to isolate the inverse cascade from the forcing in a study of two dimensional turbulence.

\chapter{S3T stability of inhomogeneous turbulent equilibria}
\label{ch:S3Tnonhom}
 
We have already seen that with homogeneous forcing the S3T system has homogeneous equilibria for any level of forcing, but also inhomogeneous equilibria in the form of zonal jets, as those obtained in chapter~\ref{ch:NLvsS3Tjas}, or non-zonal inhomogeneous equilibria in a moving frame of reference, as the traveling wave solutions that were obtained in chapter~\ref{ch:MI} (cf.~Fig.~\ref{fig:S3T_snapshots_Psi}). In chapter~\ref{ch:st3hom} we presented systematic methods that enabled us to determine the stability of the homogeneous state and we were able to predict the critical parameters for which the symmetry of the homogeneous state is broken. The stability of finite amplitude zonal jet equilibria to zonal jet perturbations has been already studied by~\textcite{Farrell-Ioannou-2003-structural} and more recently by~\textcite{Parker-Krommes-2014-generation}. Here we present more general methods for determining the stability of inhomogeneous states to zonal but also non-zonal perturbations. With this more general stability analysis we demonstrate that the phenomenon of jet merging is properly understood as an S3T instability and consequently this phenomenon is properly understood in the framework of statistical state dynamics. We also show that the transition from zonal to non-zonal turbulent states is also predicted by S3T stability analysis.

\section{Stability of finite amplitude zonal jet S3T equilibria to zonal jet mean flow perturbations\label{sec:S3Tstab_Ue_zonal}}

Consider first the equilibrium states that arise in the simpler S3Tz system~\eqref{eq:s3tz}, i.e., the S3T system in which ensemble means are interpreted as zonal means. In S3Tz by construction the mean flows are zonal jets and the associated equilibria, when they exist, can only be zonal jets. The zonal jet equilibria arise as a bifurcation of the homogeneous equilibrium state that becomes S3T unstable for energy injection rates, $\e$, that exceed a critical value, $\ecz$ (see Figs.~\ref{fig:bif_r_0p01}~and~\ref{fig:bif_r_0p1}). The resulting zonal jet equilibria have the characteristic property that the number of jets decreases as the supercriticality, $\e/\ecz$, increases\footnote{$\ecz$ is the minimum energy input rate for the instability of the homogeneous state to zonal jets.}; examples are shown in Figs.~\ref{fig:hov_IRFh_e100ec_r0p01},~\ref{fig:hov_IRFh_e10ec_S3Tb},~\ref{fig:hov_NIF_e1p5ec_r0p1_rm0p01} and~\ref{fig:hov_NIF_e20ec_r0p1_rm0p01}. We wish to study the stability of these jet equilibria in order to understand  the mechanism underlying the transition from one equilibrium state to another. In order to proceed with the stability analysis of the equilibria, we must first determine the equilibrium solutions, $\(U^e(y), C^e(x_a-x_b,y_a,y_b)\)$, with adequate accuracy in order to obtain good estimates of their stability. While stable equilibrium solutions can be in principle obtained with the required accuracy with forward time-integration of the S3T system, forward time-integration cannot determine unstable equilibria and consequently we must resort to continuation methods in order to obtain all the fixed points of the S3T equations. Both stable and unstable equilibrium states can be determined with great accuracy and ease using the continuation methods described in Appendix~\ref{app:S3Tnewton}.% that are based on Newton's iterations.

As discussed in chapter~\ref{ch:formulation}, the linear stability of S3T equilibria is studied through eigenanalysis of the operator governing the linearized S3T evolution of the perturbations $(\d Z,\d C)$ about the equilibrium state:
\begin{subequations}\begin{align}
\partial_t \,\d Z & = \Acal^e\,\d Z + \Rcal( \d C )\ ,\label{eq:s3t_pert_dZgen_2}\\
\partial_t \,\d C_{ab} & = \(\bit\Acal^e_a + \Acal^e_b \)\d C_{ab} +\(\bit\d\Acal_a + \d\Acal_b\)C^e_{ab}\ ,\label{eq:s3t_pert_dCgen_2}
\end{align}\label{eq:s3t_dZdCgen_2}\end{subequations}
with $\Acal^e\equiv\Acal(\Uv^e)$ and  $\d\Acal\equiv \Acal(\Uv^e+\d\Uv)-\Acal^e$. Note that discretized with $N$ points the dimension of the perturbation state in~\eqref{eq:s3t_dZdCgen_2} is $\Ocal(N^4)$. For example, if we use a modest discretization grid of $N_x=N_y=2^6$, the dimension of the equivalent matrix operator governing the stability of~\eqref{eq:s3t_dZdCgen_2} is $2^{24}\times 2^{24} \approx 10^7 \times 10^7$. Despite the enormity of the size of the operators the real part, $s_r$, of the maximally growing eigenvalue, $s$, and the corresponding spatial structure of the eigenfunction, $(\dZ,\dC)$, can be still obtained numerically using the power method. %Direct eigenanalysis of the above system is usually impossible since even for a discretization with the modest resolution of $N_x=N_y=64$ the system~\eqref{eq:s3t_dZdCgen_2} has a state variable of the order $10^7$.  The maximally growing eigenvalue $s$ and the corresponding eigenfunction $\(\dZ,\dC\)e^{s t}$ may be obtained numerically using a power method. Assume that we start from any random state $(\d Z,\d C)$ which is normalized so that $\|(\d Z,\d C)\|=1$, with $\|\,\bullet\,\|$ any norm, i.e., the total energy that corresponds to perturbation. %The choice of the norm used is, for example we can use \be \|(\d Z,\d C)\| \equiv \int\df^2\xv\; \d Z(\xv,t)^2 + \iint\df^2\xv_a\,\df^2\xv_b\; \d C(\xv_a,\xv_b,t)^2\ . \ee 
%We then time-integrate forward~\eqref{eq:s3t_dZdCgen_2} and at every time-step, $h$, we measure the state growth,
%\be
% \la(jh) = \frac{\log\(\bit\|(\d Z(jh),\d C(jh))\|\)}{h} \quad j=1,2,\dots\ ,\label{eq:lambda_jh}
% \ee
%and then renormalize the state so that $\|(\d Z,\d C)\|=1$ before continuing the time-inte\-gration. After sufficient time steps the growth $\la$ converges to the maximal growth rate $s_r$ and the state $(\d Z,\d C)$ converges to the maximally growing eigenfunction $(\dZ,\dC)$.
The imaginary part of the eigenvalue, $s_i$, can then be determined by solving~\eqref{eq:s3t_pert_dZgen_2}:
\be
\( s_r+\i s_i + \Acal^e\) \dZ = \Rcal( \dC )\ ,\label{eq:determine_si}
\ee
for $s_i$. This procedure is still computationally expensive because it requires time-inte\-gration of a state vector of dimension $N_xN_y+N_x^2N_y^2$. The dimension of the system can be reduced by a square root when the equilibrium states are zonal, as Bloch's theorem (cf.~Appendix~\ref{app:bloch}) requires that the spatial structure of the eigenfunction can be assumed to be of the form:\begin{subequations}
\begin{align}
\dZ(\xv) &= e^{\i n_x x}\,\dZ_{n_x}(y)\ ,\\
\dC(\xv_a,\xv_b) &=  e^{\i n_x(x_a+x_b)/2} \,\dC_{n_x}(x_a-x_b,y_a,y_b)\ ,
\end{align}\label{eq:S3Teigen_nx}\end{subequations}
and the stability of the equilibrium is determined by evolving each zonal wavenumber $n_x$ separately. More details regarding the method for determining the stability of~\eqref{eq:s3t_dZdCgen_2} are discussed in Appendix~\ref{app:S3Tlyap}. While for the case of the homogeneous equilibria the eigenfunctions are single harmonics, i.e., $\dZ_{n_x}(y)=e^{\i n_y y}$ (cf.~\eqref{eq:eig_hom}), for zonal jet S3T equilibria the eigenfunctions are single harmonic only in $x$ and have full spectrum in $y$ (cf. Fig.~\ref{fig:stab_ny5_ny7_eckhaus} and also Fig.~\ref{fig:spec_Ue_dU_nx1}). However, Bloch's theorem (cf.~Appendix~\ref{app:bloch}) restricts the meridional structure of the eigenfunction~\ref{eq:S3Teigen_nx}. For example, the mean flow perturbation must be of the form
\be
\dZ_{n_x}(y) = e^{\i q_y y} g(y)~,
\ee
with $g(y)$ any function that has the same periodicity as the equilibrium jet, $U^e(y)$, and similarly for the perturbation covariance eigenfunction, $\dC_{n_x}$ (cf.~\eqref{eq:f_xtoxma}). For a $n_y$-jet equilibrium  function $g$ is of the form: $g(y) = \sum_m e^{\i m n_y y}$. Wavenumber $q_y$ is called ``Bloch wavenumber'' and takes all values $q_y\le n_y/2$ (for our channel of length $L_x=L_y=2\pi$ wavenumber $q_y$ takes all integer values $q_y\le n_y/2$). Therefore a $q_y=0$ Bloch eigenfunction will have power at wavenumbers $0,\pm n_y,\pm2n_y,\dots$, while a  $q_y=1$ Bloch eigenfunction will have power at wavenumbers $\pm1,\pm (n_y\pm1),\pm(2n_y\pm1),\dots$.

\begin{figure}
\centering
\includegraphics[width=5.25in,trim = 2mm 2mm 2mm 0mm, clip]{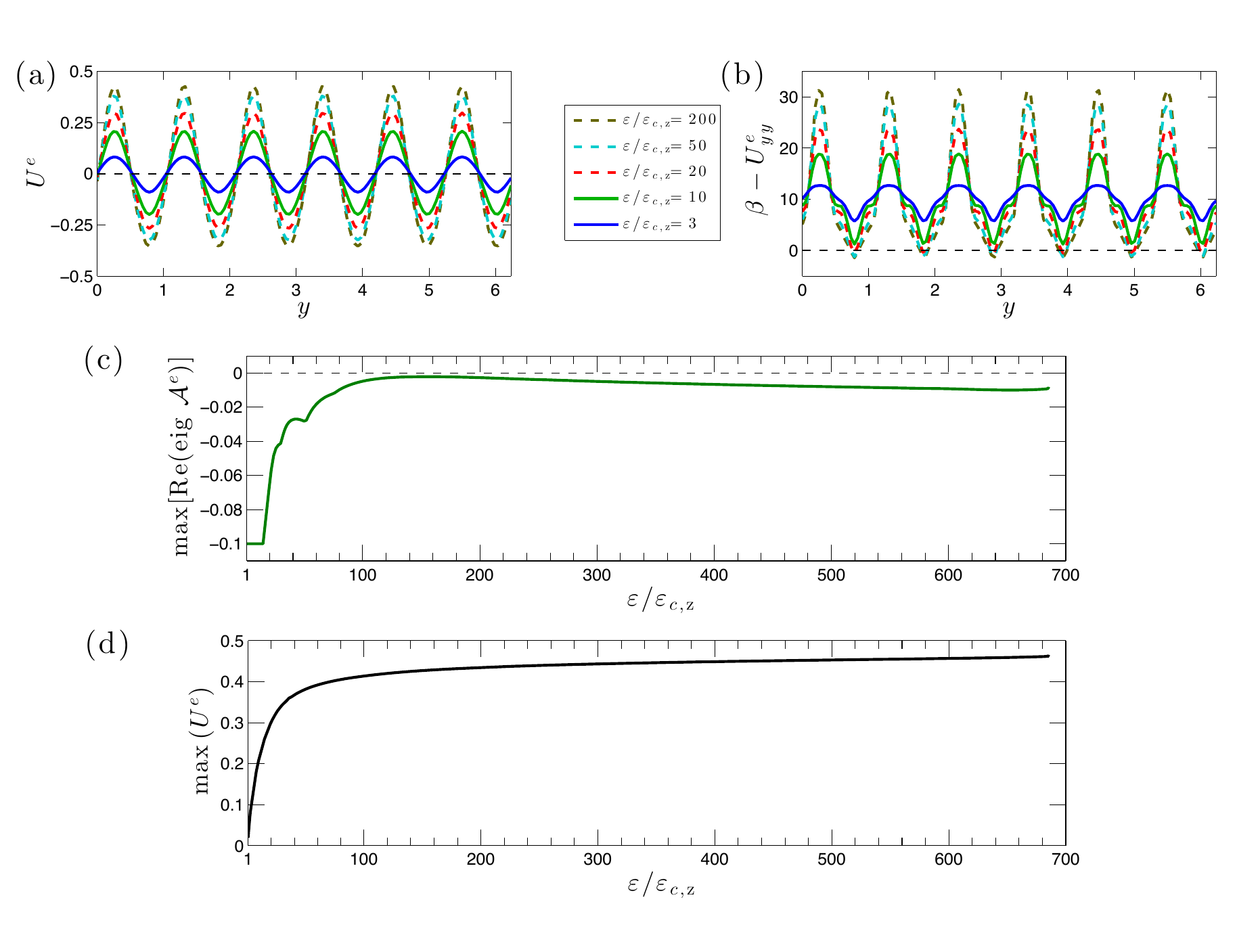}
\vspace{-1em}\caption{\label{fig:Ue_n0_6} (a) S3T zonal flow equilibria with $n_y=6$ jets for supercriticality $\e/\ecz=3,10,20,50,200$. The equilibria for $\e/\ecz=20,50$ (dashed) are S3T unstable, and those with $\e/\ecz<20$ are S3T stable. (b) The potential vorticity gradient, $\b-U^e_{yy}$ for the corresponding equilibria shown in panel (a). (c) The maximum growth rate of $\Acal(U^e)$ for the zonal flow equilibria with $n_y=6$ jets as a function of the supercriticality. All equilibria are hydrodynamically stable. S3T equilibria with $n_y=6$ jets cease to exist at $\e/\ecz=686$. (d) the amplitude of the equilibrium jets as a function of $\e/\ecz$. The amplitude of the jets does not increase substantially for $\e/\ecz>100$ and the extra energy that is imparted in the flow is absorbed by the perturbation field without being communicated to the mean flow. Parameters: NIF forcing, $\beta=10$, $r=0.1$ and $\rU=0.01$.}
\end{figure}

Using the continuation methods described in Appendix~\ref{app:S3Tnewton} we find a series of zonal jet equilibria that are characterized by different number of prograde jets, $n_y$. Consider for example the case with $n_y=6$ jets and for the parameters: NIF forcing, $\b=10$, $r=0.1$ and $\rU=0.01$ (cf.~chapter~\ref{ch:NLvsS3Tjas}, case presented in Fig.~\ref{fig:bif_r_0p1}\hyperref[fig:bif_r_0p1]{a}). For these parameter values the homogeneous equilibrium becomes first unstable to zonal jets with $n_y=6$ at $\e=\ecz$ and inhomogeneous zonal jet equilibria exist for all energy input rates $\e>\ecz$ that do not exceed $686 \ecz$. The nonexistence of zonal equilibria with $n_y=6$ for $\e >686\ecz$ is attributed to the inability to find equilibria when the flow starts supporting stable modal structures that produce strong vorticity fluxes in the neighborhood of their critical layers. Recall (cf.~Appendix~\ref{app:s3t-equil-prop}) that although all the S3T equilibria are necessarily hydrodynamically stable they may be S3T unstable. Specifically, the $n_y=6$ equilibria become S3T unstable for $\e\ge20\ecz$, that is for energy input rates substantially lower than the energy input rate at which the equilibria cease to exist. Some stable and unstable S3T equilibria, their associated planetary vorticity gradient, $\b-U^e_{yy}$, as well as the amplitude of the jets as a function of energy input rate are shown in Fig.~\ref{fig:Ue_n0_6}. Note that because of the presence of dissipation, the flow can remain stable although the mean planetary vorticity gradient changes sign and becomes slightly negative in limited regions when the jet is retrograde (cf.~Fig.~\ref{fig:Ue_n0_6}\hyperref[fig:Ue_n0_6]{c}). Note also that as the energy input rate is increased the amplitude of the mean flow grows very gradually after $\e/\ecz\approx100$ (cf.~Fig.~\ref{fig:Ue_n0_6}\hyperref[fig:Ue_n0_6]{d}) and the energy input of the stochastic forcing is absorbed mainly by the perturbation field, which indicates that the extra energy in the perturbation field is not communicated to the mean flow. This happens because for large enough energy input rates (for $\e>20\ecz$ for $n_y=6$) a nearly neutral modal structure is supported by the flow, with critical layers where $\b-U^e_{yy}$ vanishes, and this mode absorbs most of the incoming energy without producing appreciable upgradient vorticity fluxes to support a stronger flow. This modal structure eventually  develops strong critical layers and strongly localized vorticity fluxes that make the existence of an S3T equilibrium impossible.

%The flows become usually hydrodynamically unstable for large enough  energy input rates because as the forcing increases the vorticity fluxes usually increase and with them the amplitude of the zonal flow (cf.~Fig.~\ref{fig:Ue_n0_6}\hyperref[fig:Ue_n0_6]{a}), the mean deformation and the violation of the Rayleigh-Kuo stability criterion becomes greater. Note that because of the presence of dissipation the flow can remain stable although the mean potential vorticity gradient, $\beta-U_{yy}$ assumes relatively small negative values when the jets are retrograde (cf.~Fig.~\ref{fig:Ue_n0_6}\hyperref[fig:Ue_n0_6]{b}). This tendency towards neutrality is evident in Fig.~\ref{fig:Ue_n0_6}\hyperref[fig:Ue_n0_6]{c}, which shows that as the value $\e/\ecz=686$ is approached the flow becomes hydrodynamically neutrally stable and no equilibria can be located for larger energy input rates\footnote{For large enough energy input rates (for $\e>20\ecz$ for $n_y=6$) a nearly neutral mode emerges and energy can be transferred to this mode with little change in the mean flow and its stability. This can be nicely seen in the case with $n_y=2$ in which equilibria exist, as far as we could ascertain, even as $\e \rightarrow \infty$.}  Although all these S3T equilibria are hydrodynamically stable 

\begin{figure}[h]
\centering
\includegraphics[width=4.7in]{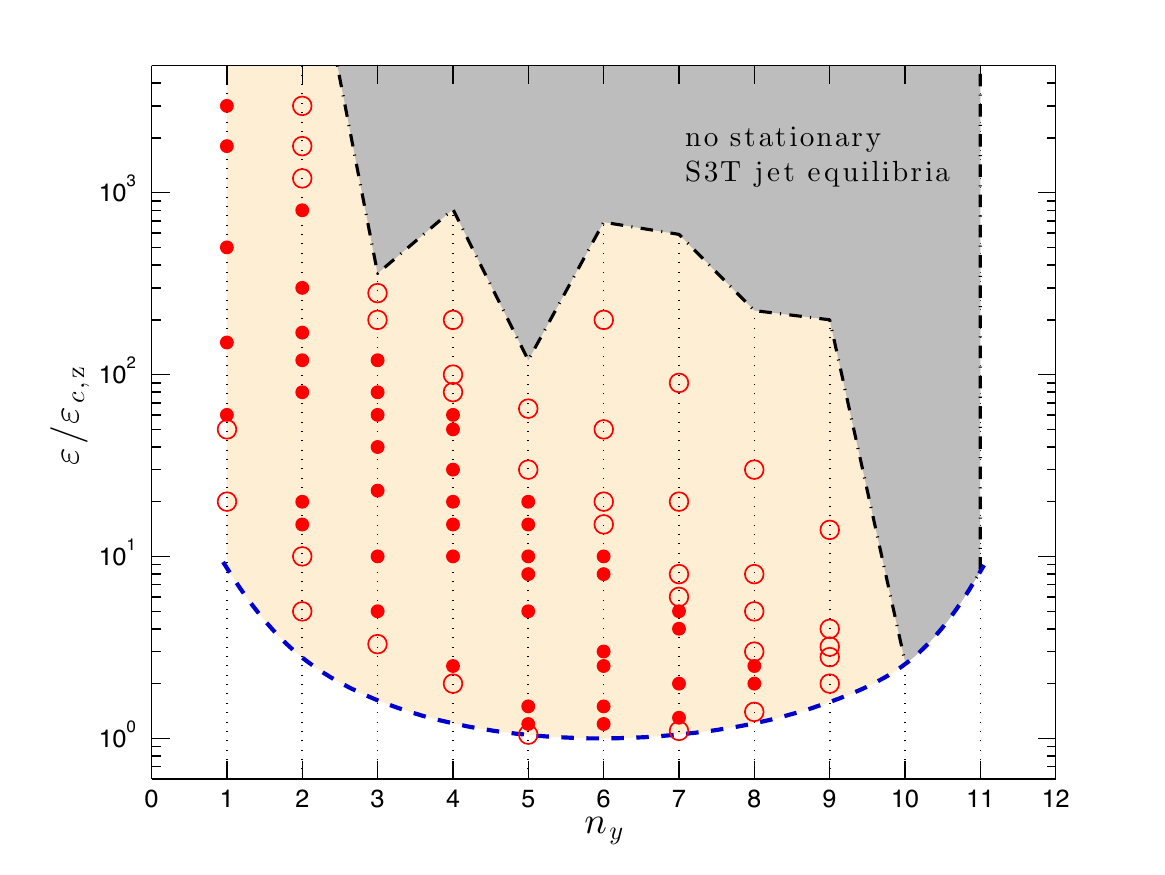}
\vspace{-.5em}\caption{\label{fig:ballonNIF} S3T zonal flow equilibria as a function of the number of prograde jets, $n_y$, and the forcing amplitude $\e/\ecz$. Jet equilibria exist in the yellow shaded region. The lower bounding curve is the curve of neutral stability of the homogeneous state (the homogeneous state is S3T stable for all $\e$ for $n_y\ge 12$). For $\e$ and $n_y$ above the upper bounding dashed-dot curve jet equilibria do not exist. S3T stable finite amplitude jet equilibria to jet perturbations are indicated with a full circle, S3T unstable equilibria to jet perturbations are indicated with an open circle. For a range of $\e$ there exist multiple S3T stable equilibria characterized by a different number of jets. Near the curve of marginal stability only the jet equilibrium that corresponds to the maximal instability of the homogeneous state, $n_y=6$ is stable, while the neighboring jet equilibria are Eckhaus unstable. Parameters: NIF forcing, $\beta=10$, $r=0.1$ and $\rU=0.01$.}
\end{figure}

A comprehensive mapping of S3T zonal jet equilibria together with their S3T stability as a function of the supercriticality, $\e/\ecz$, and the number of jets, $n_y$, is shown in the balloon diagram Fig.~\ref{fig:ballonNIF}. Jet equilibria exist in the yellow region of the diagram. For values of $\e/\ecz$ and $n_y$  below the lower bounding curve (dashed) the only S3T equilibrium is the homogeneous state with no mean flow and the dashed line is the curve of neutral S3T stability of the homogeneous state. For values of $\e$ and $n_y$ above the upper bounding curve no jet equilibria exist. The S3T stability of certain jet equilibria to zonal jet perturbations (i.e. with $n_x=0$ in~\eqref{eq:S3Teigen_nx}) is indicated with a closed circle when it is stable and with an open circle when it is unstable. %The dashed line marks the critical energy input rate value at which the homogeneous state becomes unstable to a zonal jet perturbation with $n_y=1,\dots,11$. Τhe dash-dotted line in the balloon diagram indicates the critical $\e$ for which no jet equilibria exist. 

%We have shown that all S3T equilibria are necessarily hydrodynamically stable (cf.~ Appendix~\ref{app:s3t-equil-prop}). In the presence of linear damping, if the Rayleigh-Kuo stability criterion $\b-U^e_{yy}>0$ is satisfied then the hydrodynamic stability of the zonal flow is guaranteed, while if $\b-U^e_{yy}$ changes sign the flow is unstable for appropriately small coefficient of linear damping. Consequently, the S3T equilibria we obtain satisfy the $\b-U^e_{yy}>0$ in most regions of the flow and a possibility of small excursions due to the presence of damping (cf.~Fig.~\ref{fig:kminus5}\hyperref[fig:kminus5]{e,h}). As $\e$ is increased the amplitude of the jets, $U_0$, is also increased and therefore for a fixed number of jets the stability criterion cannot be indefinitely satisfied, since $U^e_{yy}\sim n_y^2\, U_0$. Beyond that $\e$ no S3T equilibria (stable or unstable) can be found. 

The balloon diagram shows that for a range of values of  $\e$ multiple stable equilibria exist (these correspond to multiple climate states in this barotropic model). As $\e$ is increased all $n_y$-jet equilibria become eventually S3T unstable. When they become unstable the turbulent flow reorganizes, the mean flow merges and transitions to an available S3T jet stable equilibrium with fewer jets. For example, at $\e/\ecz=10$ the equilibria with $n_y=3,4,5,6$ jets are all stable (in chapter~\ref{ch:NLvsS3Tjas} we have seen that this is also reflected in the NL simulations; cf.~Fig.~\ref{fig:multiple_eq}). If we increase the supercriticality to a value $\e/\ecz>10$ the $n_y=6$ jet becomes S3T unstable and the turbulent flow reorganizes to a state with $n_y=3,4,5$ or even $n_y=2$ jets. This implies that if the energy input rate were to increase in a 6-jet equilibrium the turbulent state would transition to a state with a fewer number of jets through a process of jet mergers. %In the presence of  noise, which is absent in the infinite ensemble S3T system, turbulence is expected to settle to the S3T equilibrium with the largest attractor basin. 

\begin{figure}[h!]
\centering
\includegraphics[width=.99\textwidth,trim = 0mm 1mm 0mm 0mm, clip]{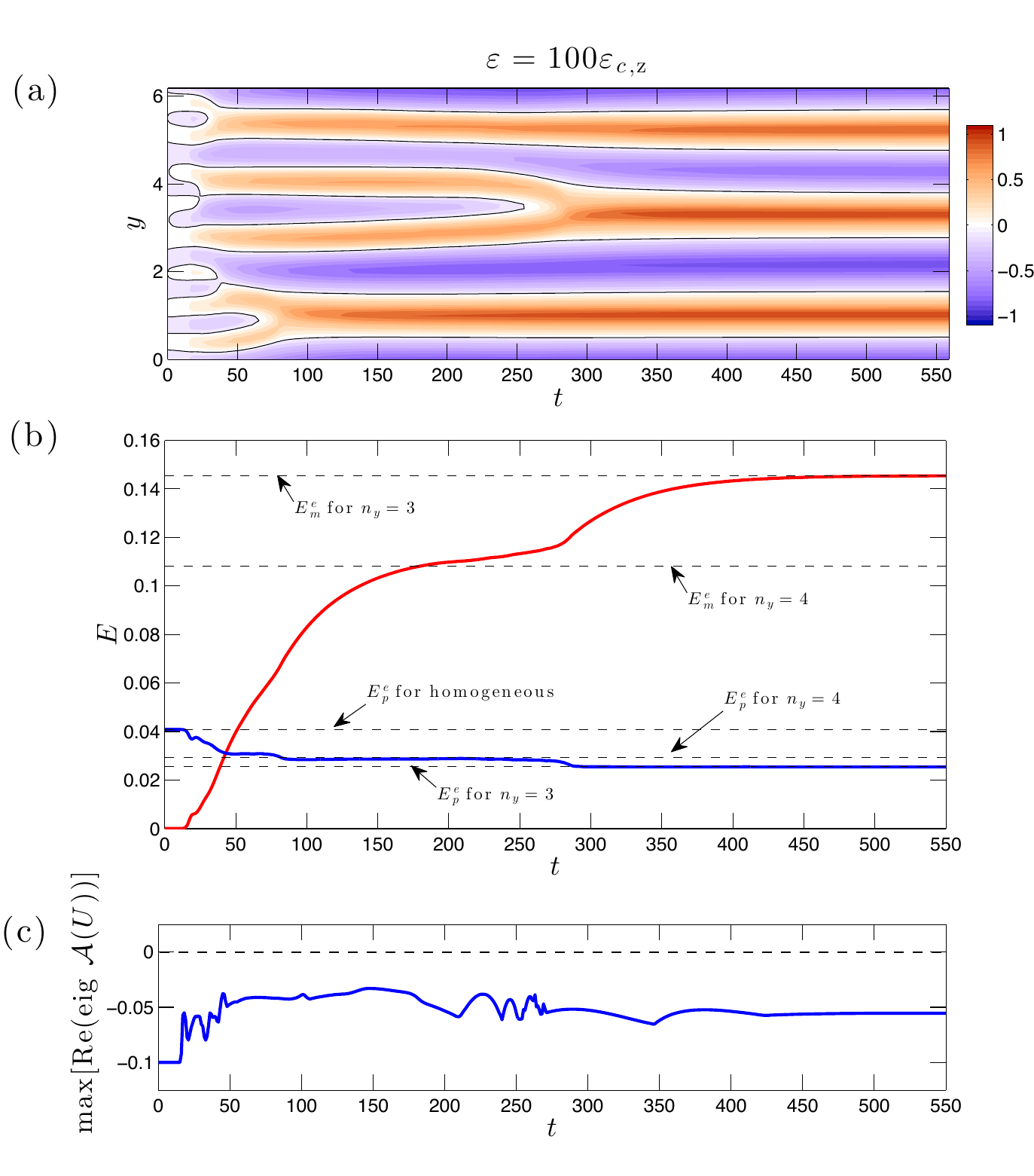}
\vspace{1em}\caption{\label{fig:merging_stab} (a) Hovm\"oller diagram of the zonal mean flow $U(y,t)$ showing a series of jet mergers leading finally to a state with 3 jets. Solid line marks the zero contour. The S3T simulation starts from the homogeneous equilibrium state perturbed by a random mean flow perturbation. (b) The corresponding evolution of the mean flow energy, $E_m$, and perturbation energy, $E_p$. Marked also are the energies of the homogeneous equilibrium and of the jet equilibria with $n_y=2,3,4$. The flow for long periods is close to S3T unstable equilibria, till it finally settles to the first available stable S3T equilibrium. For these parameters (cf. Fig.~\ref{fig:ballonNIF}) the $n_y=1,2$ jet equilibria are also stable, but both have larger/smaller mean flow/perturbation energy. (c): The evolution of the maximum growth rate of $\Acal(U)$ for the instantaneous $U(y,t)$. The flow is at every time instant hydrodynamically stable. Parameters are as in Fig.~\ref{fig:ballonNIF} with forcing at $\e=100\ecz$.}
\end{figure}

We now give an example which demonstrates in an actual S3T simulation that jet mergers do not occur because of the hydrodynamic instability of the jets but rather due to their S3T instability. In the simulation shown in Fig.~\ref{fig:merging_stab}, for $\e/\ecz = 100$. The flow evolves towards jet configurations with fewer and stronger jets though a sequence of jet mergers. Interestingly, the jet states of the flow when they do not merge evolve slowly staying close  to corresponding S3T unstable equilibria till finally they get attracted to the first available S3T stable equilibrium  (for the chosen parameters  the  first S3T stable equilibrium to jet perturbations ($n_x=0$) has $n_y=3$, all equilibria with $n_y>3$ are S3T unstable to $n_x=0$ perturbations; cf.~Fig.~\ref{fig:ballonNIF}). Because the evolution of the jets when they do not merge is slow we can interpret the jet merging process as an instability of the time evolving state of the system and because the strength of the hydrodynamic instability of a barotropic flow depends on the strength of the violation of the Rayleigh-Kuo criterion $\b-U^e_{yy}$, it is natural to attribute jet merging to the hydrodynamic instability of the flow.
We see also, that during this evolution the flow stays close to the various S3T equilibria before moving to the next equilibrium. The maximum growth rate of $\Acal(U)$ that governs the hydrodynamic stability of the flow is negative at each time indicating that jet merging can not be attributed to the hydrodynamic stability of the flow. Thus we conclude that the flow reorganization from a state with 4 jets to the state with 3 jets does not occur due to hydrodynamic instability of the zonal flow but is an inherent S3T phenomenon. In the specific example because the flow stays close to the various S3T unstable equilibria jet merging can be attributed to the S3T instability of these unstable equilibria that behave as unstable saddles of the evolving flow.

%\begin{figure}[h]
%\centering
%\includegraphics[width=.94\textwidth,trim = 2mm 1mm 2mm 0mm, clip]{merging_stab.pdf}
%\vspace{-0.5em}\caption{\label{fig:merging_stab} (a) Hovm\"oller diagram of the zonal mean flow $U(y,t)$ showing a transition for a state with 3 jets to a state with 2 jets. Solid line marks the zero contour. Dark vertical bands mark the time instants for which the zonal mean flow is hydrodynamically unstable. (b): The evolution of the maximum growth rate of $\Acal_\textrm{z}(U)$ for the instantaneous $U(y,t)$. It is clear that hydrodynamic instability only appears after the jet merger occurs. Moreover, maximum growth rates correspond to an e-folding time of 100 time units which is much longer than the time the instabilities appear. From these we conclude that jet mergers do not occur due to hydrodynamic instability. (c)-(e): Snapshots of the zonal flow, $U(y,t)$, at the time instants marked in panel (b). Parameters as in Fig.~\ref{fig:ballonNIF} with forcing at $\e=200\ecz$.}
%\end{figure}

\begin{figure}
\centering
\includegraphics[width=.93\textwidth]{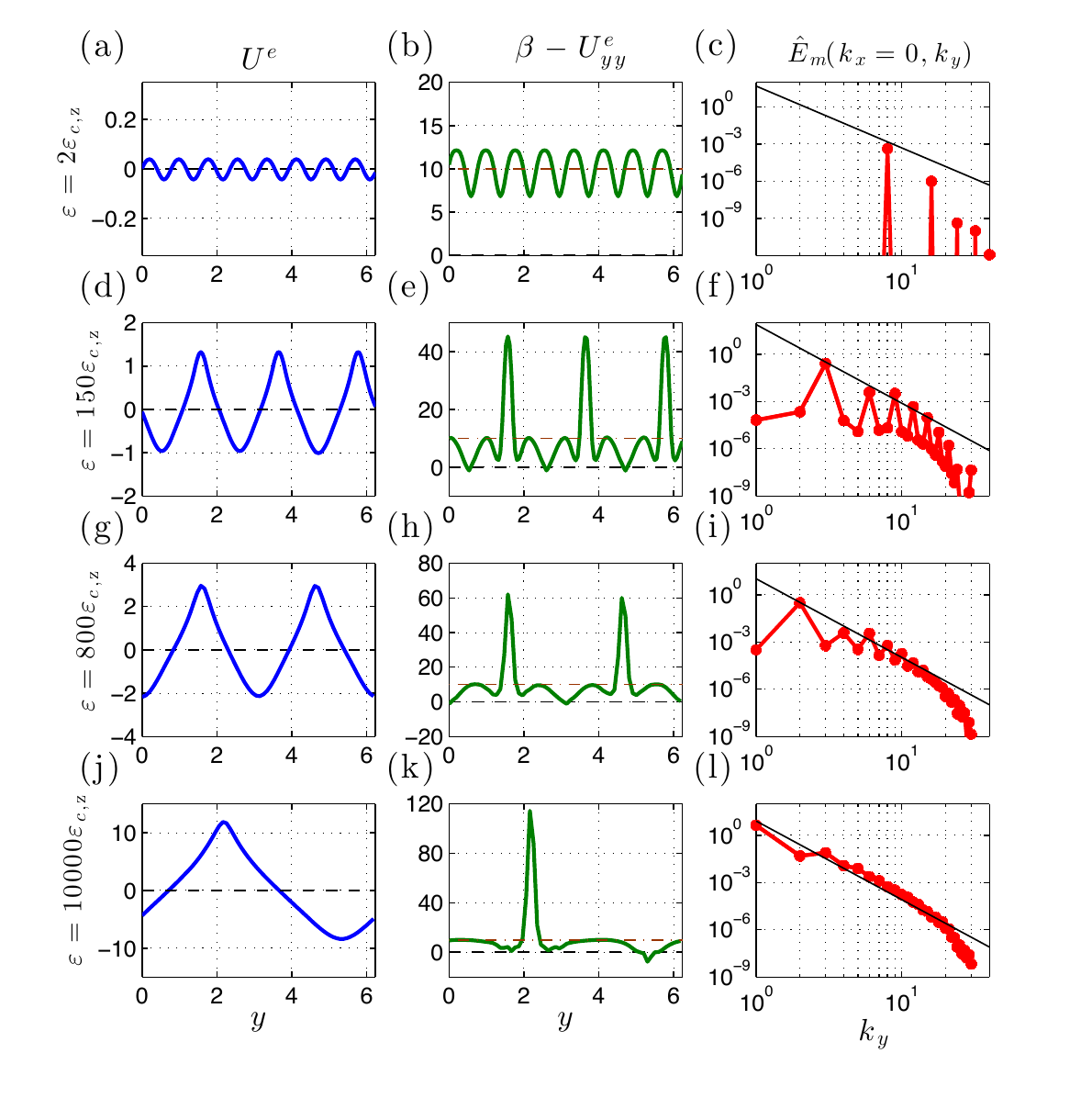}
\vspace{-1.2em} \caption{\label{fig:kminus5} Panels (a,d,g,j): The structure of the S3T stable equilibrium zonal mean flow, $U^e$, for excitation amplitudes $\e/\ecz=2,8,150,800,10^4$. Panels (b,e,h,k): The corresponding mean vorticity gradient, $\beta-U_{yy}^e$. Dash-dotted line marks the planetary vorticity gradient in the absence of mean flow, $\b=10$. Panels (c,f,i,l): The energy spectrum of the equilibrium zonal mean flow together with the $k_y^{-5}$ slope. For the highly supercritical jets the energy spectrum has approximately an $k_y^{-5}$  dependence. It is argued that in the inviscid limit this slope should approach $k_y^{-4}$ as the prograde jet becomes increasingly sharp. Other parameters as in Fig.~\ref{fig:ballonNIF}.}
\end{figure}

As we increase the energy input rate, $\e$, the stable S3T equilibria have fewer jets and the structure of the equilibrium zonal flow equilibria, $U^e$, acquires a particular shape; see Fig.~\ref{fig:kminus5}. While just above the stability boundary the jets are to a good approximation sinusoidal, $U^e\sim\sin(n_y y)$ (cf.~Fig.~\ref{fig:kminus5}\hyperref[fig:kminus5]{a}), at higher $\e$ the retrograde parts of the jets become parabolic while the prograde parts of the jets become increasingly pointed. This particular structure closely resembles the shape of the observed jets in planetary atmospheres (see the $24\deg$N jet on Jupiter, shown in Fig.~\ref{fig:JupiterJets}\hyperref[fig:JupiterJets]{c} as well as the parabolic equatorial jets in Uranus and Neptune, shown in Figs.~\ref{fig:neptune_uranus_danilov}\hyperref[fig:neptune_uranus_danilov]{a,b}). That the retrograde parts of the jets are nearly parabolic is in agreement with potential vorticity (PV) mixing or homogenization arguments \parencite{Baldwin-etal-2007,Dritschel-McIntyre-2008,Dritschel-Scott-2011,Scott-Dritschel-2012}. According to these mixing arguments the primary process in barotropic turbulence is the irreversible mixing by the strongly nonlinear processes that leads to homogenized regions of the mean PV, $\overline{q} = \b y - U_y$, that manifest as a staircase in a diagram of the PV as function of latitude (see also the discussion in section~\ref{sec:turb_theor} and Fig.~\ref{fig:staircases}). In Fig.~\ref{fig:readPVstair} we plot the PV structure for the S3T equilibrium jets shown in Fig.~\ref{fig:kminus5}. It is evident that S3T dynamics do produce a staircase structure when the forcing is strong. This shows that PV staircases are produced by S3T dynamics despite the absence of all eddy--eddy interactions. Note that the staircase structure obtained here is similar to the staircases obtained in the fully nonlinear  and nearly inviscid simulations of \textcite{Scott-Dritschel-2012} as well as those that are observed in experiments  (cf.~Fig.~\ref{fig:readPVstair}). These S3T equilibria with PV staircase structure  provide a counterexample to the necessity of wave breaking and strong nonlinearity for the formation of the staircase structure in barotropic turbulence argued by McIntyre and collaborators.

The primary mechanism responsible for the specific shape of the S3T equilibrium flows at strong forcing is that turbulence acts on the mean flow as negative diffusion (cf. section~\ref{sec:iso}) and that the flow must be hydrodynamically stable, which means that the necessary Rayleigh-Kuo criterion for stability cannot be violated (in the limit of infinitesimal friction, i.e., $\e^* = \e k_f^2/r^3 \to \infty$). The latter requirement constrains only the retrograde regions of the flow (where $U_{yy}>0$) and brings $\b-U_{yy}$ to the minimum value that does not violate Rayleigh-Kuo, i.e, brings $\b-U_{yy}$ to zero, making in this way the flow parabolic, while the former requirement leads to the formation of nearly linear prograde flows joined with sharp wedge-like peaks. As the prograde jets become sharper, the zonal mean flow energy spectrum develops a $k_y^{-5}$ slope (cf.~Fig.~\ref{fig:kminus5}), as is expected from the near discontinuity of the derivative of the prograde jets (a discontinuity would predict a $k_y^{-4}$ zonal energy spectrum; see also discussion in section~\ref{sec:turb_theor}). The absence of eddy--eddy interactions in the S3T dynamics leads us to conclude that the observed zonal energy spectrum cannot be attributed to the anisotropic and incoherent inverse turbulent energy cascade, as it was recently proposed by \textcite{Galperin-etal-04}.

\begin{figure}
\centering
\includegraphics[width=.99\textwidth,trim = 32mm 0mm 32mm 0mm, clip]{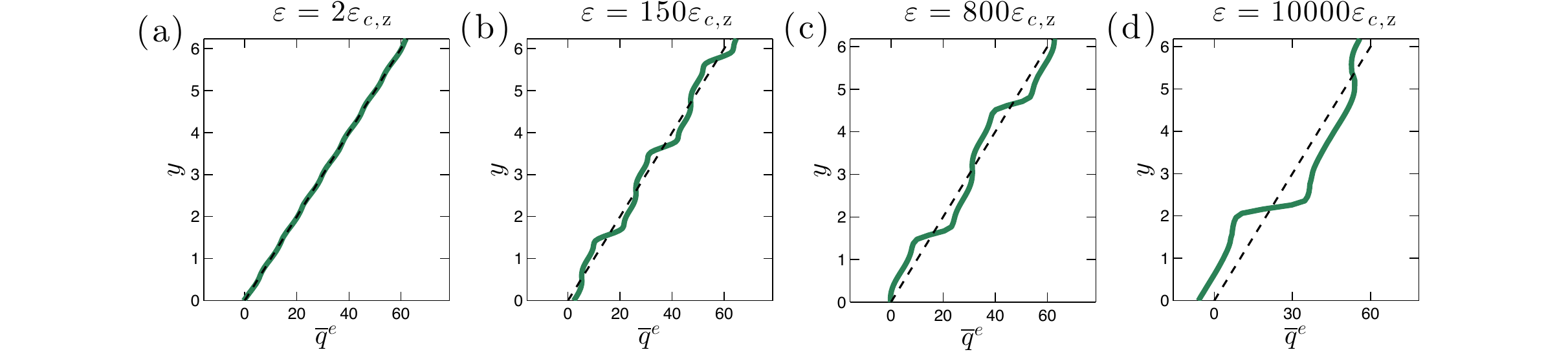}
\caption{\label{fig:stairs_Ue} The zonal mean PV, $\overline{q}^e(y) = \b y- U_y^e$, that corresponds to the equilibrium zonal flows shown in Fig.~\ref{fig:kminus5}. Dashed lines correspond to the planetary PV, $\b y$.}
\end{figure}

\begin{figure}
\centering
\includegraphics[width=.5\textwidth]{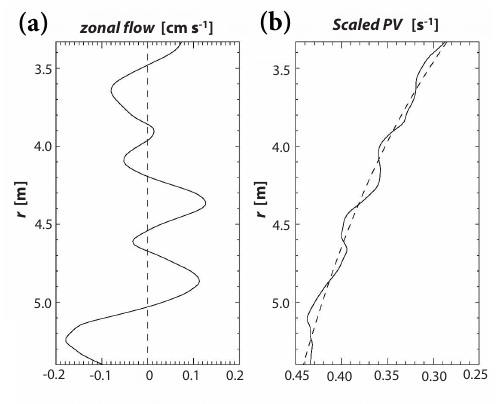}
\vspace{-0.5em}\caption{\label{fig:readPVstair} (a) The zonal mean flow and (b) the corresponding PV from the Grenoble rotating tank experiment. The PV staircases structure closely resembles the one in Fig.~\ref{fig:stairs_Ue}. (Taken from \textcite{Read-etal-2007}. The axes were flipped so that the orientation matches Fig.~\ref{fig:stairs_Ue}.)}
\end{figure}

\begin{figure}
\centering
\includegraphics[width=.98\textwidth,trim = 10mm 1mm 14mm 0mm, clip]{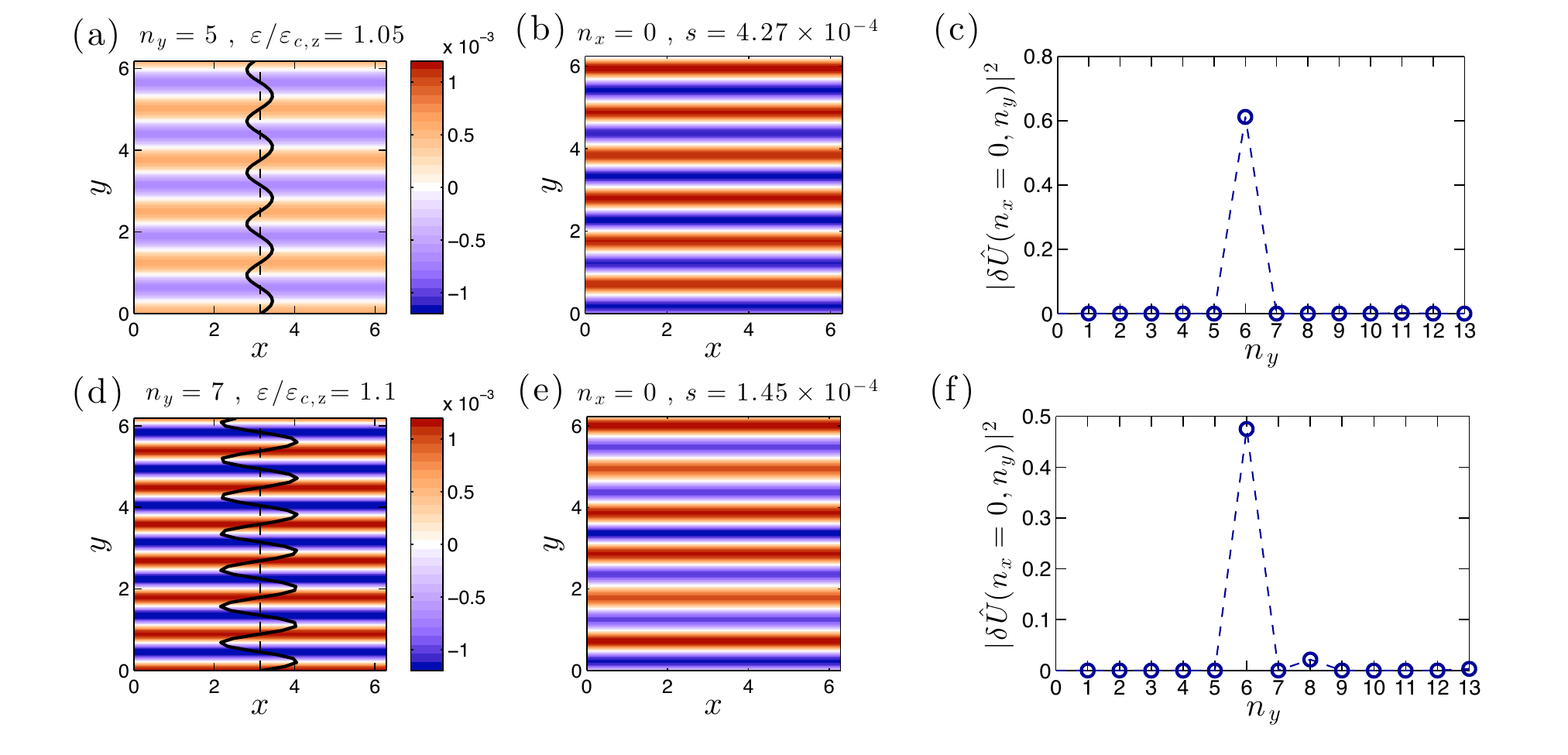}
\vspace{-.5em} \caption{\label{fig:stab_ny5_ny7_eckhaus} The jet equilibria with $n_y=5$ and $n_y=7$ near the marginal curve are both found to be Eckhaus unstable to an $n_y=6$ zonal jet perturbation. (a,d) Contour plots of the equilibrium streamfunction, $\Psi^e(y)$, together with the zonally averaged zonal velocity $U^e(y)$ (thick black line) for  jet equilibria with $n_y=5$ jets at $\e/\ecz=1.05$ and $n_y=7$ jets at $\e/\ecz=1.1$. (b,e) The structure of the mean flow streamfunction of the maximally growing S3T eigenfunctions for the jet equilibria together with its eigenvalue $\s$. (c,f) The energy spectrum of the corresponding S3T eigenfunctions as a function of the meridional wavenumber $n_y$. While both eigenfunction show maximum power at $n_y=6$ they have also non-zero power spectrum at other wavenumbers.}
\end{figure}

For very high supercriticalities only finite amplitude states with jets having the largest allowed scale, $n_y=1$, exist (this occurs in the balloon diagram Fig.~\ref{fig:ballonNIF} for $\e/\ecz>800$). Higher supercriticalities cannot sustain S3T fixed points because the periodic box does not allow jets  larger than the box size and the S3T dynamics eventually produce chaotic and non-stationary trajectories of the statistics of the turbulent flow.

We return to the equilibrium states that emerge after the homogeneous equilibrium is broken. We note that for $1<\e/\ecz<1.2$, while inhomogeneous equilibria with $n_y=5,6,7$ are found, only the equilibrium with $n_y=6$ is found to be stable. The equilibria with $n_y=5$ and $n_y=7$ are both unstable with their most unstable eigenfunction being a zonal jet with maximum power spectral power at $n_y=6$ (cf.~Fig.~\ref{fig:stab_ny5_ny7_eckhaus}). This instability is the universal Eckhaus instability that occurs near the neutral stability boundary and  attracts all finite amplitude states to the structure of the most unstable eigenfunction of the homogeneous state; in this case $n_y=6$. The nonlinear dynamics near the marginal curve obey a Ginzburg-Landau equation (cf.~\textcite{Parker-Krommes-2014-generation}). It should be noted that the accuracy of the approximation of the dynamics by the Ginzburg-Landau equation is unfortunately limited only to parameter values that are very close to the stability boundary, i.e., for $\e < 1.01\ecz$. Also the jet mergers that occur under Ginzburg-Landau dynamics are associated with the equilibration of the Eckhaus instabilities and are very different from the jet mergers that are seen in S3T simulations at higher supercriticality (cf.~\textcite{Parker-Krommes-2014-generation} and~Fig.~\ref{fig:merging_stab}).

\section{Stability of finite amplitude zonal jet S3T equilibria to non-zonal mean flow perturbations\label{sec:jet_to_nonzonal}}

The method described in the previous section for the stability of zonal jet S3T equilibria to zonal jet perturbations (i.e.~mean flow perturbations with $n_x=0$) will be employed now to determine the stability of the zonal jet equilibria that were discussed in the previous section to non-zonal mean flow perturbations. (i.e.~mean flow perturbations with $n_x \ne 0$). The question  we want to address is: can a jet be stable to jet like perturbations ($n_x=0$) but still be unstable to non-zonal large scale structures? We will demonstrate that all the stable jet equilibria of the balloon diagram Fig.~\ref{fig:ballonNIF} are unstable to non-zonal perturbations.
We note that these equilibria were obtained with NIF forcing and with the eddy flow being damped more strongly than the large-scale flow (both the jet and non-zonal large-scale components).  We have obtained similar results  when both flows were dissipated equally and the forcing was isotropic. Therefore we believe the results presented here are not pathological.  This instability of strong jet equilibria to  non-zonal perturbations is intriguing  because our theory at the level of our numerical implementation predicts  that in barotropic turbulence the statistical equilibria must have a non-zonal component, contrary for example to the observations on Jupiter which strongly suggest that the equilibria are almost purely zonal.

%We find, however, that the predicted final attracting states of the S3T theory are reconcilable with observations because S3T predicts attracting states that are mainly zonal with only a small fraction of the mean flow energy in the non-zonal component. Examples of such equilibrated states are shown in in Fig. 1.12 and Fig. 1.13 for IRFn forcing. This prediction of the theory is intriguing and it raises the possibility that the actual equilibrium states of the atmosphere have a long wave component, which arises as a bifurcation of the homogeneous turbulent state and is not forced by inhomogeneities in the forcing or the boundary conditions (like for example topography). This interpretation is supported by the study of \textcite{Bernstein-Farrell-2010} who demonstrated in a baroclinic two-layer atmosphere that the zonal equilibrium state bifurcates to non-zonal wave states.

\begin{figure}
\centering
\includegraphics[width=.95\textwidth,trim = 0mm 0mm 0mm 0mm, clip]{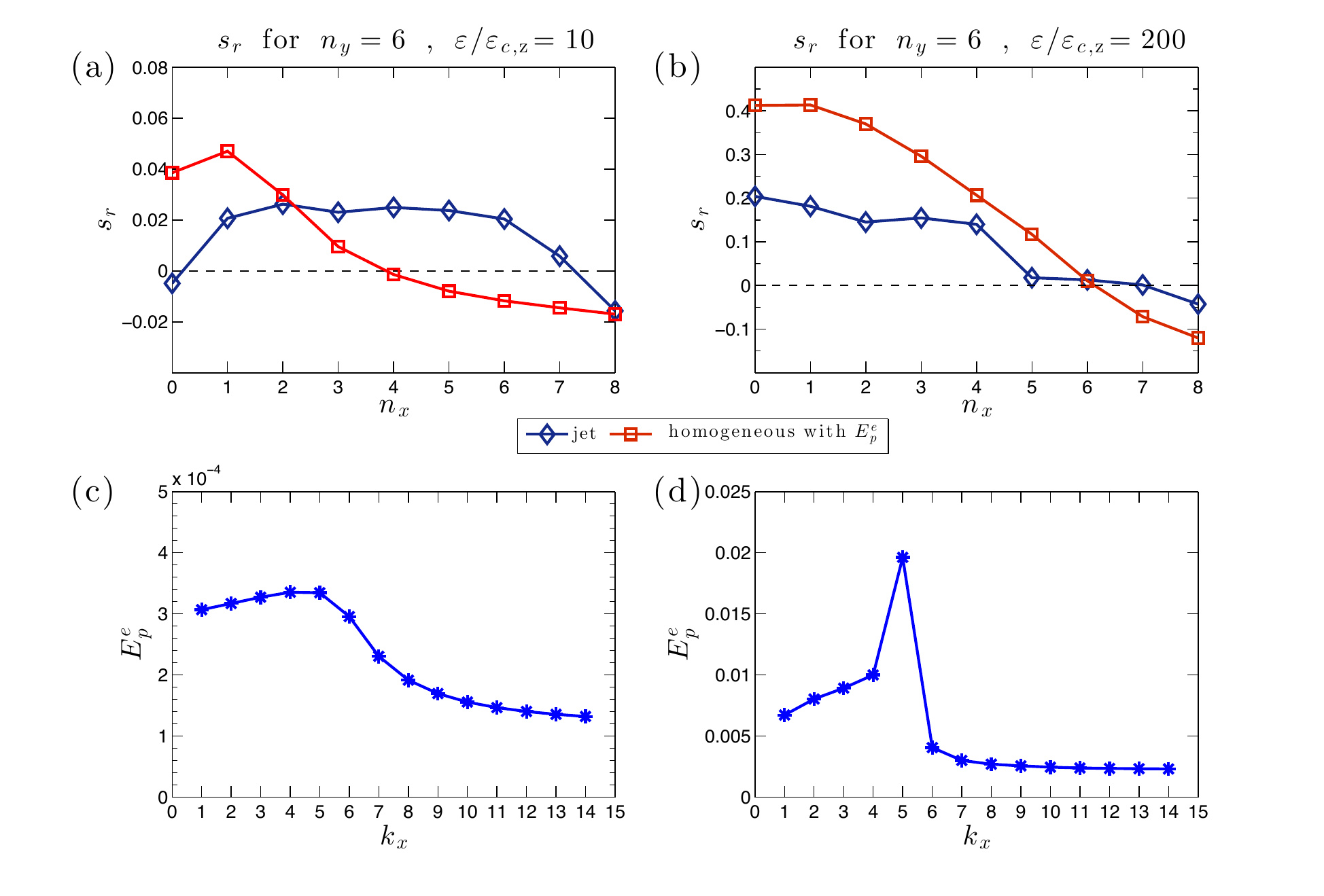}
\caption{\label{fig:sr_ny6_hom_jet} The growth rate of the maximally growing S3T eigenfunction of the 6-jet equilibrium  at supercriticality $\e/\ecz=10$ (panel (a)), and $\e/\ecz=200$ (panel (b)) as a function of the zonal wavenumber $n_x$ of the perturbation (blue line, \textemdash\hspace{-.4em}$\lozenge$\hspace{-.4em}\textemdash). In (a) the jet is S3T stable and in (b) unstable. The jet that is stable to $n_x=0$ becomes unstable for non-zonal perturbations. For comparison we also plot the growth rate of the unstable homogeneous equilibrium with perturbation energy equal to that of the inhomogeneous equilibrium (red line, \textemdash\hspace{-.4em}$\square$\hspace{-.4em}\textemdash). The equilibrium perturbation energy associated with the 6-jet equilibrium flows is shown respectively in panels (c) and (d). The presence of jets is seen to generally increase the S3T stability of the flow. Other parameters as in Fig.~\ref{fig:ballonNIF}.}
\end{figure}

\begin{figure}
\centering
\includegraphics[width=5.27in,trim = 0mm 0mm 32mm 0mm, clip]{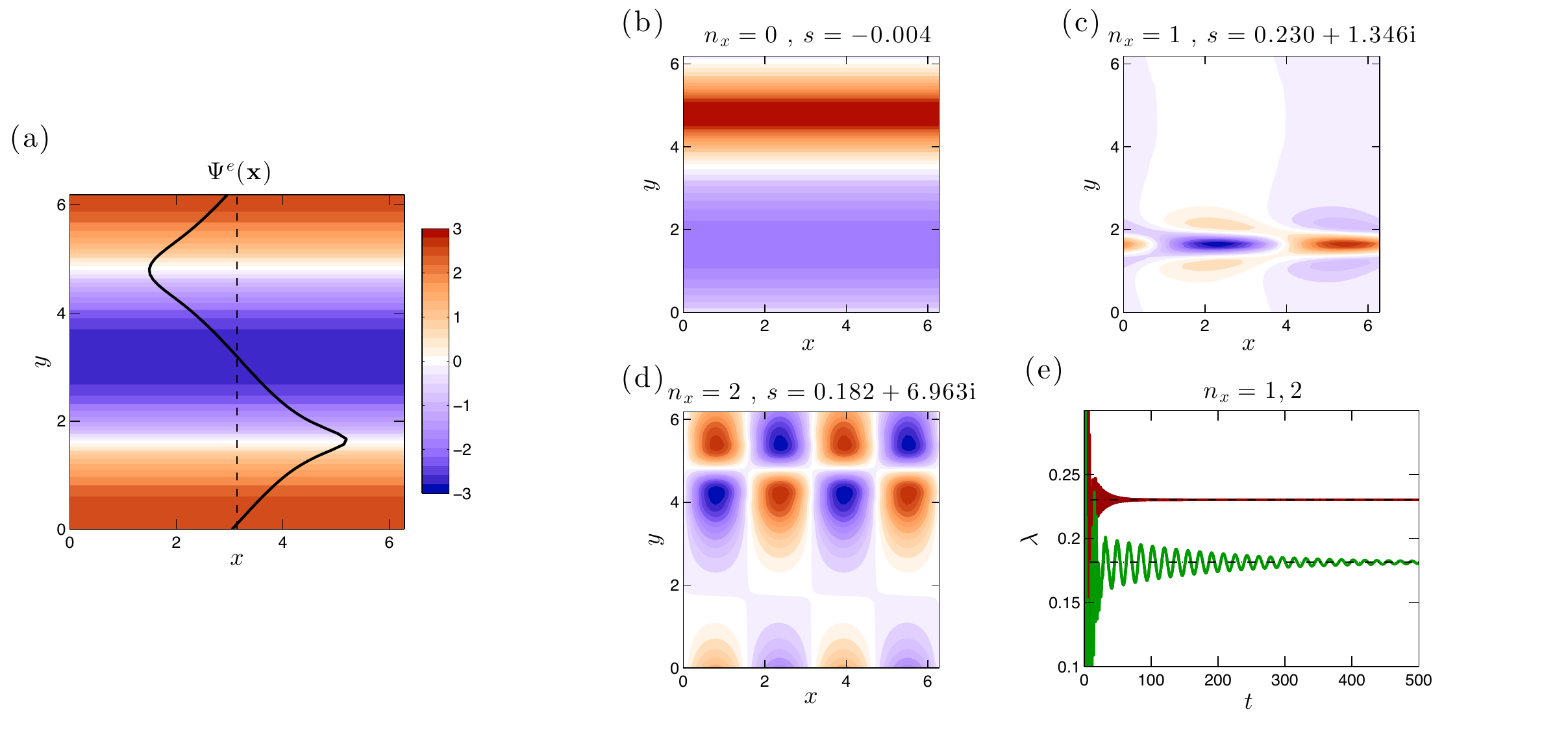}
\vspace{-2.5em} \caption{\label{fig:Psie_dPsi_ny1_e1000ecz} Stability of the zonal jet S3T equilibrium with $n_y=1$ jet at $\e/\ecz=1000$ (a): Contour plot of the equilibrium mean flow streamfunction, $\Psi^e(y)$, and a plot of the zonal velocity, $U^e(y)$, (thick black line). The jet maximum is $4.1$ and the  minimum is $-3.3$. The mean flow contains $92\%$ of the total energy of the flow. (b,c,d): The streamfunction of the most unstable mode, $\d\tilde{\Psi}$, for zonal  perturbations $(n_x=0)$ (panel (b)) and for non-zonal perturbations with   $n_x=1$ (panel (c)) and $n_x=2$ (panel (d)). The non-zonal $n_x=1$ eigenfunction is the  most unstable mode while the jet is stable to zonal jet $(n_x=0)$ perturbations.  (e):  Demonstration of the convergence of the power method iteration towards the maximum growth rate for perturbations with $n_x=1,2$. Note that the coefficient of linear damping of the mean flow (whether zonal or non-zonal) is ten times smaller than that of the perturbation field. The forcing is NIF (detailed parameter
specification can be found in Fig.~\ref{fig:ballonNIF}).}
%\end{figure}
%
%\begin{figure}
\vspace{2em}\centering
\includegraphics[width=5.27in,trim = 8mm 0mm 8mm 0mm, clip]{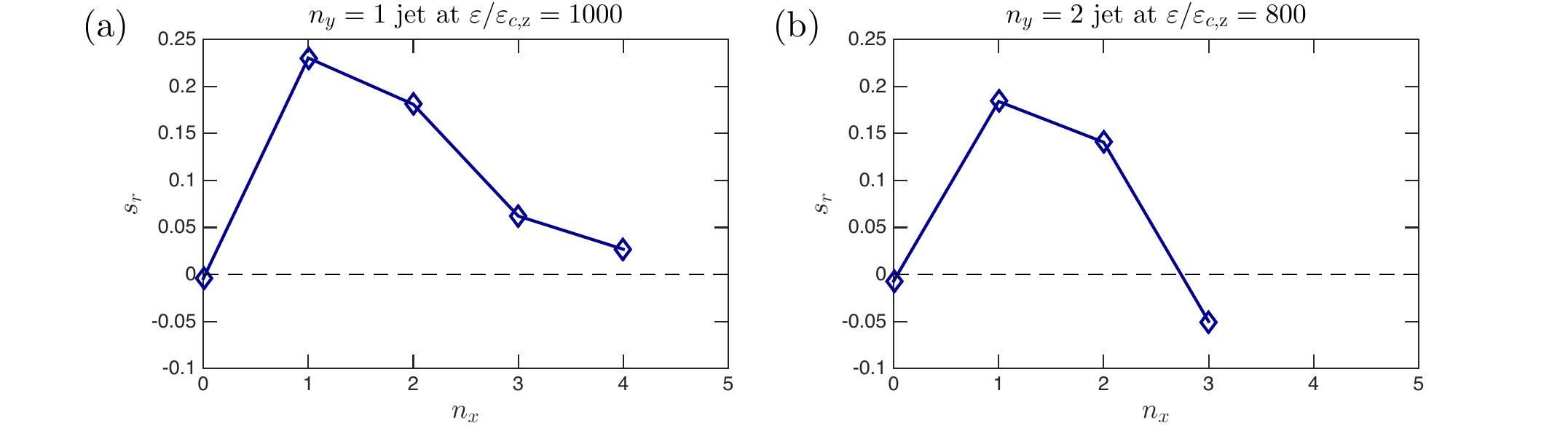}
\caption{\label{fig:S3Tgr_ny1_e1000}  The growth rate of the maximally growing S3T eigenfunction of the 1-jet equilibrium shown in Fig.~\ref{fig:Psie_dPsi_ny1_e1000ecz}\hyperref[fig:Psie_dPsi_ny1_e1000ecz]{a} as a function of the zonal wavenumber $n_x$ of the perturbation (panel (a)) and similarly for the 2-jet equilibrium at $\e/\ecz=800$ (panel (b)).  Parameters as in Fig.~\ref{fig:ballonNIF}.}
\end{figure}

Consider first the stability of two of the $n_y=6$ jet equilibria shown in Fig.~\ref{fig:Ue_n0_6}, the equilibrium for $\e/\ecz=10$, which is stable to jet perturbations, and the equilibrium for $\e/\ecz=200$, which is unstable to jet perturbations. The corresponding S3T growth rates for zonal and non-zonal perturbations are shown in Fig.~\ref{fig:sr_ny6_hom_jet}. In both cases the equilibria are unstable to non-zonal mean flow perturbations with $1\le n_x\le 7$. We have also plotted the growth rates that would obtain if  the equilibrium were homogeneous and the eddy field had the eddy energy zonal spectrum of the S3T equilibrium with jets (shown in Fig.~\ref{fig:sr_ny6_hom_jet}\hyperref[fig:sr_ny6_hom_jet]{c,d}). We see that in general the presence of the jet increases the S3T stability of the turbulent state.  However, this increase of S3T stability with jet strength does not eliminate the non-zonal instability of strong jets.
For example consider the instability of the strong jet equilibria with $n_y=1,2$ that are stable to $n_x=0$ jet perturbations. The $n_y=1$ equilibrium state at $\e/\ecz=1000$ is shown in Fig.~\ref{fig:Psie_dPsi_ny1_e1000ecz} together with the streamfunction of the least stable mode for perturbations with $n_x=0$ and the most unstable modes for perturbations with $n_x=1,2$. Maximum instability for this jet equilibrium is found at $n_x=1$, as shown in Fig.~\ref{fig:S3Tgr_ny1_e1000}\hyperref[fig:S3Tgr_ny1_e1000]{a} and similarly for the $n_y=2$ jet equilibrium at $\e/\ecz=800$ in Fig.~\ref{fig:S3Tgr_ny1_e1000}\hyperref[fig:S3Tgr_ny1_e1000]{b}.

%However, if the jet equilibrium is unstable, we have found that it stabilizes for perturbations with high enough zonal wavenumber $n_x$; for the $n_y=6$ equilibrium at $\e/\ecz=200$ this happens for $n_y>5$.
%
%The emergence of the jets has an interesting effect on the distribution of the eddy energy per zonal wavenumber, $k_x$. While for the NIF forcing used here all zonal wavenumbers are forced with the same energy input rate (cf.~Appendix~\ref{appsec:forc_spec_IRF_NIF}), the emergence of the zonal jets induces non-normality to the operator $\Acal(U)$ which leads to an anisotropic distribution of the equilibrium eddy energy into each wavenumber zonal wavenumber $k_x$. The energy distribution for the $n_y=6$ jet equilibria at $\e/\ecz=10$ and~$200$ is shown as a function of $k_x$ in Fig.~\ref{fig:sr_ny6_hom_jet}\hyperref[fig:sr_ny6_hom_jet]{c,d}. Also, the presence of the jets has a stabilizing effect on the flow. It should be expected at first order that the initial instability of the homogeneous state will be reduced after the emergence of the jets, since the jets acquire some fraction of the available flow energy. However, the presence of the jets stabilizes the flow even further. To demonstrate that, we calculate the growth rates of the homogeneous state that has an equilibrium eddy covariance with the same energy distribution per zonal wavenumber $k_x$ as the jet equilibrium; see Fig.~\ref{fig:sr_ny6_hom_jet}\hyperref[fig:sr_ny6_hom_jet]{a,b}. It is seen that even with the same available eddy energy the presence of the jets leads to less growth rates.

%{\color{Gray} Let us first we study the stability to non-zonal perturbations of all equilibria shown in Fig.~\ref{fig:ballonNIF}. Remarkably, we find that every equilibrium which is stable to $n_x=0$ perturbations is unstable to non-zonal perturbations with $n_x=1$ or $n_x=2$. This is very surprising and it argues that the attracting states in barotropic turbulence are not actually  zonal, contrary to what observations may suggest. We find, however, that the predicted final attracting states of the S3T theory are reconcilable with observations because S3T predicts attracting states that are mainly zonal with only a small fraction of the mean flow energy in the non-zonal component. Examples of such equilibrated states are shown in in Fig.~\ref{fig:S3Tmixed_a} and~Fig.~\ref{fig:S3Tmixed_b} for IRFn forcing. This prediction of the theory is intriguing and it raises the possibility that the actual equilibrium states of the atmosphere have a long wave component, which arises as a bifurcation of the homogeneous turbulent state and is not forced by inhomogeneities in the forcing or the boundary conditions (like for example topography). This interpretation is supported by the study of \textcite{Bernstein-Farrell-2010} who demonstrated in a baroclinic two-layer atmosphere that the zonal equilibrium state bifurcates to non-zonal wave states. }

\begin{figure}
\centering
\includegraphics[width=.98\textwidth,trim = 3mm 0mm 3mm 0mm, clip]{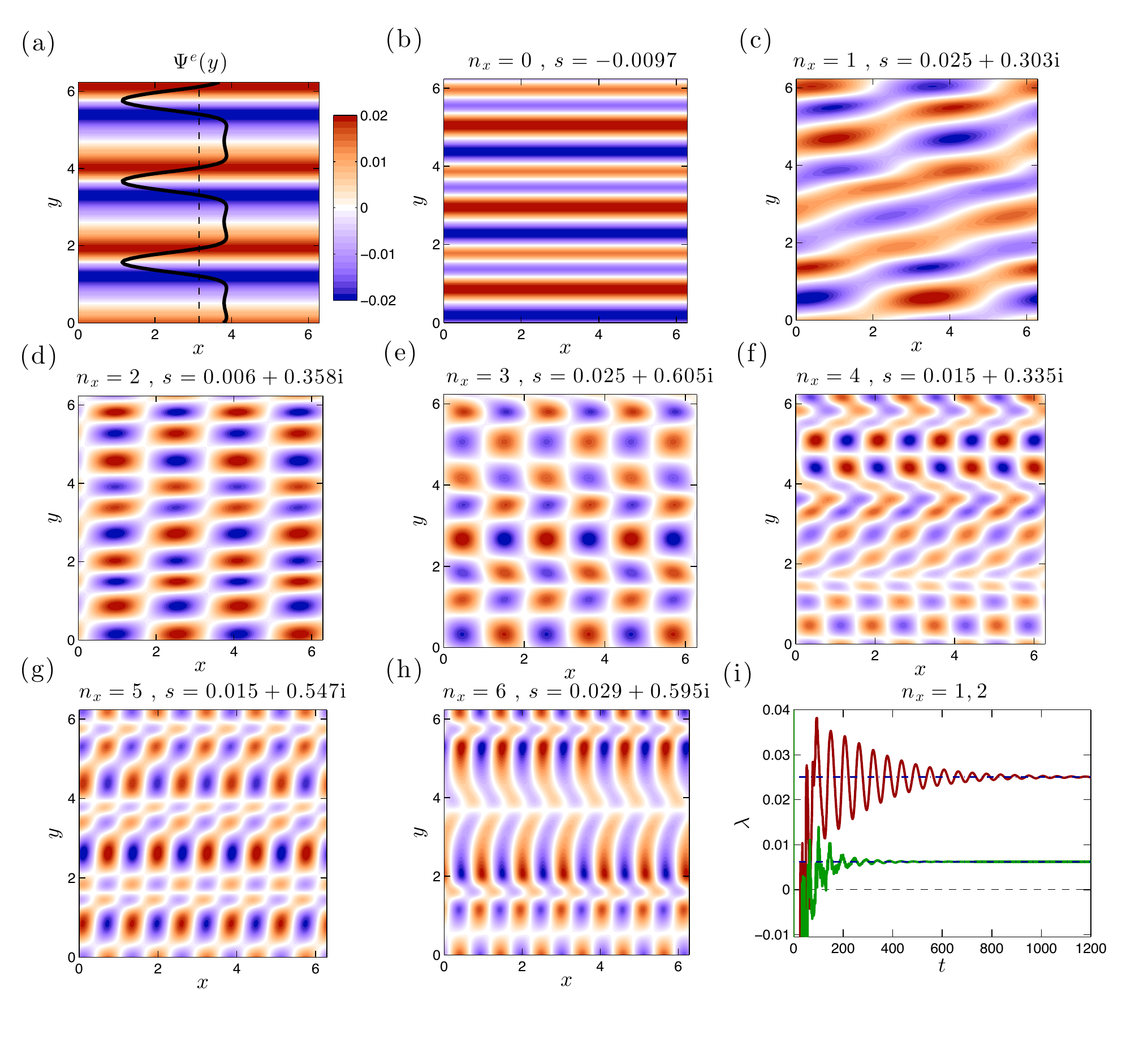}
\vspace{-1em} \caption{\label{fig:S3Teigen_S3TzMIequil} Stability of the zonal jet S3T equilibrium for the case presented in Fig.~\ref{fig:S3Tz_snapshots_psi}. (a): Contour plot of the equilibrium mean flow streamfunction, $\Psi^e(y)$, with the zonally averaged zonal velocity, $U^e(y)$, (thick black line). (b-h): The most unstable mean flow perturbation streamfunction, $\d\tilde{\Psi}$, for perturbations with $n_x=0,1,\dots,6$ together with their corresponding eigenvalues, $s$. While this zonal jet S3T equilibrium is stable to zonal jet perturbations ($n_x=0$) it is found to be unstable to non-zonal perturbations with maximum instability occurring for $n_x=1,3,6$. The maximal growing eigenfunction has the (1,4) structure that eventually emerges both in the NL and in the generalized S3T simulation (cf.~Fig.~\ref{fig:NL_stoch_snapshots_psi} and Fig.~\ref{fig:S3T_snapshots_Psi} respectively). (i): The evolution of the growth, $\la$, for $n_x=1,2$ perturbations showing the convergence to the corresponding growth rates, $s_r$ (dash-dotted line). Parameters: $\b=4.9$, linear damping coefficient $r=0.01$, stochastic forcing with single harmonics with wavenumber $(7,0)$ and energy injection rate $\e=4\times10^{-5}$.}
\vspace{.5em}
\end{figure}

\begin{figure}
\centering
\includegraphics[width=.98\textwidth,trim = 3mm 0mm 3mm 0mm, clip]{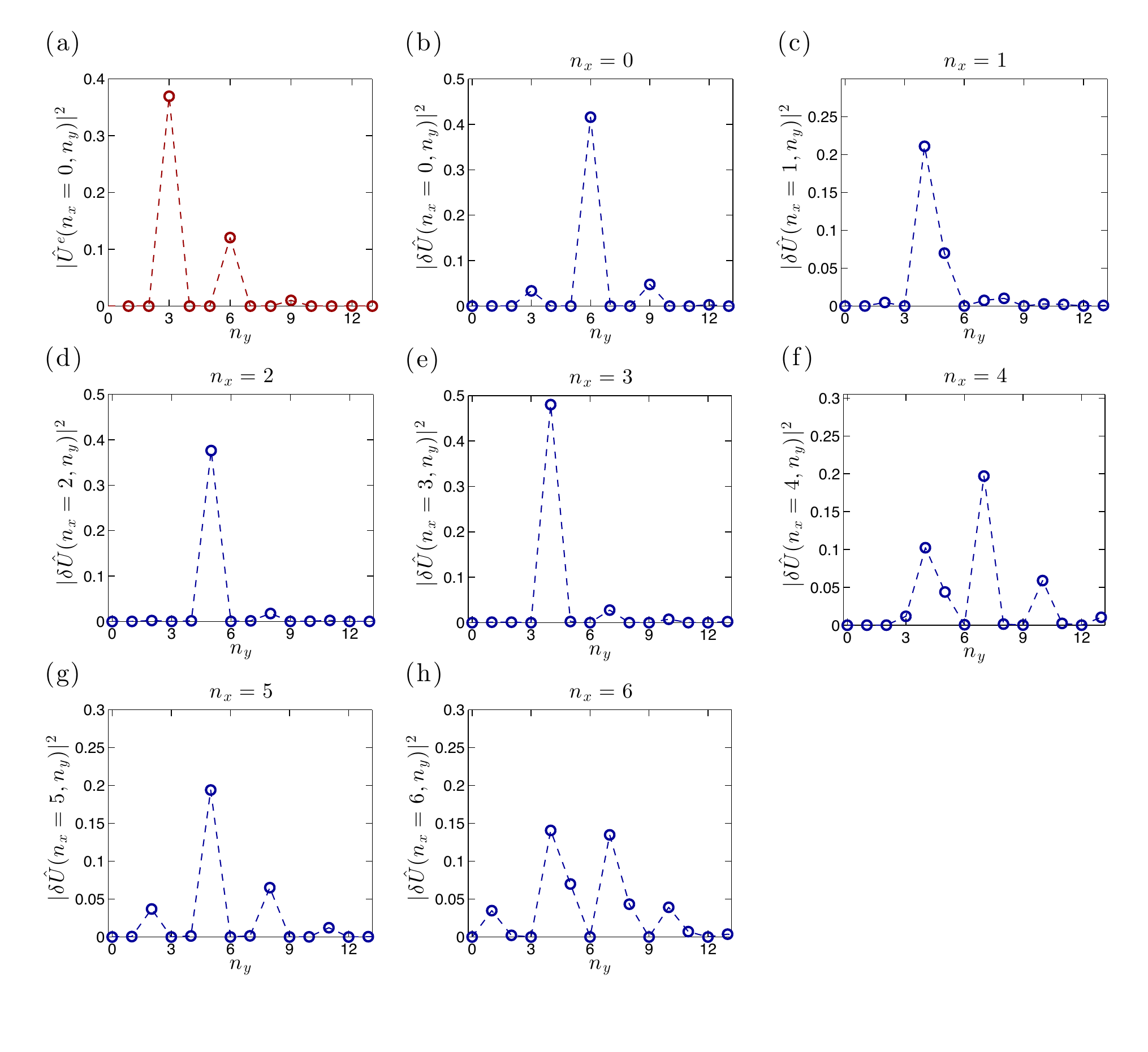}
\vspace{-1em}\caption{\label{fig:spec_Ue_dU_nx1} (a) The energy spectrum of the jet equilibrium shown in Fig.~\ref{fig:S3Teigen_S3TzMIequil}\hyperref[fig:S3Teigen_S3TzMIequil]{a} as a function of the meridional wavenumber $n_y$. (b-h) The energy spectrum of the mean flow perturbations with $n_x=0,1,\dots,6$ as a function of the meridional wavenumber $n_y$. The $n_x=0$ eigenfunction is a $q_y=0$ Bloch state, while eigenfunctions $n_x=1,\dots,6$ are $q_y=1$ Bloch states. Other parameters as in Fig.~\ref{fig:S3Teigen_S3TzMIequil}.}
\end{figure}

\begin{figure}
\centering
\includegraphics[width=2.7in]{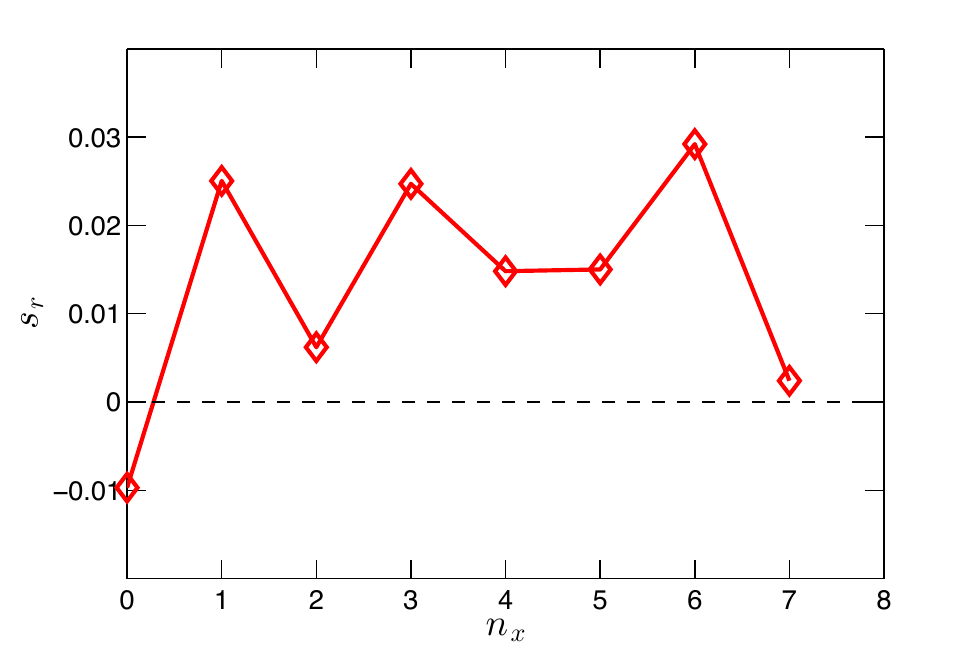}
\caption{\label{fig:S3Tgr_MI_ny3jet} The growth rate of the maximally growing S3T eigenfunction of the 3-jet equilibrium shown in Fig.~\ref{fig:S3Teigen_S3TzMIequil}\hyperref[fig:S3Teigen_S3TzMIequil]{a} as a function of the zonal wavenumber $n_x$ of the perturbation. Other parameters as in Fig.~\ref{fig:S3Teigen_S3TzMIequil}.}
\end{figure}

Consider now the stability of the zonal jet equilibrium that we obtained with non-broadband forcing in our discussion of modulational instability in chapter~\ref{ch:MI}. In that example, the fastest growing instability of the homogeneous state is a zonal 3-jet perturbation. In an S3T integration this zonal jet structure emerges and saturates to a quasi-stationary finite amplitude 3-jet flow. This flow eventually breaks giving way to a  finite amplitude traveling wave with maximal power at wavenumber $(n_x,n_y)=(1,4)$ (see Fig.~\ref{fig:S3T_snapshots_Psi}). We show here that this finite amplitude wave with maximum power at (1,4) emerges because the 3-jet state is unstable to non-zonal perturbations with the property that all the instabilities equilibrate  to a finite amplitude state with maximal power at (1,4). With the methods described in the previous section we first obtain the 3-jet equilibrium and then calculate its stability. The S3T equilibrium and the maximally growing eigenfunction for $n_x=0,\dots,6$ are shown in Fig.~\ref{fig:S3Teigen_S3TzMIequil} and the spectrum of the $n_x=0,\dots,6$ eigenfunctions is shown in Fig.~\ref{fig:spec_Ue_dU_nx1}. The growth rate of the 3-jet as function of the $n_x$ is shown in Fig.~\ref{fig:S3Tgr_MI_ny3jet}. While the 3-jet is stable to jet perturbations it is unstable to non-zonal perturbations with $n_x\le 7$. (Note that the stochastic forcing has power only at $(7,0)$). Substantial growth occurs for $n_x=1,3,6$ and maximal growth for $n_x=6$. While there is instability at both $n_x=3$ and 6 at finite amplitude all instabilities are attracted to a $n_x=1$ state with maximal power at $n_y=4$ with approximately the spectral structure of the $n_x=1$ instability. This phenomenon of saturation of the instabilities to the structure of an instability of smaller growth rate  requires further study; it appears that there is at play an Eckhaus instability for non-zonal states. We witnessed similar behavior also in cases in which the homogeneous state was more unstable to non-zonal perturbations but the instabilities saturated into zonal jets. Thus the transition from a zonal to a non-zonal flow in the S3T  simulation of section~\ref{sec:MIcompare} is caused by the secondary S3T instabilities of the finite amplitude 3-jet saddle equilibrium that emerges from the homogeneous background. This complex of phenomena predicted by S3T is remarkably reflected, with the same time sequence, in corresponding NL simulations (see~Fig.~\ref{fig:NL_stoch_snapshots_psi} in section~\ref{sec:MIcompare}).

\vfill 

%\begin{figure}[ht]
%\centering
%\includegraphics[width=4.8in]{S3Tmixed_a.pdf}
%\vspace{-1em} \caption{\label{fig:S3Tmixed_a} A mixed zonal jet--traveling wave S3T equilibrium obtained from a simulation of the S3T system~\eqref{eq:s3t}. (a) Snapshot of the finite amplitude state mean flow streamfunction, $\Psi^e$. (b) Contour plot of the non-zonal component, $\Psi^e-\overline{\Psi^e}$, of the equilibrium structure, where the overline denotes a zonal average, revealing the $(2,6)$ structure which is embedded onto the $n_y=5$ zonal jets. (c) Hovm\"oller diagram of $\Psi^e$ at a fixed latitude demonstrating that the non-zonal component is propagating westward in a coherent manner. Simulation was performed with IRFn forcing at $\b/(k_f r) = 100$ and $\e = 30\ecnz$ which for the given case corresponds to the value $\e k_f^2/r^3 =25200$. (Taken from \textcite{Bakas-Ioannou-2014-jfm}.)}
%%\end{figure}
%%
%%\begin{figure}
%%\centering
%%\vspace{1em}
%\includegraphics[width=4.8in]{S3Tmixed_b.pdf}
%\vspace{-1em} \caption{\label{fig:S3Tmixed_b} A different mixed zonal jet--travelling wave S3T equilibrium for the same parameters as Fig.~\ref{fig:S3Tmixed_a}. In this case a (1,5) wave is embedded onto a $n_y=5$ zonal jet. (Courtesy N. A. Bakas.)}
%\end{figure}

\section{Stability of finite amplitude non-zonal traveling wave S3T equilibria}

%We have seen that the S3T system does not only has solutions with large-scale flow in the form of zonal jets but also, as we have already seen in chapter~\ref{ch:MI}, the mean flow may be in the form of a large-scale traveling wave (cf. Fig.~\ref{fig:S3T_snapshots_Psi}).
%
%In chapter~\ref{ch:st3hom} we have seen that at some parameter regimes  the homogeneous state becomes first unstable to non-zonal mean flows (i.e. when $\ecnz<\ecz$, cf.~Fig.~\ref{fig:ec_RFg_z_nz}). This non-zonal mean flows emerge and reach finite amplitude and may have great effect in the emergence of jets. In chapter~\ref{ch:NLvsS3Tjas} we have seen that for those cases  the emergence of non-zonal flows alters the eddy vorticity spectrum in a way that it stabilizes the flow towards jet perturbations (cf.~section~\ref{sec:influence_eddy}).
%
%\textcite{Bakas-Ioannou-2014-jfm} have studied the bifurcation of the homogenous turbulent state as predicted by the generalized S3T~\eqref{eq:s3t}, in which ensemble means are interpreted as Reynolds averages over the intermediate time or length scale, for cases with $\ecnz<\ecz$. It was found that the homogeneity of the flow was first broken at energy input rate $\e=\ecnz$ with the emergence of large-scale finite amplitude traveling waves. Zonal jets only appeared after $\e$ exceeded another critical value, the one which was denoted as $\ecz^{(\textrm{NL})}$ in chapter~\ref{ch:NLvsS3Tjas}. At that point the finite amplitude wave solutions broke down and the flow transitioned to a state characterized by a mean flow with a finite amplitude zonal jet.

Up to now we have investigated the stability of zonal equilibria to zonal and non-zonal perturbations. We now show how to calculate the stability of non-zonal wave S3T states.
For the calculation we adopt the method for calculating the sensitivity to initial conditions
of a trajectory $\(Z(\xv,t),C(\xv_a,\xv_b,t)\)$ of the S3T system. If the state trajectory is perturbed by $(\d Z, \d C)$ then its linear stability (i.e. its Lyapunov exponent) is obtained from a simultaneous  integration of the system: \begin{subequations}
\begin{align}
\partial_t Z &+ J \(\Psi, Z+  {\bm\beta}\cdot\mathbf{x} \) = \Rcal( C )-\,Z\ ,\label{eq:s3tpert_coupl_Z}\\
\partial_t C_{ab} & = \[\bit\Acal_a(\Uv) + \Acal_b(\Uv)\]C_{ab} +\e\,Q_{ab}\ ,\label{eq:s3tpert_coupl_C}\\
\partial_t \,\d Z & = \Acal(\Uv)\,\d Z + \Rcal( \d C )\ ,\label{eq:s3tpert_coupl_dZ}\\
\partial_t \,\d C_{ab} & = \[\bit\Acal_a(\Uv) + \Acal_b(\Uv) \]\d C_{ab} +\(\bit\d\Acal_a + \d\Acal_b\)C_{ab}\ .\label{eq:s3tpert_coupl_dC}
\end{align}\label{eq:s3tpert_coupl}\end{subequations}
While with this method we determine unequivocally the largest Lyapunov exponent of a perturbation trajectory, we can also estimate the stability of a traveling wave state of the S3T system if the S3T state persists in that state despite being unstable. This often occurs because the unstable S3T states have many stable directions (they are saddles) and because the S3T equations are noiseless the unstable directions take a long time before they obtain appreciable magnitude. Under these conditions integration of the tangent linear system~\eqref{eq:s3tpert_coupl_dZ}-\eqref{eq:s3tpert_coupl_dC} produces good estimates of the growth rate of the unstable traveling wave solutions of~\eqref{eq:s3tpert_coupl_Z}-\eqref{eq:s3tpert_coupl_C}.

\begin{figure}[ht]
\centering
\includegraphics[width=3.3in]{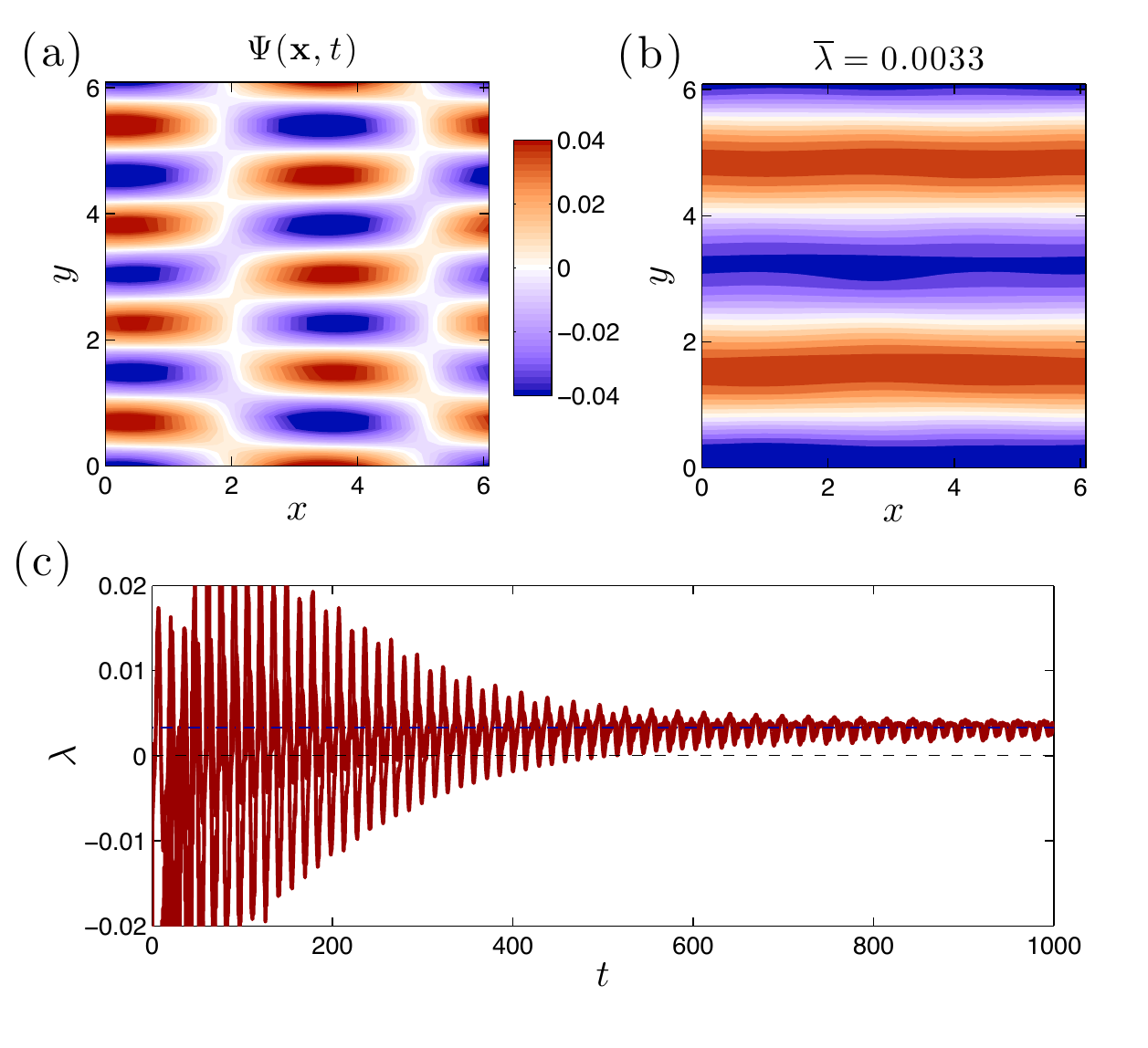}
\caption{\label{fig:S3Teigen_S3Tnz_equil} Stability of a non-zonal traveling wave quasi-equilibrium S3T state. (a): A snapshot of a finite amplitude state streamfunction, $\Psi$, that consists of a wavenumber $(1,4)$ wave traveling westwards. This state after long time (around 3000 time units) transitions to a wavenumber $(0,2)$ nearly zonal jet structure. (b): The mean flow streamfunction, $\d\tilde{\Psi}$, of the first Lyapunov vector of the state shown in (a) together with its corresponding Lyapunov exponent, $\overline{\la}$. The Lyapunov vector of the doubly periodic quasi-equilibrium is neither monochromatic in $x$ nor $y$. However, its maximum power is concentrated at $(n_x,n_y)=(0,2)$. (c) Demonstration of the convergence towards the maximum growth rate, $\overline{\la}=0.0033$. Parameters: $\b=12$, $r=0.01$, $\rU=0.005$, IRFn forcing with $k_f=6$, $\d k_f=1$ and $\e=3\ecnz=5.24\times10^{-4}$. Stability calculations were performed at resolution $N_x=N_y=32$ which results in a state variable of~\eqref{eq:s3tpert_coupl} with dimension $2\times10^6$.}
\end{figure}

With this method we consider the stability of the traveling wave S3T state shown in Fig.~\ref{fig:S3Teigen_S3Tnz_equil}. This finite traveling wave state emerged as a quasi-equilibrium
of the finite amplitude equilibration of the most unstable non-zonal eigenfunction of the homogeneous equilibrium. We show that this non-zonal state is unstable to zonal perturbations ($n_x=0$) which saturate into a  predominantly zonal state.

%
%
%
%We start with a normalized perturbation state $(\d Z,\d C)$ and at each time-step we measure the growth $\la$ (cf.~\eqref{eq:lambda_jh}) and then renormalize the perturbation state before moving to the next time-step. We initiate the integration from a state $(Z,C)$ that is close to the equilibrium solution. If the solution whose stability is in question is stable then we can integrate~\eqref{eq:s3tpert_coupl_dC} for as long as we like to obtain the desired accuracy. If however the equilibrium is unstable then we can only time-integrate the system~\eqref{eq:s3tpert_coupl_Z}-\eqref{eq:s3tpert_coupl_C} for limited amount of time before the system departs from the specific state resulting in erroneous stability calculation.
%
%Having to time-integrate forward the coupled system~\eqref{eq:s3tpert_coupl} doubles the dimension of the state variable. Also, because these kind of equilibria are not homogeneous to any direction, we cannot anticipate the form of the Lyapunov vector in a similar manner as we anticipated the eigenfunction to be of the form~\eqref{eq:S3Teigen_nx}. Therefore, we are obliged to time-integrate forward the full state without any reduction. For a discretization with resolution of $N_x=N_y=64$ the state variable of~\eqref{eq:s3tpert_coupl} has dimension of approximately $3.5\times10^7$. 
%
%By applying this method to the finite amplitude wave shown in Fig.~\ref{fig:S3Teigen_S3Tnz_equil} we obtain indication that this state is unstable to zonal jet perturbations.

%\section{Bibliographical note}
%The S3T instability of finite amplitude jet equilibria was studied by \textcite{Farrell-Ioannou-2007-structure}. 

% !TEX root = ../thesis.tex

\chapter{Conclusions\label{ch:conclusion}}

In this thesis we have presented a theory for the formation and maintenance of eddy-driven jets in planetary turbulence. The theory presented, called Stochastic Structural Stability Theory (S3T), is a non-equilibrium statistical theory which has its basis in wave--mean flow interaction theories. It studies the closed dynamics of the first two cumulants of the full statistical state dynamics of the flow, and neglects or parametrizes third and higher-order cumulants. Neglect or parametrization of third and higher order cumulants is equivalent to neglect or parametrization of the eddy--eddy interactions in the equations of motions. Thus, in the specific closure only the non-local in wavenumber space interaction between large-scale flows and the smaller scale eddies is allowed while local wavenumber interactions among eddies are not allowed. We study within this statistical closure large-scale structure formation and maintenance in stochastically forced--dissipative barotropic turbulence on a $\b$-plane.

We have performed an extensive comparison of jet formation as predicted by the S3T dynamics and as predicted by direct numerical simulations of the fully nonlinear equations. The emergence and equilibration of zonal jets in the S3T dynamics and, moreover, the remarkable agreement of their predicted structure with the jets observed in direct numerical simulations, provides a constructive proof that turbulent cascades are not required for the formation of zonal jets in $\b$-plane turbulence. Emergence and equilibration of zonal jets occurs due to cooperative quasi-linear mean flow--eddy interactions that is captured by S3T.

The S3T dynamics is autonomous and deterministic and therefore may have fixed point solutions. These solutions are statistical equilibria of the turbulent flow that describe both the large-scale structure mean flow (1st cumulant) as well as the second-order eddy statistics (2nd cumulant). S3T allows the study of the stability of such equilibrium states. Instability of a statistical equilibrium signifies transition of the turbulent regime to a different attractor (i.e.~to a different climate state). Such an equilibrium state of the statistics is the homogeneous turbulent state with no mean flow.  We have demonstrated that the homogeneous state becomes unstable at analytically predicted critical parameter values and the flow undergoes a bifurcation becoming inhomogeneous with the emergence of large-scale zonal and/or non-zonal flows. The mechanisms by which the turbulent Reynolds stresses organize to reinforce infinitesimal mean flow inhomogeneities, thus leading to this statistical instability, are extensively studied for various regimes of parameter values (planetary vorticity gradient, dissipation rate and turbulent energy injection rate). It is shown that for small and modest values of the planetary vorticity gradient, $\b$, the upgradient fluxes responsible for the large-scale structure formation instability are induced by the Orr mechanism, while for large $\b$ they are induced by resonant wave triads. The dependance of the instability on the spectrum of the stochastic excitation was also studied.

The relation between the formation of large-scale structure through modulational instability and the S3T instability of the homogeneous turbulent state was also investigated. We have demonstrated the formal equivalence between the 4MT system, that approximates well the modulational instability of coherent Rossby waves, and the S3T instability of a homogeneous turbulent state. However, we have demonstrated that the 4MT dynamical framework is inadequate for capturing the finite amplitude equilibration of the instabilities.

We also presented methods for studying the stability of inhomogeneous turbulent equilibria. We demonstrated that the phenomenon of jet merging is properly understood as an S3T instability of a finite-amplitude mean flow state and therefore is properly understood in the framework of statistical state dynamics. Also, we have shown that the transition from zonal to non-zonal turbulent states is also predicted by S3T stability analysis.

The S3T closure produces an analytical, predictive and quantitative theory for turbulence that proceeds directly from the equations of motion. It provides a way of determining turbulent statistical equilibria (climate states of our model) and moreover determining their stability.

\begin{appendices}
    % !TEX root = ../thesis.tex

\chapter{Construction of the stochastic forcing and demonstration that the energy injection rate induced by a temporally delta-correlated forcing is independent of the state of the system}
\label{app:forcing}

The stochastic equations  (i.e.~eqs.~\eqref{eq:nl} or~\eqref{eq:eql}) are nonlinear
differential equations with additive stochastic excitation of the generic form $\partial_t \phi = \Lcal(\phi) + \Ncal(\phi,\phi) +\sqrt{\varepsilon} \xi$, 
%\be
%\partial_t \phi = \Lcal(\phi) + \Ncal(\phi,\phi) +\sqrt{\varepsilon} \xi\ ,
%%\partial_t\,\phi(\xv,t) = f\(\bit\phi(\xv,t)\) + \sqrt{\varepsilon}\xi(\xv,t)\ .\label{eq:sde}
%\ee
with $\Lcal$ a linear operator and $\Ncal$ a nonlinear operator. The stochastic 
function $\xi(\xv,t)$ is taken to be a Gaussian process with zero mean,
i.e., $\<\xi(\xv,t)\> =0$, and a homogeneous function of both space and time. It  is further assumed to be delta-correlated in time but spatially correlated with spatio-temporal
covariance $ \<\xi(\xv_a,t)\xi(\xv_b,t')\> = {\protect Q(\xv_a,\xv_b) \delta(t-t')}$ 
(the averaging operator $\<\bullet\>$ denotes an average over forcing realizations).
The delta correlation in time is a very crucial assumption because it implies that the energy injection rate is constant and independent of the state the system and depends only on the amplitude factor $\varepsilon$ when $Q$ is appropriately normalized. This allows us to know the energy injection rate once we have specified the forcing without any reference to the flow that develops. If the stochastic excitation were not delta correlated, i.e., if for example it were a red-noise process, the energy injection rate would depend on the state of the system. 

\section{Construction of the stochastic forcing}

We show first that the spatio-temporal homogeneity of the random field $\xi$ implies that $\<\xi(\xv_a,t)\xi(\xv_b,t')\>=F(\xv_a-\xv_b,t-t')$, which in turn implies that the spatial covariance of the forcing, $Q$, is a function of the difference coordinate $\xv_a-\xv_b$, i.e., $Q(\xv_a,\xv_b)=Q(\xv_a-\xv_b)$.%= and determines its structure from the specification of the random $\xi$. 

Since $\xi$ is chosen to be homogeneous its statistical properties are invariant under translations, that is $\<\xi(\xv_a,t)\xi(\xv_b,t')\>=\<\xi(\xv_a+\alphav,t)\xi(\xv_b+\alphav,t')\>$ for any $\alphav$. By expanding $\xi$ in Fourier series,
\be
\xi(\xv,t) = \int \frac{\df^2 \kv}{(2\pi)^2} \hat{\xi}(\kv,t)\,e^{\i\kv\cdot\xv}\ ,\label{eq:xi_fourier}
\ee
with $\hat{\xi}(-\kv,t)=\hat{\xi}(\kv,t)^*$ in order that $\xi$ be real valued, we have that:\begin{subequations}
\begin{align}
\<\xi(\xv_a,t)\xi(\xv_b,t')\>&=\iint \frac{\df^2 \kv}{(2\pi)^2} \frac{\df^2 \kv'}{(2\pi)^2}\, \<\hat{\xi}(\kv,t)\hat{\xi}(\kv',t')\> e^{\i\kv\cdot\xv_a} e^{\i\kv'\cdot\xv_b}\ ,\label{eq:exp1}
\end{align}
while
\begin{align}
\<\xi(\xv_a+\alphav,t)\xi(\xv_b+\alphav,t')\>&=\iint \frac{\df^2 \kv}{(2\pi)^2} \frac{\df^2 \kv'}{(2\pi)^2}\, \<\hat{\xi}(\kv,t)\hat{\xi}(\kv',t')\> e^{\i\kv\cdot\xv_a} e^{\i\kv'\cdot\xv_b}\,e^{\i(\kv+\kv')\cdot\alphav}\ .\label{eq:exp2}
\end{align}\end{subequations}
For~\eqref{eq:exp1} and \eqref{eq:exp2} to be equal we must have $e^{\i(\kv+\kv')\cdot\alphav}=1$ for every $\alphav$, which requires that $\kv'=-\kv$ and which in turn implies that the wavenumber covariance of the field is of the form: $\<\hat{\xi}(\kv,t)\hat{\xi}(\kv',t')\>=(2\pi)^2\<\hat{\xi}(\kv,t)\hat{\xi}(-\kv,t')\>\d(\kv+\kv')$. Consequently,
\begin{align}
\<\xi(\xv_a,t)\xi(\xv_b,t')\>&=\int \frac{\df^2 \kv}{(2\pi)^2} \, \<\hat{\xi}(\kv,t) \hat{\xi}(-\kv,t')\> e^{\i\kv\cdot(\xv_a-\xv_b)}\ , \label{eq:xiat_xibt'}
\end{align}
which is a function only of the difference $\xv_a-\xv_b$. Similarly, we can show that temporal homogeneity implies that the $\<\hat{\xi}(\kv,t) \hat{\xi}(-\kv,t')\> = f(\kv,t-t')$.

Note if the random field is homogeneous only in the zonal $x$ direction, then the above argument would require that $e^{\i (k_x+k_x') \a_x}=1$ for every $\a_x$, which would imply $k_x=-k_x'$ and as a result the spatial covariance of a random field homogeneous in $x$ should be a function of $x_a-x_b$, $y_a$ and $y_b$. 

\vspace{1em}

To construct the delta-correlated Gaussian stochastic excitation we choose the Fourier amplitude of~\eqref{eq:xi_fourier} at each wavenumber $\kv$ and at each instant to be:
\be
\hat{\xi}(\kv,t) = w(\kv)\,\eta_{\kv,t}\ ,\label{eq:xihat_choice}
\ee
with $w(\kv)$ the amplitude at wavenumber $\kv$, satisfying $w(-\kv)=w(\kv)^*$,  and $\eta_{\kv,t}$ a complex valued Gaussian random variable of $\kv$ and $t$ satisfying $\eta_{-\kv,t} = \eta_{\kv,t}^*$, with zero mean and delta-correlated covariance both in wavenumber and time: $\<\eta_{\kv,t} \eta_{\kv',t'}^*\>=\d(t-t')\d(\kv-\kv')$. Equation~\eqref{eq:xihat_choice} implies that $\<\xi(\xv,t)\>=0$ and, moreover~\eqref{eq:xiat_xibt'} can be written as,
\begin{align}
\<\xi(\xv_a,t)\xi(\xv_b,t')\> &= \iint \frac{\df^2 \kv}{(2\pi)^2} \frac{\df^2 \kv'}{(2\pi)^2}\, w(\kv)w(\kv')\,\<\eta_{\kv,t}\,\eta_{\kv',t'}\> e^{\i\kv\cdot\xv_a} e^{\i\kv'\cdot\xv_b}\nonumber\\
&= \iint \frac{\df^2 \kv}{(2\pi)^2} \frac{\df^2 \kv'}{(2\pi)^2}\, w(\kv)w(\kv')^*\,\<\eta_{\kv,t}\,\eta_{\kv',t'}^*\> e^{\i\kv\cdot\xv_a} e^{-\i\kv'\cdot\xv_b}\nonumber\\
&= \d(t-t')\int \frac{\df^2 \kv}{(2\pi)^4}  |w(\kv)|^2\, e^{\i\kv\cdot(\xv_a-\xv_b)}\ ,
\end{align}
(in the second equality we changed the integrating variables $\kv'\to-\kv'$). Therefore the spatial structure of the covariance, $Q$, is given by:
\be
Q(\xv_a-\xv_b) = \int \frac{\df^2 \kv}{(2\pi)^2}  \frac{|w(\kv)|^2}{(2\pi)^2}\, e^{\i\kv\cdot(\xv_a-\xv_b)} = \int \frac{\df^2 \kv}{(2\pi)^2} \hat{Q}(\kv)\, e^{\i\kv\cdot(\xv_a-\xv_b)}\ .\label{eq:appQhat}
\ee
The spatial covariance $Q$ turns out to be an even function of its argument, since $|w(-\kv)|=|w(\kv)|$. This is physically expected due to the homogeneity of $\xi$, which implies the exchange symmetry $\<\xi(\xv_a,t)\xi(\xv_b,t')\>=\<\xi(\xv_b,t)\xi(\xv_a,t')\>$. Moreover, equation~\eqref{eq:appQhat} shows that the Fourier transform of $Q$ besides being real is also non-negative, as is demanded by Wiener-Khinchin theorem. The statement that homogeneous covariances have real  and positive Fourier transforms is often also referred to as Bochner's theorem.

That the covariance $\<\xi(\xv_a,t)\xi(\xv_b,t')\>$ turns out to be a function only of the difference coordinates $\xv_a-\xv_b$ may be alternatively seen as a result of the fact that each of the Fourier components of the stochastic forcing corresponds to a different random variable which is uncorrelated to the others, except for the pairs $\kv$ and $-\kv$.

We construct a discrete representation of $\xi$ in time as follows.
Discretize time with time steps $h$. The complex random variable $\eta_{\kv,t}$ at time  $t=jh$, $j=0,1,\dots$, is taken to be
\be
\eta_{\kv,t} =\frac1{\sqrt{2h}} \( X_{\kv,t} + \i Y_{\kv,t} \)\ ,\label{eq:Wkt_discr}
\ee
with $X$ and $Y$ real valued random numbers taken from a Gaussian number 
generator with zero mean and unit variance; we further set $X_{-\kv,t}=X_{\kv,t}$ and $Y_{-\kv,t}=-Y_{\kv,t}$. The $\sqrt{h}$ in the denominator ensures that~\eqref{eq:Wkt_discr} is delta-correlated in time satisfying as $h\to 0$ the discrete expression of the delta function requirement, $\int_{-\infty}^{\infty} \<\eta_{\kv,t} \eta_{\kv,t'}^*\>\,\df t'=1$. Note that because $X$ and $Y$ are normally distributed, the probability density function of  $\eta_{\kv,t}$, $P(\eta_{\kv,t})$, is only a function of the amplitude of $\eta_{\kv,t}$, as
\begin{align}
P(\eta_{\kv,t}) &=P(X_{\kv,t},Y_{\kv,t}) \nonumber\\
& =  P(X_{\kv,t})P(Y_{\kv,t}) \nonumber\\
&= \frac1{\sqrt{2\pi}}e^{-\frac1{2}X_{\kv,t}^2}\,\frac1{\sqrt{2\pi}}e^{-\frac1{2}Y_{\kv,t}^2}\nonumber\\
%&= \frac1{2\pi}e^{-\frac1{2}\(X_{\kv,t}^2+Y_{\kv,t}^2\)}\nonumber\\
&=  \frac1{2\pi}e^{-\frac1{2}|\eta_{\kv,t}|^2}\ ,
\end{align}
which is only a function of the absolute value of the $\eta_{\kv,t}$.

%
%\section{Numerical integration of stochastic partial differential equation}
%
%Having constructed the stochastic forcing we next turn to the way we integrate~\eqref{eq:sde}. Assuming that at the $n$-th time step the field is $\phi(\xv,nh)$, we can formally integrate~\eqref{eq:sde} to obtain:
%\be
%\phi(\xv,(n+1)h) = \int_{nh}^{(n+1)h} f\(\bit\phi(\xv,t)\) \,\df t + \sqrt{\varepsilon}\int_{nh}^{(n+1)h} \xi(\xv,t) \,\df t \ .
%\ee
%We approximate the first integral by a fourth order Runge-Kutta (RK4) time step of the deterministic part of~\eqref{eq:sde},
%\be
%\int_{nh}^{(n+1)h} f\(\bit\phi(\xv,t)\) \,\df t \approx \phi(\xv,nh) + \Df f\(\bit\phi(\xv,nh),h\) \ ,
%\ee
%where $\Df f\(\bit\phi(\xv,nh),h\)$ is the RK4 time step, while for the the stochastic integral we take
%\be
% \int_{nh}^{(n+1)h} \xi(\xv,t) \,\df t \approx h\,\xi(\xv,nh)\ .
%\ee
%

\section[Proof of the relation $\<\xi_a \z'_b+\z'_a \xi_b\>=\sqrt{\varepsilon}\,Q_{ab}$]{Proof of the relation $\<\xi(\xv_a,t)\z'(\xv_b,t)+\z'(\xv_a,t)\xi(\xv_b,t)\>=\sqrt{\varepsilon}\,Q(\xv_a-\xv_b)$\label{sec:proof_xizeta}}

In this section we calculate the term $\<\xi(\xv_a,t)\z'(\xv_b,t)+\z'(\xv_a,t)\xi(\xv_b,t)\>$ in~\eqref{eq:s3t_C_xizeta}, where $\z'$ satisfies the NL eddy equation~\eqref{eq:enl_pert}.

We should note that because the stochastic forcing is additive (it does not depend on the state of the system) there is no need to distinguish between the mathematically convenient It\^o interpretation of the stochastic differential equations, in which the state at $t$ is uncorrelated with the stochastic forcing at the same time $t$, i.e.,  $\<\psi(\xv,t) \xi(\xv,t)\> =0$, and the physically relevant Stratonovich interpretation in which state and forcing are correlated at the same time. Both interpretations lead to the same results (\citet[][p.~35]{Oksendal-2000}). Throughout this thesis we have adopted the Stratonovich interpretation.

To calculate $\<\xi(\xv_a,t)\z'(\xv_b,t)+\z'(\xv_a,t)\xi(\xv_b,t)\>$ we write the solution of~\eqref{eq:nl} in integral form as
\begin{align}
\z(\xv,t) &= \z(\xv,0) - \int_{0}^{t} \[\bit  J\(\psi(\xv,s),\zeta(\xv,s)+\bv\cdot\xv\bit\) -r\zeta(\xv,s)\]\,\df s+ \sqrt{\varepsilon} \int_{0}^{t} \xi(\xv,s)\,\df s \ .\label{eq:nlsol_integral}
\end{align}
from which it follows that the eddy vorticity is given as:
\begin{align}
\z'(\xv,t) &= \z'(\xv,0) - \int_{0}^{t} \left\{\vphantom{\frac{1}{2}} J\(\psi(\xv,s),\zeta(\xv,s)+\bv\cdot\xv\bit\)  \right.\nonumber\\
&\hspace{3em}-\left. \vphantom{\frac{1}{2}} \tav{J\(\psi(\xv,s),\zeta(\xv,s)+\bv\cdot\xv\bit\)} -r\z'(\xv,s)\right\}\,\df s+ \sqrt{\varepsilon} \int_{0}^{t} \xi(\xv,s)\,\df s \ .\label{eq:nlpertsol_integral}
\end{align}
We want to calculate $\<\xi(\xv_a,t)\z'(\xv_b,t)\>$. From~\eqref{eq:nlpertsol_integral} we have:
%\begin{align}
%&\<\xi(\xv_a,t)\z'(\xv_b,t)\>= \<\xi(\xv_a,t)\z'(\xv_b,0)\> -\int_{0}^{t} \<\left\{\vphantom{\frac1{2}} \xi(\xv_a,t) J\(\psi(\xv_b,s),\zeta(\xv_b,s)+\bv\cdot\xv\bit\)  \right.\right.\nonumber\\
%&\hspace{2.5em}-\left.\left. \vphantom{\frac1{2}}  \xi(\xv_a,t) \tav{J\(\psi(\xv_b,s),\zeta(\xv_b,s)+\bv\cdot\xv\bit\)} -r\xi(\xv_a,t)\z'(\xv_b,s)\right\}\>\,\df s\nonumber\\
%&\hspace{2.5em}+ \sqrt{\varepsilon} \int_{0}^{t} \<\xi(\xv_a,t)\xi(\xv_b,s)\>\,\df s \ .
%\end{align}
\begin{align}
&\<\xi(\xv_a,t)\z'(\xv_b,t)\>= \nonumber\\
&\quad= \<\xi(\xv_a,t)\z'(\xv_b,0)\> - \int_{0}^{t} \<\left\{\bit\xi(\xv_a,t) J\(\psi(\xv_b,s),\zeta(\xv_b,s)+\bv\cdot\xv\bit\)  \right.\right.\nonumber\\
&\hspace{2.5em}-\left.\left. \bit \xi(\xv_a,t)\, \tav{J\(\psi(\xv_b,s),\zeta(\xv_b,s)+\bv\cdot\xv\bit\)} -r\xi(\xv_a,t)\z'(\xv_b,s)\right\}\>\,\df s\nonumber\\
&\hspace{2.5em}+ \sqrt{\varepsilon} \int_{0}^{t} \<\xi(\xv_a,t)\xi(\xv_b,s)\>\,\df s \ .
\end{align}
The initial state of the system is clearly uncorrelated with $\xi$ at time $t$, $ \<\xi(\xv_a,t)\z'(\xv_b,0)\>=0$. The first integral on the r.h.s. does not contribute to the correlation since i) there is no correlation between stochastic forcing at time $t$, $\xi(\xv,t)$, and state of the system at any time $t'< t$ and ii) also because the integral is unchanged if calculated over the interval $s\in[0,t)$ instead over $s\in[0,t]$ since at time $t$, $\<\xi(\xv_a,t)\z'(\xv_b,t)\><\infty$. The excluded point $s=t$ is only a point of measure zero. Therefore, the only non-zero contribution to $\< \xi(\xv_a,t)\z'(\xv_b,t)\>$ is the last integral,
\begin{align}
\<\xi(\xv_a,t)\z'(\xv_b,t)\> &= \sqrt{\varepsilon}\,\int_{0}^{t} \<\xi(\xv_a,t)\xi(\xv_b,s)\>\,\df s\nonumber\\
&= \sqrt{\varepsilon}\,\int_{0}^{t} Q(\xv_a-\xv_b)\,\d(t-s)\,\df s\nonumber\\
&=\frac{\sqrt{\varepsilon}}{2} Q(\xv_a-\xv_b)\ ,\label{eq:xia_zb}
\end{align}
where the integration of the delta function gives $1/2$. Similarly, $\<\z'(\xv_a,t)\xi(\xv_b,t)\>= (\sqrt{\varepsilon}/2)Q(\xv_a-\xv_b)$, leading to 
\be
\<\xi(\xv_a,t)\z'(\xv_b,t)+\z'(\xv_a,t)\xi(\xv_b,t)\> = \sqrt{\varepsilon}\,Q(\xv_a-\xv_b)\ .\label{eq:xiazb_xibza}
\ee
%(If $\z'$ obeys the QL eddy equation~\eqref{eq:eql_pert} instead of~\eqref{eq:enl_pert} the relation~\eqref{eq:xiazb_xibza} still holds, since in this case only the first integral on the r.h.s. of~\eqref{eq:nlpertsol_integral} is different and that does not affect the result.)

\section{Energy injection rate in the NL and QL systems by the stochastic excitation}

The energy of the flow is defined as $\tilde{E} \equiv \int \df^2 \xv\;\frac1{2}|\uv|^2$ and the time rate of change of $\tilde{E}$, assuming periodic boundary conditions, 
is given after an integration by parts as
\begin{align}
 \frac{\df \tilde{E}}{\df t}  &=%\frac{\df}{\df t} \int \df^2 \xv\;\frac1{2}|{\bm\nabla}\psi|^2 %\nonumber\\
\frac{\df}{\df t}\int \df^2 \xv\;\frac1{2}(-\psi\,\Del\psi)  %\nonumber\\ 
% &= -\int \df^2 \xv\;\frac1{2}\[(\partial_t\psi)\Del\psi+\psi(\partial_t\Del\psi)\bit\]\nonumber\\ 
 =-\int \df^2 \xv\;\psi\,\partial_t\z \ .
\end{align}

%\begin{align}
% \frac{\df E}{\df t}  &=\frac{\df}{\df t} \frac{\int \df^2 \xv\;\frac1{2}|{\bm\nabla}\psi|^2}{\int \df^2 \xv} \nonumber\\
%&=\frac{\df}{\df t}\frac{\int \df^2 \xv\;\frac1{2}(-\psi\,\Del\psi)}{\int \df^2 \xv}  \nonumber\\ 
%% &= -\int \df^2 \xv\;\frac1{2}\[(\partial_t\psi)\Del\psi+\psi(\partial_t\Del\psi)\bit\]\nonumber\\ 
% &=-\frac{\int \df^2 \xv\;\psi\,\partial_t\z}{\int \df^2 \xv} \ .
%\end{align}

We calculate the contribution of each term of $\partial_t\z$ from~\eqref{eq:nl}.
After integration by parts, the Jacobian term gives
\be
 \int \df^2 \xv~\psi\, J(\psi,\z+\bv\cdot\xv) = 0\ ,\label{eq:energ_J}
\ee
as expected because the Jacobian terms merely redistributes the energy among the flow scales. The rate of energy due to the dissipation is 
\be
r\int \df^2 \xv~\psi \z  = -r \int \df^2 \xv~|{\bm\nabla}\psi|^2 =-2r  \tilde{E} < 0\ .\label{eq:energ_r}
\ee
The ensemble average energy injection rate from the stochastic forcing is
\be
-\sqrt{\varepsilon}\int \df^2 \xv\;\<\psi \,\xi\>\ .\label{eq:energ_input}
\ee
Since $\xi$ is a stochastic variable we seek to determine the ensemble average energy injection rate  over all forcing realizations. To obtain this estimate we proceed as in~\eqref{eq:nlsol_integral}. As argued above, the first integral on the r.h.s. of~\eqref{eq:nlsol_integral} does not contribute at all to the correlation $\<\psi(\xv,t)\xi(\xv,t)\>$ and the only contribution comes from the last term in~\eqref{eq:nlsol_integral} so that,
\begin{align}
\<\vphantom{\z'} \xi(\xv,t)\psi(\xv,t)\>&=\left.\<\vphantom{\z'} \xi(\xv_a,t)\psi(\xv_b,t)\>\vphantom{\frac1{2}}\right|_{\xv_a=\xv_b} \nonumber\\
&=\frac{\sqrt{\varepsilon}}{2} \left. \Del^{-1}_a Q(\xv_a-\xv_b) \vphantom{\frac1{2}}\right|_{\xv_a=\xv_b}  \nonumber\\
&=\frac{\sqrt{\varepsilon}}{2} \left. \int \frac{\df^2 \kv}{(2\pi)^2}\;  \frac{\hat{Q}(\kv)}{-k^2}e^{\i\kv\cdot(\xv_a-\xv_b)} \right|_{\xv_a=\xv_b} \nonumber\\
&=-\sqrt{\varepsilon} \int \frac{\df^2 \kv}{(2\pi)^2}\;  \frac{\hat{Q}(\kv)}{2k^2}\ . 
\end{align}
From~\eqref{eq:energ_input} we have that the energy injection rate is
\begin{align}
-\sqrt{\varepsilon}\<\int \df^2 \xv\;\psi(\xv,t) \,\xi(\xv,t) \>&= \varepsilon\(\int \frac{\df^2 \kv}{(2\pi)^2}\;  \frac{\hat{Q}(\kv)}{2k^2}\)\(\int \df^2 \xv\)\ .\label{eq:energ_xi}\end{align}
Therefore, if the covariance spectrum is normalized according to
\be
\int \frac{\df^2 \kv}{(2\pi)^2}\;  \frac{\hat{Q}(\kv)}{2k^2}=1\ ,\label{eq:Qhat_norm}
\ee
then the energy injection rate per unit area is $\varepsilon$.

From~\eqref{eq:energ_J}, \eqref{eq:energ_r}, \eqref{eq:energ_xi} and \eqref{eq:Qhat_norm} we conclude that the domain and ensemble averaged total energy of the flow, $\<E\> \equiv (\int\df^2\xv)^{-1}  \tilde{E}$, satisfies:
\begin{align}
\frac{\df \<E\>}{\df t}  = \varepsilon - 2r \<E\>\ ,
\end{align}
and this implies the total energy  the flow (the sum of the eddy and mean flow energy) will always approach with time the constant value:
\be 
\<E\>_{\infty} = \frac{\varepsilon }{2r}\ .\label{eq:Einfinity}
\ee

\chapter{General properties of S3T equilibrium solutions}
\label{app:s3t-equil-prop}

Assume that $(Z^e,C^e)$ is an equilibrium solution of the S3T statistical state dynamics governed by~\eqref{eq:s3t} satisfying:
\begin{subequations}\begin{align}
J \(\Psi^e, Z^e+  {\bm\beta}\cdot\mathbf{x} \) - \Rcal( C^e )+\,Z^e&=0\ ,\\
\[\bit\Acal_a(\Uv^e) + \Acal_b(\Uv^e)\]C_{ab}^e +\varepsilon\,Q_{ab} &=0\ ,\label{eq:ex2}
\end{align}\label{eq:s3tequilibr}\end{subequations}
with $\Uv^e = \zhat\times\nablav\Del^{-1}Z^e$.

We show first that when $\Acal^e\equiv\Acal(\Uv^e)$ is stable then there exists a unique Hermitian and positive definite solution  $C^e$ of~\eqref{eq:s3tequilibr}.

When the operator $(\Acal_a^e + \Acal_b^e )$ is invertible then~\eqref{eq:ex2} has a unique solution. The spectrum of $( \Acal_a^e + \Acal_b^e )$ is $\mu_i+\mu_j^*$, $i,j=1,2,\dots$ where $\mu_i$ are the eigenvalues of $\Acal^e$. If $\Acal^e$ is stable then $\real(\mu_i)<0$, which implies that $\real(\mu_i+\mu_j^*)<0$ for every $i,j$ and hence $( \Acal_a^e + \Acal_b^e )$ is invertible. So a unique $C^e$ that solves~\eqref{eq:s3tequilibr} exists. We now show that it is a physically realizable solution, i.e., it is Hermitian and positive definite, by giving the explicit expression of the solution. We verify that the solution of~\eqref{eq:s3tequilibr} is  
\begin{align}
C^e(\xv_a,\xv_b) &=\varepsilon \lim_{t\to+\infty} \int_0^t e^{\Acal_a^e\,s}e^{\Acal_b^e\,s} Q(\xv_a-\xv_b)\,\df s\ ,\label{eq:Cinfty}
\end{align}
since
\begin{align}
\(\bit\Acal_a^e + \Acal_b^e\)C_{ab}^e&=\varepsilon \lim_{t\to+\infty} \int_0^t \(\bit\Acal_a^e + \Acal_b^e\) e^{\Acal_a^e\,s}e^{\Acal_b^e\,s} Q_{ab}\,\df s\nonumber\\
&=\varepsilon \lim_{t\to+\infty} \int_0^t  \frac{\df}{\df s} \( e^{\Acal_a^e\,s}e^{\Acal_b^e\,s} Q_{ab}\)\,\df s \nonumber\\
&=\varepsilon \lim_{t\to+\infty} \[ e^{\Acal_a^e\,s}e^{\Acal_b^e\,s} Q_{ab}\]_{s=0}^t\nonumber\\
&=-\varepsilon\, Q_{ab}  + \varepsilon \lim_{t\to+\infty}  e^{\Acal_a^e\,t}e^{\Acal_b^e\,t} Q_{ab}\nonumber\\
&=-\varepsilon\,Q_{ab}\ ,\label{eq:proofCe}
\end{align}
as $\lim_{t\to+\infty}  e^{\Acal_a^e\,t}e^{\Acal_b^e\,t} Q_{ab}$ vanishes when $\Acal^e$ is stable.

From~\eqref{eq:Cinfty} it is clear because $Q(\xv)=Q(-\xv)$ that $C$ is Hermitian, i.e., it satisfies the exchange symmetry $C_{ab}=C_{ba}$. It is only left to prove that $C^e$ is positive definite. A real function $F(\xv_a,\xv_b)$ is positive definite if and only if for every complex function $f$,
\be
\iint \df\xv_a\,\df\xv_b\; F(\xv_a,\xv_b)\,f(\xv_a) f(\xv_b)^*\;>0\ .\label{eq:Fxaxb_posdef}
\ee
%We may define the inner product between any two functions $g(\xv)$ and $h(\xv)$ as $\(\bit g,h\)\equiv\int\df^2\xv\;g(\xv)^*h(\xv)$ as well as the inner product between any two function with two arguments $G(\xv_a,\xv_b)$ and $H(\xv_a,\xv_b)$ as $\(\bit G(\xv_a,\xv_b),H(\xv_a,\xv_b) \)_2 = \iint \df\xv_a\,\df\xv_b\; G(\xv_a,\xv_b)^* H(\xv_a,\xv_b)$. With this definition~\eqref{eq:Fxaxb_posdef} may be rewritten as $\(f(\xv_a),\bit\)_2>0$.
Consider a positive definite homogeneous covariance, like the forcing covariance $Q$. Then for all functions (or even distributions) $f$ we must have: 
\begin{align}
&\iint \df\xv_a\,\df\xv_b\; Q(\xv_a-\xv_b)\,f(\xv_a) f(\xv_b)^* =\nonumber\\
&\qquad=\int\frac{\df^2\kv}{(2\pi)^2}\iint \df\xv_a\,\df\xv_b\; \hat{Q}(\kv)\,e^{\i\kv\cdot(\xv_a-\xv_b)}\,f(\xv_a) f(\xv_b)^*\nonumber\\
&\qquad=\int\frac{\df^2\kv}{(2\pi)^2}\;\hat{Q}(\kv)\(\int \df\xv_a\; e^{\i\kv\cdot\xv_a}\,f(\xv_a)\) \(\int \df\xv_b\; e^{\i\kv\cdot\xv_b}\,f(\xv_b)\)^*\nonumber\\
&\qquad=\int\frac{\df^2\kv}{(2\pi)^2}\;\hat{Q}(\kv)\,\left|\hat{f}(\kv)\right|^2>0\ ,
\end{align}
This implies that $\hat{Q}(\kv)\ge 0$ given that this integral must be positive for all functions $f$. (This is Bochner's theorem.) We now show that $C^e$ is also positive definite.  Indeed from~\eqref{eq:Cinfty} we have for all functions $f$:
\begin{align}
&\iint \df\xv_a\,\df\xv_b\; C^e(\xv_a,\xv_b)\,f(\xv_a) f(\xv_b)^*=\nonumber\\
&\quad=\varepsilon \iint \df\xv_a\,\df\xv_b\;\int_0^{\infty}\df s\; \[e^{\Acal_a^e\,s}e^{\Acal_b^e\,s} \;  Q(\xv_a-\xv_b)\]f(\xv_a) f(\xv_b)^*\nonumber\end{align}\begin{align}%\\
&\quad=\varepsilon \int\!\frac{\df^2\kv}{(2\pi)^2}\!\iint \df\xv_a\,\df\xv_b\,\int_0^{\infty}\!\df s\[ e^{\Acal_a^e\,s}e^{\Acal_b^e\,s} \, \hat{Q}(\kv) e^{\i\kv\cdot(\xv_a-\xv_b)}\]f(\xv_a) f(\xv_b)^*\nonumber\\
%&\qquad=\varepsilon \int_0^{+\infty}\df s \int\frac{\df\kv}{(2\pi)^2}\,\hat{Q}(\kv)\iint \df\xv_a\,\df\xv_b\;f(\xv_a) f(\xv_b)^*\, e^{\Acal_a^e\,s}e^{\Acal_b^e\,s} \; e^{\i\kv\cdot(\xv_a-\xv_b)}=\nonumber\\
&\quad=\varepsilon \int_0^{\infty}\df s \int\frac{\df^2\kv}{(2\pi)^2}\,\hat{Q}(\kv) \left|\int \df\xv \;\( e^{\Acal^e s}\,e^{\i\kv\cdot\xv}\)f(\xv)\right|^2>0\ .
\end{align}

Note that,
\be
C(\xv_a,\xv_b,t) =e^{\Acal^e_a\,t}e^{\Acal^e_b\,t} C(\xv_a,\xv_b,0) + \varepsilon \int_0^t e^{\Acal^e_a\,s}e^{\Acal^e_b\,s} Q(\xv_a-\xv_b)\,\df s\ ,\label{eq:Ct}
\ee
solves the time dependent S3T covariance equation~\eqref{eq:s3t_C},
\begin{align}
\partial_t C_{ab} & = \(\bit\Acal^e_a + \Acal^e_b\)C_{ab} +\varepsilon\,Q_{ab}~.\label{eq:st3C_again}
\end{align}
To show that make the change of variable  $s\to t-s$ in~\eqref{eq:Ct} and write the covariance in the form
\begin{align}
C(\xv_a,\xv_b,t) &=e^{\Acal^e_a\,t}e^{\Acal^e_b\,t} C(\xv_a,\xv_b,0) + \varepsilon \int_0^t e^{\Acal^e_a\,(t-s)}e^{\Acal^e_b\,(t-s)} Q(\xv_a-\xv_b)\,\df s\ .\label{eq:B10}
\end{align}
which satisfies~\eqref{eq:st3C_again}. From~\eqref{eq:Ct} we see that if $C(\xv_a,\xv_b,0)$ is positive definite, then $C(\xv_a,\xv_b,t)$ is positive definite at all times $t$, irrespectively of whether $\Acal^e$ is stable or unstable and when $\Acal^e$ is hydrodynamically unstable $C(\xv_a,\xv_b,t)$ diverges as $t \to \infty$. This implies that no S3T equilibrium exists when $\Uv^e$ is hydrodynamically unstable.

It should be noted that the time independent Lyapunov equation~\eqref{eq:ex2} has always formally a solution unless $\Acal^e$ has a neutral eigenvalue, i.e. unless there is an eigenvalue with $\real(\mu_i)=0$. However, the formal solution of~\eqref{eq:ex2} for $\Acal^e$ unstable is not positive definite and consequently is not physically realizable given that all physically realizable states produce positive definite covariances (or  from the previous argument this non-positive steady state covariance solution is unreachable from any physical initial covariance which generates through~\eqref{eq:B10} only positive definite covariances). In order to see that consider 
%
%
%%define inner product, say that they are normalized, no iint
%%
%%
%%
%%\noindent\emph{Corollaries}:
%%
%%\begin{enumerate}
%%
%%\item[a.]
%%If $\Uv^e$ is hydrodynamically unstable, i.e. $\Acal^e$ has at least one eigenvalue with $\real(\mu)>0$, a physically realizable S3T equilibrium does not exist because $C^e$, even if it solves~\eqref{eq:ex2}, is not positive definite and therefore not a physically realizable covariance.   Hence, if $(Z^e,C^e)$ is an S3T equilibrium (irrespectively of whether it is an S3T stable or unstable equilibrium) then 
%%$\Uv^e$ is by necessity hydrodynamically stable. So the existence of S3T equilibria requires the hydrodynamic stability of $\Uv^e$. However, the hydrodynamic stability of $\Uv^e$ does not imply the S3T stability
%%of the equilibrium $(\Uv^e,C^e)$.
%
%
%
%\noindent\emph{Proof}:\\
the eigenrelation of $\Acal^e$ and its adjoint ${\Acal^e}^\dag$,\footnote{The adjoint ${\Acal^e}^\dag$ of operator $\Acal^e$ is defined as the operator that satisfies $\int\df^2\xv\; g(\xv)^*\[\bit\Acal^e\, h(\xv)\] = \int\df^2\xv\; \[\bit{\Acal^e}^\dag g(\xv)\]^* h(\xv)$, for every functions $g$, $h$.} 
\begin{align}
\Acal^e u_n  = \mu_n u_n \ ,\ \ {\Acal^e}^\dag v_n= \mu_n^* v_n \ ,\ \ n=1,2,\dots\ .
\end{align} 
The eigenvalues of ${\Acal^e}^\dag$ are the complex conjugate of the eigenvalues of $\Acal^e$ and in general the eigenfuctions $u_n$ of $\Acal^e$ and eigenfuctions $v_n$ of ${\Acal^e}^\dag$ are not the same.\footnote{Eigenfunctions $u_n$ and $v_n$ coincide only when the two operators commute, i.e., $\Acal^e{\Acal^e}^\dag={\Acal^e}^\dag\Acal^e$, in which case the operator $\Acal^e$ is called normal.} Moreover, they satisfy the following relations:\begin{subequations}
\begin{align}
\(\bit u_m,v_n\) &= \d_{mn}\,\(\bit u_m,v_m\)\quad\textrm{(orthogonality) ,}\label{eq:orthogonality}\\
\sum_n \frac{u_n(\xv_a) v_n^*(\xv_b)}{\(\bit v_n,u_n\)}&=\delta(\xv_a-\xv_b)\quad\textrm{(completeness) ,}\label{eq:completeness}
\end{align}\end{subequations}
where $\(\bullet,\bullet\)$ denotes the inner product between two functions $g$ and $h$, taken here as $\(\bit g,h\)\equiv\int\df^2\xv\;g(\xv)^*h(\xv)$. 

Assume that $\Acal^e$ has an eigenfunction, say $u_1$, with eigenvalue with $\real(\mu_1)>0$ with associated adjoint eigenfunction $v_1$. We will show that the solution $C^e$ of~\eqref{eq:ex2} is not positive definite. Consider the integral
\begin{align}
&\int\df^2\xv_a\int\df^2\xv_b\; \[\bit\Acal^e_a C^e(\xv_a,\xv_b)\] v_1^*(\xv_a) v_1(\xv_b) =\nonumber\\
&\qquad= \int\df^2\xv_a\int\df^2\xv_b\int\df^2\xv'\;   \Acal^e_a \sum_n \frac{u_n(\xv_a) v_n^*(\xv')}{\(\bit v_n,u_n\)} \, C^e(\xv',\xv_b)  \,v_1^*(\xv_a) v_1(\xv_b)\nonumber\\
&\qquad= \int\df^2\xv_b\int\df^2\xv'\;  \sum_n \mu_n  \frac{\(\bit v_1,u_n\)}{\(\bit v_n,u_n\)}\, C^e(\xv',\xv_b)  \,v_n^*(\xv') v_1(\xv_b)\nonumber\\
%&\qquad= \sum_n\mu_n\int\df^2\xv_b\df^2\xv'\; v_n^*(\xv')\delta_{1,n} C^e(\xv',\xv_b)  \,v_1(\xv_b)\nonumber\\
&\qquad= \mu_1\int\df^2\xv_a\int\df^2\xv_b\; C^e(\xv_a,\xv_b)  \, v_1^*(\xv_a)v_1(\xv_b) \ ,\label{eq:AaCe}
\end{align} 
and similarly,
\begin{align}
&\int\df^2\xv_a\int\df^2\xv_b\; \[\bit\Acal^e_b C^e(\xv_a,\xv_b)\] v_1^*(\xv_a) v_1(\xv_b) =\nonumber\\
%&\qquad= \iint\df^2\xv_a\df^2\xv_b\int \df^2\xv'\;   \Acal^e_b \sum_n \frac{v_n(\xv') u_n^*(\xv_b)}{\int v_n(\xv) u_n^*(\xv)\,\df^2\xv} C^e(\xv_a,\xv')  \,v_1^*(\xv_a) v_1(\xv_b)\nonumber\\
%&\qquad= \sum_n\iint\df^2\xv_a\df^2\xv_b\int \df^2\xv'\;   \mu_n^*  \frac{v_n(\xv') u_n^*(\xv_b)}{\int v_n(\xv) u_n^*(\xv)\,\df^2\xv} C^e(\xv_a,\xv')  \,v_1^*(\xv_a) v_1(\xv_b)\nonumber\\
%&\qquad= \sum_n\mu_n^*\iint\df^2\xv_a\df^2\xv'\; v_n(\xv')\delta_{1,n} C^e(\xv_a,\xv')  \,v_1^*(\xv_a)\nonumber\\
&\qquad\quad=  \mu_1^*\int\df^2\xv_a\int\df^2\xv_b\; C^e(\xv_a,\xv_b)  \, v_1^*(\xv_a)v_1(\xv_b) \ .\label{eq:AbCe}
\end{align} 
%Adding~\eqref{eq:AaCe} and~\eqref{eq:AbCe} we obtain that
%\begin{align}
%&\int\df^2\xv_a\int\df^2\xv_b\; \[\bit(\Acal^e_a+\Acal^e_b) C^e(\xv_a,\xv_b)\] v_1^*(\xv_a) v_1(\xv_b) =\nonumber\\
%&\qquad= 2\real(\mu_1)\,\int\df^2\xv_a\int\df^2\xv_b\; C^e(\xv_a,\xv_b)  \, v_1^*(\xv_a) v_1(\xv_b)  \ .\label{eq:B15}
%\end{align} 
Using~\eqref{eq:AaCe},~\eqref{eq:AbCe} and~\eqref{eq:ex2} we have that
\begin{align}
&\int\df^2\xv_a\int\df^2\xv_b\; \[\bit(\Acal^e_a+\Acal^e_b) C^e(\xv_a,\xv_b)\] v_1^*(\xv_a) v_1(\xv_b) =\nonumber\\
&\qquad\quad= 2\real(\mu_1)\,\int\df^2\xv_a\int\df^2\xv_b\; C^e(\xv_a,\xv_b)  \, v_1^*(\xv_a) v_1(\xv_b)  \nonumber\\
&\qquad\quad= -\varepsilon\int\df^2\xv_a\int\df^2\xv_b\; Q(\xv_a-\xv_b)  \, v_1^*(\xv_a) v_1(\xv_b)  <0\ ,
\end{align} 
hence, because $\real(\mu_1)>0$, we conclude that 
\be
\int\df^2\xv_a\int\df^2\xv_b\; C^e(\xv_a,\xv_b)  \, v_1^*(\xv_a) v_1(\xv_b)  <0\ ,
\ee
which shows that $C^e$ is not positive definite and hence not realizable when $\Acal^e$ is unstable.

\vspace{1em}

Further, in the absence of forcing, i.e., $Q=0$, the equilibrium solution $(Z^e,C^e=0)$ is hydrodynamically stable if and only if it is S3T stable. To see that consider the S3T perturbation equations~\eqref{eq:s3t_dZdCgen} about $(Z^e,C^e=0)$ which for this equilibrium simplify to:
\begin{subequations}\begin{align}
\partial_t \,\d Z &=\Acal^e\,\d Z+ \Rcal( \d C )\ ,\label{eq:pertZeCe0_dZ}\\
\partial_t \,\d C_{ab} & = \(\bit\Acal^e_a + \Acal^e_b \)\d C_{ab} \ .\label{eq:pertZeCe0_dC}
\end{align}\label{eq:pertZeCe0}\end{subequations}
From the above equations we infer the equivalence between the stability of $\Acal^e$ and the S3T stability of $(Z^e,C^e)$, which is governed by the upper-diagonal operator
\be
\left(
\begin{array}{cc}
 \Acal^e &    \Rcal \\
 0  & \Acal^e_a  +\Acal^e_b   
\end{array}
\right)\ .
\ee

    % !TEX root = ../thesis.tex

\chapter{Numerical integration of the ~NL, ~QL and ~S3T systems}
\label{app:numerical_method}

In this appendix we present details of the numerical method used to integrate the NL system~\eqref{eq:nl}, the QL system~\eqref{eq:eql} and~\eqref{eq:qlz} and the S3T system~\eqref{eq:s3t} and~\eqref{eq:s3tz}. The flow domain is taken to be a rectangle of size $L_x\times L_y$, with periodic boundary conditions on its boundary and discretized with $N_x\times N_y$ points, $N_x$ in the zonal and $N_y$ in the meridional direction. The allowed wavenumbers in this domain, for $N_x$, $N_y$ even, are:
\begin{align}
k_j = \frac{2\pi}{L_j}\times\(0,\pm 1,\pm2,\dots,\pm (N_j/2-1),-N_j/2\bit\)\ \ ,\ j=x,y\ ,\label{eq:allowed_k}
\end{align}
and moreover, the continuous Fourier transforms become discrete Fourier transforms:
\be
\phi(x) = \int \frac{\df k_x}{2\pi}\hat{\phi}(k_x) e^{\i k_x x} \longrightarrow \sum_{k_x=-N_x/2}^{N_x/2-1} \hat{\phi}_{k_x} e^{\i k_x x}\ ,
\ee
with the continuous and the discrete Fourier amplitudes $\hat{\phi}(k_x)$ and $\hat{\phi}_{k_x}$ related by
\be
\hat{\phi}(k_x) = 2\pi \sum_{k_x'=-N_x/2}^{N_x/2-1} \hat{\phi}_{k_x'}\,\d(k_x-k_x')\ .
\ee
To simplify notation we denote:
\be
\sum_{\l_j}\equiv\sum_{\substack{\l_j=-N_j/2\\k_j\ne0}}^{N_j/2-1}\ \ ,\ \ \ \sum_{\substack{\l_j\\\l_j>0}}\equiv\sum_{\l_j=1}^{N_j/2-1}\ \ ,\ \ \ j=x,y \ ,\label{eq:def_sumpos_posneg}
\ee
for any variable $\l$.

\section{Numerical integration of NL and QL system}

Both NL and QL systems are a stochastic partial differential equations of the form:
\be
\partial_t \phi(\xv,t) = \Lcal\(\bit\phi(\xv,t)\) + \Ncal\(\bit\phi(\xv,t),\phi(\xv,t)\) +\sqrt{\varepsilon}\, \xi(\xv,t)\ ,\label{eq:sde}
\ee
with $\Lcal$ a linear operator and $\Ncal$ a bilinear operator. We use pseudospectral methods to evaluate  $\Lcal(\phi)$ and $\Ncal(\phi,\phi)$. Time is discretized in equal steps, $h$, and the deterministic part of~\eqref{eq:sde} is advanced in time by $h$ using a fourth-order Runge-Kutta (RK4) scheme. After the completion of each RK4 step, the stochastic excitation term, $h\sqrt{\e} \xi(x,nh)$, is added to the state of the system \parencite{Godunov-1959}.

%
%To calculate the terms $\Lcal\(\bit\phi(\xv,nh),h\)$ and $\Ncal\(\bit\phi(\xv,nh),h\)$ we use a pseudospectral scheme, i.e., we calculate the derivatives of the fields spectrally and the product of two fields in physical space. We consider the excitation to be constant during a time step and therefore we approximate the second integral in the r.h.s. of~\eqref{eq:time_step} as,
%\be
% \int\limits_{nh}^{(n+1)h} \xi(\xv,t) \,\df t \approx h\,\xi(\xv,nh)\ .
%\ee
%

\section{Numerical integration of the S3T system}

In this section we discuss the numerical integration of the S3T system. While the state of the NL and QL has dimension $N_x N_y$, the S3T system has the far larger dimension $N_x N_y + N_x^2N_y^2$. This makes the numerical integration of the S3T system extremely costly and special techniques need to be developed in order to obtain resolved simulations. 

\subsection{Numerical integration zonal S3T system\label{sec:numerical_S3Tz}}

The S3T system which is computationally more easier to integrate is the zonal mean S3T system~\eqref{eq:s3tz} because of the homogeneity in the $x$ direction. In that case the dimension of the mean flow is $N_y$ and  the  dimension of the covariance is, as we will show, at most $N_x N_y^2$.

\vspace{1em}

We first use the homogeneity in $x$ to split the covariance equation~\eqref{eq:s3tz_C} into a system of decoupled equations for the zonal Fourier components $\tilde{C}_{k_x}$ of $C$, defined
as 
\begin{align}
C(\xv_a,\xv_b,t) %&= \int \frac{\df k_x}{2\pi} \tilde{C}(k_x,y_a,y_b,t)\,e^{\i k_x(x_a-x_b)}\nonumber\\
&= \sum_{k_x} \tilde{C}_{k_x}(y_a,y_b,t)\,e^{\i k_x(x_a-x_b)}\ ,\label{eq:C_Ckab}
\end{align}
while the spatial forcing covariance is similarly expanded as
\begin{align}
Q(\xv_a-\xv_b) %&= \int \frac{\df k_x}{2\pi} \tilde{Q}(k_x,y_a-y_b)\,e^{\i k_x(x_a-x_b)}\nonumber\\
&=  \sum_{k_x} \tilde{Q}_{k_x}(y_a-y_b)\,e^{\i k_x(x_a-x_b)}\ .\label{eq:Q_Qab}
\end{align}
Note that the zonal Fourier component  $\tilde{Q}_{k_x}$ is related to the forcing spectrum $\hat{Q}(\kv)= \int\df^2(\xv_a-\xv_b)\; Q(\xv_a-\xv_b)\,e^{-\i \kv\cdot(\xv_a-\xv_b)}$, with $\kv=(k_x,k_y)$, by
\be
\tilde{Q}_{k_x}(y_a-y_b) = \sum_{k_y} \hat{Q}(\kv) \,e^{\i k_y(y_a-y_b)}\ .
\ee
Note also that if we take the $x$ Fourier transform of $\z'$
\begin{align}
\z'(\xv,t) &= \sum_{k_x} \tilde{\z}_{k_x}(y,t)\,e^{\i k_x x}\ ,\label{eq:z_zka}
\end{align}
the eddy vorticity covariance, being homogeneous in $x$,  can be written as
\begin{align}
C(\xv_a,\xv_b,t) &=\< \z'(\xv_a,t)\z'(\xv_b,t) \> \nonumber\\
&= \sum_{k_x,k_x'} \< \tilde{\z}_{k_x}(y_a,t)\tilde{\z}_{k'_x}(y_b,t)\>\,e^{\i k_x x_a}e^{\i k'_x x_b} \nonumber\\
&= \sum_{k_x}  \<\tilde{\z}_{k_x}(y_a,t)\tilde{\z}_{k_x}(y_b,t)^*\>\,e^{\i k_x (x_a-x_b)}\ ,
\end{align}
which implies that
\be
\tilde{C}_{k_x}(y_a,y_b,t) = \<\tilde{\z}_{k_x}(y_a,t)\tilde{\z}_{k_x}(y_b,t)^*\>\ .\label{eq:Ckab_zkazkb}
\ee

We can write:
\begin{align}
&\Acal_{\textrm{z},a}(U)\,C(\xv_a,\xv_b,t) =\nonumber\\
&\quad=\sum_{k_x} \[\bit -\i k_x U_a - \i k_x  \(\bit \b-\partial^2_{y_ay_a}U_a\)\Del_a^{-1} -1\]\tilde{C}_{k_x}(y_a,y_b,t)\,e^{\i k_x(x_a-x_b)}\ ,\nonumber\\
&\quad\equiv\sum_{k_x} \Acal_{\textrm{z},k_x,a}(U) \,\tilde{C}_{k_x}(y_a,y_b,t)\,e^{\i k_x(x_a-x_b)}\ ,
\label{eq:AaC_AkCkab}\end{align}
where
\be
\Acal_{\textrm{z},k_x}(U) = -\i k_x U - \i k_x  \(\bit \b-\partial^2_{yy}U\)\Del^{-1} -1\ ,
\ee
and similarly
\begin{align}
\Acal_{\textrm{z},b}(U)\,C(\xv_a,\xv_b,t) &= \sum_{k_x}\Acal_{\textrm{z},k_x,b}(U)^*\tilde{C}_{k_x}(y_a,y_b,t)\,e^{\i k_x(x_a-x_b)}\ .
\label{eq:AbC_AkCkab}\end{align}
Combining~\eqref{eq:C_Ckab},~\eqref{eq:Q_Qab},~\eqref{eq:AaC_AkCkab} and~\eqref{eq:AbC_AkCkab}  we write equation~\eqref{eq:s3tz_C} as:
\begin{align}
&\sum_{k_x} \left\{\vphantom{\frac1{2}} \partial_t\tilde{C}_{k_x}(y_a,y_b,t) -\[\bit\Acal_{\textrm{z},k_x,a}(U)+\Acal_{\textrm{z},k_x,b}(U)^*\]\tilde{C}_{k_x}(y_a,y_b,t)\right.\nonumber\\
&\hspace{16em}\left.\vphantom{\frac1{2}} -\varepsilon \,\tilde{Q}_{k_x}(y_a-y_b) \right\}e^{\i k_x(x_a-x_b)}=0\ .\label{eq:sumCk}
\end{align}
 which implies that  the covariance equation decouples to the $N_x$ equations
\begin{align}
&\partial_t\tilde{C}_{k_x}(y_a,y_b,t) =\nonumber\\
&\qquad=\[\bit\Acal_{\textrm{z},k_x,a}(U)+\Acal_{\textrm{z},k_x,b}(U)^*\]\tilde{C}_{k_x}(y_a,y_b,t)+\varepsilon \,\tilde{Q}_{k_x}(y_a-y_b)\ .\label{eq:C_kx_cont}
\end{align}
Covariance $C$ is real valued, implying that $\tilde{C}_{-k_x}(y_a,y_b,t)=\tilde{C}_{k_x}^*(y_a,y_b,t)$, and therefore we only need to solve for $k_x\ge0$. Note also that the exchange  symmetry $C(\xv_a,\xv_b,t)=C(\xv_b,\xv_a,t)$ implies $\tilde{C}_{-k_x}(y_a,y_b,t)=\tilde{C}_{k_x}(y_a,y_b,t)$ and therefore 
\be
 \tilde{C}_{k_x}(y_a,y_b,t)=\tilde{C}_{k_x}(y_b,y_a,t)^*\ .\label{eq:symm_Ck}
 \ee
 Moreover, from~\eqref{eq:C_kx_cont} we can see that we can integrate only the components
$k_x$ for which $\tilde{Q}_{k_x}$ has power (as from uniqueness if $\tilde{Q}_{k_x} =0$ and initially $\tilde{C}_{k_x} =0$ then $\tilde{C}_{k_x}=0$ for all times). By removing the wavenumber components  for which $\tilde{Q}_{k_x} =0$ we may remove  also  self-sustained turbulent states, i.e., the possibility that a non-zero covariance component is maintained in the absence of forcing (which can happen when $U$ produces a neutrally stable $\Acal_{\textrm{z},k_x}(U)$, when $U$ is time independent, or a $\Acal_{\textrm{z},k_x}(U(t))$ with zero top Lyapunov exponent when $U$ is time-dependent). These self-sustained states are singular in barotropic turbulence because barotropic turbulence does not self-sustain and including them may result  in  singular states that disappear with the introduction of even the slightest forcing (see for example the states in~\cite{Marston-etal-2008}). However, there are particularly interesting self-sustained states in 3D turbulence (cf.~\cite{Farrell-Ioannou-2012,Constantinou-etal-Madrid-2014}). We do not include in the S3T calculations wavenumber components that are not externally forced. In this way we consider in~\eqref{eq:C_kx_cont} only the $k_x=1,2,\dots,N_k$ for which $\tilde{Q}_{k_x}$ is non-zero.

\vspace{1em}

We now formulate~\eqref{eq:C_kx_cont} on the discretized domain.

The Fourier coefficients $\tilde\z_{k_x}(y,t)$ on the discretized domain become a vector $\zv_{k_x}(t)$ with elements $[\zv_{k_x}(t)]_j=\tilde{\z}_{k_x}(y_j,t)$, while the Fourier coefficients  $\tilde{C}_{k_x}(y_a,y_b,t)$ and $\tilde{Q}_{k_x}(y_a-y_b)$ become matrices $\C_{k_x}(t)$ and $\Q_{k_x}$ with elements $[\C_{k_x}(t)]_{ij} = \tilde{C}_{k_x}(y_i,y_j,t)$ and $[\Q_{k_x}]_{ij} = \tilde{Q}_{k_x}(y_i-y_j)$ respectively. Also ,the mean flow $U(y,t)$ becomes a vector ${\bm U}(t)$ with elements $[{\bm U}(t)]_j = U(y_j,t)$. The symmetry property~\eqref{eq:symm_Ck} requires that the $\C_k$ and $\Q_k$ matrices are hermitian, i.e., $\C_k^\dag=\C_k$. Relation~\eqref{eq:Ckab_zkazkb} implies that $\[\C_{k_x}\]_{ij} = \<[\zv_{k_x}]_i\,[\zv_{k_x}]_j^*\>$, or simply $\C_{k_x} = \<\zv_{k_x}\,\zv_{k_x}^\dag\>$, demonstrating that the $\C_{k_x}$ matrices are also positive definite.

We express the S3T covariance equations~\eqref{eq:C_kx_cont} in terms of matrices $\C_{k_x}$ and $\Q_{k_x}$. Term $\[\bit\Acal_{\textrm{z},k_x,a}(U)+\Acal_{\textrm{z},k_x,b}(U)^*\]\tilde{C}_{k_x}(y_a,y_b,t)$ becomes in matrix notation
\begin{align}
\[\Acal_{\textrm{z},k_x,a}(U)+\Acal_{\textrm{z},k_x,b}(U)^*\] \tilde{C}_{k_x}(y_a,y_b,t) &= \[ \A_{k_x}({\bm U})\,\C_{k_x}(t)+\C_{k_x}(t)\, \A_{k_x}({\bm U})^\dag \]_{ab}\ ,
\end{align}
where $\A_{k_x}({\bm U})$ is the time-depended matrix
\be
\A_{k_x}({\bm U})= -\i k_x \U  - \i k_x \[ \bit\b\I- (\U_{yy})\] \DDel_{k_x}^{-1} -\I\ ,
\ee
with $\I$ the identity, $\DDel_{k_x}$ the matrix approximation of $\partial^2_{yy}-k_x^2$, $\DDel_{k_x}^{-1}$ its inverse and $\U=\diag(U)$, $\U_{yy}=\diag( \partial^2_{yy} U)$, where $\diag(\bullet)$ denotes the diagonal matrix with diagonal elements the values of its argument at the $y$ collocation points. We thus end up with $N_k$ equations for $(N_y\times N_y)$ sized matrices:
\be
\frac{\df }{\df t}\C_{k_x}(t) = \A_{k_x}({\bm U})\,\C_{k_x}(t)+\C_{k_x}(t)\, \A_{k_x}({\bm U})^\dag  + \varepsilon \,\Q_{k_x}\ ,\ \ \text{for }k_x=1,\dots,N_k\ .
\ee

Turning to the mean flow equation~\eqref{eq:s3tz_mean} we express the Reynolds stress divergence term, $\[ \frac1{2}(\Del^{-1}_a\partial_{x_a}\!+\!\Del^{-1}_b\partial_{x_b}) C_{ab}\]_{\xv_a=\xv_b}$ in terms of $\C_{k_x}$. The Reynolds stress evaluated at $y=y_j$ is,
\begin{align}
&\left.\[ \frac1{2}\(\Del^{-1}_a\partial_{x_a}\!+\!\Del^{-1}_b\partial_{x_b}\) C_{ab}\]_{\xv_a=\xv_b}\right|_{y=y_j}=  \nonumber\\
%&\qquad=\int \frac{\df k_x}{2\pi} \[\frac1{2} \( \i k_x\, \Del_{k_x,a}^{-1} -\i k_x\, \Del_{k_x,b}^{-1}\) \tilde{C}(k_x,y_a,y_b,t) \,e^{\i k_x(x_a-x_b)} \]_{\xv_a=\xv_b}\nonumber\\
&\qquad=  \sum_{k_x}\left.\[\frac1{2} \( \i k_x\, \Del_{k_x,a}^{-1} -\i k_x\, \Del_{k_x,b}^{-1}\) \tilde{C}_{k_x}(y_a,y_b,t) \,e^{\i k_x(x_a-x_b)} \]_{\xv_a=\xv_b}\right|_{y=y_j}\nonumber\\
&\qquad=  \sum_{k_x}\left.\left\{\frac1{2} \[ \i k_x\, \DDel_{k_x}^{-1} \C_{k_x}(t) +  \C_{k_x}(t) \(\i k_x\, \DDel_{k_x}^{-1}\)^\dag \]_{ab} \,e^{\i k_x(x_a-x_b)} \right\}_{\xv_a=\xv_b}\right|_{y=y_j}\nonumber\\
&\qquad= \sum_{k_x}\, \real\left\{\[ \(\i k_x\, \DDel_{k_x}^{-1}\) \C_{k_x}(t) +  \C_{k_x}(t) \(\i k_x\, \DDel_{k_x}^{-1}\)^\dag \]_{jj}\right\} \nonumber\\
&\qquad= \sum_{\substack{k_x\\k_x>0}} 2 \real\left\{\[ \i k_x\, \DDel_{k_x}^{-1}\, \C_{k_x}(t) \]_{jj}\right\} \ .\label{eq:RC_discrete_z}
\end{align}

The discrete S3T system in the zonal mean--eddy decomposition takes the form:
\begin{subequations}
\begin{align}
\frac{\df}{\df t} {\bm U}(t) &=\sum_{\substack{k_x=1}}^{N_k}2 \real\[\vphantom{\frac1{2}}\vecd{\(\i k_x\, \DDel_{k_x}^{-1} \C_{k_x}(t) \bit\)}\]- {\bm U}(t)\ ,\label{eq:s3tz_discrete_mean}\\
\frac{\df }{\df t}\C_{k_x}(t) &= \A_{k_x}({\bm U})\,\C_{k_x}(t)+\C_{k_x}(t)\, \A_{k_x}({\bm U})^\dag  + \varepsilon \,\Q_{k_x}\ ,\quad\text{for }k_x=1,\dots,N_k\ ,
\end{align}\label{eq:S3Tz_discrete}\end{subequations}
where $\vecd(\M)$ denotes the vector with elements the diagonal elements of matrix $\M$. The homogeneity in the zonal direction has resulted in reducing the dimension of the S3T system from $N_x N_y+N_x^2N_y^2$ to $N_y+2N_k N_y^2$ (the factor 2 comes up because $\C_{k_x}$ are complex valued). This is a tremendous reduction, i.e., for $N_x=N_y=128$ and $N_k=15$ this gives around 550-fold decrease in the dimension of the S3T state variable.

In previous S3T studies (for example in \cite{Farrell-Ioannou-2003-structural,Farrell-Ioannou-2007-structure,Constantinou-etal-2014}) the Reynolds stress divergence term~\eqref{eq:RC_discrete_z} appears with a factor 1/2 instead of a factor 2. The reason is that in those studies Fourier transforms are defined as $\phi'=\sum_{\substack{k_x\\k_x>0}} \real\(\tilde{\phi}_{k_x}e^{\i k_xx}\)$. This results in a factor 1/2 difference in the Fourier coefficients of $\z'$ and in turn a $1/4$ difference in $\tilde{C}_{k_x}$.

To integrate~\eqref{eq:S3Tz_discrete} a RK4 time stepping scheme is used.

\subsection{Numerical integration of the generalized S3T system\label{sec:numS3Tgen}}

The generalized S3T system~\eqref{eq:s3t} does not a priori posses any homogeneity in $x$, so the dimension of the S3T state is $N_x N_y + N_x^2N_y^2$ instead of $N_y+2N_k N_y^2$, when the flow is homogeneous in $x$ and is represented by $N_k$ zonal waves.

On the discretized domain the mean flow streamfunction, $\Psi(\xv,t)$, is represented by an $(N_xN_y)$-column vector with elements:
\be
[\Psiv(t)]_{P} = \Psi(\xv_P,t)\ .
\ee
where index $P$ runs through $1,2,\dots,N_xN_y$ covering all the $(x,y)$ points of the domain. All other mean flow fields can be expressed in terms of $\Psi$. The eddy vorticity covariance $C(\xv_a,\xv_b,t)$ is represented by an $(N_xN_y)\times(N_x N_y)$ matrix, $\C$, with elements:
\be
[\C(t)]_{PQ} = C(\xv_P,\xv_Q,t)\ ,\quad\text{for }P,Q=1,\dots,N_xN_y\ .
\ee
Similarly, the spatial forcing covariance is represented with matrix $\Q$ with elements:
\be
[\Q]_{PQ} = Q(\xv_P-\xv_Q)\ ,\quad\text{for }P,Q=1,\dots,N_xN_y\ .
\ee

In matrix form $C(\xv_a,\xv_b,t)= \<\z'(\xv_a,t)\z'(\xv_b,t)\>$ is
\be
\C(t)= \<\zv(t)\,\zv(t)^\transp\>\ ,
\ee
where $\zv$ is the $(N_xN_y)$-column vector with elements
\be
[\zv(t)]_P = \z'(\xv_P,t)\ .
\ee

We express the S3T covariance equation~\eqref{eq:s3t_C} in terms of matrices $\C$ and $\Q$. For example, term $(U_a\partial_{x_a}+U_b\partial_{x_b}) C(\xv_a,\xv_b,t)$ becomes in matrix notation:
\begin{align}
\[U(\xv_a,t)\partial_{x_a}+U(\xv_b,t)\partial_{x_b}\] C(\xv_a,\xv_b,t) &= \left\{ \[\U(t)\,\D_x\]\C(t) + \C(t)\[\U(t)\,\D_x\]^\transp\right\}_{ab}\ ,
\end{align}
with $\D_x$ the matrix approximation of $\partial_x$ and $\U=\diag(U)$ the diagonal matrix with elements the values of $U(\xv,t)$ at the $(x,y)$ collocation points. Similarly, we transcribe in matrix form all other terms in~\eqref{eq:s3t_C}, which becomes
\be
\frac{\df }{\df t}\C(t)  = \A(\Uv)\, \C(t)+ \C(t)\,\A(\Uv)^\transp +\varepsilon\, \Q\ ,\ee
with $\A(\Uv)$ is the time-depended matrix
%\be
%\A(\Uv) = -(\U\D_x+\V\D_y) + \[\bit (\DDel\U)\D_x + (\DDel\V)\D_y+\( \b_x\D_y-\b_y\D_x\)\]\DDel^{-1} -\I~,
%\ee
\be
\A(\Uv) = -(\U\D_x+\V\D_y) + \[ ( \DDel\U-\b_y\I )\D_x +  ( \DDel\V+\b_x\I )\D_y\]\DDel^{-1} -\I\ ,\label{eq:def_Amatr}
\ee
and $\DDel$ the matrix approximation of the Laplacian operator $\Del$ and $\DDel^{-1}$ of its inverse.

In the mean flow equation~\eqref{eq:s3t_mean}, we express the Reynolds stress divergence term, $\Rcal( C )$ in terms of the matrix $\C$. At point $\xv_P$ we have:
\begin{align}
\left.\vphantom{\frac1{2}} \Rcal( C ) \right|_{\xv=\xv_P}&= -\left.\vphantom{\frac1{2}} \nablav\cdot\[\frac1{2}\hat{\mathbf{z}}\times\(\nablav_a \Del_{a}^{-1}+\nablav_b \Del_{b}^{-1}\) C_{ab}\]_{\xv_a=\xv_b} \right|_{\xv=\xv_P} \nonumber\\
%&=-\left.\vphantom{\frac1{2}} \nablav\cdot\[\frac1{2} \(-\partial_{y_a}\Del_a^{-1} -\partial_{y_b}\Del_b^{-1} ,\partial_{x_a}\Del_a^{-1}+ \partial_{x_b}\Del_b^{-1}\) C_{ab}\]_{\xv_a=\xv_b} \right|_{\xv=\xv_P} \nonumber\\
&=\frac1{2}\left.\nablav\cdot\[ \vphantom{\frac1{2}} \(\partial_{y_a}\Del_a^{-1} +\partial_{y_b}\Del_b^{-1} ,-\partial_{x_a}\Del_a^{-1}- \partial_{x_b}\Del_b^{-1}\) C_{ab}\]_{\xv_a=\xv_b} \right|_{\xv=\xv_P} \nonumber\\
%&=\frac1{2}\left.\nablav\cdot\[ \vphantom{\frac1{2}} \(\[\(\D_y\DDel^{-1}\)\C + \C\(\D_y\DDel^{-1}\)^\transp\]_{ab} ,-\[\(\D_x\DDel^{-1}\)\C + \C\(\D_x\DDel^{-1}\)^\transp\]_{ab}\)\]_{\xv_a=\xv_b} \right|_{\xv=\xv_P} \nonumber\\
&=\frac1{2}\left.\D_x\[ \vphantom{\frac1{2}} \[\(\D_y\DDel^{-1}\)\C + \C\(\D_y\DDel^{-1}\)^\transp\]_{ab} \]_{\xv_a=\xv_b} \right|_{\xv=\xv_P}  \nonumber\\
&\qquad+\frac1{2}\left.\D_y\[ \vphantom{\frac1{2}} \[-\(\D_x\DDel^{-1}\)\C + \C\(\D_x\DDel^{-1}\)^\transp\]_{ab} \]_{\xv_a=\xv_b} \right|_{\xv=\xv_P} \nonumber\\
&=\[ \vphantom{\frac1{2}}\D_x\, \vecd\[  \(\D_y\DDel^{-1}\)\C \]  + \D_y  \,\vecd\[-\(\D_x\DDel^{-1}\)\C \] \]_P \ .\label{eq:RC_discrete}
\end{align}
Therefore the discrete S3T system becomes in matrix form:
\begin{subequations}
\begin{align}
\frac{\df}{\df t} \Zv(t) &+ J\(\bit\Psiv(t),\Zv(t)+\bv\cdot\xv\)=\R\(\C(t)\) - \Zv(t)\ ,\label{eq:s3t_discrete_mean}\\
\frac{\df }{\df t}\C(t)  &= \A(\Uv)\, \C(t)+ \C(t)\,\A(\Uv)^\transp +\varepsilon\, \Q\ ,
\end{align}\label{eq:S3T_discrete}\end{subequations}
where $\R$ produces the Reynolds stress divergence for covariance $\C$ and is defined as 
\be
\R(\M)\equiv\D_x\, \vecd\[  \(\D_y\DDel^{-1}\)\M \]  + \D_y  \,\vecd\[-\(\D_x\DDel^{-1}\)\M \]\ .\label{eq:def_Rvec}
\ee

The Jacobian in~\eqref{eq:s3t_discrete_mean} is calculated pseudo-spectrally. To integrate~\eqref{eq:S3T_discrete} a RK4 time stepping scheme is used.

    % !TEX root = ../thesis.tex

\chapter{Bloch's theorem}
\label{app:bloch}

In this Appendix we will use Bloch's theorem to prove~\eqref{eq:eigendZdC} and~\eqref{eq:dz}.

\vspace{2em}

Consider the eigenvalue problem:
\be
\Acal(\xv)\,f_\s(\xv)= \s\,f_\s(\xv)\ ,\label{eq:Af_sf}
\ee
where $\Acal$ is the linear differential operator,
\be
\Acal(\xv) =b(\xv) + \sum_j b_j(\xv) \partial_{x_j} +  \sum_{j,k} b_{jk}(\xv) \partial^2_{x_jx_k} +\dotsb\ .
\ee
Suppose that the coefficients of $\Acal$ are invariant under the translation $\xv\to\xv + \av$, where $\av$ is any integer multiple of the constant vector $\av_0$. This implies that if $\f_\s(\xv)$ is an eigenfunction of $\Acal$ so is $\f_\s(\xv+\av)$. To see that set $\xv\to\xv+\av$ in~\eqref{eq:Af_sf} and use $\Acal(\xv+\av)=\Acal(\xv)$ to obtain $\Acal(\xv) f_\s(\xv+\av)= \s\,f_\s(\xv+\av)$.

Define $T_\rv$ the translation operator:
\be
T_\rv\,f(\xv) \equiv f(\xv+\rv)\ .
\ee
Operators $\Acal$ and $T_\av$ commute, i.e., $\Acal\,T_\av = T_\av\, \Acal$. This can be established by considering, without loss of generality, the action of $T_\av\,\Acal$ on an eigenfunction of $\Acal$: $\[ T_\av\,\Acal(\xv) \] f_\s(\xv) = T_\av \[ \Acal(\xv)\,f_\s(\xv)\] = \s\, T_\av\, f_\s(\xv) = \s f_\s (\xv+\av) = \Acal(\xv+\av)\, f_\s(\xv+\av) = \Acal(\xv) \[ T_\av  f_\s(\xv)\] = \[\Acal(\xv) T_\av\] f_\s(\xv)$.
The commutation of $\Acal$ and $T_\av$ implies that the eigenbasis of $\Acal$ and $T_\av$ can be chosen to be common.

%{\color{green}
%\begin{align}
%T_{\av} e^{\i\qv\cdot\xv}& = e^{\i\qv\cdot\av}\,e^{\i\qv\cdot\xv}
%\end{align}
%Therefore all function $e^{\i\qv\cdot\xv}$ with $\qv=\nv+m p_0$ will correspond to eigenvalue $e^{\i\nv\cdot\av}$
%}

This is not automatically achieved. For
example all plane waves  of the form
$e^{\i\qv\cdot\xv}$ are eigenfunctions of $T_\av$ 
but not necessarily of $\Acal$. From Fourier analysis we know that because every function can be written as a superposition of plane waves, the eigenfunctions of $\Acal$ will be of the general form $f_\s(\xv)=\int \df^2\qv\; c(\qv)\,e^{\i\qv\cdot\xv}$.
To determine the constraints imposed by the periodicity of $\Acal$ on the eigenfunction it is only required, because $\Acal$ and $T_\av$  commute, to render $f_\s(\xv)$ an eigenfunction of $T_\av$. Since $\av$ is an integer multiple of $\av_0$, to obtain the most general eigenfuction satisfying all symmetries of $\Acal$ we need to render $f_\s(\xv)$ an eigenfunction of $T_{\av_0}$. Then, since
\begin{align}
T_{\av_0} f_\s(\xv)& = T_{\av_0}\[ \int\df^2\qv\;c(\qv)\,e^{\i\qv\cdot\xv}\] = \int\df^2\qv\;c(\qv)\,e^{\i\qv\cdot\xv} e^{\i\qv\cdot\av_0} \ ,\label{eq:Ta_eigen}
\end{align}
if $f_\s(\xv)$ is to be an eigenfunction of $T_{\av_0}$ then $e^{\i\qv\cdot\av_0}$ cannot depend on $\qv$ and should be of the form  $e^{\i\qv\cdot\av_0}= e^{\i\phi}$. This holds when $\qv = \nv + m\pv_0$ with $m$ any integer, $\pv_0$ the vector:\footnote{In solid state physics $\pv_0$ is referred to as the fundamental vector of the reciprocal lattice.}
\be
\pv_0=2\pi\av_0/|\av_0|^2\ ,
\ee
that has the property that $\pv_0\cdot\av_0=2\pi$ and $\nv$ a constant vector satisfying $|\nv|\le|\pv_0|/2=\pi/|\av_0|$. This restricts the Fourier spectrum of the eigenfunction $f_\s(\xv)$ allowing only power at discrete wavenumbers. By writing $c(\qv)$ in the form
\be
c(\qv) = \sum_m\, C_m \,\d\(\qv-\nv-m\pv_0\)\ ,
\ee
we obtain that a general eigenfunction of $T_\av$ is:
\be
f(\xv) = \sum_m C_m \,e^{\i(\nv+m\pv_0)\cdot\xv} = e^{\i\nv\cdot\xv} \underbrace{\sum_m C_m \,e^{\i m\pv_0\cdot\xv}}_{g(\xv)}\ .\label{eq:f_xtoxma}
\ee

This implies that the general eigenfunction of $T_{\av_0}$ and $\Acal$ is of the form $e^{\i\nv\cdot\xv} g(\xv)$, for any $\nv$ satisfying $|\nv|\le\pi/|\av_0|$, and for periodic functions $g(\xv)$ with period $\av_0$, i.e., $g(\xv+\av_0)=g(\xv)$. This the content of Bloch's theorem about the structure of the eigenfunctions of operators with periodic coefficients.

In the special case that $\Acal$ is homogeneous, i.e., its coefficients do not depend on $\xv$, $
\Acal$ commutes with $T_\av$ for \emph{every} vector $\av$. From~\eqref{eq:f_xtoxma} we see that in this case $g(\xv)$ must be constant and therefore the only common eigenfunction of $\Acal$ and of all the $T_\av$'s is the single harmonic 
\be
f_\s(\xv) = C_0 \,e^{\i\nv\cdot\xv} \ ,\label{eq:f_xtoxa}
\ee
with no restriction on $\nv$ since $|\av_0|$ may be taken infinitely small.

\vspace{1em}

Another way at arriving at the same result is to note that  eigenvalue $e^{\i \nv\cdot \av_0}$ of $T_{\av_0}$ is degenerate and the degenerate eigenfunctions form a subspace  $G_{\av_0}$ spanned by $e^{\i\qv\cdot\xv}$ with $\qv = \nv + m \pv_0$, $m=0,\pm1,\dots$, $\pv_0\cdot\av_0=2\pi$ and $|\nv|\le\pi/|\av_0|$, as previously.  The common eigenfunctions of $\Acal$ and $T_{\av_0}$ will thus be indexed by $\nv$ and  for each $\nv$ will be linear combinations of basis of $G_{\av_0}$ (as in~\eqref{eq:f_xtoxma}). To arrive at the homogeneous result~\eqref{eq:f_xtoxa} consider a decreasing sequence of $\av_0$, i.e., $\av_0,\av_0/2,\av_0/2^2,\dots$, each associated with its own degenerate subspace for an $\nv$. The eigenfunctions of $\Acal$ for that $\nv$ will belong to $\bigcap_{s=1}^\infty G_{\av_0/2^s}$, which has the only element: $e^{\i\nv\cdot\xv}$, with $\nv$ unrestricted.

\chapter{Derivation of the eigenvalue relation for the S3T stability of a homogeneous equilibrium}
\label{app:S3Tgr}

\section{Eigenvalue relation for homogeneous S3T equilibrium} 
\label{appsec:eigenvalue}

Here we derive an analytic expression satisfied by  the eigenvalue $\s$  that determines the stability of the homogeneous equilibrium~\eqref{eq:ZeCe}. The eigenvalue problem to be solved is~\eqref{eq:s3t_dZdC_s}, %\textcolor{red}{\textbf{(Petro the equations do not change with the inclusion of hyperdiffusion, only $\Acal^e$ changes. Do you think I should repeat them anyway?)}}
\begin{subequations}\begin{align}
s \,\dZ &=\Acal^e\,\dZ+ \Rcal( \dC )\ ,\label{eq:s3t_pert_dZ_s_app}\\
s \,\dC_{ab} & = \(\bit\Acal^e_a + \Acal^e_b \)\dC_{ab} +\(\bit\d\tilde{\Acal}_a + \d\tilde{\Acal}_b\)C^e_{ab}\ .\label{eq:s3t_pert_dC_s}
\end{align}\label{eq:s3t_dZdC_s_app}\end{subequations}
We have included in~\eqref{eq:s3t_dZdC_s_app} also hyperdiffusive damping of the form: $-(-1)^h\nu_{2h} \Delta^{h}\,\z$. For $h=1$ this is normal diffusion, while for $h>1$ is hyperdiffusion of order $2h$. With this damping the operator $\Acal^e$ becomes:
\be
\Acal^e =\zhat\cdot\( \bv\times\nablav\)\Del^{-1} -1 + (-1)^h\nu_{2h} \Delta^{h}\ .
\ee
Equations~\eqref{eq:s3t_dZdC_s_app} is a linear system in $\dZ$ and $\dC$ and can be written symbolically as:\footnote{To find explicitly the form of $\Lcal$ one needs to manipulate $\d\tilde{\Acal}_a\,C^e_{ab}$ in the following manner:
\begin{align}
\d\tilde{\Acal}_a \,C^e_{ab} &   =-\d\tilde{\Uv}_a\cdot \nablav_a C^e_{ab} + (\Del_a\d\tilde{\Uv}_a)\cdot \nablav_a\Del_a^{-1}C^e_{ab} \nonumber\\
&   =-\(\zhat\times\nablav_a\d\tilde{\Psi}_a\) \cdot\nablav_a C^e_{ab} + \(\zhat\times\nablav_a\dZ_a\) \cdot \nablav_a \Del_a^{-1}C^e_{ab} \nonumber\\
&   =\(\zhat\times\nablav_a C^e_{ab}\) \cdot \nablav_a\d\tilde{\Psi}_a-\(\zhat\times\nablav_a \Del_a^{-1}C^e_{ab}\) \cdot\nablav_a\dZ_a \nonumber\\
&   =\[ \(\zhat\times\nablav_a C^e_{ab}\) \cdot \nablav_a\Del_a^{-1}-\(\zhat\times\nablav_a \Del_a^{-1}C^e_{ab}\) \cdot\nablav_a\]\dZ_a\ .\nonumber
\end{align}
Similarly for $\d\tilde{\Acal}_b\,C^e_{ab}$.}
\be
s\,\[\dZ,\dC\]^\transp = \Lcal(C^e)\,\[\dZ,\dC\]^\transp\ .
\ee
Because the only spatial dependence of the coefficients of $\Lcal$ is through $C^e(\xv_a-\xv_b)$ it is advantageous to consider the eigenvalue problem in terms of the relative coordinate $\xv_a-\xv_b$ and centroid coordinate $(\xv_a+\xv_b)/2$. Since~\eqref{eq:s3t_pert_dZ_s_app} is homogeneous in $\xv$ and~\eqref{eq:s3t_dZdC_s_app} homogeneous in $(\xv_a+\xv_b)/2$ the eigenfunctions $\dZ$ and $\dC$ will be single harmonics of their homogeneous coordinate (cf. Appendix~\ref{app:bloch}) and can be assumed to be of the form:
\begin{subequations}\begin{align}
\dZ_\nv(\xv) &= e^{\i\nv \cdot\xv}\ ,\\
\dC_\nv(\xv_a,\xv_b) &=\tilde{C}_\nv^{(\textrm{h})}\(\bit\xv_a-\xv_b\)\,e^{\i\nv\cdot(\xv_a+\xv_b)/2}\ .
\end{align}\label{eq:eig_hom}\end{subequations}
The homogeneous part of the perturbation covariance eigenfunction, $\tilde{C}_\nv^{(\textrm{h})}$, is expanded as
\be
C_\nv^{\textrm{(h)}}(\xv_a-\xv_b) = \int\frac{\df^2\kv}{(2\pi)^2}\hat{C}^{\textrm{(h)}}(\kv)\,e^{\i\kv\cdot(\xv_a-\xv_b)}\ ,\label{eq:Cnv_Cnvhat}
\ee
and introducing~\eqref{eq:Cnv_Cnvhat} to the perturbation covariance equation~\eqref{eq:s3t_pert_dC_s}, we obtain:
\begin{subequations}\begin{align}
%s \,\dC_{ab}  & = \int\frac{\df^2\kv}{(2\pi)^2}(\s-\i\om_{\nv})\hat{C}^{\textrm{(h)}}(\kv)\,e^{\i\kv\cdot(\xv_a-\xv_b)}\ ,\\
\(\bit\Acal^e_a + \Acal^e_b \)\dC_{ab}  & = -\int\frac{\df^2\kv}{(2\pi)^2}\[ 2+\nu_{2h} (k^{2h}_+ + k^{2h}_-) + \i\(\om_{\kv_+}-\om_{\kv_-}\)\]\hat{C}^{\textrm{(h)}}(\kv)\,e^{\i\kv\cdot(\xv_a-\xv_b)}\ ,\end{align}
with $\kv_{\pm} \equiv \kv\pm\nv/2$. We also have that \begin{align}
\d\tilde{\Acal}_a \,C^e_{ab} &=\int\!\frac{\df^2\kv}{(2\pi)^2}\,\hat{\mathbf{z}}\cdot(\nv\times\kv)\(\frac1{k^2}-\frac1{n^2}\)\,\hat{C}^e(\kv)\,e^{\i(\kv+\nv)\cdot\xv_a-\i\kv\cdot\xv_b}\nonumber\\
&= e^{\i\nv\cdot(\xv_a+\xv_b)/2} \int\!\frac{\df^2\kv}{(2\pi)^2}\,\hat{\mathbf{z}}\cdot(\nv\times\kv_-)\(\frac1{k_-^2}-\frac1{n^2}\)\hat{C}^e(\kv_-)\,e^{\i\kv\cdot(\xv_a-\xv_b)}\ ,
\end{align}
and similarly,
\begin{align}\d\tilde{\Acal}_b \,C^e_{ab} &=\int\!\frac{\df^2\kv}{(2\pi)^2}\,\hat{\mathbf{z}}\cdot(\kv\times\nv)\(\frac1{k^2}-\frac1{n^2}\)\,\hat{C}^e(\kv)\,e^{\i\kv\cdot\xv_a-\i(\kv+\nv)\cdot\xv_b}\nonumber\\
&=- e^{\i\nv\cdot(\xv_a+\xv_b)/2} \int\!\frac{\df^2\kv}{(2\pi)^2}\,\hat{\mathbf{z}}\cdot(\nv\times\kv_+)\(\frac1{k_+^2}-\frac1{n^2}\)\hat{C}^e(\kv_+)\,e^{\i\kv\cdot(\xv_a-\xv_b)}\ .
\end{align}\label{eq:dCpert_terms}\end{subequations}
From~\eqref{eq:s3t_pert_dC_s} we then obtain that
\begin{align}
\hat{C}^{\textrm{(h)}}(\kv)&=\dfrac{\hat{\mathbf{z}}\cdot(\nv\times\kv_-) \(\dfrac1{k^2_{-}}-\dfrac1{n^2}\)\,\hat{C}^e(\kv_-)-\hat{\mathbf{z}}\cdot(\nv\times\kv_+)\(\dfrac1{k^2_{+}}-\dfrac1{n^2}\)\,\hat{C}^e(\kv_+)}{\s+2 + \nu_{2h} (k^{2h}_+ + k^{2h}_-) +\i \(  \om_{\kv_+}- \om_{\kv_-} - \om_\nv\)}e^{\i\nv\cdot(\xv_a+\xv_b)/2}\nonumber\\
  &\equiv \[\bit F(\kv_-)-F(\kv_+)\]e^{\i\nv\cdot(\xv_a+\xv_b)/2} \ ,\label{eq:Chk}
\end{align}
with $\s\equiv s+\i\om_\nv$. Further,
\begin{align}
&\Rcal (\dC_\nv)     =\nonumber\\
&\hspace{.3em}  =-\nablav\cdot\[\frac{\hat{\mathbf{z}}}{2}\!\times\!\(\nablav_a \Del_{a}^{-1}+\nablav_b \Del_{b}^{-1}\) \dC\]_{\xv_a=\xv_b}\nonumber\\
%&  =-\nablav\cdot\[\frac{\hat{\mathbf{z}}}{2}\times\(\nablav_a \Del_{\la,a}^{-1}+\nablav_b \Del_{\la,b}^{-1}\) \int\frac{\df^2\kv}{(2\pi)^2} \hat{C}^{\textrm{(h)}}(\kv) e^{\i\kv\cdot(\xv_a-\xv_b)} \]_{\xv_a=\xv_b}\nonumber\\
&\hspace{.3em}  =-\nablav\cdot\[\frac{\hat{\mathbf{z}}}{2}\!\times\!\(\nablav_a \Del_{a}^{-1}+\nablav_b \Del_{b}^{-1}\) \int\!\frac{\df^2\kv}{(2\pi)^2} \[\bit F(\kv_-)-F(\kv_+)\] e^{\i\nv\cdot(\xv_a+\xv_b)/2}e^{\i\kv\cdot(\xv_a-\xv_b)} \]_{\xv_a=\xv_b}\nonumber\\
%&  =-\nablav\cdot\[\frac{\hat{\mathbf{z}}}{2}\times \iint\frac{\df^2\kv}{(2\pi)^2} \(\frac{\i \kv_+}{-k^2_{\la+}}-\frac{\i \kv_-}{-k^2_{\la-}} \)\[\bit F(\kv_-)-F(\kv_+)\] e^{\i (\kv_+\cdot \xv_a-\kv_-\xv_b)} \]_{\xv_a=\xv_b}\nonumber\\
&\hspace{.3em}  =-\nablav\cdot \int\!\frac{\df^2\kv}{(2\pi)^2} \frac{\hat{\mathbf{z}}}{2}\!\times\!\(\frac{\i \kv_-}{k^2_{-}}-\frac{\i \kv_+}{k^2_{+}} \)\[\bit F(\kv_-)-F(\kv_+)\] e^{\i \nv\cdot\xv}\ ,\label{eq:app8}
\end{align}
and since under the transformation $\kv\to-\kv$ we have that $\kv_+ \to- \kv_-$, $k^2_{+} \to k^2_{-}$ and $F(\kv_+)\to -F(\kv_-)$,~\eqref{eq:app8} becomes
\begin{align}
\Rcal (\dC_\nv)    &   =-\nablav\cdot \int\!\frac{\df^2\kv}{(2\pi)^2}\, \hat{\mathbf{z}}\times\(\frac{\i \kv_-}{k^2_{-}}-\frac{\i \kv_+}{k^2_{+}} \)\, F(\kv_-)  e^{\i \nv\cdot\xv}~.\label{eq:RdC_prefinal}
\end{align}
Using $\nablav\cdot\( \hat{\mathbf{z}}\times\pv\,e^{\i\nv\cdot\xv}\) = -\i\,\hat{\mathbf{z}}\cdot(\nv\times\pv)\,e^{\i\nv\cdot\xv}$, we then obtain,
\begin{align}
\Rcal (\dC_\nv)  &   = \int\!\frac{\df^2\kv}{(2\pi)^2} \hat{\mathbf{z}}\cdot(\nv\times\kv)\(\frac1{k^2_{+}} -\frac1{k^2_{-}}\)\, F(\kv_-)  e^{\i \nv\cdot\xv} \nonumber\\
%& = \d\tilde Q\iint\frac{\df^2\kv}{(2\pi)^2} \hat{\mathbf{z}}\cdot(\nv\times\kv)\(\frac1{k^2_{s}} -\frac1{k^2}\)\, F(\kv)   \nonumber\\
&   =\dZ \int\!\frac{\df^2\kv}{(2\pi)^2} \dfrac{|\nv\times\kv|^2 \(\dfrac1{k^2_{s}}-\dfrac1{k^2} \)\(\dfrac1{k^2}-\dfrac1{n^2}\)}{\s+2+ \nu_{2h} (k^{2h}_s + k^{2h}) +\i \(  \om_{\kv+\nv}- \om_{\kv} - \om_\nv\)}  \frac{\e\hat{Q}(\kv)}{2(1+\nu_{2h} k^{2h})}\ ,\label{eq:RdC_efdZ}
\end{align}
with $\kv_s\equiv \kv+\nv$ and $k_s\equiv|\kv_s|$. The last equality was obtained with the substitution $\kv\to\kv-\nv/2$. For $\nu_{2h}=0$~\eqref{eq:RdC_efdZ} gives~\eqref{eq:f_factor}. Finally, using~\eqref{eq:s3t_pert_dZ_s} we obtain the dispersion relation for the stability of the homogeneous equilibrium:
\begin{align}
\s + 1 + \nu_{2h} n^{2h} &=\e \int\frac{\df^2\kv}{(2\pi)^2} \dfrac{|\nv\times\kv|^2 \(\dfrac1{k^2_{s}}-\dfrac1{k^2} \)\(\dfrac1{k^2}-\dfrac1{n^2}\)}{\s+2+ \nu_{2h} (k^{2h}_s + k^{2h}) +\i \(  \om_{\kv+\nv}- \om_{\kv}  - \om_\nv\)}  \frac{\hat{Q}(\kv)}{2(1+\nu_{2h} k^{2h})} \ .\label{eq:s3tgr}
%&= \iint\frac{\df^2\kv}{(2\pi)^2} \dfrac{|\nv\times\kv|^2 \(k^2_{s}-k^2\)\(k^2-n^2\)\,\hat{C}^e(\kv)}{k_{s}^2k^4 n^2 \[\s+2r + \nu_{2h} (k^{2h}_s + k^{2h}) -\i \, \( \om_{\kv} - \om_{\kv+\nv}\)\bit\]}  
\end{align}

Equation~\eqref{eq:s3tgr} can be written in terms of the real and
imaginary part of the eigenvalue $s$ as:\begin{subequations}
\begin{align}
s_r&=-(1+\nu_{2h} n^{2h})+\varepsilon\,\real[f(\s)] ~,\label{eq:s3t_slanted_real}\\
s_i&=-\om_\nv+\varepsilon\,\imag[f(\s)]~.\label{eq:s3t_slanted_imag}
\end{align}\label{eq:s3t_slanted}\end{subequations}
The real part of the eddy feedback contributes to the growth rate of the mean flow and the imaginary part determines the departure of the phase speed of the mean flow from the Rossby wave frequency. For $\b\gg1$ the marginally unstable eigenfunctions have $\s_i=-\om_\nv$ as it can be readily shown that $f_i$ is at most of $\mathcal{O}(1)$ and therefore produces only a small correction to the Rossby phase speed which is of $\mathcal{O}(\b)$.

Equation~\eqref{eq:s3tgr} is solved numerically for $\s$ for a given $\nv$, $\b$, $\e$ and $\hat{Q}(\kv)$. However, in some special cases $\s$ can be solved in closed form. Such an example is when $\b=0$ and with $\nu_{2h}=0$. Then~\eqref{eq:s3tgr} takes the form:
\be
(s + 1)(s+2) =\e \underbrace{\int \frac{\df^2 \kv}{(2\pi)^2}\;\frac{|\mathbf{k}\times{\mathbf{n}}|^2\, (k^2_s-k^2) (k^2-n^2) }{k^4 k^2_s n^2 }\frac{\hat{Q}(\kv)}{2}}_{J}\equiv\e\, J\ ,\label{eq:s3tgr_b0}
\ee
and the eigenvalues are $s=-3/2\pm\(1/4+\e J\)^{1/2}$. The value of the integral, $J$, depends only on the wavevector $\nv$ and the forcing spectrum. For the ring forcing spectrum~\eqref{eq:Qhat_deltaG} by using the definitions of Fig.~\ref{fig:einx_angles} we have
\begin{align}
|\mathbf{k}\times{\mathbf{n}}|^2 &= n^2\cos^2\thet\ ,\ \ k_s^2 = 1+n^2+2n\sin\thet\ ,\label{eq:defs_inangles}
\end{align}
and therefore $J$ takes the form:
\begin{align}
J&=  \int\limits_{0}^{2\pi}\frac{\df \thet}{2\pi} \frac{2n\(1-n^2\)\,\cos^2{\thet} (\sin\thet+n/2) }{1 +n^2+2n\sin{\thet} }   \,\Gcal(\thet-\varphi)\nonumber\\
&=\frac{\mu}{4} n^2(1-n^2)\,\cos(2\varphi)\ ,\ \ \text{for }n\le 1\ .
\end{align}
For isotropic forcing $J=0$ and there is no instability since $\s=-1$ or $\s=-2$. When $\mu\ne0$ we have instability when $\e\ge 32/[\mu\cos(2\varphi)]>0$, which occurs either when $\mu>0$ and $0\le \varphi<45\deg$ or when $\mu<0$ and $45\deg<\varphi\le 90\deg$.

\section{Derivation of expression for $\fr$ for the ring forcing spectrum}
\label{appsec:fr_theta_phi_n}

In this section we will derive the expression~\eqref{eq:fr} for $\fr\equiv\real\[f(0)\]$ for the ring forcing spectrum~\eqref{eq:Qhat_deltaG}. For $\nu_{2h}=0$, $\s=0$ and for spectrum~\eqref{eq:Qhat_deltaG} we have that
\begin{align}
 f(0) &= \int\frac{k\,\df k\,\df \thet}{(2\pi)^2} \frac{|\mathbf{k}\times{\mathbf{n}}|^2\, (k^2_s-k^2) (k^2-n^2) }{k^4 k^2_s n^2  \[ 2 +\i\( \om_{\kv+\nv}- \om_{\nv} - \om_{\kv} \)\bit \] }  \,2\pi\,\Gcal(\thet-\varphi)\,\d(k-1)\nonumber\\
 &= \int\limits_{0}^{2\pi}\frac{\df \thet}{2\pi} \frac{|\mathbf{k}\times{\mathbf{n}}|^2\, (k^2_s-1) (1-n^2) }{k^2_s n^2  \[ 2 +\i\( \om_{\kv+\nv}- \om_{\nv} - \om_{\kv} \)\bit \] }   \,\Gcal(\thet-\varphi)\ .
\end{align}
Using the definitions of Fig.~\ref{fig:einx_angles} we have:
\begin{align}
\om_\nv = -\b \sin\varphi/n \ ,\ \ \om_\kv = -\b \cos(\thet-\varphi) \ ,\ \ \om_{\kv+\nv} = -\b \frac{n\sin\varphi+\cos(\thet-\varphi)}{1+n^2+2n\sin\thet}\ .
\end{align}
and together with~\eqref{eq:defs_inangles} we get
\begin{align}
\fr &=\real\[\f(0)\bit\]  = \real\(\spa\right.\int\limits_{0}^{2\pi} { \frac{\Ncal}{\Dcal_0 +\i\b\,\Db }}\df \thet\left.\spa \)\nonumber\\
&=\int\limits_{0}^{2\pi}\frac{\Ncal\,\Dcal_0}{ \Dcal_0^2 + \b^2\,\Db^2 }\,\df \thet\equiv\int\limits_{0}^{2\pi} F(\thet,\nv)\,\df \thet\ ,\label{eq:dvz}
\end{align}
with $F(\thet,\nv)$ defined by~\eqref{eq:defF} and
\begin{subequations}
\begin{align}
\Dcal_0(\thet,\nv)&=2(1 +n^2+2n\sin{\thet})~,\\
%\Db(\thet,n)&=\sin{\varphi}/n+(1+n^2)\cos{(\thet-\varphi)} - \cos{(\thet+\varphi)} + 2n\sin\thet\cos(\thet-\varphi)~,\\
%\Db(\thet,n)&=\sin{\varphi}/n+\cos{(\thet-\varphi)}\(1+n^2+2n\sin{\thet}\) - \cos{(\thet+\varphi)}~,\\
\Db(\thet,\nv)&=(1 +n^2+2n\sin{\thet})\sin{\varphi}/n+n^2\cos{(\thet-\varphi)} + n \sin{(2\thet-\varphi)}~,\\
%\Db(\thet,n)&=(1 +n^2+2n\sin{\thet})\[\frac{\sin{\varphi}}{n} + \cos{(\thet-\varphi)} - \frac{\cos{(\thet-\varphi)} + n\sin{\varphi}}{1 +n^2+2n\sin{\thet}}\]~,\\
%N(\thet)&=4\pi n\(1-n^2\)\,\cos^2{\thet} (\st+n/2) \left\{ 1-\mu \cos[2(\thet-\varphi)] \bit\right\}\,D_0(\thet)~,
\Ncal(\thet,\nv)&= \frac1{\pi}\, n\(1-n^2\)\,\cos^2{\thet} (\sin\thet+n/2) \, \mathcal{G}(\thet-\varphi) ~.\label{eq:defNcal}\end{align}\label{eq:defD0DbN}\end{subequations}

Note that $\b\Db$ is 
\be
\b\,\Db(\thet,\nv)= k^2_s (\om_{\kv+\nv}-\om_{\nv}-\om_{\kv})~,\label{eq:D2zero}
\ee
and therefore vanishing of $\Db$ occurs when the resonant condition is satisfied:
\be
\om_{\nv}+\om_{\kv}=\om_{\kv+\nv}~.\label{eq:res}
\ee 

Also note that the $F$ defined by~\eqref{eq:defF} remains unchanged when the angle $\varphi$ is shifted by
$180\deg$ ($\varphi\to180\deg+\varphi$) or when there is a simultaneous shift of $\varphi\to180\deg-\varphi$
and $\thet\to180\deg-\thet$. As a result, it suffices to only consider cases with $0\le\varphi\le 90\deg$.

\section{Asymptotic expression for the eigenvalue $s_r$ at high supercriticality}

At high supercriticality, i.e., as  $\varepsilon \to \infty$, the maximal growth rate, $s_r$, of the large-scale structure with wavevector $\nv$ scales with $\sqrt{\varepsilon}$ while the frequency of this eigenstructure, $s_i$, asymptotes to values $s_i\approx -\om_\nv$.

Specifically, $s_r$ asymptotes to:
\begin{align}
s_r^2 &=\e \int\frac{\df^2\kv}{(2\pi)^2} |\nv\times\kv|^2 \(\dfrac1{k^2_{s}}-\dfrac1{k^2} \)\(\vphantom{\dfrac1{k^2_{s}}}\dfrac1{k^2}-\dfrac1{n^2}\)  \frac{\hat{Q}(\kv)}{2} \ .\label{eq:asymptote_sr_e}
\end{align}
This asymptotic expression for the growth rate and phase speed of the large-scale structure is useful for tracing the maximal growth rates as a function of supercriticality using Newton's iterations.  

\begin{figure}[ht]
\centering\includegraphics[width=24pc,trim=7mm 0mm 7mm 0mm,clip]{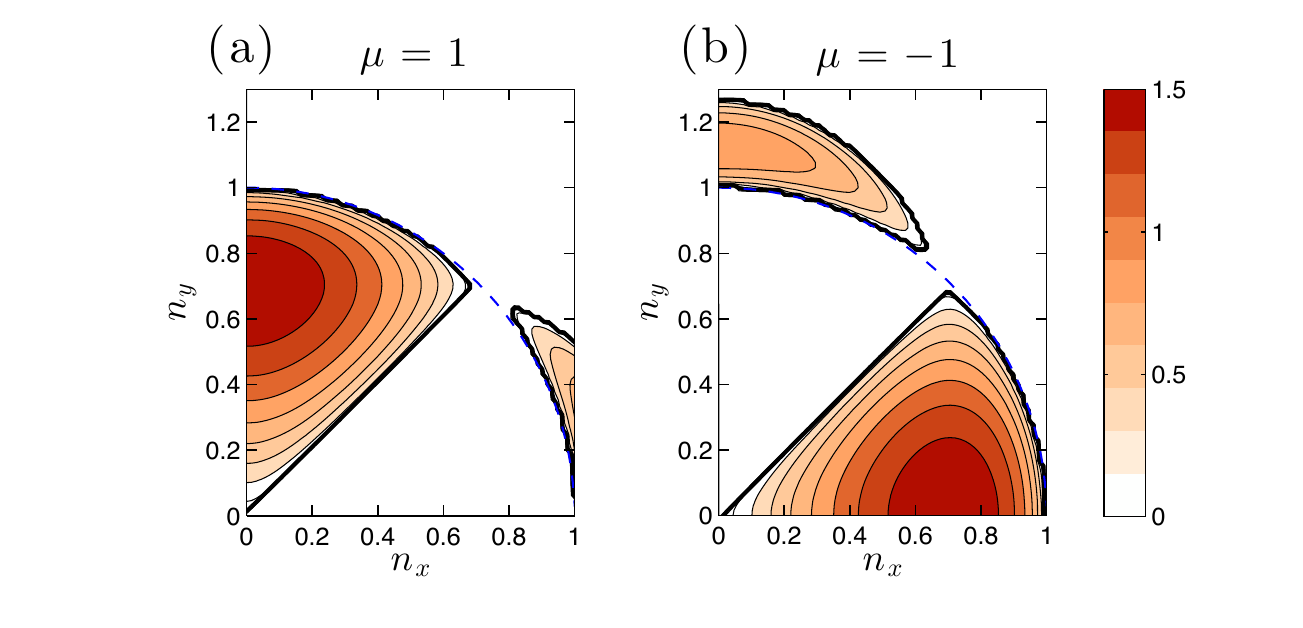}
\caption{The $\varepsilon\to+\infty$ asymptotic maximal growth rate $s_r$ scaled by $\sqrt{\varepsilon}$ as a function of the wavenumbers $(n_x,n_y)$ of the S3T eigenfunction. Shown are contours for $s_r \ge 0$ and the zero contour is marked with a thick solid line. The asymptotic growth rate is independent of $\bv$ and dissipation and depends only the forcing spectrum. Shown are the asymptotic growth rates for forcing~\eqref{eq:Qhat_deltaG} for (a) $\mu=1$ and (b) $\mu=-1$. For $\mu>0$ maximal instability occurs for jet structures, while for $\mu<0$ maximal instability occurs for non-zonal structures.}
  \label{fig:asymptotic}
\end{figure}

The asymptotic growth rates depend only on the forcing distribution and for the forcing spectrum~\eqref{eq:Qhat_deltaG} are shown in Fig.~\ref{fig:asymptotic} for $\mu>0$ and $\mu<0$.  It can be shown that the asymptotic growth rate vanishes for exactly isotropic forcing. Asymptotically the growth rates do not depend on the damping rate of the mean flow, $\rU$ (cf.~chapter~\ref{ch:NLvsS3Tjas}). 

\chapter{Asymptotic expressions for eddy feedback}
\label{app:fR}

In this Appendix we calculate in closed form asymptotic expressions for the eddy feedback induced by a mean flow perturbation with wavevector $\nv$, in the cases $\b\ll1$ and $\b\gg1$.

\section{Case $\b\ll 1$ \label{sec:asympt_smallb}}

When $\b\ll1$ and for $n$ satisfying $\b /n\ll1$, we expand
$\Fcal(\thet,\nv) =F(\thet,\nv)+F(180\deg+\thet,\nv)$ in~\eqref{eq:fr} in powers of $\b$. Since $\Fcal$ is a function of
$\b^2$ we have the expansion:
\be
\Fcal=\Fcal_{0} + \b^2\, \Fcal_2 + \Ocal(\b^4)~,
\ee
with $\Fcal_2 = \frac1{2}\left . \bit \partial^2_{\b\b} \Fcal \right|_{\b=0}$.
%\Fcal_0(\thet,\nv) = F_0(\thet,\nv) + F_0(\pi+\thet,\nv)$, $\Fcal_2(\thet,\nv) = F''_\b(\thet,\nv) + F''_\b(\pi+\thet,\nv)$ and
%\be
%F_0 \equiv \left. F\bit\right|_{\b=0} -\frac{\Ncal}{\Dcal_0}~~,~~F''_\b \equiv\left. \partial^2_{\b\b} F\bit\right|_{\b=0}= -\frac{\Ncal\,%\Db^2}{\Dcal_0^3}~.
%\ee
%where $\Dcal_0$, $\Db$ and $\Ncal$ as defined in~\eqref{eq:defD0DbN}.and
The leading order term is:
\begin{align}
\Fcal_0 & = \frac{1}{\pi}\, n^2\(1-n^2\)\,\Gcal (\thet-\varphi)\frac{1+n^2-4\sin^2\thet }{ (1 + n^2)^2-4n^2\sin^2{\thet}}\cos^2{\thet} ~,\label{eq:F0}
\end{align}
due to the property $\Gcal(180\deg+\thet)=\Gcal(\thet)$. Positive values
of $\Fcal_0$ indicate that the stochastically forced waves with phase
lines inclined at angle $\thet$ with respect to the wavevector $\nv$ (cf.~Fig.~\ref{fig:einx_angles})
induce upgradient vorticity fluxes to a mean flow with wavenumber $n$
when $\b=0$. Given that $n<1$ and $\Gcal>0$, $\Fcal_0$ is positive for
any forcing distribution, only in the sector shown in
Fig.~\ref{fig:smallb_F}\hyperref[fig:smallb_F]{a} in which
$4\sin^2\thet<1+n^2$. Specifically, in the absence of $\b$ all waves
with $|\thet|\le 30\deg$ reinforce mean flows with $n<1$. Note that
the condition $4\sin^2\thet<1+n^2$ is also the necessary condition for
modulational instability of a Rossby wave with wavevector components
$(\cos \thet, \sin \thet)$ to any mean flow (zonal or non-zonal) of
total wavenumber $n$ for $\b\ll 1$~\citep{Gill-1974}.

\vspace{1em}

The total vorticity flux feedback $\fr$ for
% since each integral involving $\Fcal_j$, $j=0,2,\dots$ can be calculated in closed form.
$\Gcal(\thet-\varphi) = 1+\mu\cos{[2(\thet-\varphi)]}$ is at leading order:
\be
\fr =\frac{\mu}{8} n^2 \left(1-n^2\right) \cos (2 \varphi ) + \Ocal(\b^2)~,\label{eq:dvz_b0}
\ee
which is proportional to the anisotropy factor, $\mu$. The maximum feedback factor is in this case
\be
\f_{r,\max} = \frac{|\mu|}{32} ~,\label{eq:frmax_smallb_mu}
\ee
and is achieved for mean flows with $n=1/\sqrt{2}$. This maximum is achieved for
zonal jets ($\varphi=0\deg$) if $\mu>0$ and for meridional jets ($\varphi=90\deg$) if $\mu<0$.
This implies that for $\b\ll 1$ the first structures to become unstable are zonal jets if
$\mu>0$ and meridional jets if $\mu<0$, as shown in Fig.~\ref{fig:frmax}\hyperref[fig:frmax]{c}.

\begin{figure}
\centerline{\includegraphics[width=6in]{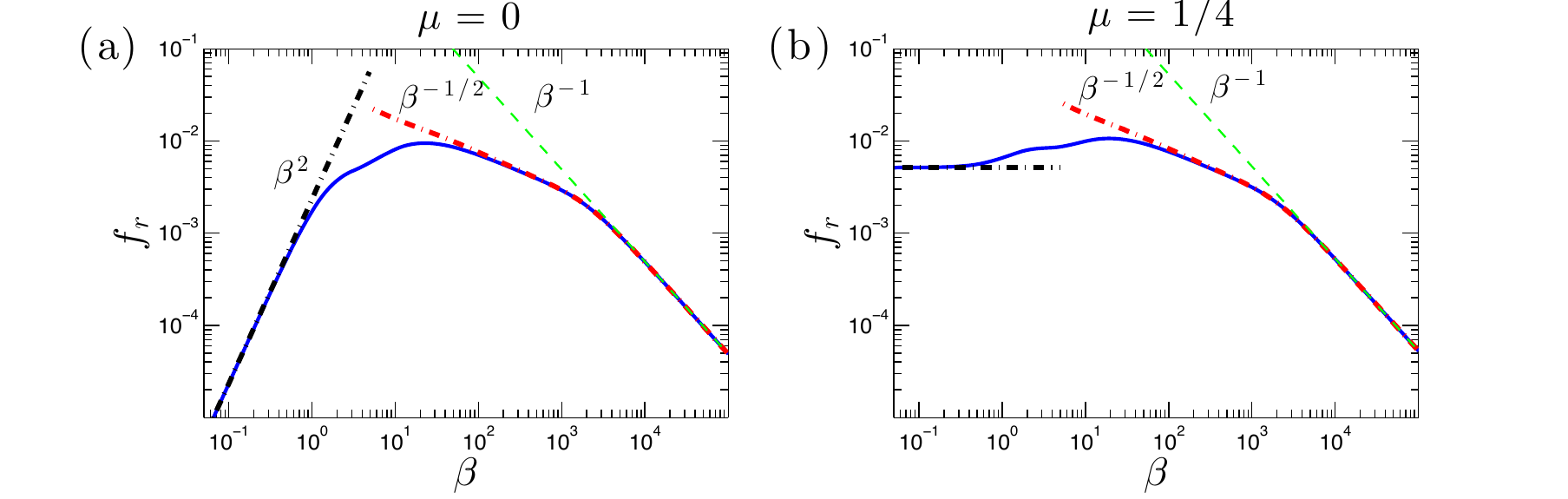}}
\caption{Feedback factor $\fr$ for a non-zonal perturbation with $n=0.4751$ and $\varphi=10^\circ$ (which belongs in region~A of Fig.~\ref{fig:R_Fcal}\hyperref[fig:R_Fcal]{a}) (solid lines) in the case of a forcing covariance with (a) $\mu=0$ and (b) $\mu=1/4$. Also shown are asymptotic expressions for $\b\ll1$ (\eqref{eq:frmax_smallb_mu0} in (a) and~\eqref{eq:frmax_smallb_mu} in (b)) and the resonant contribution~ \eqref{eq:frR} for $\b\gg1$ (dash-dot). For $\b\gg1$, expression~\eqref{eq:fR_b} is also plotted (dashed). It can be seen that only~\eqref{eq:frR} can captures the $\b^{-1/2}$ decrease of $\fr$.}\label{fig:Rdvz}
\end{figure}

For isotropic forcing ($\mu=0$), the leading order term is zero and $\fr$ depends quadratically on $\b$:
\begin{align}
\fr &= \b^2\frac{n^4}{64} \[\bit 2 + \cos (2 \varphi)\]+ \Ocal(\b^4)~ ~~{\textrm{for}~~~n<1}\ ,\label{eq:dvz_b0_isotr}
\end{align}
producing upgradient fluxes for $n<1$. Note that for the delta function ring forcing
$\int_0^{2\pi}\Fcal_2\,\df\thet$ is discontinuous at $n=1$, with positive values for $n=1^-$ and negative values
for $n=1^+$. The accuracy of these asymptotic expressions is shown in Fig.~\ref{fig:Rdvz}.
%
%and the $\Ocal(\b^2)$ term is %In the limit $\b\ll1$, $n\ll 1$ with $\b/n\ll1$ we get that the $\dvz$ are:
%\begin{align}
%f_{r,2} &= \int\limits_{0}^{\pi} \Fcal_2 (\thet, n)\,\df\thet \nonumber\\
%&=\frac{3n^4}{64} \(\bit\cos^2\varphi+\frac1{3}\sin^2\varphi\)+\nonumber\\
%&\qquad\qquad+\mu \left\{\bit -4\cos(2\varphi)\sin^2\varphi + 4n^2 \sin^2\varphi\,[3+4\cos(2\varphi)] - n^4[3-2\cos\varphi+2\cos(2\varphi)] - n^6 [2\cos(2\varphi)+\cos(4\varphi)] \right\}\nonumber\\
%%&=+\frac{\b^2 }{128}\left\{ -4\mu \sin ^2\varphi \cos (2 \varphi ) +4n^2 \mu \sin ^2\varphi \[3 \cos (2 \varphi )+2\bit\] +\bit\right.\nonumber\\
%%&\qquad+\left. n^4 \[\bit4-3 \mu+2 \mu \cos (4 \varphi )-2 (\mu -1) \cos (2 \varphi ) \] -\mu n^6\[\bit2\cos(2\varphi)+\cos(4\varphi)\]\bit\right\} \;+\Ocal(\b^4)~.
%\label{eq:dvq_b_n_phi}
%\end{align}
%
%
%
%\begin{align}
%\dvz = &\frac{\mu}{8} n^2 \left(1-n^2\right) \cos (2 \varphi ) \,\d U+\nonumber\\
%&~-\i\frac{\b\mu}{16}\sin \varphi\left\{ n \cos (2 \varphi )- n^3 \[\bit2 \cos (2 \varphi )+1\]+\mathcal{O}(n^5)\bit\right\} \,\d U+\nonumber\\
%&~+\frac{\b^2 }{32}\left\{ -\mu \sin ^2\varphi \cos (2 \varphi ) +n^2 \mu \sin ^2\varphi \[3 \cos (2 \varphi )+2\bit\] +\bit\right.\nonumber\\
%&\qquad\qquad\qquad\qquad\qquad+\left. \frac{n^4}{4} \[\bit4-3 \mu+2 \mu \cos (4 \varphi )-2 (\mu -1) \cos (2 \varphi ) \] +\mathcal{O}(n^5)\bit\right\} \,\d U~.
%\label{eq:dvq_b_n_phi}
%\end{align}
%From~\eqref{eq:dvz_b0} we can see that maximum feedback gain $\fr$ is obtained for zonal jet perturbations if $\mu>0$ and for non-zonal perturbations with $\varphi=90\deg$ if $\mu<0$.
%
The maximum feedback factor, shown in Fig.~\ref{fig:frmax}\hyperref[fig:frmax]{a}, is
\begin{align}
\f_{r,\max} = & \frac{3\b^2}{64} ~,\label{eq:frmax_smallb_mu0}
\end{align}
and is attained by zonal jets ($\varphi=0\deg$) with wavenumber
$n\to1^{-}$ as $\b\rightarrow 0$, a result that was previously derived by \citet{Srinivasan-Young-2012}. The
accuracy of~\eqref{eq:frmax_smallb_mu} and~\eqref{eq:frmax_smallb_mu0} extends to $\b\approx 0.1$, as shown
in Fig.~\ref{fig:frmax}\hyperref[fig:frmax]{a}.

\section{Case $\b\gg 1$\label{sec:asympt_largeb}}

When $\b\gg1$, we write~\eqref{eq:dvz} in the form:
\begin{align}
\label{eq:B7}
\f_r &=\frac{I} {\b^2}~,~~{\textrm{with}}~~I = \int\limits_{0}^{2\pi} F_\chi(\thet,\nv) \,\df\thet~,
\end{align}
where
\begin{align}
F_\chi(\thet,\nv) = \frac{\Ncal\,\Dcal_0}{\chi^2 \Dcal_0^2 + \Db^2 }~,\label{eq:refFchi}
\end{align}
and $\chi\equiv1/\b$. When $\Db\sim\mathcal{O}(1)$ for all angles $\thet$, then the feedback factor
is $\fr\sim\mathcal{O}(\b^{-2})$. However, if $\Db\sim\mathcal{O}(\b^{-1})$ for some angle $\thet$, then
as we will show in this Appendix, $\fr$ decays as $\mathcal{O}(\b^{-1})$ or as $\mathcal{O}(\b^{-1/2})$. This
is illustrated in Fig.~\ref{fig:Rdvz} showing the feedback factor $\fr$ as a function of $\b$ in cases
in which $\Db$ vanishes.
%\begin{figure}
%\centerline{\includegraphics[width=.9\textwidth,trim = 30mm 1mm 30mm 0mm, clip]{Db_circ_3.eps}}
%\caption{\label{fig:Db} Zero contours of $\Db(\thet,\nv)$ for (a): jet perturbations ($\varphi=0$) and (b), (c) non-zonal perturbations ($\varphi=15\deg$, $\varphi=60\deg$ respectively) in a $(\thet,\nv)$ polar plot. Shaded area marks $n\le 1$. Light shade corresponds to $(\thet,\nv)$ satisfying $\sin\thet>-n/2$ for which constructive resonances occur, while dark areas correspond to $\sin\thet<-n/2$ for which destructive resonances occur. The radial grid interval is $\Delta n=0.25$.}
%\end{figure}

$\Db$ can have at most 4 roots, $0\deg \le \thet_j \le 360\deg$ ($j=1,2,3,4$), for any given $(n,\varphi)$. At these angles
the resonance condition (\ref{eq:res}) is satisfied. To calculate asymptotic approximations to the integral $I$,
we split the range of integration to a small range close to the roots of $\Db$ for which we have resonance, $I^{(\textrm{R})}$, and to a range away from the roots of $\Db$, $I^{\textrm{(NR)}}$:
\be
I = \sum_{j=1}^{N_r}\[ \spb \right. \underbrace{\int\limits_{\thet_{j-1}+\d\thet}^{\thet_j-\d\thet} \!\!
F_\chi(\thet,\nv)\,\df\thet}_{I^{(\textrm{NR})}_j} + \underbrace{\int
\limits_{\thet_{j}-\d\thet}^{\thet_j+\d\thet} \!\! F_\chi(\thet,\nv)\,\df \thet}_{I^{(\textrm{R})}_j}
\left.\spb \]~,\label{eq:sumI}
\ee
where $N_r$ is the total number of the roots of $\Db$ and $\thet_0\equiv\thet_{N_r}$. Asymptotic approximations to the integral over the two ranges are then found separately using a proper rescaling for the regions close to the roots of $\Db$ (cf.~\citet{Hinch-1991}).

When the distance between two consecutive roots
is $|\thet_j-\thet_{j-1}|>\sqrt{\chi}$, as in the examples shown in
Figs.~\ref{fig:R_Fcal}\hyperref[fig:R_Fcal]{c,e}, then the dominant
contribution to the integral comes from the $\mathcal{O}(\chi)$ regions close to
the roots $\thet_j$, since $F_\chi(\thet,\nv)$ close to $\thet_j$ is approximately a Lorentzian
of half-width $\mathcal{O}(\chi)$.
Therefore, choosing the range $\d \thet$ close to the roots to be
$\sqrt{\chi}\ll\delta\thet\ll1$, Taylor expanding $F_\chi(\thet,\nv)$
close to $\thet_j$ and rescaling $\thet=\thet_j+\chi u$ we obtain:
\be
I^{(\textrm{R})}_j=\frac1{\chi}\int\limits_{-\delta\thet/\chi}^{\delta\thet/\chi}
\frac{\Ncal_j\,\Dcal_{0,j}\,\df u}{\Dcal_{0,j}^2+\Dbj'^{\,2}\, u^2}+\mathcal{O}(\chi^{-3})~,
\ee
where $\Db'\equiv\partial_\thet\Db$ and the subscript $j$ denotes the value at
$\thet_j$. In the limit $\d \thet/\chi\rightarrow\infty$ we obtain:
\be
I^{(\textrm{R})}_j = \frac1{\chi}\frac{ \pi\,\Ncal_j}{|\Dbj'|}~,\label{eq:fR_b}
\ee
and as a result, the resonant contribution produces the asymptotic approximation:
\be
\fr^{(\textrm{R})} = \frac{1}{\b}\sum_{j=1}^{N_r}\frac{ \pi\,\Ncal_j}{|\Dbj'|}~.\label{eq:fR_b}
\ee

\begin{figure}[t]
\centering\includegraphics[width=2.7in]{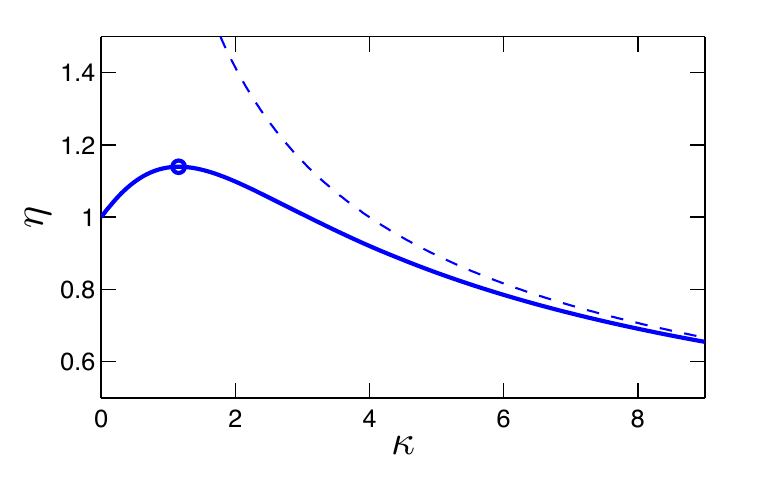}
\caption{\label{fig:xifactor} The factor $\eta = 2\, (\kappa^2+4)^{-3/4} \, \csc{\[\frac1{2}\arctan{(2/\kappa)}\]}$ as a function of $\kappa$ that is a measure of the distance between two consecutive resonant angles. The maximum value of $\eta$ marked with an open circle (and consequently of the feedback gain that is proportional to $\eta$) is $\eta =3^{3/4}/2\approx 1.14$ and it is achieved at $\kappa=2/\sqrt{3}\approx1.16$. Also shown is the asymptote $\eta=2/\sqrt{\kappa}$ that $\eta$ follows for $\kappa\gg1$ (dashed). This suggests that the resonant contribution is maximum when the two roots are very close to each other ($\kappa\approx 1$) but not on top of each other ($\kappa\ll1$).}
%matlab code for xi
%k = linspace(0,4,301);x =1/2* (k.^4+4).^(3/4).*(sin(1/2 *atan(2./k.^2)));figure(1);set(gcf,'Renderer','Painters','color','w');plot(k,x,'b','LineWidth',2);hold on;plot(k,k/2,'--b','LineWidth',1);plot((4/3)^(1/4),2/3^(3/4),'ob','LineWidth',2);hold off;ylim([1 4]/2);xlabel('$\kappa$','FontSize',16,'Interpreter','latex');ylabel('$\eta$','FontSize',16,'Interpreter','latex');set(gca,'FontSize',12);set(gca,'TickLength',[.025 .025]);

%matlab code for 1/xi
%k = linspace(0,4,301);x =1/2* (k.^4+4).^(3/4).*(sin(1/2 *atan(2./k.^2)));x=1./x;figure(1);set(gcf,'Renderer','Painters','color','w');plot(k,x,'b','LineWidth',2);hold on;plot(k,2./k,'--b','LineWidth',1);plot((4/3)^(1/4),1/(2/3^(3/4)),'ob','LineWidth',2);hold off;ylim([1 3]/2);xlabel('$\kappa$','FontSize',16,'Interpreter','latex');ylabel('$\eta$','FontSize',16,'Interpreter','latex');set(gca,'FontSize',12);set(gca,'TickLength',[.025 .025]);set(gca,'YTick',[0:.2:2]);

%matlab code for eta defined with k=kappa^2
%k = linspace(0,10,201);x =1/2* (k.^2+4).^(3/4).*(sin(1/2 *atan(2./k)));x=1./x;figure(1);set(gcf,'Renderer','Painters','color','w');plot(k,x,'b','LineWidth',2);hold on;plot(k,2./sqrt(k),'--b','LineWidth',1);plot((4/3)^(1/2),1/(2/3^(3/4)),'ob','LineWidth',2);hold off;ylim([1 3]/2);xlabel('$\kappa_j$','FontSize',16,'Interpreter','latex');ylabel('$\eta_j$','FontSize',16,'Interpreter','latex');set(gca,'FontSize',12);set(gca,'TickLength',[.025 .025]);set(gca,'YTick',[0:.2:2]);
\end{figure}

However, special attention should be given to the case in which two consecutive roots are
close to each other. When $|\thet_j-\thet_{j-1}|\sim\mathcal{O}(\sqrt{\chi})$
then $\Dbj' \sim \mathcal{O}(\sqrt{\chi})$ and $\fr^{\textrm{(R)}}$ scales as
$1/\sqrt{\b}$ instead of $1/\b$ for $\b\gg 1$. Indeed,
when $F_\chi$ is double peaked, as in Fig.~\ref{fig:R_Fcal}\hyperref[fig:R_Fcal]{d}, the
dominant contribution comes from the whole range between the two resonant angles which are a distance
$\mathcal{O}(\sqrt{\chi})$ apart. The proper scaling for the angles close to $\thet_j$ is therefore
$\thet=\thet_j+\sqrt{\chi}u$. Taylor expanding the denominator under this scaling we obtain:
\be
\label{eq:B13}
\chi^2\Dcal_0^2+\Db^2=\chi^2\Dcal^2_{0,j}+\chi\Dbj'^{\,2}\, u^2+
\chi^{3/2}\Dbj'\Dbj'' u^3 + \chi^2\left(\frac1{4}\Dbj''^{\,2}+\frac1{3}\Dbj'\Dbj'''\right)u^4+\Ocal{(\chi^{5/2})}~,%\mbox{h.o.t}~,
\ee
where $\Dcal_{2}''\equiv \partial_{\thet\thet}^2\Db$ and $\Dcal_{2}'''\equiv
\partial_{\thet\thet\thet}^3\Db$. When $\Dbj' \sim \mathcal{O}(\sqrt{\chi})$
all the terms in (\ref{eq:B13}) are $\Ocal ( \chi^2)$ and
writing $\Dbj'=\sqrt{\x}\,d(n,\thet_j)\equiv\sqrt{\x}\,d_j$, where $d$ is of $\mathcal{O}(1)$,
the leading order resonant contribution is:
%Since $\Dcal_0$ is multiplied with $\x^2$ we have to keep up to $\mathcal{O}(\x^2)$ in $\Db$ and to do that we need to expand $\Db$ up to fourth order.
%Each of the terms of $F$ can be expanded as:
%\begin{subequations}
%\begin{align}
%\Ncal&=\Ncal_j + \Ncal'_j \sqrt{\x}u + \mathcal{O}\(\x \)~,\\
%\Dcal_0^2&=\Dcal_{0,j}^{\,2}+2 \Dcal_{0,j} \Dcal_{0,j}'\sqrt{\x}u + \mathcal{O}\( \x\)~,\\
%\Db^2&=\Dbj'^{\,2}\,\x u^2 +\Dbj'D_{\b,j}'' \,\x^{3/2} u^3 + \frac1{12}\(3\Dbj''^{\,2}+4\Dbj'\,\Dbj'''\) \x^2 u^4+ \mathcal{O}\( \x^{5/2}\)~.
%\end{align}\end{subequations}
%where primes denote partial differentiation with respect to $\thet$ and subscript $j$ at $\Ncal$, $\Dcal_0$ and $\Dcal_\b$ and their derivatives denotes their corresponding value calculated at $F_\chi$. Now the integrand $f$ takes the form:
%\begin{align}
%\frac{\Ncal}{\chi^2 \Dcal_0^2 + \Dcal_\b^2} &=\chi^{-2}\frac{\Ncal_j}{ \Dcal_{0,j}^2 + d_{j}^{\,2}\, u^2 + d_{j}\la_j u^3 + \frac1{4} \la_j^{2} u^4 } +\mathcal{O}(\chi^{-3/2})&
%\end{align}
%where $\la_j\equiv \Dbj''$ and integral $I_{\textrm{r},j}$ becomes:{\color{red}
\begin{align}
I^{(\textrm{R})}_j &=\chi^{-3/2}\!\!\int\limits_{-\d\thet/\sqrt{\x}}^{\d\thet/\sqrt{\x}} \frac{\Ncal_j\,\Dcal_{0,j}\; \df u}{ \Dcal_{0,j}^2 + d_{j}^{\,2}\, u^2 + d_{j}\rho_j u^3 + \frac1{4} \rho_j^{2} u^4 } + \mathcal{O}(\x^{-1})\ ,
\end{align}
where $\rho_j\equiv \Dbj''$. In the limit $\d\thet/\sqrt{\x}\to\infty$ the integral can be evaluated
from the residues from two of the four poles of the integrand. The two poles are at $
%\begin{align}
u = - d_{j}/\rho_j \pm |z_j|^{1/2}\sgn{(\rho_j) }\, e^{\pm\i w_j/2} $ ,
%\end{align}
where $|z_j| = \Dcal_{0,j} |\rho_j |^{-1} (\kappa_j^2+4)^{1/2}$, $w_j=\arctan(2/\kappa_j)$ and
$\kappa_j\equiv d_{j}^2 \Dcal_{0,j}^{-1} |\rho_j|^{-1}$ is an increasing function of the
distance between the two roots of $\Db$. Therefore:
\begin{align}
I^{(\textrm{R})}_j & =\chi^{-3/2} \frac{ \pi \,\Ncal_j\,\eta_j}{\Dcal_{0,j}^{1/2}|\rho_j|^{1/2}}+
\mathcal{O}(\chi^{-1})~,\label{eq:IRj}
\end{align}
and
%\be
%\fr^{\textrm{(R)}} = \frac{1}{\sqrt{\b}}\sum_{j=1}^{N_r}\frac{\pi \, \Ncal_j\,\eta_j}{ 2\,\Dcal_{0,j}^{1/2}|\la_j|^{1/2}}~,\label{eq:dvz_approx_sqrtb}
%\ee
\be
\fr^{\textrm{(R)}} = \frac{1}{\b^2}\sum_{j=1}^{N_r}\frac1{2}\,I^{(\textrm{R})}_j= \frac{1}{\sqrt{\b}}\sum_{j=1}^{N_r}\frac{\pi \, \Ncal_j\,\eta_j}{ 2\,\Dcal_{0,j}^{1/2}|\rho_j|^{1/2}}~,\label{eq:dvz_approx_sqrtb}
\ee
which is exactly~\eqref{eq:frR}. The factor $1/2$ in~\eqref{eq:dvz_approx_sqrtb}
arises because the range of integration includes both angles and~\eqref{eq:IRj} must be divided by 2, in order to avoid double counting. The resonant response is proportional to
\be
\eta = 2\, (\kappa ^2+4)^{-3/4} \csc{\[\frac1{2}\arctan{(2/\kappa)}\]}~,\label{eq:def_eta}
\ee
which is always positive, because $\kappa >0$ as $\Dcal_{0} >0$. The factor $\eta$ is shown
as a function of $\kappa$ (which is a rough measure of the distance between the roots) in
Fig.~\ref{fig:xifactor}. We observe that the maximum value is attained at $\kappa =2/\sqrt{3}\approx1.16$,
that is when the roots are at a distance $\mathcal{O}(\chi^{1/2})$ apart. Note also that by taking the limit
of the resonant angles being away from each other, that is by taking
the limit $\kappa\gg1$, $\eta\sim 2/ \sqrt{\kappa}$ and~\eqref{eq:dvz_approx_sqrtb} reduces to~\eqref{eq:fR_b}. Consequently, \eqref{eq:dvz_approx_sqrtb} is a valid asymptotic expression regardless of the distance between the roots $\thet_j$. The accuracy of~\eqref{eq:fR_b} and~\eqref{eq:dvz_approx_sqrtb} in comparison with the numerically obtained integral is shown in Fig.~\ref{fig:Rdvz}.

\begin{figure}
%\centerline{\includegraphics[width=.9\textwidth,trim = 30mm 1mm 30mm 0mm, clip]{Db_circ_2.eps}}
\centerline{\includegraphics[width=.95\textwidth,trim = 45mm 0mm 40mm 0mm, clip]{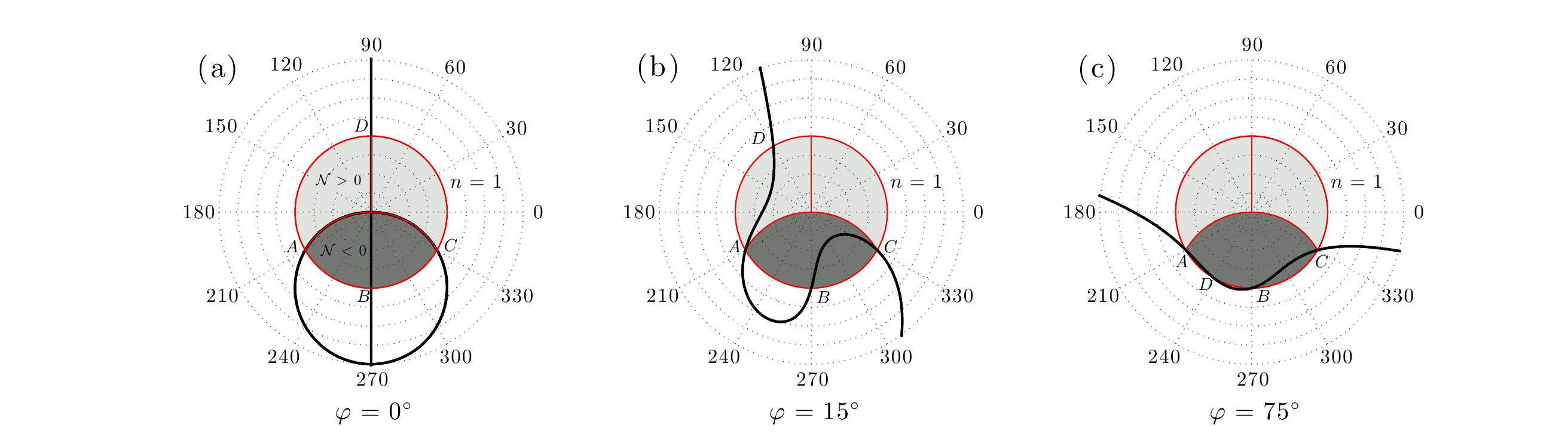}}
\caption{Zero contours of $\Db(\thet,\nv)$ for (a) zonal jet perturbations ($\varphi=0\deg$), (b) non-zonal perturbations with $\varphi=15\deg$ and (c) non-zonal perturbations with $\varphi=75\deg$ in a $(\thet,n)$ polar plot. Shaded areas mark $n\le 1$. Light shade corresponds to $(\thet,n)$ satisfying $\sin\thet>-n/2$ for which we have positive resonant contributions ($\Ncal>0$), while dark areas correspond to $\sin\thet<-n/2$ for which we have negative resonant contributions ($\Ncal<0$). Points of intersection of the $\Db=0$ curve with the unit circle are marked with $A$, $B$, $C$, $D$. The radial grid interval is $\Delta n=0.25$. The curve $\Db=0$ does not enter the $\Ncal>0$ area for $60\deg\le\varphi\le120\deg$.}\label{fig:Db}
\end{figure}

The sign of the resonant contribution depends only on the sign of $\Ncal$. From~\eqref{eq:defNcal}
we see that $\Ncal>0$ when $\sin\thet > -n/2 $ for $n<1$; this region is highlighted with light shading in Fig.~\ref{fig:Db}. It should be noted that for the important case of zonal jet perturbations ($\varphi=0\deg$) the resonant contribution is exactly zero because $\Ncal_j=0$, as shown in Fig.~\ref{fig:Db}\hyperref[fig:Db]{a}. The asymptotic behavior of the feedback factor for this case is found from the non-resonant part of the integral. Expanding in this case the integrand for $\chi\ll1$, we obtain to leading order:
\be
\f_{r} \approx \f_{r}^{(\textrm{NR})}=(1-n^2)(2+\mu) \b^{-2} + \Ocal(\b^{-4})~,\label{eq:dvzR_largeb}
\ee
with the maximum feedback gain
\be
\f_{r,\textrm{max}} = (2+\mu) \b^{-2} + \Ocal(\b^{-4})~,\label{eq:sr_largeb_max}
\ee
occurring for $n\to0$.

Consider now non-zonal perturbations $(\varphi\ne0\deg)$. There is a large region in the $(n,\varphi)$ plane (region~D in Fig.~\ref{fig:R_Fcal}\hyperref[fig:R_Fcal]{a}) in which $\Db$ has no roots and $\fr = \Ocal(\b^{-2})$.
For larger values of $n$ (region~B in Fig.~\ref{fig:R_Fcal}\hyperref[fig:R_Fcal]{a}), and for any given $\varphi$, $\Db=0$ for exactly two $\thet_j$ that satisfy the inequality $\sin\thet_j<-n/2$. Consequently, $\Ncal_j<0$ and
the resonant contribution from these roots is negative. For even larger values of $n$ (regions A and C in
Fig.~\ref{fig:R_Fcal}\hyperref[fig:R_Fcal]{a}), $\Db$ has exactly 4 roots. Only two of the roots in region A produce positive resonant contributions. Note also that region A extends to $\varphi < 60\deg$ and $\varphi>120\deg$.\footnote{It can be shown that fluxes from the resonant contributions for $n<1$ are necessarily downgradient (negative)
for $60\deg\le\varphi\le120\deg$.
%Proof: There are 4 roots of $\Db=0$ on $n=1$: $\thet=210\deg$, $270\deg$, $330\deg$ and $\thet=90\deg+2\varphi$ and consequently for a given $\varphi$ the locus in the plane $(\thet,\nv)$ of the roots of $\Db$
%in order to produce positive resonant contribution, which is predicated on $\Ncal>0$,
%must enter the region with $\Ncal>0$. The bounding curve $\Ncal=0$ separating positive from negative
%contributions, shown in Fig.~\ref{fig:Db}, crosses $n=1$ at the angles $210\deg$ and $330 \deg$
%that are also roots of $\Db$ on $n=1$. If all the roots of $\Db =0$ on $n=1$ are in the sector
%$210\deg \le \thet \le 330\deg$, implying that $60 \deg \le \varphi \le 120 \deg$, the
%continuation of the curve $\Db=0$ into $n<1$
%will remain in $\Ncal <0$ and can not cross into $\Ncal >0$
%(since $\Db=0$ and $\Ncal=0$ can intersect only at $n=1$ and for these $\varphi$ the
%curve $\Db=0$ at $\thet= 210\deg$ and $330\deg$ continues into $\Ncal<0$ for $n<1$).
Proof: A positive contribution is produced when the $\Db=0$ curve enters into the $\Ncal>0$, highlighted with light grey
in Fig.~\ref{fig:Db}. There are 4 roots of $\Db$ on the unit circle $n=1$ (on which also $\Ncal=0$), at angles:
$\thet=210\deg$, $270\deg$, $330\deg$ and $\thet=90\deg+2\varphi$ (marked with $A$, $B$, $C$ and $D$ respectively). The $\Db=0$ curve can cross the curve $AOC$, which separates positive from negative $\Ncal$, only at points $A$ and $C$, since $\Db=0$ only at these points on $AOC$.
 %, and $\Db\ne 0$ on the rest of $AOC$ for $n<1$.
Therefore, the $\Db=0$ curve can enter the $\Ncal>0$ region i) through $D$, if it lies outside the arc $ABC$, and/or ii) through $A$, $C$. However, for $60\deg\le \varphi \le 120\deg$ point $D$ lies within the arc $ABC$ and moreover, the gradient ${\bm\nabla} \Db$ at points $A$ and $C$ is oriented in such way that does not allow the $\Db=0$ curve to enter $\Ncal>0$, as $\partial_n\Db<0$ and $\partial_\thet \Db\le 0$ ($\partial_\thet \Db\ge 0$) at point $A$ (point $C$).}

The maximum response, which is $\Ocal(\b^{-1/2})$, arises in region A close to the curve separating regions~A and C where $\kappa \approx 1.16$. While the roots of $\Db$ are independent of $\b$, the location and the size of the region of maximum response depends on $\b$ through the dependence of $\kappa$ on $\b$. However, as $\b$ increases this dependence is weak and as $\b \rightarrow \infty$ the maximum response occurs in a narrow region near $n \approx 0.5$ and $\varphi \approx 10\deg$, marked with a star in Fig.~\ref{fig:R_Fcal}\hyperref[fig:R_Fcal]{a}. The width of this region decreases with $\b$, making it exceedingly hard to locate for large $\b$, and the asymptotic approach of $(n,\varphi)$ to $(0.5,10\deg)$ is shown in Figs.~\ref{fig:frmax}\hyperref[fig:frmax]{b,c}.

    % !TEX root = ../thesis.tex

\chapter{Formal equivalence between the S3T instability of a homogeneous equilibrium with the modulational instability of a corresponding basic flow}
\label{app:MI}

 In this Appendix we  demonstrate the formal equivalence between the modulational instability  (MI) of any solution of the barotropic equation, which may be in general time dependent but has stationary power spectrum, with the S3T instability of the homogeneous state with the same power spectrum.
Consider a  solution $\psi_G(\xv,t)$, with vorticity $\z_G = \Delta \psi_G$, of the inviscid and unforced nonlinear barotropic equation~\eqref{eq:nl} with time-independent power spectrum. Because $J(\psi_G, \z_G)=0$,
$\z_G$ satisfies the equation
\begin{align}
\partial_t\,\z_G& =\Lcal^{\textrm{(h)}} \z_G\,\label{eq:psiG}\ ,
\end{align}
with $\Lcal^{\textrm{(h)}} =\zhat\cdot\( \bv\times\nablav\)\Del^{-1}$. Linear perturbations $\d\z$ to this solution evolve according to the equation:
\begin{align}
\partial_t\,\d\z& =\Lcal\,\d\z\ ,\label{eq:dpsi_lorenz}
\end{align}
where
\begin{align}
\Lcal &= \underbrace{-\uv_G \cdot\nablav + (\Del\uv_G)\cdot\nablav\Del^{-1}}_{\Lcal_G'}+\underbrace{\zhat\cdot\( \bv\times\nablav\)\Del^{-1}}_{\Lcal^{\textrm{(h)}}}=\Lcal_G' + \Lcal^{\textrm{(h)}}\ ,\label{eq:C4}
\end{align}
is the time-dependent linear operator about $\z_G$ that has been decomposed into a spatially homogeneous
operator, $\Lcal^{\textrm{(h)}}$, that governs the evolution of $\z_G$ and the inhomogeneous operator
$\Lcal_G'$ that depends on $\z_G$.  The hydrodynamic instability of  $\z_G$ is ascertained when
the largest Lyapunov exponent of (\ref{eq:dpsi_lorenz}) is positive.

We proceed with the study of the MI by decomposing the perturbation into a mean $\d Z=\<\d\z\>$ and deviations from the mean $\d \z'=\d \z-\d Z$, where $\<\;\bullet\;\>$ is an averaging operation. The averaging operation in MI is projection to the eigenstructure with wavenumber $\nv$, which is orthogonal to $\z_G$, because only orthogonal eigenstructures to $\zeta_G$ could become unstable. With this averaging  operator $\< \zeta_G \> =0$, and therefore $\z_G=\z'_G$, whereas the perturbations has a non-zero mean, $\d Z$, and a deviation and is expressed as $\d \z = \d Z +\d \z'$.
For example, if $\psi_G$ is  a sum of Rossby waves as in~\eqref{eq:C1} the perturbation field from Bloch's theorem comprises Fourier components with wavenumbers $\nv,~\nv\pm \pv_j,~\nv \pm2\pv_j,~\nv\pm3\pv_j,\dots$ for all the $\pv_j$. In this case $\d Z$ is a plane wave with wavenumber $\nv$ and $\d \z'$ comprises the remaining Fourier components. With these definitions~\eqref{eq:dpsi_lorenz} is equivalently written as:
\begin{align}
\partial_t\(\d Z+\d\z'\)& =\Lcal'_G \d Z + \Lcal^{\textrm{(h)}}\d\z' + \Lcal'_G\, \d\z' + \Lcal^{\textrm{(h)}} \d Z\ ,\label{eq:dZ+dz}
\end{align}
where $\Lcal'_G $ is primed in order to stress that the operator linearly depends on the
deviation quantity $\z'_G$. Equation~\eqref{eq:dZ+dz} is then separated to form an equivalent system of equations for the evolution of the mean perturbation, $\d Z$, and the deviation perturbation, $\d\z'$:
\begin{subequations}
\label{eq:system}
\begin{align}
\partial_t\,\d Z& = \Lcal^{\textrm{(h)}} \d Z + \< \Lcal'_G\d\z' \>\ ,\label{eq:MIdZ}\\
\partial_t\,\d\z'& = \Lcal^{\textrm{(h)}}\d\z' +\Lcal'_G\,\d Z +\Lcal'_G\,\d\z'- \< \Lcal'_G\,\d\z' \>\ .\label{eq:MIdz}
\end{align}\end{subequations}
The stability equation (\ref{eq:dpsi_lorenz}) and the stability equations (\ref{eq:system}) for $\d Z$ and $\d \z'$ are equivalent.

In MI studies usually the term $\Lcal'_G\,\d\z'- \< \Lcal'_G\,\d\z' \>$ in~\eqref{eq:MIdz} is neglected and the stability of the following simpler system is studied:
\begin{subequations}
\label{eq:systemMI}
\begin{align}
\partial_t\,\d Z& = \Lcal^{\textrm{(h)}} \d Z + \< \Lcal'_G\d\z' \>\ , \label{eq:MI1}\\
\partial_t\,\d\z'& = \Lcal^{\textrm{(h)}}\d\z' +\Lcal'_G\,\d Z\ .\label{eq:MI2}
\end{align}\label{eq:MIdz_trunc}\end{subequations}
For example, if $\psi_G$ is in the form of~\eqref{eq:C1} the  neglected term comprises waves with wavevectors $\nv\pm2\pv_j,~\nv\pm3\pv_j,\dots$ and the truncated system~\eqref{eq:MIdz_trunc} allows only interaction between the primary finite amplitude waves $\pv_j$, the perturbation $\nv$
and the waves $\nv\pm\pv_j$. If $\z_G$ is a single wave $\pv$ (as in MI studies), \eqref{eq:systemMI} is referred to as the 4 mode truncation or ``4MT'' system since it comprises only modes $\pv$, $\nv$ and $\nv\pm\pv$.

However, instead of studying the MI stability of $\d Z$ and $\d \z'$ using the approximate equations~\eqref{eq:systemMI}, we can equivalently study the stability of $\d Z$ and
\begin{align}
\d C(\xv_a,\xv_b,t) &= \<\bit\right. \z'_{G}(\xv_a,t)\,\d\z'(\xv_b,t)+\z'_{G}(\xv_b,t)\,\d\z'(\xv_a,t)\bit \left.\bit\> \nonumber\\
&\equiv \<\bit\right.\z'_{G,a}\,\d\z'_b+\z'_{G,b}\,\d\z'_a \left.\bit\>\ .
\end{align}
With these definitions we obtain from~\eqref{eq:psiG} and~\eqref{eq:MI2} the evolution equation for $\d C$:
\begin{align}
\partial_t \d C &= \< (\partial_t\z'_{G,a})\,\d\z'_b + (\partial_t\z'_{G,b})\,\d\z'_a + \z'_{G,a}\,(\partial_t\d\z'_b)+ \z'_{G,b}\,(\partial_t\d\z'_a) \bit \>\nonumber\\
 &= \(\bit\right.\Lcal^{\textrm{(h)}}_a +\Lcal^{\textrm{(h)}}_b\left.\bit\)\d C+ \<\bit\right. \z'_{G,a}\, \Lcal'_{G,b}\,\d Z_b + \z'_{G,b}\, \Lcal'_{G,a}\,\d Z_a \bit \left.\bit\>\ .\label{eq:c8}
 \end{align}
We note from the definition of $\Lcal'_G$ (cf.~\eqref{eq:C4}) that:
%\begin{align}
%\Lcal'_{G}\,\d Z &=(\partial_y\psi'_G)(\partial_x\d Z) -(\partial_x\psi'_G)(\partial_y\d Z) - (\partial_y\z'_G)(\partial_x\d\Psi)+(\partial_x\z'_G)(\partial_y\d\Psi) \nonumber\\
%&=(\Del\,\d V)(\partial_y\psi'_G) +(\Del \,\d U)(\partial_x\psi'_G) - \d V\,(\partial_y\z'_G)-\d U\,(\partial_x\z'_G) \nonumber\\
% %& = \[-\uv'_\pv\cdot{\bm\nabla} + (\Del\uv'_\pv)\cdot{\bm\nabla}\Del^{-1}\]\d Z \nonumber\\
%% & = \varepsilon_{jk}(\partial_{x_k}\psi'_G) (\partial_{x_j} \d Z) - \Del (\varepsilon_{jk} \partial_{x_k}\psi'_G)(\partial_{x_j} \d\Psi) \nonumber\\
%% & = -\varepsilon_{kj}(\Del \partial_{x_j} \d \Psi)(\partial_{x_k}\psi'_\pv) + \varepsilon_{kj} (\partial_{x_j} \d \Psi) (\partial_{x_k}\z'_\pv)\nonumber\\
% & = (\Del\,\d\Uv)\cdot({\bm\nabla} \psi'_G) - (\d\Uv)\cdot({\bm\nabla}\z'_G) = \d\Acal \;\z'_G\ .
%\end{align}
\begin{align}
\Lcal'_{G}\,\d Z %&= -\(\zhat\times{\bm\nabla\psi'_G}\) \cdot\nablav \d Z+ \[\Del\(\zhat\times{\bm\nabla\psi'_G}\)\]\cdot\nablav\Del^{-1}\d Z\nonumber\\
&= -\(\zhat\times{\bm\nabla\psi'_G}\) \cdot\nablav \d Z+ \(\zhat\times{\bm\nabla\z'_G}\)\cdot\nablav\d \Psi\nonumber\\
&= \(\bit\zhat\times{\bm\nabla\d Z}\) \cdot\nablav \psi'_G - \(\bit\zhat\times{\bm\nabla\d\Psi}\)\cdot\nablav \z'_G\nonumber\\
 & = (\Del\,\d\Uv)\cdot(\nablav \psi'_G) - (\d\Uv)\cdot(\nablav\z'_G) = \d\Acal \;\z'_G\ ,
\end{align}
where $\d\Uv=\zhat\times\nablav\d\Psi$ is the velocity field associated with $\d Z$ and $\d\Acal = -\d\Uv\!\cdot\!\nablav + \[(\Del\;\d\Uv)\!\cdot\!\nablav\]\Del^{-1}$ is the operator that also
appears in~\eqref{eq:s3t_pert_dC}. As a result,~\eqref{eq:c8} becomes:
\be
\partial_t \d C =\(\bit\right.\Lcal^{\textrm{(h)}}_a +\Lcal^{\textrm{(h)}}_b\left.\bit\)\d C + \( \d\Acal_a +\d\Acal_b\bit\) C^G\ ,
\ee
where $C^G= \< \z'_{G,a} \z'_{G,b} \>$. Returning now to~\eqref{eq:MI1} we note that $\< \Lcal'_G\d\z' \> = \Rcal(\d C) $, where $\Rcal(\d C)$ is defined in~\eqref{eq:def_Rcal}, as:
%\begin{align}
%\Rcal( \d C ) &=-\partial_x\,\[- \frac1{2}(\Del^{-1}_a\partial_{y_a}\!+\!\Del^{-1}_b\partial_{y_b}) \< \bit \z'_{G,a}\,\d\z'_b+\z'_{G,b}\,\d\z'_a \>\]_{\xv_a=\xv_b}-\nonumber\\
%&\qquad\qquad\qquad-\partial_y\,\[ \frac1{2}(\Del^{-1}_a\partial_{x_a}\!+\!\Del^{-1}_b\partial_{x_b}) \< \bit\z'_{G,a}\,\d\z'_b+\z'_{G,b}\,\d\z'_a \>\]_{\xv_a=\xv_b} \nonumber\\
% &=\< -\partial_x\,\( \bit u'_G\,\d\z' + \z'_G\,\d u'\) -\partial_y\,\( \bit v'_G\,\d\z' + \z'_G\,\d v'\) \>\nonumber\\
% &=\<\bit -u'_G\,\partial_x\d\z'-v'_G\,\partial_y\d\z' -\(\partial_x\,\z'_G\)\d u'-\(\partial_y\,\z'_G\)\d v' \> = \< \Lcal'_G\,\d\z \>\ .
%\end{align}
\begin{align}
\Rcal( \d C ) &=-\nablav\cdot\[ \frac{\zhat}{2}\times(\nablav_a\Del^{-1}_a+\nablav_b\Del^{-1}_b) \< \bit \z'_{G,a}\,\d\z'_b+\z'_{G,b}\,\d\z'_a \>\]_{\xv_a=\xv_b}\nonumber\\
&=-\nablav\cdot\left\{\vphantom{\frac1{2}} \zhat\times\< \bit (\nablav\psi'_{G})\d\z'  + (\nablav\d\psi') \z'_G\>\right\}\nonumber\\
&=-\nablav\cdot\< \bit \uv'_G\,\d\z'  + \d\uv'\, \z'_G\>\nonumber\\
&=\<\bit  -\uv'_G\cdot\nablav\,\d\z'  +  (\Del\uv'_G)\cdot\nablav \d\psi'\> = \< \Lcal'_G\,\d\z' \>\ .
\end{align}

Consequently, the MI of $\z'_G$ in the approximation~\eqref{eq:MIdz_trunc} is equivalently determined from the stability of the system: \begin{subequations}\begin{align}\partial_t\,\d Z& = \Lcal^{\textrm{(h)}} \d Z + \Rcal(\d C)\ , \label{eq:M1}\\
\partial_t \d C & = \(\bit\right.\Lcal^{\textrm{(h)}}_a +\Lcal^{\textrm{(h)}}_b\left.\bit\)\d C + \(\d\Acal_a +\d\Acal_b\bit\) C^G\ ,
\end{align}\label{eq:system1}\end{subequations}
 which is identical to equations~\eqref{eq:s3t_dZdC} that determine the S3T stability of the homogeneous equilibrium with zero mean flow, $\Uv^e=0$, and
equilibrium covariance $C^e=C^G$ under the ergodic assumption that ensemble averages are equal to averages under operation $\< \;\bullet\; \>$.

For example, consider the nonlinear solution
\be
\psi(\xv,t) = \int\limits_0^{2 \pi} a(\thet) \cos ( \pv \cdot \xv - \om_{\pv}t) \,\df \thet\ ,
\label{eq:C10}
\ee
with wavevectors $\pv=(\cos\thet,\sin\thet)$ on the unit circle ($p=1$) and take $\bv=(0,\b)$. Expanding the plane waves into cylindrical waves:
\be
e^{\i\[ ( x+\b t)\cos\thet +y\sin\thet \]} = \sum_{m=-\infty}^{+\infty} \i^m J_m(R) e^{\i m (\phi-\thet)}\ ,
\ee
with $R^2 = (x+\b t)^2 + y^2$, $\phi = \arctan\[y/(x+\b t)\bit\]$ and $J_m$
the $m$-th Bessel function of the first kind, this can be shown to be the non-dispersive
structure
\be
\psi (x+\b t,y) = \real\[ \sum_{m=-\infty}^{+\infty} \gamma_m\,J_m(R) e^{\i m \phi} \]\ ,
\ee
propagating westward with velocity $\b$, where $\gamma_m = \int_0^{2\pi} a(\thet)\,e^{-\i m \thet}\,\df\thet$. The results in this Appendix show that the MI of the propagating structure~\eqref{eq:C10} in the approximation~\eqref{eq:MIdz_trunc} is equivalent to the S3T instability of the homogeneous equilibrium with covariance $C^e$ prescribed by power spectrum $\hat{C}^e(\kv) = (2\pi)^2 \left| a(\thet)\right|^2\,\d(k-1)$. Note that this S3T equilibrium is also an exact  homogeneous statistical equilibrium of the nonlinear barotropic equations without approximation.

%Note that  the vanishing of  $J(\psi,\Del\psi)=0$ for~\eqref{eq:C1} and~\eqref{eq:C10} implies that waves forced exactly on a ring can not advect each other and do not lead to a cascade. Further, waves stochastically forced with forcing covariance with spectrum $\hat{Q}(\kv)=4\pi\,\Gcal(\thet) \delta(k-1)$ produce a homogeneous equilibrium of the full nonlinear equations, the stability of which is governed, because of the absence of eddy--eddy interactions, exactly and not approximately by the S3T stability equations. %{\color{green}\bf Consequently, the critical $\varepsilon_c$ for transition to an inhomogeneous state is predicted exactly and without approximation by the S3T dynamics in the case of delta-function ring forcing.}
%

    % !TEX root = ../thesis.tex

\chapter{Specification of the forcing structures used in chapter~\ref{ch:NLvsS3Tjas}}
\label{appsec:forc_spec_IRF_NIF}

In chapter~\ref{ch:NLvsS3Tjas} three spatial structure of  stochastic forcing are used in the investigation of the correspondence among S3T, QL and NL dynamics: a forcing with narrow isotropic ring spectrum (IRFn), a forcing with wide isotropic ring spectrum (IRFw) and a forcing with non-isotropic spectrum (NIF).

For the IRFn we take the forcing spectrum
\be
\hat{Q}_{\kv}  = \left\{
	\begin{array}{ll}
		c  & \mbox{if\ \ } |k-k_f| \le  \d k_f  \\
		0 & \mbox{if\ \ } |k-k_f| >  \d k_f \mbox{ or } k_x=0\ .
	\end{array}
\right.\label{eq:spec_IRFn}
\ee
The constant $c$ is chosen so that $\hat{Q}_{\kv}$ satisfies
\be
\sum_{k_x,k_y} \frac{\hat{Q}_{\kv}}{2 k^2} =1\ .\label{eq:Qhat_norm_discrete}
\ee
and therefore, according to~\eqref{eq:Qhat_norm}, the energy injection rate by $\hat{Q}_{\kv}$
is a unit. For the narrow ring forcing we choose $k_f=14$ and $\d k_f=1$.

For the IRFw we take the forcing spectrum to be
\be
\hat{Q}_{\kv}  = \left\{
	\begin{array}{ll}
		 c\exp{\[-{\(k-k_f\)^2}/({2\,\d k_f^2})\]}  & \mbox{if } k_x\ne0  \\
		0 &  \mbox{if } k_x=0  
	\end{array}
\right.\label{eq:spec_IRFw}
\ee
Again $c$ is chosen so that~\eqref{eq:Qhat_norm_discrete} is satisfied. For the wide ring forcing we choose $k_f=14$ and $\d k_f=8$.

For NIF we force the zonal wavenumbers $k_x=1,\dots,N_k$ with power:
\be
\hat{Q}_{\kv} = c_{k_x} d^2 e^{-k_y^2 d^2}\ ,\label{eq:spec_NIF}
\ee
with constants $c_{k_x}$ chosen in such manner so that for a fixed zonal wavenumber $k_x$,
\be
\sum_{k_y} \frac{\hat{Q}_{\kv}}{2(k_x^2 + k_y^2)} = \frac1{N_k}\ ,
\ee
so that all zonal wavenumbers $k_x$ inject the same energy input rate and the total injection rate by $\hat{Q}_{\kv}$ is unity. For the anisotropic forcing we force wavenumbers $k=1,\dots,14$ and $d=1/5$.

Comparison of the forcing spectra structure~\eqref{eq:spec_IRFn}, \eqref{eq:spec_IRFw} and~\eqref{eq:spec_NIF}  as well all forcing realizations that they induce are shown in Fig.~\ref{fig:Qkl_Fxy_NIF_IRF}.

\chapter{Determination of inhomogeneous zonal jet S3T equilibrium solutions using Newton's iteration}
\label{app:S3Tnewton}

In this appendix we present a method for determining stationary equilibrium solutions $\(U^e(y),C^e(x_a-x_b,y_a,y_b)\bit\)$ of the S3T system~\eqref{eq:s3t}.
Remember that for solutions of the form $\(\bit U(y,t),C(x_a-x_b,y_a,y_b,t)\)$ and for the usual orientation of $\bv=(0,\b)$ the S3T system~\eqref{eq:s3t} collapses to the simpler S3Tz system~\eqref{eq:s3tz} and therefore it is sufficient to determine stationary equilibrium solutions of the S3Tz system, or equivalently (cf.~Appendix~\ref{sec:numerical_S3Tz}), stationary solutions $\({\bm U}^e,\C^e_{1},\dots,\C^e_{N_k}\)$ of~\eqref{eq:S3Tz_discrete} that satisfy the matrix equations:\begin{subequations}
\begin{align}
0 &=\sum_{\substack{k_x=1}}^{N_k}2 \real\[\vphantom{\frac1{2}}\vecd{\(\i k_x\, \DDel_{k_x}^{-1} \C^e_{k_x} \bit\)}\]- {\bm U}^e\ ,\label{eq:s3tz_discrete_mean_N}\\
0 &= \A_{k_x}\({\bm U}^e\)\,\C^e_{k_x}+\C^e_{k_x}\, \A_{k_x}\({\bm U}^e\)^\dag  + \varepsilon \,\Q_{k_x}\ ,\quad\text{for }k_x=1,\dots,N_k~.\label{eq:s3tz_discrete_pert_N}\end{align}\label{eq:S3Tz_discrete_N}\end{subequations}

We determine the equilibrium solution satisfying~\eqref{eq:s3tz_discrete_mean_N} through Newton's iterations. However instead of iterating  both the ${\bm U}$ and $\C_{k_x}$ towards the solution, which is computationally very costly,  for each  flow iteration ${\bm U}$ we determine the $\C_{k_x}$ that satisfy~\eqref{eq:s3tz_discrete_pert_N} and treat with this understanding the $\C_{k_x}$  as  linear functions of ${\bm U}$, writing  $\C_{k_x} = \Lcal_{k_x}\!\({\bm U}\)$. This approach has  the disadvantage that we constrain the iteration to the manifold of hydrodynamically stable flows for which there exist physical steady covariances that satisfy~\eqref{eq:s3tz_discrete_pert_N}. Under these conditions we can obtain the equilibrium state by solving for ${\bm U}^e$  the linear equation:
\be
 {\bm G}\( {\bm U}^e \) \equiv\sum_{\substack{k_x=1}}^{N_k} 2 \real\[\vphantom{\frac1{2}}\vecd{\(\i k_x\, \DDel_{k_x}^{-1} \,\Lcal_{k_x}\!\({\bm U}^{e}\)\bit\)}\]- {\bm U}^e= 0\ .\label{eq:G(U)_0}
\ee
In order to solve~\eqref{eq:G(U)_0} with Newton's method we start the iteration by selecting a hydrodynamically stable flow ${\bm U}^0$. If ${\bm G}\({\bm U}^{(0)}\)=0$ no iteration is needed. If not, assume that the equilibrium ${\bm U}$ is nearby, and hence to first order it must satisfy:
\be
0=G_i\({\bm U}\) \approx G_i\({\bm U}^{(0)}\) +\[\left.\frac{\partial{\bm G}}{\partial{\bm U}}\right|_{{\bm U}^{(0)}}\]_{ij}\(U_j-U_j^{(0)}\)\ ,\label{eq:Gexpansion}
\ee
where $\partial{\bm G}/\partial{\bm U}$ is  is the Jacobian matrix of ${\bm G}$ with elements $\[\partial{\bm G}/\partial{\bm U}\]_{ij} = \partial G_i / \partial U_j$, $U_i$ and $G_i$ are the $N_y$ elements of ${\bm U}$ and ${\bm G}$ respectively and subscript ${{\bm U}^{(0)}}$ denotes that the matrix elements are evaluated at ${\bm U}={\bm U}^{(0)}$. The iteration is continued by  taking as the new iterant the ${\bm U}$ that satisfies~\eqref{eq:Gexpansion}, i.e.,
\be
{\bm U}^{(1)} = {\bm U}^{(0)} -\left.\(\frac{\partial{\bm G}}{\partial{\bm U}}\)^{-1}\right|_{{\bm U}^{(0)}} \,{\bm G}\({\bm U}^{(0)}\)\ ,\label{eq:newton_nextU}
\ee
where $\( \partial{\bm G}/\partial{\bm U} \)^{-1}$ is inverse of the Jacobian matrix $\partial{\bm G}/\partial{\bm U}$. We approximate the elements of $\left.\partial{\bm G}/\partial{\bm U}\right|_{{\bm U}={\bm U}^{(0)}}$ with 
\be
\left.\frac{\partial G_i}{\partial U_j}\right|_{{\bm U}^{(0)}} \approx \frac{G_i\({\bm U}^{(0)}+h\,{\bm e}_j\) - G_i\({\bm U}^{(0)}-h\,{\bm e}_j\)}{2h}\ ,
\ee
where ${\bm e}_j$ is the unit-vector in the $j$-th direction with elements $[{\bm e}_j]_i=\d_{ij}$ and $h>0$ is sufficiently small. After ${\bm U}^{(1)}$ is calculated from~\eqref{eq:newton_nextU} we repeat the iteration. If the initial guess ${\bm U}^0$ is close to the equilibrium ${\bm U}^e$ the iteration converges rapidly to the equilibrium solution. We consider the iterations converged if 
\be
\dfrac{\sum_j \[G_j\({\bm U}^{(n)}\)\]^2}{\sum_j \[ U_j^{(n)}\]^2} < 10^{-14}\ .
\ee

\chapter{Stability of inhomogeneous ~S3T ~equilibrium ~solutions}
\label{app:S3Tlyap}

%In this appendix we describe the method by which we calculate the S3T stability of inhomogeneous states stationary or non-stationary S3T states. %To study the stability of a time-depended solution done by determining the first Lyapunov exponent and its corresponding first Lyapunov vector. For stationary solutions the first Lyapunov vector collapses to the maximally growing eigenfunction and the first Lyapunov exponent to its corresponding growth.

Consider a solution $(Z,C)$ of the S3T system:\begin{subequations}
\begin{align}
\partial_t Z &+ J \(\Psi, Z+  {\bm\beta}\cdot\mathbf{x} \) = \Rcal( C )-\,Z\ ,\label{eq:s3tpert_coupl_Z_v2}\\
\partial_t C_{ab} & = \[\bit\Acal_a(\Uv) + \Acal_b(\Uv)\]C_{ab} +\e\,Q_{ab}\ .\label{eq:s3tpert_coupl_C_v2}
\end{align}\label{eq:s3t_v2}\end{subequations}
with $\Acal$ defined in~\eqref{eq:def_Acal}. Linear perturbations $(\d Z,\d C)$ about this solution obey:\begin{subequations}
\begin{align}
\partial_t \,\d Z & = \Acal(\Uv)\,\d Z + \Rcal( \d C )\ ,\label{eq:s3tpert_coupl_dZ_v2}\\
\partial_t \,\d C_{ab} & = \[\bit\Acal_a(\Uv) + \Acal_b(\Uv) \]\d C_{ab} +\(\bit\d\Acal_a + \d\Acal_b\)C_{ab}\ ,\label{eq:s3tpert_coupl_dC_v2}
\end{align}\label{eq:s3tpert_v2}\end{subequations}
with $\d\Acal = \Acal(\Uv+\d\Uv)-\Acal(\Uv)$. The stability of $(Z,C)$ is determined by the top Lyapunov exponent of~\eqref{eq:s3tpert_v2}, which is calculated numerically using the power method. We initiate the integration of~\eqref{eq:s3tpert_v2} with a normalized perturbation state $(\d Z,\d C)$, using with any norm. At every time-step, $h$, we calculate the perturbation state growth,
\be
\la(jh) = \frac{\log\(\bit\|(\d Z(jh),\d C(jh))\|\)}{h} \ ,\quad j=1,2,\dots\ ,\label{eq:lambda_jh}
\ee
and then renormalize the perturbation state before moving to the next time-step. After sufficient time-steps the growth $\la$ converges to the top Lyapunov exponent and the state $(\d Z,\d C)$ to the first Lyapunov vector. 

The time-integration of the discretized~\eqref{eq:s3tpert_v2} proceeds as described in Appendix~\ref{sec:numS3Tgen} and system~\eqref{eq:s3tpert_v2} takes the form:\begin{subequations}
\begin{align}
\frac{\df}{\df t} \d \Zv &= \A(\Uv)\,\d\Zv + \R(\d\C)\ ,\\
\frac{\df}{\df t} \d\C  &= \A(\Uv) \,\d\C+ \d\C\,\[\A(\Uv)\]^\transp + \d\A\,\C  +\C\,\(\d\A\)^\transp\ ,\end{align}\label{eq:s3tpert_discrete}\end{subequations}
where $\R$ is defined in~\eqref{eq:def_Rvec}, $\A$ is defined in~\eqref{eq:def_Amatr} and $\d\A = \A(\Uv+\d\Uv)-\A(\Uv)$. In~\eqref{eq:s3tpert_discrete} $\d\Zv$ is an $(N_xN_y)$-column vectors while $\d\C$ is an $(N_xN_y)\times(N_x N_y)$ matrix. %This sums up to a state of $N_xN_y+N_x^2N_y^2$ variables.

%For stationary equilibria $(Z^e,C^e)$ we use the equilibrium state $(Z^e,C^e)$ in~\eqref{eq:s3tpert_v2}. %Therefore stability of stationary equilibria requires in general time-integration of a system of $N_xN_y+N_x^2N_y^2$ variables.

\section{Stability of zonal jet equilibria}

The homogeneity in $x$ of zonal jet equilibria enables us to seek S3T eigenfunctions in the form%~\eqref{eq:S3Teigen_nx} (cf.~Appendix~\ref{app:bloch}). In a similar manner we expand the S3T perturbation state as:
\begin{subequations}
\begin{align}
\d Z(\xv,t) &= e^{\i n_x x} \int\df n_y\; \a_{n_x}(n_y,t)\,e^{\i n_y y}\ ,\\
\d C(\xv_a,\xv_b,t) &=  e^{\i n_x(x_a+x_b)/2} \int\df k_x\; \d\tilde{C}_{n_x}(k_x,y_a,y_b,t)\,e^{\i k_x(x_a-x_b)}\ .
\end{align}\label{eq:S3Tpert_nx}\end{subequations}
i.e., the perturbation mean flow is a single harmonic in $x$ with wavenumber $n_x$ and the inhomogeneous part of the perturbation covariance is also a single harmonic with the same wavenumber. We will show that by using~\eqref{eq:S3Tpert_nx} the S3T perturbation system~\eqref{eq:s3tpert_v2} results in a smaller system of decoupled equations for Fourier components $\a_{n_x}(n_y,t)$ and $\d\tilde{C}_{n_x}(k_x,y_a,y_b,t)$. %The discrete version of each of these systems for a fixed $n_x$ has dimension $N_y+N_x N_y^2$, i.e., reduced $N_x$ times compared to the initial $N_xN_y+N_x^2N_y^2$.

In an infinite domain the wavenumbers $n_x$, $n_y$ and~$k_x$ assume continuous values. However, system~\eqref{eq:s3tpert_coupl_dC_v2} is solved in a channel of dimension $L_x\times L_y$ with periodic boundary conditions. For a box with $L_x=L_y=2\pi$, the periodic boundary conditions on $\d Z$ impose that  $n_x$ and $n_y$ take only integer values and the periodicity on $\d C$ imposes that
\begin{align}
k_x= \left\{
	\begin{array}{ll}
		  \text{integer}  &\hspace{-.5em} \textrm{,\ \ for $n_x$ even\ \ ,} \\
		 \\
		  \textrm{half-integer}  &\hspace{-1em} \textrm{,\ \ for $n_x$ odd\ \ .}\\
	\end{array}\right.
\end{align}
(A number $m$ is called half-integer when $m+1/2\in\mathbb{Z}$, which implies that $m=(2\kappa+1)/2$ for $\kappa\in\mathbb{Z}$.) Therefore, after redefining $k_x\to k_x+n_x/2$ so that $k_x$ assumes integer values for any value of $n_x$, the eigenfunction~\eqref{eq:S3Tpert_nx} takes the form:
\begin{subequations}\begin{align}
\d Z(\xv,t) &= e^{\i n_x x} \sum_{n_y}  \a_{n_x,n_y}(t)\,e^{\i n_y y}\ ,\label{eq:dZ_a_nxny}\\
\d C(\xv_a,\xv_b,t) &=  e^{\i n_x(x_a+x_b)/2} \sum_{k_x}  \d\tilde{C}_{k_x,n_x}(y_a,y_b,t) e^{\i (k_x-n_x/2)(x_a-x_b)}\ ,
\end{align}\label{eq:S3Tpert_nx_discrete}\end{subequations}
where notation~\eqref{eq:def_sumpos_posneg} is used in the summation. Notice that we have chosen to define $\d\tilde{C}_{k_x,n_x}$ as the Fourier coefficient that corresponds to $e^{\i(k_x-n_x/2)(x_a-x_b)}$. The equilibrium covariance $C^e$ is also expanded as in Appendix~\ref{sec:numerical_S3Tz},
\begin{subequations}\begin{align}
C^e(\xv_a,\xv_b) &= \sum_{k_x} \tilde{C}^e_{k_x}(y_a,y_b) e^{\i k_x(x_a-x_b)}\ .
%&= \frac1{2} \sum_{k_x=1}^{N_k} \left\{ \hat{C}^{E,R}_{k_x}(y_a,y_b) \cos{ \[k(x_a-x_b) \]}-\hat{C}^{E,I}_{k_x}(y_a,y_b) \sin{ \[k(x_a-x_b) \]} \right\}
\end{align}\end{subequations}

We want to write now the S3T perturbation system~\eqref{eq:s3tpert_v2} in terms of the Fourier coefficients $\a_{n_x,n_y}(t)$ and $\d\tilde{C}_{k_x,n_x}(y_a,y_b,t)$. Consider first a single harmonic of~\eqref{eq:dZ_a_nxny}  with  vorticity: $\d Z_{\nv} = -n^2\,e^{\i\nv\cdot\xv}$. Then:\begin{subequations}
\begin{align}
\[\d \Psi_{\nv}, \d \Uv_{\nv} , \nablav\d Z_{\nv} \bit\] &= \[\bit -1/n^2,-(\zhat\times\nablav)/n^2,\i \nv\]\d Z_\nv\ ,
%\[\d \Psi_{n_x}, \d \Uv_{n_x} , \nablav\d Z_{n_x} \bit\] &= \int\df n_y\,\[\bit -1/n^2,-(\zhat\times\nablav)/n^2,\i \nv\]\d Z_\nv\ ,
\end{align}\end{subequations}
and\begin{subequations}\begin{align}
\partial_{x_a} \d C  &= e^{\i n_x(x_a+x_b)/2}\sum_{k_x} \,\i k_x\, \d\tilde{C}_{k_x,n_x}(y_a,y_b) \,e^{\i (k_x-n_x/2)(x_a-x_b)} \ ,\\
\partial_{x_b} \d C &= e^{\i n_x(x_a+x_b)/2}\sum_{k_x} -\i\(k_x-n_x\) \d\tilde{C}_{k_x,n_x}(y_a,y_b) \,e^{\i (k_x-n_x/2)(x_a-x_b)} \ ,\\
\Del_{a} \d C  &= e^{\i n_x(x_a+x_b)/2}\sum_{k_x} \( \partial^2_{y_a} - k_x^2 \bit\)\d\tilde{C}_{k_x,n_x}(y_a,y_b) \,e^{\i (k_x-n_x/2)(x_a-x_b)} \ ,\\
\Del_{b} \d C  &= e^{\i n_x(x_a+x_b)/2}\sum_{k_x} \[ \partial^2_{y_b} - \(k_x-n_x\)^2 \]\d\tilde{C}_{k_x,n_x}(y_a,y_b) \,e^{\i (k_x-n_x/2)(x_a-x_b)} \ ,
\end{align}\end{subequations}
imply that:
\begin{subequations}\begin{align}
 -(&U^e_a \partial_{x_a} +U^e_b \partial_{x_b}) \,\d C_{n_x} = \nonumber\\
 &=e^{\i n_x(x_a+x_b)/2}\sum_{k_x}  \[  -\i k_x U^e_a\,\d\tilde{C}_{k_x,n_x}(y_a,y_b)  \right.\nonumber\\
&\hspace{6em}\left.+  \i (k_x-n_x) U^e_b\,\d\tilde{C}_{k_x,n_x}(y_a,y_b) \]\,e^{\i (k_x-n_x/2)(x_a-x_b)} \ ,\end{align}\begin{align}
 -(&\d U_a \partial_{x_a} + \d U_b \partial_{x_b} )  C^e = \nonumber\\
 &=-\sum_{k_x} \[\d U_a\, \i k_x \,\tilde{C}^e_{k_x}(y_a,y_b) + \d U_b \(-\i k_x\) \tilde{C}^e_{k_x}(y_a,y_b) \] \,e^{\i k_x(x_a-x_b)}\nonumber\\
%  &=\i n \sum_{k_x}  \[ e^{\i m x_a} e^{\i n y_a}\, \i k\,  \tilde{C}^e_{k_x}(y_a,y_b) + e^{\i m x_b} e^{\i n y_b} \(-\i k\) \tilde{C}^e_{k_x}(y_a,y_b) \] e^{\i k(x_a-x_b)}~\nonumber\\
%  &=\i n\, e^{\i n_x(x_a+x_b)/2} \sum_{k_x}  \[ e^{\i m (x_a-x_b)/2} e^{\i n y_a}\, \i k\,  \tilde{C}^e_{k_x}(y_a,y_b) + e^{-\i m (x_a-x_b)/2} e^{\i n y_b} \(-\i k\) \tilde{C}^e_{k_x}(y_a,y_b) \] e^{\i k(x_a-x_b)}\nonumber\\
  &=\i n_y \, e^{\i n_x(x_a+x_b)/2} \sum_{k_x} \left\{  \vphantom{\frac1{2}} \[  \(\i k_x\,\d E_{n_y,a}\)   \tilde{C}^e_{k_x}(y_a,y_b) \] \,e^{\i (k_x+n_x/2) (x_a-x_b)}\right.\nonumber\\
 &\hspace{9em}\left. + \[   \tilde{C}^e_{k_x}(y_a,y_b) \(-\i k_x\,\d E_{n_y,b}\) \] \,e^{\i (k_x-n_x/2) (x_a-x_b)}\right\}\nonumber\\
 &=\i n_y \, e^{\i n_x(x_a+x_b)/2} \sum_{k_x} \left\{  \vphantom{\frac1{2}} \[  \i (k_x-n_x)\,\d E_{n_y,a} \,\tilde{C}^e_{k_x-n_x}(y_a,y_b) \] \right.\nonumber\\
 &\hspace{9em}\left. + \[   \tilde{C}^e_{k_x}(y_a,y_b) \(-\i k_x\,\d E_{n_y,b}\) \]  \vphantom{\frac1{2}}\right\}e^{\i (k_x-n_x/2) (x_a-x_b)}\ ,\label{eq:dUadUbCE} \end{align}\end{subequations}
with the notation $\d E_{n_y} \equiv  e^{\i n_y y}$.  We see that each component of~\eqref{eq:s3tpert_coupl_dC_v2} has as a common factor the term $e^{\i n_x(x_a+x_b)/2}$. For any mean flow perturbation~\eqref{eq:dZ_a_nxny} we can rewrite~\eqref{eq:s3tpert_coupl_dC_v2} as a system of matrix equations for the perturbation covariances matrices $\d\C_{k_x,n_x}$, with elements $\[\d\C_{k_x,n_x}(t)\]_{ab} =\d\tilde{C}_{k_x,n_x}(y_a,y_b,t)$, and equilibrium covariance matrices $\C^e_{k_x}$ with elements $\[\C^e_{k_x}\]_{ab}=\tilde{C}^e_{k_x}(y_a,y_b)$. Consequently,  for the usual orientation $\bv=(0,\b)$, each $\d\C_{k_x,n_x}$ satisfies:\begin{align}
\frac{\df}{\df t}\d \C_{k_x,n_x} &=  \A^{e}_{k_x}\,\d\C_{k_x,n_x} + \d\C_{k_x,n_x}\(\A^{e}_{n_x-k_x}\)^\transp \nonumber\\
&\qquad+ \sum_{n_y}  \a_{n_x,n_y} \[\vphantom{\frac1{2}}\d\A_{k_x-n_x,n_x,n_y}\C^e_{k_x-n_x} + \C^e_{k_x}\(\d\A_{-k_x,n_x,n_y}\)^\transp\]\ ,
\label{eq:pert_dC_s_matr}\end{align}
with:\begin{subequations}\begin{align}
\A^e_{k_x} &\equiv -\i k_x \U^e  -\i k_x \(\bit \beta\I -\U^e_{yy} \) \DDel^{-1}_{k_x}  - \I \ ,\\
%\A^{e,\textrm{R}}_{k_x} &\equiv +\i k_x\U^e  + \i k_x \(\bit \beta\I -\U^e_{yy} \) \DDel^{-1}_{k_x}  - \I \ ,\\
\d\A_{k_x,n_x,n_y} &\equiv -\d\mathsf{E}_{n_y}\(+ n_yk_x\I+\i n_x \D_y\)\(\I+n^2\DDel^{-1}_{k_x}\)\ ,%\\
%\d\A^{\textrm{R}}_{k_x,n_x,n_y} &\equiv -\d\mathsf{E}_{n_y}\(- n_yk_x\I+\i n_x \D_y\)\(\I+n^2\DDel^{-1}_{k_x}\)\ ,
\end{align}\label{eq:def_AkRL_dAkRL}\end{subequations}
and $\d\mathsf{E}_{n_y}\equiv \diag(e^{\i n_y})$.

%From definitions~\eqref{eq:def_AkRL_dAkRL} it can be seen that the following symmetries hold:
%\begin{subequations}\begin{align}
%\A^{e,\textrm{L}}_{-k_x} &= \A^{e,\textrm{R}}_{k_x}\ ,\\
%\d\A^{\textrm{L}}_{-k_x,n_x,n_y} &= \d\A^{\textrm{R}}_{k_x,n_x,n_y} \ ,\\
%\d\A^{\textrm{L}}_{k_x,-n_x,n_y} &= -\d\A^{\textrm{R}}_{k_x,n_x,n_y} \ .
%\end{align}
%\end{subequations}
Because $\d C$ and $C^e$ are symmetric to the exchange $\xv_a\leftrightarrow\xv_b$ and further because $C^e$ is real, we have that\begin{subequations}\begin{align}
&\C^e_{k_x} = \(\C^e_{-k_x}\)^\transp= \(\C^e_{k_x}\)^\dag\ ,\label{eq:symm_Cek}\\
&\d\C_{-k_x,n_x} = \(\d\C_{k_x+n_x,n_x}\bit\)^\transp \ .\label{eq:symm_dCk}
\end{align}\end{subequations} From~\eqref{eq:symm_Cek} we see that is only necessary to determine $\C^e_{k_x}$ for non-negative $k_x$. In general, if $C^e$ has non-zero $\C^e_{k_x}$ only for $k_x=1,\dots,N_k$ then for a fixed $n_x$ we only need to solve for $2(N_k+n_x)+1$ perturbation covariance matrices $\d\C_{k_x,n_x}$, since perturbation covariances with $|k_x|>N_k+n_x$ are necessarily zero because they do not couple with any of the non-zero $\C^e_{k_x}$. In practice however, we only solve for the first $N_k+2n_x$ matrices $\d\C_{k_x,n_x}$ and deduce from them the rest $N_k+1$ using the symmetry~\eqref{eq:symm_dCk}. Note that for $n_x=0$ perturbations, in the case that $\C_{k_x=0}^e\ne0$ we see that we need to solve for $N_k+1$ perturbation covariances $\d\C_{k_x,0}$, and not for $N_k$ as it is claimed above. However, it can be proven that $\d\C_{k_x=0,n_x=0}=0$, even if $\C_{k_x=0}^e\ne0$ and therefore only $N_k$ perturbation covariances are non-zero.

Turning now to the mean flow perturbation equation~\eqref{eq:s3tpert_coupl_dZ_v2} we want to express the Reynolds stress divergence associated with perturbation covariance $\d C$ in terms of the covariances $\d\tilde{C}_{k_x,n_x}$. At point $\xv_m$ the Reynolds stress divergence is  \begin{align}
\left. \vphantom{\frac1{2}} \Rcal(\d C) \right|_{\xv=\xv_m}&=  \left. \vphantom{\frac1{2}}\frac1{2} \nablav\cdot\[ \vphantom{\frac1{2}} \(\partial_{y_a}\Del_a^{-1} +\partial_{y_b}\Del_b^{-1} ,-( \partial_{x_a}\Del_a^{-1}+ \partial_{x_b}\Del_b^{-1} )\) \d C_{ab}\]_{\xv_a=\xv_b}  \right|_{\xv=\xv_m}\ .
\end{align}
Using~\eqref{eq:S3Tpert_nx} and  expressing everything in term of the matrices $\d\C_{k_x,n_x}$ we get
\begin{align}
&\left. \vphantom{\frac1{2}}\partial_x\left\{ \left. \bit \frac1{2}\(\partial_{y_a} \Del_a^{-1}+\partial_{y_b} \Del_b^{-1}\) \d C_{ab} \right|_{\xv_a=\xv_b} \right\} \right|_{\xv=\xv_m}= \nonumber\\
&\quad=\left. \vphantom{\frac1{2}}\partial_x\left\{ \frac1{2}  \left. \bit \sum_{k_x=-(N_k+n_x)}^{N_k+n_x}  \(\partial_{y_a} \Del_a^{-1}+\partial_{y_b} \Del_b^{-1}\) \right.\right.\right.\nonumber\\
&\hspace{9em}\left.\left.\left. \[e^{\i n_x(x_a+x_b)/2} \,\d\tilde{C}_{k_x,n_x} (y_a,y_b) \,e^{\i (k_x-n_x/2)(x_a-x_b)} \]\vphantom{\frac1{2}}\right|_{\xv_a=\xv_b}  \vphantom{\sum_{k_x=-(N_k+n_x)}^{N_k+n_x} }\right\} \right|_{\xv=\xv_m}\nonumber\end{align}\begin{align}%\\
&\quad=\left. \vphantom{\frac1{2}}\partial_x\left\{ \frac1{2}  \left. \bit e^{\i n_x(x_a+x_b)/2} \sum_{k_x=-(N_k+n_x)}^{N_k+n_x}  \[\vphantom{\(\D_y\)^\transp}\D_y \DDel^{-1}_{k_x} \,\d\C_{k_x,n_x} \right. \right.\right.\right.\nonumber\\
&\hspace{9em}\left.\left.\left.\left. + \d\C_{k_x,n_x} \(\DDel^{-1}_{k_x-n_x}\)^\transp\(\D_y\)^\transp\]_{ab}  e^{\i (k_x-n_x/2)(x_a-x_b)} \vphantom{\sum_{k_x=-\infty}^{+\infty} }\right|_{\xv_a=\xv_b}  \right\} \right|_{\xv=\xv_m}\nonumber\\
&\quad=\frac1{2} \i n_x \,  e^{\i n_x x_m} \sum_{k_x=-(N_k+n_x)}^{N_k+n_x}  \[\D_y \DDel^{-1}_{k_x} \,\d\C_{k_x,n_x} + \d\C_{k_x,n_x} \(\DDel^{-1}_{k_x-n_x}\)^\transp\(\D_y\)^\transp\]_{mm} \ .\label{eq:duzk}
\end{align}
Similarly,
\begin{align}
&-\left. \vphantom{\frac1{2}}\partial_y\left\{ \left. \bit \frac1{2}\(\partial_{x_a} \Del_a^{-1}+\partial_{x_b} \Del_b^{-1}\) \d C_{ab} \right|_{\xv_a=\xv_b} \right\} \right|_{\xv=\xv_m}= \nonumber\\
&\quad=-\frac1{2}   e^{\i n_x x_m} \left\{\D_y\,\vecd\[ \sum_{k_x=-(N_k+n_x)}^{N_k+n_x} \hspace{-1em}  \i k_x \DDel^{-1}_{k_x} \,\d\C_{k_x,n_x} + \d\C_{k_x,n_x} \(\DDel^{-1}_{k_x-n_x}\)^\transp\[-\i (k_x-n_x)\]\]\right\}_m .\label{eq:dvzk}
\end{align}
From~\eqref{eq:duzk} and~\eqref{eq:dvzk} it can be seen that the $e^{\i n_x x}$ dependance factors out and therefore the Reynolds stress divergence takes the form $\Rcal(\d C)=e^{\i n_x x}\,\Rcal_{n_x}(\d C_{n_x})$. After discretization $\Rcal_{n_x}(\d C_{n_x})$ is approximated by the column vector $\d\Rv_{n_x}$:%\begin{align}
%&\d\Rv_{n_x} =\frac{1}{2}\left\{ \i n_x \, \vecd\[ \sum_{k_x=-(N_k+n_x)}^{N_k+n_x} \D_y \DDel^{-1}_{k_x} \,\d\C_{k_x,n_x} + \d\C_{k_x,n_x} \(\D_y\DDel^{-1}_{k_x-n_x}\)^\transp\] \right.\nonumber\\
%&\qquad\qquad+\left. \D_y \vecd\[ \sum_{k_x=-(N_k+n_x)}^{N_k+n_x}   \i k_x \DDel^{-1}_{k_x} \,\d\C_{k_x,n_x} + \d\C_{k_x,n_x} \(\DDel^{-1}_{k_x-n_x}\)^\transp\[-\i (k_x-n_x)\]\] \right\}\ .\label{eq:dRv_nx}
%\end{align}
\begin{align}
&\d\Rv_{n_x} =\frac{1}{2}\sum_{k_x=-(N_k+n_x)}^{N_k+n_x} \left\{ \vphantom{\sum_i^j} \i n_x \, \vecd\[ \D_y \DDel^{-1}_{k_x} \,\d\C_{k_x,n_x} + \d\C_{k_x,n_x} \(\D_y\DDel^{-1}_{k_x-n_x}\)^\transp\] \right.\nonumber\\
&\qquad\qquad\hspace{0.8em}+\left. \D_y \vecd\[ \i k_x \DDel^{-1}_{k_x} \,\d\C_{k_x,n_x} + \d\C_{k_x,n_x} \(\DDel^{-1}_{k_x-n_x}\)^\transp\[-\i (k_x-n_x)\]\] \vphantom{\sum_i^j}\right\}\ .\label{eq:dRv_nx}
\end{align}
(For a homogeneous S3T equilibrium, i.e., in the absence of mean flow $\Uv^e=0$, the Reynolds stress $\Rcal_{n_x}(\d C_{n_x})$ becomes proportional to $e^{\i n_y y}$ or $\d\Rv_{n_x}$ is proportional to $\vecd[\d\mathsf{E}_{n_y}]$.)

Note that the operator $\D_y\DDel^{-1}_{k_x}$ appearing in~\eqref{eq:duzk} is ill-defined  for $k_x=0$,  since then $\DDel_{k_x}^{-1} $ is the non-invertible operator  $(\D_y^{-1} )^2$. In this case we calculate $\D_y  \DDel^{-1}_{k_x}$ as the pseudoinverse of $\D_y$ using its SVD decomposition and then by removing the zero singular values of $\D_y$.

After all these consideration the discrete version of~\eqref{eq:s3tpert_v2} is:
\begin{subequations}
\begin{align}
\frac{\df}{\df t} \d \Zv_{n_x} &= \A^{e}_{n_x} \,\d\Zv_{n_x} + \d\Rv_{n_x}\ ,\\
\frac{\df}{\df t}\d \C_{k_x,n_x} &=  \A^{e}_{k_x}\,\d\C_{k_x,n_x} + \d\C_{k_x,n_x}\(\A^{e}_{n_x-k_x}\)^\transp \nonumber\\
&\qquad+ \sum_{n_y=-N_y/2}^{N_y/2-1}  \a_{n_x,n_y} \[\vphantom{\frac1{2}}\d\A_{k_x-n_x,n_x,n_y}\C^e_{k_x-n_x} + \C^e_{k_x}\(\d\A_{-k_x,n_x,n_y}\)^\transp\]\ ,\nonumber\\
&\hspace{16em}\text{for }k_x=1,\dots,N_k+2n_x\ ,
\end{align}\label{eq:s3tpert_discrete_kx}\end{subequations}
where $\d \Zv_{n_x}$ is the $N_y$-column vector with elements $[\d \Zv_{n_x}(t)]_a= \sum_{n_y}  \a_{n_x,n_y}(t)\,e^{\i n_y y_a}$.  This system has a state of $2N_y + 2(N_k+2n_x)N_y^2$ real variables. For $N_x=N_y=128$, $N_k=15$ and $n_x=2$ this gives an order of 400-fold decrease in the dimension of the state variable compared to the full S3T perturbation system~\eqref{eq:s3tpert_discrete}.

An example demonstrating the convergence of the growth $\la$ to the top Lyapunov exponent using~\eqref{eq:s3tpert_discrete_kx} is shown in Fig.~\ref{fig:S3Teigen_S3TzMIequil}\hyperref[fig:S3Teigen_S3TzMIequil]{e}.

\end{appendices}

\singlespacing

% the back matter

%\bibliography{references}
%\addcontentsline{toc}{chapter}{References}
%\bibliographystyle{apalike2}

% \bibliographystyle{ametsoc2014}
% \bibliography{references}

\begin{thebibliography}{148}
\providecommand{\natexlab}[1]{#1}
\providecommand{\url}[1]{\texttt{#1}}
\renewcommand{\UrlFont}{\rmfamily}
\providecommand{\urlprefix}{URL }
\expandafter\ifx\csname urlstyle\endcsname\relax
  \providecommand{\doi}[1]{doi:\discretionary{}{}{}#1}\else
  \providecommand{\doi}{doi:\discretionary{}{}{}\begingroup
  \urlstyle{rm}\Url}\fi
\providecommand{\eprint}[2][]{\url{#2}}

\bibitem[{Andrews and McIntyre(1976)Andrews, and
  McIntyre}]{Andrews-McIntyre-1976-I}
Andrews, D.~G., and M.~E. McIntyre, 1976: Planetary waves in horizontal and
  vertical shear: {The} generalized {Eliassen-Palm} relation and the mean zonal
  acceleration. \textit{J. Atmos. Sci.}, \textbf{33~(11)}, 2031--2048,
  \doi{10.1175/1520-0469(1976)033<2031:PWIHAV>2.0.CO;2}.

\bibitem[{Atkinson et~al.(1997)Atkinson, Ingersoll,, and
  Seif}]{Atkinson-etal-1997}
Atkinson, D.~H., A.~P. Ingersoll, and A.~Seif, 1997: Deep winds on {Jupiter} as
  measured by the {Galileo} probe. \textit{Nature}, \textbf{388}, 649--650,
  \doi{10.1038/41718}.

\bibitem[{Bakas et~al.(2015)Bakas, Constantinou,, and
  Ioannou}]{Bakas-etal-2014}
Bakas, N.~A., N.~C. Constantinou, and P.~J. Ioannou, 2015: {S3T} stability of
  the homogeneous state of barotropic beta-plane turbulence. \textit{J. Atmos.
  Sci.}, \textbf{72~(5)}, 1689--1712, \doi{10.1175/JAS-D-14-0213.1}.

\bibitem[{Bakas and Ioannou(2011)Bakas, and Ioannou}]{Bakas-Ioannou-2011}
Bakas, N.~A., and P.~J. Ioannou, 2011: Structural stability theory of
  two-dimensional fluid flow under stochastic forcing. \textit{J. Fluid Mech.},
  \textbf{682}, 332--361, \doi{10.1017/jfm.2011.228}.

\bibitem[{Bakas and Ioannou(2013{\natexlab{a}})Bakas, and
  Ioannou}]{Bakas-Ioannou-2013-prl}
Bakas, N.~A., and P.~J. Ioannou, 2013{\natexlab{a}}: Emergence of large scale
  structure in barotropic $\beta$-plane turbulence. \textit{Phys. Rev. Lett.},
  \textbf{110}, 224\,501, \doi{10.1103/PhysRevLett.110.224501}.

\bibitem[{Bakas and Ioannou(2013{\natexlab{b}})Bakas, and
  Ioannou}]{Bakas-Ioannou-2013-jas}
Bakas, N.~A., and P.~J. Ioannou, 2013{\natexlab{b}}: On the mechanism
  underlying the spontaneous emergence of barotropic zonal jets. \textit{J.
  Atmos. Sci.}, \textbf{70~(7)}, 2251--2271, \doi{10.1175/JAS-D-12-0102.1}.

\bibitem[{Bakas and Ioannou(2014)Bakas, and Ioannou}]{Bakas-Ioannou-2014-jfm}
Bakas, N.~A., and P.~J. Ioannou, 2014: A theory for the emergence of coherent
  structures in beta-plane turbulence. \textit{J. Fluid Mech.}, \textbf{740},
  312--341, \doi{10.1017/jfm.2013.663}.

\bibitem[{Bakas and Ioannou(2019)Bakas, and Ioannou}]{Bakas-Ioannou-2015-book}
Bakas, N.~A., and P.~J. Ioannou, 2019: Emergence of non-zonal coherent
  structures. \textit{Zonal jets: {Phenomenology}, genesis, and physics},
  B.~Galperin, and P.~L. Read, Eds., Cambridge University Press, chap.~27,
  419--436, \doi{10.1017/9781107358225.028}.

\bibitem[{Baldwin et~al.(2007)Baldwin, Rhines, Huang,, and
  McIntyre}]{Baldwin-etal-2007}
Baldwin, M.~P., P.~B. Rhines, H.-P. Huang, and M.~E. McIntyre, 2007: The
  jet-stream conundrum. \textit{Science}, \textbf{39}, 467--468,
  \doi{10.1126/science.1131375}.

\bibitem[{Balk(1991)}]{Balk-1991}
Balk, A.~M., 1991: A new invariant for {Rossby} wave systems. \textit{Phys.
  Lett. A}, \textbf{155}, 20--24, \doi{10.1016/0375-9601(91)90501-X}.

\bibitem[{Balk et~al.(1991)Balk, Nazarenko,, and Zakharov}]{Balk-etal-1991}
Balk, A.~M., S.~V. Nazarenko, and V.~E. Zakharov, 1991: New invariant for drift
  turbulence. \textit{Phys. Lett. A}, \textbf{152}, 276--280,
  \doi{10.1016/0375-9601(91)90105-H}.

\bibitem[{Balk and Yoshikawa(2009)Balk, and Yoshikawa}]{Balk-etal-2009}
Balk, A.~M., and T.~Yoshikawa, 2009: The {Rossby} wave extra invariant in the
  physical space. \textit{Physica D}, \textbf{238}, 384--394,
  \doi{10.1016/j.physd.2008.11.008}.

\bibitem[{Batchelor(1969)}]{Batchelor-1969}
Batchelor, G.~K., 1969: Computation of the energy spectrum in homogeneous
  two-dimensional turbulence. \textit{Phys. Fluids}, \textbf{12}, II233--II239.

\bibitem[{Benjamin(1967)}]{Benjamin-1967}
Benjamin, T.~B., 1967: Instability of periodic wavetrains in nonlinear
  dispersive systems. \textit{Proc. R. Soc. Lond. A}, \textbf{299}, 59--76,
  \doi{10.1098/rspa.1967.0123}.

\bibitem[{Benjamin and Feier(1967)Benjamin, and Feier}]{Benjamin-Feir-1967}
Benjamin, T.~B., and J.~E. Feier, 1967: The disintegration of wave trains on
  deep water. {Part 1. Theory}. \textit{J. Fluid Mech.}, \textbf{27~(3)},
  417--430, \doi{10.1017/S002211206700045X}.

\bibitem[{Berloff et~al.(2009)Berloff, Kamenkovich,, and
  Pedlosky}]{Berloff-etal-2009a}
Berloff, P., I.~Kamenkovich, and J.~Pedlosky, 2009: A mechanism of formation of
  multiple zonal jets in the oceans. \textit{J. Fluid Mech.}, \textbf{628},
  395--425, \doi{10.1017/S0022112009006375}.

\bibitem[{Berloff et~al.(2011)Berloff, Karabasov, Farrar,, and
  Kamenkovich}]{Berloff-etal-2011}
Berloff, P., S.~Karabasov, J.~Farrar, and I.~Kamenkovich, 2011: On latency of
  multiple zonal jets in the oceans. \textit{J. Fluid Mech.}, \textbf{686},
  534--567, \doi{10.1017/jfm.2011.345}.

\bibitem[{Bernstein and Farrell(2010)Bernstein, and
  Farrell}]{Bernstein-Farrell-2010}
Bernstein, J., and B.~F. Farrell, 2010: Low frequency variability in a
  turbulent baroclinic jet: {Eddy}--mean flow interactions in a two-level
  model. \textit{J. Atmos. Sci.}, \textbf{67~(2)}, 452--467,
  \doi{10.1175/2009JAS3170.1}.

\bibitem[{Boffetta and Musacchio(2010)Boffetta, and
  Musacchio}]{Boffetta-Musacchio-2010}
Boffetta, G., and S.~Musacchio, 2010: Evidence for the double cascade scenario
  in two-dimensional turbulence. \textit{Phys. Rev. E}, \textbf{82}, 016\,307,
  \doi{10.1103/PhysRevE.82.016307}.

\bibitem[{Bouchet et~al.(2013)Bouchet, Nardini,, and
  Tangarife}]{Bouchet-etal-2013}
Bouchet, F., C.~Nardini, and T.~Tangarife, 2013: Kinetic theory of jet dynamics
  in the stochastic barotropic and 2{D} {N}avier-{S}tokes equations. \textit{J.
  Stat. Phys.}, \textbf{153~(4)}, 572--625, \doi{10.1007/s10955-013-0828-3}.

\bibitem[{Bouchet and Sommeria(2002)Bouchet, and
  Sommeria}]{Bouchet-Sommeria-2002}
Bouchet, F., and J.~Sommeria, 2002: Emergence of intense jets and {Jupiter's
  Great Red Spot} as maximum-entropy structures. \textit{J. Fluid Mech.},
  \textbf{464}, 165--207, \doi{10.1017/S0022112002008789}.

\bibitem[{Bouchet and Venaille(2012)Bouchet, and
  Venaille}]{Bouchet-Venaile-2012}
Bouchet, F., and A.~Venaille, 2012: Statistical mechanics of two-dimensional
  and geophysical flows. \textit{Phys. Rep.}, \textbf{515~(5)}, 227--295,
  \doi{10.1016/j.physrep.2012.02.001}.

\bibitem[{Bouchet and Venaille(2019)Bouchet, and
  Venaille}]{Bouchet-Venaille-2014-book}
Bouchet, F., and A.~Venaille, 2019: Zonal flows as statistical equilibria.
  \textit{Zonal jets: {Phenomenology}, genesis, and physics}, B.~Galperin, and
  P.~L. Read, Eds., Cambridge University Press, chap.~22, 347--359,
  \doi{10.1017/9781107358225.023}.

\bibitem[{Boyd(1976{\natexlab{a}})}]{Boyd-1976}
Boyd, J.~P., 1976{\natexlab{a}}: The noninteraction of waves with the zonally
  averaged flow on a spherical {Earth} and the interrelationships on eddy
  fluxes of energy, heat and momentum. \textit{J. Atmos. Sci.},
  \textbf{33~(12)}, 2285--2291,
  \doi{10.1175/1520-0469(1976)033<2285:TNOWWT>2.0.CO;2}.

\bibitem[{Boyd(1976{\natexlab{b}})}]{Boyd-1976-thesis}
Boyd, J.~P., 1976{\natexlab{b}}: Planetary waves and the semiannual wind
  oscillation in the tropical upper stratosphere. Ph.D. thesis, Harvard
  University.

\bibitem[{Carnevale and Martin(1982)Carnevale, and
  Martin}]{Carnevale-Martin-1982}
Carnevale, G.~F., and P.~C. Martin, 1982: Field theoretic techniques in
  {Statistical Fluid Dynamics}: with application to nonlinear wave dynamics.
  \textit{Geophys. Astrophys. Fluid Dyn.}, \textbf{20}, 131--164.

\bibitem[{Charney and Drazin(1961)Charney, and Drazin}]{Charney-Drazin-1961}
Charney, J.~G., and P.~G. Drazin, 1961: Propagation of planetary-scale
  disturbances from the lower into the upper atmosphere. \textit{J. Geophys.
  Res.}, \textbf{66~(1)}, 83--109, \doi{10.1029/JZ066i001p00083}.

\bibitem[{Chekhlov et~al.(1996)Chekhlov, Orszag, Sukoriansky, Galperin,, and
  Staroselsky}]{Chekhlov-etal-1996}
Chekhlov, A., S.~Orszag, S.~Sukoriansky, B.~Galperin, and I.~Staroselsky, 1996:
  The effect of small-scale forcing on large-scale structures in two
  dimensional flows. \textit{Physica D}, \textbf{96}, 321--334.

\bibitem[{Cho and Polvani(1996{\natexlab{a}})Cho, and
  Polvani}]{Cho-Polvani-1996pof}
Cho, J. Y.-K., and L.~M. Polvani, 1996{\natexlab{a}}: The emergence of jets and
  vortices in freely evolving, shallow-water turbulence on a sphere.
  \textit{Phys. Fluids}, \textbf{8~(6)}, 1531--1552, \doi{10.1063/1.868929}.

\bibitem[{Cho and Polvani(1996{\natexlab{b}})Cho, and
  Polvani}]{Cho-Polvani-1996}
Cho, J. Y.-K., and L.~M. Polvani, 1996{\natexlab{b}}: The morphogenesis of
  bands and zonal winds in the atmospheres on the giant outer planets.
  \textit{Science}, \textbf{273~(5273)}, 335--337,
  \doi{10.1126/science.273.5273.335}.

\bibitem[{Connaughton et~al.(2010)Connaughton, Nadiga, Nazarenko,, and
  Quinn}]{Connaughton-etal-2010}
Connaughton, C.~P., B.~T. Nadiga, S.~V. Nazarenko, and B.~E. Quinn, 2010:
  Modulational instability of {Rossby} and drift waves and generation of zonal
  jets. \textit{J. Fluid Mech.}, \textbf{645}, 207--231,
  \doi{10.1017/S0022112010000510}.

\bibitem[{Constantinou et~al.(2014{\natexlab{a}})Constantinou, Farrell,, and
  Ioannou}]{Constantinou-etal-2014}
Constantinou, N.~C., B.~F. Farrell, and P.~J. Ioannou, 2014{\natexlab{a}}:
  Emergence and equilibration of jets in beta-plane turbulence: applications of
  {Stochastic Structural Stability Theory}. \textit{J. Atmos. Sci.},
  \textbf{71~(5)}, 1818--1842, \doi{10.1175/JAS-D-13-076.1}.

\bibitem[{Constantinou et~al.(2014{\natexlab{b}})Constantinou, Lozano-Dur\'an,
  Nikolaidis, Farrell, Ioannou,, and Jim\'enez}]{Constantinou-etal-Madrid-2014}
Constantinou, N.~C., A.~Lozano-Dur\'an, M.-A. Nikolaidis, B.~F. Farrell, P.~J.
  Ioannou, and J.~Jim\'enez, 2014{\natexlab{b}}: Turbulence in the highly
  restricted dynamics of a closure at second order: comparison with {DNS}.
  \textit{J. Phys. Conf. Ser.}, \textbf{506}, 012\,004,
  \doi{10.1088/1742-6596/506/1/012004}.

\bibitem[{Danilov and Gurarie(2004)Danilov, and Gurarie}]{Danilov-04}
Danilov, S., and D.~Gurarie, 2004: Scaling, spectra and zonal jets in
  beta-plane turbulence. \textit{Phys. Fluids}, \textbf{16}, 2592--2603,
  \doi{10.1063/1.1752928}.

\bibitem[{DelSole(2001)}]{DelSole-01a}
DelSole, T., 2001: A simple model for transient eddy momentum fluxes in the
  upper troposphere. \textit{J. Atmos. Sci.}, \textbf{58}, 3019--3035,
  \doi{10.1175/1520-0469(2001)058<3019:ASMFTE>2.0.CO;2}.

\bibitem[{DelSole and Farrell(1996)DelSole, and Farrell}]{DelSole-Farrell-1996}
DelSole, T., and B.~F. Farrell, 1996: The quasi-linear equilibration of a
  thermally mantained stochastically excited jet in a quasigeostrophic model.
  \textit{J. Atmos. Sci.}, \textbf{53}, 1781--1797,
  \doi{10.1175/1520-0469(1996)053<1781:TQLEOA>2.0.CO;2}.

\bibitem[{Dritschel and McIntyre(2008)Dritschel, and
  McIntyre}]{Dritschel-McIntyre-2008}
Dritschel, D.~G., and M.~E. McIntyre, 2008: Multiple jets as {PV} staircases:
  {The} {Phillips} effect and the resilience of eddy-transport barriers.
  \textit{J. Atmos. Sci.}, \textbf{65}, 855--874, \doi{10.1175/2007JAS2227.1}.

\bibitem[{Dritschel and Scott(2011)Dritschel, and Scott}]{Dritschel-Scott-2011}
Dritschel, D.~G., and R.~K. Scott, 2011: Jet sharpening by turbulent mixing.
  \textit{Phil. Trans. R. Soc. A}, \textbf{369}, 754--770,
  \doi{10.1098/rsta.2010.0306}.

\bibitem[{Eliassen and Palm(1961)Eliassen, and Palm}]{Eliassen-Palm-1961}
Eliassen, A., and E.~Palm, 1961: On the transfer of energy in stationary
  mountain waves. \textit{Geofys. Publ.}, \textbf{XXII~(3)}, 1--23.

\bibitem[{Espa et~al.(2010)Espa, Di~Nitto,, and Cenedese}]{Espa-etal-2010}
Espa, S., G.~Di~Nitto, and A.~Cenedese, 2010: The emergence of zonal jets in
  forced rotating shallow water turbulence: A laboratory study. \textit{EPL},
  \textbf{92}, 34\,006, \doi{10.1209/0295-5075/92/34006}.

\bibitem[{Eyink(1996)}]{Eyink-1996}
Eyink, G.~L., 1996: Exact results on stationary turbulence in {2D}:
  consequences of vorticity conservation. \textit{Physica D}, \textbf{91},
  97--142, \doi{10.1016/0167-2789(95)00250-2}.

\bibitem[{Eyink and Sreenivasan(2006)Eyink, and
  Sreenivasan}]{Eyink-Sreenivasan-2006}
Eyink, G.~L., and K.~Sreenivasan, 2006: Onsager and the theory of hydrodynamic
  turbulence. \textit{Rev. Mod. Phys.}, \textbf{78~(1)}, 87--135,
  \doi{10.1103/RevModPhys.78.87}.

\bibitem[{Farrell and Ioannou(1993)Farrell, and
  Ioannou}]{Farrell-Ioannou-1993d}
Farrell, B.~F., and P.~J. Ioannou, 1993: Stochastic dynamics of baroclinic
  waves. \textit{J. Atmos. Sci.}, \textbf{50}, 4044--4057,
  \doi{10.1175/1520-0469(1993)050<4044:SDOBW>2.0.CO;2}.

\bibitem[{Farrell and Ioannou(1994)Farrell, and
  Ioannou}]{Farrell-Ioannou-1994a}
Farrell, B.~F., and P.~J. Ioannou, 1994: A theory for the statistical
  equilibrium energy spectrum and heat flux produced by transient baroclinic
  waves. \textit{J. Atmos. Sci.}, \textbf{51}, 2685--2698,
  \doi{10.1175/1520-0469(1994)051<2685:ATFTSE>2.0.CO;2}.

\bibitem[{Farrell and Ioannou(1995)Farrell, and Ioannou}]{Farrell-Ioannou-1995}
Farrell, B.~F., and P.~J. Ioannou, 1995: Stochastic dynamics of the midlatitude
  atmospheric jet. \textit{J. Atmos. Sci.}, \textbf{52}, 1642--1656,
  \doi{10.1175/1520-0469(1995)052<1642:SDOTMA>2.0.CO;2}.

\bibitem[{Farrell and Ioannou(2003)Farrell, and
  Ioannou}]{Farrell-Ioannou-2003-structural}
Farrell, B.~F., and P.~J. Ioannou, 2003: Structural stability of turbulent
  jets. \textit{J. Atmos. Sci.}, \textbf{60}, 2101--2118,
  \doi{10.1175/1520-0469(2003)060<2101:SSOTJ>2.0.CO;2}.

\bibitem[{Farrell and Ioannou(2007)Farrell, and
  Ioannou}]{Farrell-Ioannou-2007-structure}
Farrell, B.~F., and P.~J. Ioannou, 2007: Structure and spacing of jets in
  barotropic turbulence. \textit{J. Atmos. Sci.}, \textbf{64}, 3652--3665,
  \doi{10.1175/JAS4016.1}.

\bibitem[{Farrell and Ioannou(2008)Farrell, and
  Ioannou}]{Farrell-Ioannou-2008-baroclinic}
Farrell, B.~F., and P.~J. Ioannou, 2008: Formation of jets by baroclinic
  turbulence. \textit{J. Atmos. Sci.}, \textbf{65}, 3353--3375,
  \doi{10.1175/2008JAS2611.1}.

\bibitem[{Farrell and Ioannou(2009{\natexlab{a}})Farrell, and
  Ioannou}]{Farrell-Ioannou-2009-equatorial}
Farrell, B.~F., and P.~J. Ioannou, 2009{\natexlab{a}}: Emergence of jets from
  turbulence in the shallow-water equations on an equatorial beta plane.
  \textit{J. Atmos. Sci.}, \textbf{66}, 3197--3207,
  \doi{10.1175/2009JAS2941.1}.

\bibitem[{Farrell and Ioannou(2009{\natexlab{b}})Farrell, and
  Ioannou}]{Farrell-Ioannou-2009-plasmas}
Farrell, B.~F., and P.~J. Ioannou, 2009{\natexlab{b}}: A stochastic structural
  stability theory model of the drift wave-zonal flow system. \textit{Phys.
  Plasmas}, \textbf{16}, 112\,903, \doi{10.1063/1.3258666}.

\bibitem[{Farrell and Ioannou(2009{\natexlab{c}})Farrell, and
  Ioannou}]{Farrell-Ioannou-2009-closure}
Farrell, B.~F., and P.~J. Ioannou, 2009{\natexlab{c}}: A theory of baroclinic
  turbulence. \textit{J. Atmos. Sci.}, \textbf{66}, 2444--2454,
  \doi{10.1175/2009JAS2989.1}.

\bibitem[{Farrell and Ioannou(2012)Farrell, and Ioannou}]{Farrell-Ioannou-2012}
Farrell, B.~F., and P.~J. Ioannou, 2012: Dynamics of streamwise rolls and
  streaks in turbulent wall-bounded shear flow. \textit{J. Fluid Mech.},
  \textbf{708}, 149--196, \doi{10.1017/jfm.2012.300}.

\bibitem[{Fj{\o}rtoft(1953)}]{Fjortoft-1953}
Fj{\o}rtoft, R., 1953: On the changes in the spectral distribution of kinetic
  evergy for twodimensional, nondivergent flow. \textit{Tellus}, \textbf{5},
  120--140, \doi{10.1111/j.2153-3490.1953.tb01051.x}.

\bibitem[{Galperin et~al.(2004)Galperin, Nakano, Huang,, and
  Sukoriansky}]{Galperin-etal-04}
Galperin, B., H.~Nakano, H.-P. Huang, and S.~Sukoriansky, 2004: The ubiquitous
  zonal jets in the atmospheres of giant planets and {Earth's} oceans.
  \textit{Geophys. Res. Lett.}, \textbf{31}, 13\,303--13\,308,
  \doi{10.1029/2004GL019691}.

\bibitem[{Galperin et~al.(2014)Galperin, Young, Sukoriansky, Dikovskaya, Read,
  Lancaster,, and Armstrong}]{Galperin-etal-2014}
Galperin, B., R.~M.~B. Young, S.~Sukoriansky, N.~Dikovskaya, P.~L. Read, A.~L.
  Lancaster, and D.~Armstrong, 2014: {Cassini} observations reveal a regime of
  zonostrophic macroturbulence on {Jupiter}. \textit{Icarus}, \textbf{229},
  295--320, \doi{10.1016/j.icarus.2013.08.030}.

\bibitem[{Gierasch et~al.(2000)}]{Gierasch-etal-2000}
Gierasch, P.~J., and Coauthors, 2000: Observation of moist convection in
  {Jupiter's} atmosphere. \textit{Nature}, \textbf{403~(6)}, 628--630,
  \doi{10.1038/35001017}.

\bibitem[{{Gill}(1974)}]{Gill-1974}
{Gill}, A.~E., 1974: The stability of planetary waves on an infinite
  beta-plane. \textit{Geophys. Astrophys. Fluid Dyn.}, \textbf{6}, 29--47,
  \doi{10.1080/03091927409365786}.

\bibitem[{Godunov(1959)}]{Godunov-1959}
Godunov, S.~K., 1959: A difference scheme for numerical solution of
  discontinuous solution of hydrodynamic equations. \textit{Math. Sb. (N.S.)},
  \textbf{47(89)~(3)}, 271--306, (translated US Joint Publ. Res. Service, JPRS
  7226, 1969).

\bibitem[{Hasselmann(1966)}]{Hasselmann-1966}
Hasselmann, K., 1966: Feynman diagrams and interaction rules of wave--wave
  scattering processes. \textit{Rev. Geophys.}, \textbf{4~(1)}, 1--32,
  \doi{10.1029/RG004i001p00001}.

\bibitem[{Hasselmann(1967)}]{Hasselmann-1967}
Hasselmann, K., 1967: Nonlinear interactions treated by the methods of
  theoretical physics (with application to the generation of waves by wind).
  \textit{Proc. R. Soc. Lond. A}, \textbf{299~(1456)}, 77--103,
  \doi{10.1098/rspa.1967.0124}.

\bibitem[{Held(1999)}]{Held-1999}
Held, I.~M., 1999: The macroturbulence of the troposphere. \textit{Tellus},
  \textbf{51~(1)}, 59--70, \doi{10.1034/j.1600-0870.1999.t01-1-00006.x}.

\bibitem[{Held and Hoskins(1985)Held, and Hoskins}]{Held-Hoskins-1985}
Held, I.~M., and B.~J. Hoskins, 1985: Large-scale eddies and the general
  circulation of the troposphere. \textit{Adv. Geophys.}, \textbf{28A}, 3--31,
  \doi{10.1016/S0065-2687(08)60218-6}.

\bibitem[{Held and {Hou}(1980)Held, and {Hou}}]{Held-Hou-1980}
Held, I.~M., and A.~Y. {Hou}, 1980: Nonlinear axially symmetric circulations in
  a nearly inviscid atmosphere. \textit{J. Atmos. Sci.}, \textbf{37~(3)},
  515--533, \doi{10.1175/1520-0469(1980)037<0515:NASCIA>2.0.CO;2}.

\bibitem[{Hinch(1991)}]{Hinch-1991}
Hinch, E.~J., 1991: \textit{Perturbation Methods}. Cambridge University Press.

\bibitem[{Hopf(1952)}]{Hopf-1952}
Hopf, E., 1952: Statistical hydromechanics and functional calculus. \textit{J.
  Ration. Mech. Anal.}, \textbf{1}, 87--123, \doi{10.1512/iumj.1952.1.51004}.

\bibitem[{Hoskins(1983)}]{Hoskins-1983}
Hoskins, B.~J., 1983: Modelling of the transient eddies and their feedback on
  the mean flow. \textit{Large-scale dynamical processes in the atmosphere},
  B.~J. Hoskins, and R.~Pearce, Eds., 169--199.

\bibitem[{Huang and Robinson(1998)Huang, and Robinson}]{Huang-Robinson-98}
Huang, H.-P., and W.~A. Robinson, 1998: Two-dimensional turbulence and
  persistent zonal jets in a global barotropic model. \textit{J. Atmos. Sci.},
  \textbf{55}, 611--632, \doi{10.1175/1520-0469(1998)055<0611:TDTAPZ>2.0.CO;2}.

\bibitem[{Ingersoll(1990)}]{Ingersoll-90}
Ingersoll, A.~P., 1990: Atmospheric dynamics of the outer planets.
  \textit{Science}, \textbf{248}, 308--315, \doi{10.1126/science.248.4953.308}.

\bibitem[{Ingersoll et~al.(1981)Ingersoll, Beebe, Mitchell, Garneau, Yagi,, and
  M{\"u}ller}]{Ingersoll-etal-1981}
Ingersoll, A.~P., R.~F. Beebe, J.~L. Mitchell, G.~W. Garneau, G.~M. Yagi, and
  J.-P. M{\"u}ller, 1981: Interaction between eddies and mean zonal flow on
  {Jupiter} as inferred from {Voyager 1} and {Voyager 2} images. \textit{J.
  Geophys. Res.}, \textbf{86~(A10)}, 8733--8743, \doi{10.1029/JA086iA10p08733}.

\bibitem[{Ioannou and Lindzen(1986)Ioannou, and Lindzen}]{Ioannou-Lindzen-1986}
Ioannou, P.~J., and R.~S. Lindzen, 1986: Baroclinic instability in the presence
  of barotropic jets. \textit{J. Atmos. Sci.}, \textbf{43}, 2999--3014,
  \doi{10.1175/1520-0469(1986)043<2999:BIITPO>2.0.CO;2}.

\bibitem[{James(1987)}]{James-1987}
James, I.~N., 1987: Suppression of baroclinic instability in horizontally
  sheared flows. \textit{J. Atmos. Sci.}, \textbf{44~(24)}, 3710--3720,
  \doi{10.1175/1520-0469(1987)044<3710:SOBIIH>2.0.CO;2}.

\bibitem[{Jeffreys(1926)}]{Jeffreys-1926-winds}
Jeffreys, H., 1926: On the dynamics of geostrophic winds. \textit{Q. J. Roy.
  Meteor. Soc.}, \textbf{52}, 85--104, \doi{10.1002/qj.49705221708}.

\bibitem[{Kaneda et~al.(2003)Kaneda, Ishihara, Yokokawa, Itakura,, and
  Uno}]{Kaneda-etal-2003}
Kaneda, Y., T.~Ishihara, M.~Yokokawa, K.~Itakura, and A.~Uno, 2003: Energy
  dissipation rate and energy spectrum in high resolution direct numerical
  simulations of turbulence in a periodic box. \textit{Phys. Fluids},
  \textbf{15}, L21, \doi{10.1063/1.1539855}.

\bibitem[{{Kaspi}(2013)}]{Kaspi-2013}
{Kaspi}, Y.~H., 2013: Inferring the depth of the zonal jets on {Jupiter} and
  {Saturn} from odd gravity harmonics. \textit{Geophys. Res. Lett.},
  \textbf{40~(4)}, 1--5, \doi{10.1029/2012GL053873}.

\bibitem[{Kaspi et~al.(2013)Kaspi, Showman, Hubbard, Aharonson,, and
  Helled}]{Kaspi-etal-2013}
Kaspi, Y.~H., A.~P. Showman, W.~B. Hubbard, O.~Aharonson, and R.~Helled, 2013:
  Atmospheric confinement of jet streams on {Uranus} and {Neptune}.
  \textit{Nature}, \textbf{497}, 344--347, \doi{10.1038/nature12131}.

\bibitem[{Kraichnan(1967)}]{Kraichnan-1967}
Kraichnan, R.~H., 1967: Inertial ranges in two-dimensional turbulence.
  \textit{Phys. Fluids}, \textbf{10}, 1417--1423, \doi{10.1063/1.1762301}.

\bibitem[{Kuo(1951)}]{Kuo-1951}
Kuo, H.-L., 1951: Dynamical aspects of the general circulaiton and the
  stability of zonal flow. \textit{Tellus}, \textbf{3~(4)}, 268--284,
  \doi{10.1111/j.2153-3490.1951.tb00809.x}.

\bibitem[{Lee and Smith(2003)Lee, and Smith}]{Lee-Smith-2003}
Lee, Y., and L.~M. Smith, 2003: Stability of {Rossby} waves in the
  $\beta$-plane approximation. \textit{Physica D}, \textbf{179}, 53--91,
  \doi{10.1016/S0167-2789(03)00010-1}.

\bibitem[{Leith(1968)}]{Leith-1968}
Leith, C.~E., 1968: Diffusion approximation for two-dimensional turbulence.
  \textit{Phys. Fluids}, \textbf{11}, 671--673, \doi{10.1063/1.1691968}.

\bibitem[{Lewis(2003)}]{Lewis-2003}
Lewis, J.~M., 2003: Ooishi's observation: {Viewed} in the context of jet stream
  discovery. \textit{Bull. Am. Met. Soc.}, \textbf{84~(3)}, 357--369,
  \doi{10.1175/BAMS-84-3-357}.

\bibitem[{Lighthill(1978)}]{Lighthill-1978}
Lighthill, M.~J., 1978: Acoustic streaming. \textit{J. Sound Vib.},
  \textbf{61~(3)}, 391--418, \doi{10.1016/0022-460X(78)90388-7}.

\bibitem[{Lilly(1969)}]{Lilly-1969}
Lilly, D.~K., 1969: Numerical simulation of two-dimensional turbulence.
  \textit{Phys. Fluids}, \textbf{12}, II240--II249, \doi{10.1063/1.1692444}.

\bibitem[{Limaye(1986)}]{Limaye-1986}
Limaye, S.~S., 1986: {Jupiter}: new estimates of the mean zonal flow at the
  cloud level. \textit{Icarus}, \textbf{65}, 335--352,
  \doi{10.1016/0019-1035(86)90142-9}.

\bibitem[{Lindzen(1993)}]{Lindzen-1993}
Lindzen, R.~S., 1993: Baroclinic neutrality and the tropopause. \textit{J.
  Atmos. Sci.}, \textbf{50~(8)}, 1148--1151,
  \doi{10.1175/1520-0469(1993)050<1148:BNATT>2.0.CO;2}.

\bibitem[{Lindzen and Hou(1988)Lindzen, and Hou}]{Lindzen-Hou-1988}
Lindzen, R.~S., and A.~Y. Hou, 1988: Hadley circulations for zonally averaged
  heating centered off the equator. \textit{J. Atmos. Sci.}, \textbf{45~(17)},
  2416--2427, \doi{10.1175/1520-0469(1988)045<2416:HCFZAH>2.0.CO;2}.

\bibitem[{Little et~al.(1999)Little, Anger, Ingersoll, Vasavada, Shenke,
  Breneman, Borucki,, and {The Galileo {SSI} Team}}]{Little-etal-1999}
Little, B., C.~D. Anger, A.~P. Ingersoll, A.~R. Vasavada, D.~A. Shenke, H.~H.
  Breneman, W.~J. Borucki, and {The Galileo {SSI} Team}, 1999: Galileo images
  of lightning on {Jupiter}. \textit{Icarus}, \textbf{142~(2)}, 306--323,
  \doi{10.1006/icar.1999.6195}.

\bibitem[{Longuet-Higgins(1964)}]{Longuet-Higgins-1964}
Longuet-Higgins, M.~S., 1964: Planetary waves on a rotating sphere.
  \textit{Proc. R. Soc. Lond. A}, \textbf{279~(1)}, 446--473,
  \doi{10.1098/rspa.1964.0116}.

\bibitem[{Lorenz(1967)}]{Lorenz-67}
Lorenz, E.~N., 1967: \textit{The nature and theory of the general circulation
  of the atmosphere}. World Meteorological Organization, 161 pp.

\bibitem[{Lorenz(1972)}]{Lorenz-1972}
Lorenz, E.~N., 1972: Barotropic instability of {Rossby} wave motion. \textit{J.
  Atmos. Sci.}, \textbf{29}, 258--269,
  \doi{10.1175/1520-0469(1972)029<0258:BIORWM>2.0.CO;2}.

\bibitem[{Manfroi and Young(1999)Manfroi, and Young}]{Manfroi-Young-99}
Manfroi, A.~J., and W.~R. Young, 1999: Slow evolution of zonal jets on the beta
  plane. \textit{J. Atmos. Sci.}, \textbf{56}, 784--800,
  \doi{10.1175/1520-0469(1999)056<0784:SEOZJO>2.0.CO;2}.

\bibitem[{Marcinkiewicz(1939)}]{Marcinkiewicz-1939}
Marcinkiewicz, J., 1939: Sur une propri{\'e}t{\'e} de la loi de {G}auss.
  \textit{Mathematische Zeitschrift}, \textbf{44~(1)}, 612--618,
  \doi{10.1007/BF01210677}.

\bibitem[{Marston(2012)}]{Marston-2012}
Marston, J.~B., 2012: Atmospheres as nonequilibrium condensed matter.
  \textit{Annu. Rev. Condens. Matter Phys.}, \textbf{3}, 285--310,
  \doi{10.1146/annurev-conmatphys-020911-125114}.

\bibitem[{Marston et~al.(2008)Marston, Conover,, and
  Schneider}]{Marston-etal-2008}
Marston, J.~B., E.~Conover, and T.~Schneider, 2008: Statistics of an unstable
  barotropic jet from a cumulant expansion. \textit{J. Atmos. Sci.},
  \textbf{65~(6)}, 1955--1966, \doi{10.1175/2007JAS2510.1}.

\bibitem[{Marston et~al.(2014)Marston, Qi,, and Tobias}]{Marston-etal-2014}
Marston, J.~B., W.~Qi, and S.~M. Tobias, 2014: Direct statistical simulation of
  a jet. \textit{Zonal jets: Phenomenology, genesis, physics}, B.~Galperin, and
  P.~L. Read, Eds., Cambridge University Press, chap.~5, (submitted,
  arXiv:1412.0381).

\bibitem[{McEwan et~al.(1980)McEwan, Thompson,, and Plumb}]{McEwan-etal-1980}
McEwan, A.~D., R.~O. R.~Y. Thompson, and R.~A. Plumb, 1980: Mean flows driven
  by weak eddies in rotating systems. \textit{J. Fluid Mech.}, \textbf{99~(3)},
  656--672, \doi{10.1017/S002211208000081X}.

\bibitem[{Miller(1990)}]{Miller-1990}
Miller, J., 1990: Statistical mechanics of {Euler} equations in two dimensions.
  \textit{Phys. Rev. Lett.}, \textbf{65}, 2137--2140,
  \doi{10.1103/PhysRevLett.65.2137}.

\bibitem[{Nastrom and Gage(1985)Nastrom, and Gage}]{Nastrom-Gage-1985}
Nastrom, G.~D., and K.~S. Gage, 1985: A climatology of atmospheric wavenumber
  spectra of wind and temperature observed by commercial aircraft. \textit{J.
  Atmos. Sci.}, \textbf{42~(9)}, 950--060,
  \doi{10.1175/1520-0469(1985)042<0950:ACOAWS>2.0.CO;2}.

\bibitem[{Nazarenko and Quinn(2009)Nazarenko, and Quinn}]{Nazarenko-09}
Nazarenko, S.~V., and B.~E. Quinn, 2009: Triple cascade behavior in
  quasigeostrophic and drift turbulence and generation of zonal jets.
  \textit{Phys. Rev. Lett.}, \textbf{103}, 118\,501,
  \doi{10.1103/PhysRevLett.103.118501}.

\bibitem[{Nozawa and Yoden(1997)Nozawa, and Yoden}]{Nozawa-and-Yoden-97}
Nozawa, T., and Y.~Yoden, 1997: Formation of zonal band structure in forced
  two-dimensional turbulence on a rotating sphere. \textit{Phys. Fluids},
  \textbf{9}, 2081--2093, \doi{10.1063/1.869327}.

\bibitem[{O'Gorman and Schneider(2007)O'Gorman, and
  Schneider}]{OGorman-Schneider-2007}
O'Gorman, P.~A., and T.~Schneider, 2007: Recovery of atmospheric flow
  statistics in a general circulation model without nonlinear eddy-eddy
  interactions. \textit{Geophys. Res. Lett.}, \textbf{34}, L22\,801,
  \doi{10.1029/2007GL031779}.

\bibitem[{{\O}ksendal(2000)}]{Oksendal-2000}
{\O}ksendal, B., 2000: \textit{{S}tochastic {D}ifferential {E}quations}.
  {S}pringer-{V}erlag, {B}erlin.

\bibitem[{Onsager(1949)}]{Onsager-1949}
Onsager, L., 1949: Statistical hydrodynamics. \textit{Nuovo Cimento},
  \textbf{6}, 249--286.

\bibitem[{Parker and Krommes(2014)Parker, and
  Krommes}]{Parker-Krommes-2014-generation}
Parker, J.~B., and J.~A. Krommes, 2014: Generation of zonal flows through
  symmetry breaking of statistical homogeneity. \textit{New J. Phys.},
  \textbf{16~(3)}, 035\,006, \doi{10.1088/1367-2630/16/3/035006}.

\bibitem[{Parker and Krommes(2019)Parker, and
  Krommes}]{Parker-Krommes-2014-book}
Parker, J.~B., and J.~A. Krommes, 2019: Zonal flow as pattern formation.
  \textit{Zonal jets: {Phenomenology}, genesis, and physics}, B.~Galperin, and
  P.~L. Read, Eds., Cambridge University Press, chap.~26, 401--418.

\bibitem[{Peix{\'o}to and Oort(1984)Peix{\'o}to, and Oort}]{Peixoto-Oort-1984}
Peix{\'o}to, J.~P., and A.~H. Oort, 1984: Physics of climate. \textit{Rev. Mod.
  Phys.}, \textbf{56~(3)}, 365--429, \doi{10.1103/RevModPhys.56.365}.

\bibitem[{Phillips(1972)}]{Phillips-1972}
Phillips, O.~M., 1972: Turbulence in a strongly stratified fluid -- is it
  unstable? \textit{Deep-Sea Res.}, \textbf{19~(1)}, 79--81,
  \doi{10.1016/0011-7471(72)90074-5}.

\bibitem[{Phillips(1977)}]{Phillips-77}
Phillips, O.~M., 1977: \textit{The Dynamics of the Upper Ocean}. Cambridge
  University Press, Cambridge.

\bibitem[{Porco et~al.(2003)}]{Porco-etal-2003}
Porco, C.~C., and Coauthors, 2003: Cassini imaging of {Jupiter's} atmosphere,
  satellites and rings. \textit{Science}, \textbf{299~(5612)}, 1541--1547,
  \doi{10.1126/science.1079462}.

\bibitem[{Rayleigh(1896)}]{Rayleigh-1896}
Rayleigh, L., 1896: \textit{Theory of sound (two volumes)}. 2nd ed., Dover
  Publications.

\bibitem[{Read(2013)}]{Read-2013}
Read, P.~L., 2013: Plumbing the depths of {Uranus} and {Neptune}.
  \textit{Nature}, \textbf{497}, 323--324, \doi{10.1038/497323a}.

\bibitem[{Read et~al.(2004)Read, Yamazaki, Lewis, Williams, Miki-Yamazaki,
  Sommeria, Didelle,, and Fincham}]{Read-etal-2004}
Read, P.~L., Y.~H. Yamazaki, S.~R. Lewis, P.~D. Williams, K.~Miki-Yamazaki,
  J.~Sommeria, H.~Didelle, and A.~Fincham, 2004: Jupiter's and {Saturn's}
  convectively driven banded jets in the laboratory. \textit{Geophys. Res.
  Lett.}, \textbf{31}, {L}22\,701, \doi{10.1029/2004GL020106}.

\bibitem[{Read et~al.(2007)}]{Read-etal-2007}
Read, P.~L., and Coauthors, 2007: Dynamics of convectively driven banded jets
  in the laboratory. \textit{J. Atmos. Sci.}, \textbf{64}, 4031--4052,
  \doi{10.1175/2007JAS2219.1}.

\bibitem[{Reynolds(1883)}]{Reynolds-1883}
Reynolds, O., 1883: An experimental investigation of the circumstances which
  determine whether the motion of water shall be direct or sinuous, and of the
  law of resistance in parallel channels. \textit{Phil. Trans. R. Soc. Lond.},
  \textbf{174}, 935--982, \doi{10.1098/rstl.1883.0029}.

\bibitem[{Rhines(1975)}]{Rhines-1975}
Rhines, P.~B., 1975: Waves and turbulence on a beta-plane. \textit{J. Fluid
  Mech.}, \textbf{69}, 417--433, \doi{10.1017/S0022112075001504}.

\bibitem[{Riehl(1962)}]{Riehl-1962}
Riehl, H., 1962: Jet streams of the atmosphere. Tech. Rep.~32, Colorado State
  University.

\bibitem[{Robert and Sommeria(1991)Robert, and Sommeria}]{Robert-Sommeria-1991}
Robert, R., and J.~Sommeria, 1991: Statistical equilibrium states for
  two-dimensional flows. \textit{J. Fluid Mech.}, \textbf{229}, 291--310,
  \doi{10.1017/S0022112091003038}.

\bibitem[{Roe and Lindzen(1996)Roe, and Lindzen}]{Roe-Lindzen-1996}
Roe, H.~H., and R.~S. Lindzen, 1996: Baroclinic adjustment in a two-level model
  with barotropic shear. \textit{J. Atmos. Sci.}, \textbf{53~(18)}, 2749--2754,
  \doi{10.1175/1520-0469(1996)053<2749:BAIATL>2.0.CO;2}.

\bibitem[{Rossby and Collaborators(1939)Rossby, and
  Collaborators}]{Rossby-1939}
Rossby, C.-G., and Collaborators, 1939: Relation between variations in the
  intensity of the zonal circulation of the atmosphere and the displacements of
  the semi-permanent centers of action. \textit{J. Mar. Res.}, \textbf{2},
  38--55.

\bibitem[{Rossby and collaborators(1947{\natexlab{a}})Rossby, and
  collaborators}]{StaffMembers-1947}
Rossby, C.-G., and collaborators, 1947{\natexlab{a}}: On the circulation of the
  atmosphere in middle latitudes. \textit{Bull. Am. Met. Soc.}, \textbf{28},
  255--280.

\bibitem[{Rossby and collaborators(1947{\natexlab{b}})Rossby, and
  collaborators}]{Rossby-1947}
Rossby, C.-G., and collaborators, 1947{\natexlab{b}}: On the distribution of
  angular velocity in gaseous envelopes under the influence of large-scale
  horizontal mixing processes. \textit{Bull. Am. Met. Soc.}, \textbf{28},
  53--68.

\bibitem[{Rutgers(1998)}]{Rutgers-1998}
Rutgers, M., 1998: Forced {2D} turbulence: {Experimental} evidence of
  simultaneous inverse energy and forward enstrophy cascades. \textit{Phys.
  Rev. Lett.}, \textbf{8~(1)}, 2244--2247, \doi{10.1103/PhysRevLett.81.2244}.

\bibitem[{Salmon(1998)}]{Salmon-1998}
Salmon, R., 1998: \textit{Lectures on Geophysical Fluid Dynamics}. Oxford
  University Press.

\bibitem[{Salyk et~al.(2006)Salyk, Ingersoll, Lorre, Vasavada,, and
  Del~Genio}]{Salyk-etal-2006}
Salyk, C., A.~P. Ingersoll, J.~Lorre, A.~Vasavada, and A.~D. Del~Genio, 2006:
  Interaction between eddies and mean flow in {Jupiter's} atmosphere:
  {Analysis} of {Cassini} imaging data. \textit{Icarus}, \textbf{185},
  430--442, \doi{10.1016/j.icarus.2006.08.007}.

\bibitem[{S\'anchez-Lavega et~al.(2008)}]{Sanchez-etal-2008}
S\'anchez-Lavega, A., and Coauthors, 2008: Depth of a strong {Jovian} jet from
  a planetary-scale disturbance driven by storms. \textit{Nature},
  \textbf{451~(7177)}, 437--440, \doi{10.1038/nature06533}.

\bibitem[{Schneider and Lindzen(1977)Schneider, and
  Lindzen}]{Schneider-Lindzen-1977}
Schneider, E.~K., and R.~S. Lindzen, 1977: Axially symmetric steady state
  models of the basic state of instability and climate studies. {Part I:
  Linearized} calculations. \textit{J. Atmos. Sci.}, \textbf{34}, 263--279,
  \doi{10.1175/1520-0469(1977)034<0263:ASSSMO>2.0.CO;2}.

\bibitem[{Schneider and Walker(2006)Schneider, and
  Walker}]{Schneider-Walker-2006}
Schneider, T., and C.~C. Walker, 2006: Self-organization of atmospheric
  macroturbulence into critical states of weak nonlinear eddy-eddy
  interactions. \textit{J. Atmos. Sci.}, \textbf{63~(6)}, 1569--1586,
  \doi{10.1175/JAS3699.1}.

\bibitem[{Schoeberl and Lindzen(1984)Schoeberl, and
  Lindzen}]{Schoeberl-Lindzen-84}
Schoeberl, M.~R., and R.~S. Lindzen, 1984: A numerical simulation of barotropic
  instability. {Part I}: {Wave}-mean flow interaction. \textit{J. Atmos. Sci.},
  \textbf{41~(8)}, 1368--1379,
  \doi{10.1175/1520-0469(1984)041<1368:ANSOBI>2.0.CO;2}.

\bibitem[{Scott and Dritschel(2012)Scott, and Dritschel}]{Scott-Dritschel-2012}
Scott, R.~K., and D.~G. Dritschel, 2012: The structure of zonal jets in
  geostrophic turbulence. \textit{J. Fluid Mech.}, \textbf{711}, 576--598,
  \doi{10.1017/jfm.2012.410}.

\bibitem[{Scott and Polvani(2007)Scott, and Polvani}]{Scott-Polvani-2007}
Scott, R.~K., and L.~M. Polvani, 2007: Forced-dissipative shallow-water
  turbulence on the sphere and the atmospheric circulation of the giant
  planets. \textit{J. Atmos. Sci.}, \textbf{64}, 3158--3176,
  \doi{10.1175/JAS4003.1}.

\bibitem[{Shepherd(1987)}]{Shepherd-1987}
Shepherd, T.~G., 1987: A spectral view of nonlinear fluxes and stationary
  transient interaction in the atmosphere. \textit{J. Atmos. Sci.},
  \textbf{44}, 1166--1178,
  \doi{10.1175/1520-0469(1987)044<1166:ASVONF>2.0.CO;2}.

\bibitem[{Smith and Waleffe(1999)Smith, and Waleffe}]{Smith-Waleffe-1999}
Smith, L.~M., and F.~Waleffe, 1999: Transfer of energy to two-dimensional large
  scales in forced, rotating three-dimensional turbulence. \textit{Phys.
  Fluids}, \textbf{11}, 1608, \doi{10.1063/1.870022}.

\bibitem[{Srinivasan and Young(2012)Srinivasan, and
  Young}]{Srinivasan-Young-2012}
Srinivasan, K., and W.~R. Young, 2012: Zonostrophic instability. \textit{J.
  Atmos. Sci.}, \textbf{69~(5)}, 1633--1656, \doi{10.1175/JAS-D-11-0200.1}.

\bibitem[{Srinivasan and Young(2014)Srinivasan, and
  Young}]{Srinivasan-Young-2014}
Srinivasan, K., and W.~R. Young, 2014: Reynold stress and eddy difusivity of
  $\beta$-plane shear flows. \textit{J. Atmos. Sci.}, \textbf{71~(6)},
  2169--2185, \doi{10.1175/JAS-D-13-0246.1}.

\bibitem[{Starr(1953)}]{Starr-1953}
Starr, V.~P., 1953: Note concerning the nature of the large-scale eddies in the
  atmosphere. \textit{Tellus}, \textbf{5~(4)}, 494--498,
  \doi{10.1111/j.2153-3490.1953.tb01079.x}.

\bibitem[{Sukoriansky et~al.(2008)Sukoriansky, Dikovskaya,, and
  Galperin}]{Sukoriansky-etal-2008}
Sukoriansky, S., N.~Dikovskaya, and B.~Galperin, 2008: Nonlinear waves in
  zonostrophic turbulence. \textit{Phys. Rev. Lett.}, \textbf{101~(1)},
  178\,501, \doi{10.1103/PhysRevLett.101.178501}.

\bibitem[{Thompson(1971)}]{Thompson-1971}
Thompson, R. O. R.~Y., 1971: Why there is an intense eastward current in the
  {North Atlantic} but not in the {South Atlantic}? \textit{J. Phys.
  Oceanogr.}, \textbf{1~(3)}, 235--237,
  \doi{10.1175/1520-0485(1971)001<0235:WTIAIE>2.0.CO;2}.

\bibitem[{Thompson(1980)}]{Thompson-1980}
Thompson, R. O. R.~Y., 1980: A prograde jet driven by {Rossby} waves.
  \textit{J. Atmos. Sci.}, \textbf{37~(6)}, 1216--1226,
  \doi{10.1175/1520-0469(1980)037<1216:APJDBR>2.0.CO;2}.

\bibitem[{Tobias and Marston(2013)Tobias, and Marston}]{Tobias-Marston-2013}
Tobias, S.~M., and J.~B. Marston, 2013: Direct statistical simulation of
  out-of-equilibrium jets. \textit{Phys. Rev. Lett.}, \textbf{110~(10)},
  104\,502, \doi{10.1103/PhysRevLett.110.104502}.

\bibitem[{Tung and Orlando(2003{\natexlab{a}})Tung, and
  Orlando}]{Tung-Orlando-2003}
Tung, K.-K., and W.~W. Orlando, 2003{\natexlab{a}}: The $k^{-3}$ and $k^{-5/3}$
  energy spectrum of atmospheric turbulence: quasigeostrophic two-level model
  simulation. \textit{J. Atmos. Sci.}, \textbf{50}, 824--835.

\bibitem[{Tung and Orlando(2003{\natexlab{b}})Tung, and
  Orlando}]{Tung-Orlando-2003-2DQG}
Tung, K.-K., and W.~W. Orlando, 2003{\natexlab{b}}: On the differences between
  {2D} and {QG} turbulence. \textit{Discrete Cont. Dyn.-B}, \textbf{3~(2)},
  145--162, \doi{10.3934/dcdsb.2003.3.145}.

\bibitem[{Vallis and Maltrud(1993)Vallis, and Maltrud}]{Vallis-Maltrud-93}
Vallis, G.~K., and M.~E. Maltrud, 1993: Generation of mean flows and jets on a
  beta-plane and over topography. \textit{J. Phys. Oceanogr.}, \textbf{23},
  1346--1362, \doi{10.1175/1520-0485(1993)023<1346:GOMFAJ>2.0.CO;2}.

\bibitem[{Williams(1978)}]{Williams-78}
Williams, G.~P., 1978: Planetary circulations: 1. {Barotropic} representation
  of {Jovian} and terrestrial turbulence. \textit{J. Atmos. Sci.},
  \textbf{35~(8)}, 1399--1426,
  \doi{10.1175/1520-0469(1978)035<1399:PCBROJ>2.0.CO;2}.

\bibitem[{{Wunsch}(2003)}]{Wunsch-2003}
{Wunsch}, C., 2003: {Greenland}--{Antarctic} phase relations and millennial
  time-scale climate fluctuations in the {Greenland} ice-cores.
  \textit{Quaternary Sci. Rev.}, \textbf{22}, 1631--1646,
  \doi{10.1016/S0277-3791(03)00152-5}.

\bibitem[{Yuen and Lake(1980)Yuen, and Lake}]{Yuen-Lake-1980}
Yuen, H.~C., and B.~M. Lake, 1980: Instabilities of waves on deep water.
  \textit{Ann. Rev. Fluid Mech.}, \textbf{12}, 303--304,
  \doi{10.1146/annurev.fl.12.010180.001511}.

\bibitem[{Zakharov(1965)}]{Zakharov-1965}
Zakharov, V.~E., 1965: Weak turbulence in media with a decay spectrum.
  \textit{J. Appl. Mech. Tech. Phy.}, \textbf{6~(4)}, 22--24,
  \doi{10.1007/BF01565814}.

\bibitem[{Zakharov et~al.(1992)Zakharov, L'Vov,, and Falkovich}]{Zakharov-1992}
Zakharov, V.~E., V.~S. L'Vov, and G.~Falkovich, 1992: \textit{Kolmogorov
  spectra of turbulence 1. {Wave} turbulence}. Springer.

\bibitem[{Zakharov and Ostrovsky(2009)Zakharov, and
  Ostrovsky}]{Zakharov-Ostrovsky-2009}
Zakharov, V.~E., and L.~A. Ostrovsky, 2009: Modulation instability: {The}
  beginning. \textit{Physica D}, \textbf{238}, 540--548,
  \doi{10.1016/j.physd.2008.12.002}.

\bibitem[{Zhang and Held(1999)Zhang, and Held}]{Zhang-Held-99}
Zhang, Y., and I.~M. Held, 1999: A linear stochastic model of a {GCM's}
  midlatitude storm tracks. \textit{J. Atmos. Sci.}, \textbf{56}, 3416--3435,
  \doi{10.1175/1520-0469(1999)056<3416:ALSMOA>2.0.CO;2}.

\end{thebibliography}

%\printbibliography[heading=bibintoc,title=References]

%\include{endmatter/colophon}

\end{document}